\documentclass{emulateapj}

\usepackage{graphicx}
\usepackage[]{natbib}
\usepackage{placeins}
\usepackage{lineno}
\usepackage{ulem}
\usepackage{rotating}
\usepackage{adjustbox}
\usepackage{float}

\usepackage[caption=false]{subfig}    

\usepackage{color}

\pdfoutput=1

\newcommand{\ltsima} {$\; \buildrel < \over \sim \;$}

\newcommand{\gtsima} {$\; \buildrel > \over \sim \;$}
\newcommand{\lta} {\lower.5ex\hbox{\ltsima}}
\newcommand{\gta} {\lower.5ex\hbox{\gtsima}}

\shorttitle{\it{CGRO}/BATSE Data Support the New Paradigm for GRB Prompt Emission}
\shortauthors{S.Guiriec}

\begin{document}

\title{\it{CGRO}/BATSE Data Support the New Paradigm for GRB Prompt Emission and the New L$_\mathrm{\lowercase{i}}^\mathrm{\lowercase{n}T\lowercase{h}}$--E$_\mathrm{\lowercase{peak,i}}^\mathrm{\lowercase{n}T\lowercase{h,rest}}$ relation}

\author{S. Guiriec\altaffilmark{1,2,3,4}, M. M. Gonzalez\altaffilmark{5}, J. R. Sacahui\altaffilmark{6}, C. Kouveliotou\altaffilmark{7}, N. Gehrels\altaffilmark{1} \& J. McEnery\altaffilmark{1}}


\altaffiltext{1}{NASA Goddard Space Flight Center, Greenbelt, MD 20771, USA}
\altaffiltext{2}{Department of Physics and Department of Astronomy, University of Maryland, College Park, MD 20742, USA}
\altaffiltext{3}{Center for Research and Exploration in Space Science and Technology (CRESST)}
\altaffiltext{4}{NASA Postdoctoral Program}
\altaffiltext{5}{Instituto de Astronom'a, UNAM, MŽxico 04510, Mexico}
\altaffiltext{6}{Instituto Nacional de Pesquisas Espaciais -- INPE, Avenida dos Astronautas 1758, 12227-010, S‹o JosŽ dos Campos-SP, Brazil}
\altaffiltext{7}{Department of Physics, The George Washington University, Washington, DC 20052}



\email{sylvain.guiriec@nasa.gov}


\begin{abstract}

The paradigm for the Gamma Ray Burst (GRB) prompt emission is changing. Since early in the {\it Compton Gamma Ray Observatory} ({\it CGRO}) era, the empirical Band function has been considered a good description of the keV-MeV $\gamma$-ray prompt emission spectra despite the fact that its shape was very often inconsistent with the theoretical predictions especially these expected in the pure synchrotron emission scenarios. 
We have recently established a new observational model analyzing data of the NASA {\it Fermi Gamma-ray Space Telescope}. In this model, GRB prompt emission would be a combination of three main emission components: (i) a thermal-like component that we interpreted so far as emission from the jet photosphere, (ii) a non-thermal component that we interpreted so far as synchrotron radiation from the propagating and accelerated charged particles within the jet, and (iii) an additional non-thermal (cutoff) power-law extending from low to high energies in $\gamma$-rays and most likely of inverse Compton origin. In this article we reanalyze some of the bright GRBs observed with the Burst And Transient Source Experiment (BATSE) on-board {\it CGRO} with the new model, namely GRBs 941017, 970111 and 990123. We conclude that BATSE data are fully consistent with the recent results obtained with {\it Fermi}: some bright BATSE GRBs exhibit three separate components during the prompt phase with similar spectral parameters as these reported from {\it Fermi} data. In addition, the analysis of the BATSE GRBs with the new prompt emission model results in a relation between the time-resolved energy flux of the non-thermal component, F$_\mathrm{i}^\mathrm{nTh}$, and its corresponding $\nu$F$_\nu$ spectral peak energy, E$_\mathrm{peak,i}^\mathrm{nTh}$ (i.e., F$_\mathrm{i}^\mathrm{nTh}$--E$_\mathrm{peak,i}^\mathrm{nTh}$) that has a similar index---when fitted to a power law---as the one initially derived from {\it Fermi} data. For GRBs with known redshift (z) this results in a possible universal relation between the luminosity of the non-thermal component, L$_\mathrm{i}^\mathrm{nTh}$, and its corresponding $\nu$F$_\nu$ spectral peak energy in the rest frame, E$_\mathrm{peak,i}^\mathrm{NT,rest}$ (i.e., L$_\mathrm{i}^\mathrm{nTh}$--E$_\mathrm{peak,i}^\mathrm{NT,rest}$). We estimated the redshifts of GRBs~941017 and~970111 using GRB~990123---with z=1.61---as a reference. The estimated redshift for GRB~941017 is typical for long GRBs and the estimated redshift for GRB~970111 is right in the range of the expected values for this burst.
\break
\end{abstract}


\section{Introduction}

During the past years, a new paradigm has emerged for Gamma Ray Burst (GRB) prompt emission\footnote{See \citet{Peer:2015} for a recent review on GRB prompt emission.} in which multiple component models seem to be favored compared to single component ones \citep[see for instance][]{Guiriec:2015a}. Until recently, the empirical Band function---a smoothly broken power law (PL) described with four free parameters: $\alpha_\mathrm{Band}$ and $\beta_\mathrm{Band}$ for the indices of the low and high energy PLs, respectively, $E_\mathrm{peak}^\mathrm{Band}$ for the $\nu$F$_\nu$ spectral peak energy \citep{Gehrels:1997}, and an amplitude parameter \citep{Band:1993,Greiner:1995}---was considered as an adequate description of the keV--MeV $\gamma$-ray prompt emission spectra. However, an additional PL was required in a few cases to account for spectral deviations at high energies \citep{Gonzalez:2003,Gonzalez:2012,Guiriec:2010,Guiriec:2015a,Abdo:2009:GRB090902B,Ackermann:2010:GRB090510,Ackermann:2011:GRB090926A}. Moreover, despite their compatibility with non-thermal processes, the values of $\alpha_\mathrm{Band}$ were usually too high to be consistent with the predictions from the pure synchrotron emission scenarios from electrons in the slow and fast cooling regimes that require index values $<$-2/3 and $<$-3/2, respectively \citep{Cohen:1997,Crider:1997,Preece:1998,Ghisellini:2000}, as expected in the framework of the popular fireball model \citep{Cavallo:1978,Paczynski:1986,Goodman:1986,Shemi:1990,Rees:1992,Rees:1994,Meszaros:1993}.

The recent discovery of a thermal-like component, C$_\mathrm{Th}$, together with a non-thermal one, C$_\mathrm{nTh}$, in the prompt emission of both long \citep{Guiriec:2011a,Guiriec:2015a,Guiriec:2015b,Axelsson:2012:GRB110721A} and short \citep{Guiriec:2013a} GRBs observed with the {\it Fermi Gamma-ray Space Telescope} (hereafter, {\it Fermi}), challenges the well established Band function paradigm. The C$_\mathrm{Th}$ spectral shape is usually compatible with a Planck function---based on the quality of the {\it Fermi} data. Although a broader spectral shape more compatible with the GRB jet photospheric models was recently reported in~\citet{Guiriec:2015b} who analyzed the spectra of GRB 131014A and showed that it exhibited an intense C$_\mathrm{Th}$ component. Indeed, while a pure Planck function is well approximated with a cutoff PL (CPL) with an index $\alpha_\mathrm{Th}$=+1, the thermal-like component of GRB 131014A is best described with a CPL with an index $\alpha_\mathrm{Th}$$\approx$+0.6. The C$_\mathrm{nTh}$ spectral shape is adequately described with a Band function but with spectral parameters usually very different from those resulting from fits to a Band function alone. For instance, $\alpha_\mathrm{nTh}$ is systematically lower than $\alpha_\mathrm{Band}$ and, therefore, more compatible with the synchrotron emission scenarios, $\beta_\mathrm{nTh}$ is $<\sim$-3 and compatible with an exponential cutoff (i.e., the Band function can be replace with a CPL with no change in the fit statistics), and $E_\mathrm{peak}^\mathrm{nTh}$ is systematically shifted to higher values compared to $E_\mathrm{peak}^\mathrm{Band}$ \citep{Guiriec:2011a,Guiriec:2013a,Guiriec:2015a,Guiriec:2015b}. In the case of short GRB 120323A, in fits to C$_\mathrm{nTh}$+C$_\mathrm{Th}$ the $\alpha$ indices drop from positive values---in fits to a Band function alone---down to values low enough to be compatible with the pure fast-cooling synchrotron scenario (i.e., $\sim$-1.2)\citep{Guiriec:2013a}. Conversely to the genuine fireball model, which predicts a thermal-like component from the jet photosphere that overpowers the non-thermal one \citep{Zhang:2009}, the observed C$_\mathrm{Th}$ is usually energetically subdominant compared to C$_\mathrm{nTh}$ as predicted by \citet{Daigne:2002}, \citet{Nakar:2005}, \citet{Zhang:2009} and \citet{Hascoet:2013} in the case of highly magnetized outflows. The diversity of C$_\mathrm{Th}$ relative contribution to the total energy from burst to burst indicates that the magnetization parameter can vary over a large range of values \citep{Guiriec:2011a,Guiriec:2013a,Guiriec:2015a,Guiriec:2015b}. Although we interpreted C$_\mathrm{nTh}$ as synchrotron emission, it is also possible that this component corresponds to a comptonized photosphere~\citep{Peer:2008,Beloborodov:2010,Beloborodov:2014,Vurm:2015}.

\citet{Guiriec:2011b,Guiriec:2013a,Guiriec:2015a} went a step further by identifying simultaneously the three known components---namely C$_\mathrm{nTh}$, C$_\mathrm{Th}$ and the additional (cutoff) PL (i.e., C$_\mathrm{nTh}$+C$_\mathrm{Th}$+(C)PL)---in the prompt emission of some bright {\it Fermi} GRBs. This completely changed the view we had of GRB 080916C considered before as adequately fitted to a single Band function from 8 keV up to GeVs \citep{Abdo:2009:GRB080916C}. While the spectral and temporal behaviors of GRBs 080916C and 090926A were very different when fitted to a Band function alone for the former \citep{Abdo:2009:GRB080916C} and to a combination of a Band function and a additional CPL for the latter \citep{Ackermann:2011:GRB090926A}, the two GRBs are like ``twins'' in the context of the C$_\mathrm{nTh}$+C$_\mathrm{Th}$+(C)PL model. We note here, however, that the three components of the C$_\mathrm{nTh}$+C$_\mathrm{Th}$+(C)PL model are not systematically present or detectable in all GRBs.

\citet{Guiriec:2013a,Guiriec:2015a,Guiriec:2015b} reported a strong correlation between the time-resolved energy flux of C$_\mathrm{nTh}$, F$_\mathrm{i}^\mathrm{nTh}$, and its corresponding $\nu$F$_\nu$ spectral peak energy, E$_\mathrm{peak,i}^\mathrm{nTh}$ intrinsic to each burst (hereafter, F$_\mathrm{i}^\mathrm{nTh}$--E$_\mathrm{peak,i}^\mathrm{nTh}$ relation where ``i'' counts time intervals). Interestingly, the F$_\mathrm{i}^\mathrm{nTh}$--E$_\mathrm{peak,i}^\mathrm{nTh}$ relations for all GRBs are adequately described by PLs with very similar index values indicating a possible common physical process in jets of both short and long GRBs to explain the C$_\mathrm{nTh}$ component. Moreover, when accounting for the redshift and the K-correction, a strong correlation appears between the time-resolved luminosity of C$_\mathrm{nTh}$, L$_\mathrm{i}^\mathrm{nTh}$, and its corresponding $\nu$F$_\nu$ spectral peak energy in the GRB central engine rest frame, E$_\mathrm{peak,i}^\mathrm{C_{nTh},rest}$ (hereafter, L$_\mathrm{i}^\mathrm{nTh}$--E$_\mathrm{peak,i}^\mathrm{C_{nTh},rest}$ relation); this relation is not only intrinsic to each GRB but it is also common to all GRBs and points, therefore, towards a possible universal physical mechanism intrinsic to all GRB jets, which may eventually be used, for instance, (i) to measure the distance of GRBs using solely their $\gamma$-ray prompt emission and (ii) to constrain the cosmological parameters complementary to the type IA supernova sample.

While all the parameters of the Band function are highly variable---within each burst and from burst to burst---when a Band function alone is fitted to $\gamma$-ray prompt emission data, $\alpha_\mathrm{nTh}$, $\beta_\mathrm{nTh}$, $\alpha_\mathrm{Th}$ and the index of the additional (C)PL, $\alpha_\mathrm{(C)PL}$, do not vary much with time or from burst to burst when fitting either C$_\mathrm{nTh}$+C$_\mathrm{Th}$ or C$_\mathrm{nTh}$+C$_\mathrm{Th}$+(C)PL to the data \citep{Guiriec:2013a,Guiriec:2015a}. Indeed, $\alpha_\mathrm{nTh}$ indices have values that are either $\sim$-0.7 or $\sim$-1.2 depending on the burst \citep{Guiriec:2013a,Guiriec:2015a,Guiriec:2015b}, $\beta_\mathrm{nTh}$ index values are usually $<$-3.5 and compatible with exponential cutoffs in many cases, $\alpha_\mathrm{Th}$ indices are usually $\sim$+0.6, and $\alpha_\mathrm{(C)PL}$ values are usually $\sim$-1.5. Therefore, by freezing some parameters of the C$_\mathrm{nTh}$+C$_\mathrm{Th}$ and C$_\mathrm{nTh}$+C$_\mathrm{Th}$+(C)PL models to their typical values, we reduce the number of free parameters to 4 and 5, respectively, without degrading significantly the fit quality, and the new models become statistically competitive with the Band function in terms of free parameters. Moreover, thanks to the F$_\mathrm{i}^\mathrm{nTh}$--E$_\mathrm{peak,i}^\mathrm{nTh}$ and L$_\mathrm{i}^\mathrm{nTh}$--E$_\mathrm{peak,i}^\mathrm{C_{nTh},rest}$ relations it is possible to reduce again the number of free parameters to 3 and 4 for C$_\mathrm{nTh}$+C$_\mathrm{Th}$ and C$_\mathrm{nTh}$+C$_\mathrm{Th}$+(C)PL, respectively.

In this article we reanalyzed three famous bright long GRBs (namely, GRBs~941017,~970111 and~990123) observed with the Burst And Transient Source Experiment (BATSE) on-board the {\it Compton Gamma Ray Observatory} ({\it CGRO})---that was operating in low Earth orbit from 1991 to 2000---in the context of the new paradigm to verify its consistency with these archival data. In Sections~ \ref{sec:data} and~\ref{sec:analysis} we discuss the data selection criteria and our analysis methodology. Finally in Section~\ref{sec:results} we present our results. We will see how the view we had of those GRBs---that attracted a lot of attention in the scientific community---may be completely refreshed in the framework of the new model.

\begin{figure}[ht!]
\begin{center}
\includegraphics[totalheight=0.31\textheight, clip]{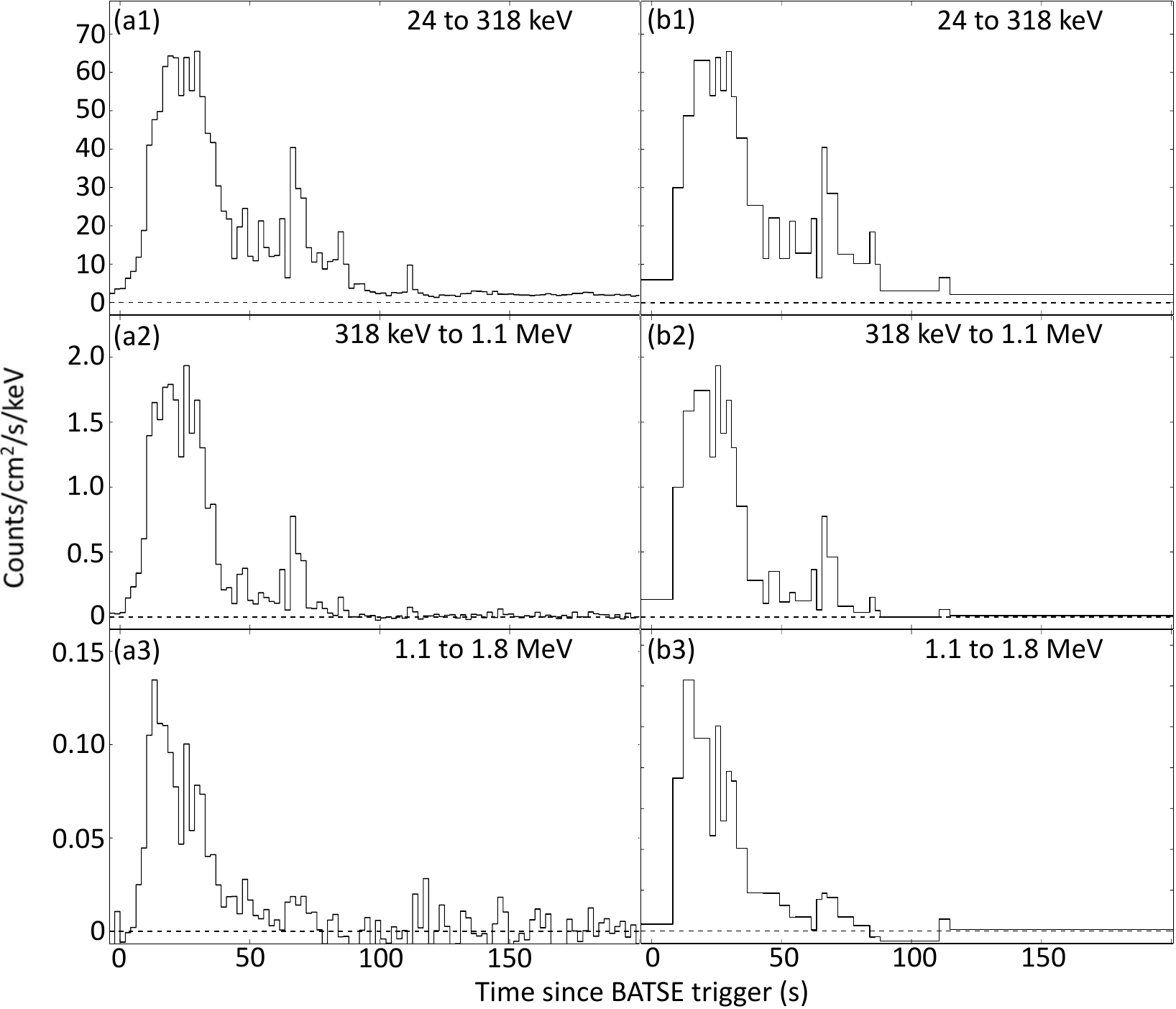}
\caption{\label{fig01}GRB 941017 count light curves as observed with BATSE LAD4 ($\sim$20 keV--2 MeV). (a1--3) 2.048 s time-resolved light curves. (b1--3) Light curves with the binning used for our time-resolved spectral analysis.}
\end{center}
\end{figure}
\begin{figure}[ht!]
\begin{center}
\includegraphics[totalheight=0.31\textheight, clip]{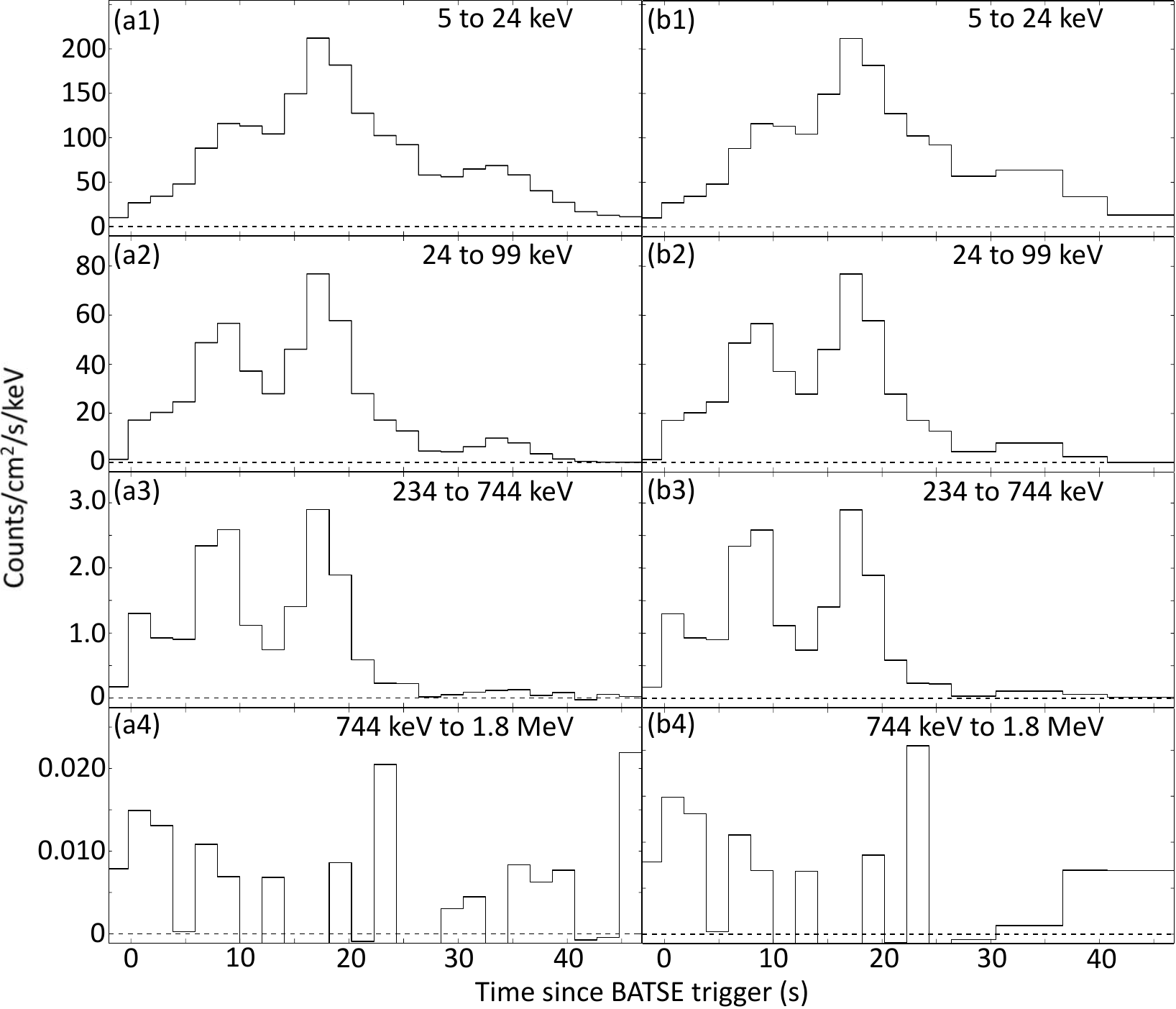}
\caption{\label{fig02}GRB 970111 count light curves as observed with BATSE LAD0 ($\sim$20 keV--2 MeV). (a1--4) 2.048 s time-resolved light curves. (b1--4) Light curves with the binning used for our time-resolved spectral analysis.}
\end{center}
\end{figure}
\begin{figure}[ht!]
\begin{center}
\includegraphics[totalheight=0.31\textheight, clip]{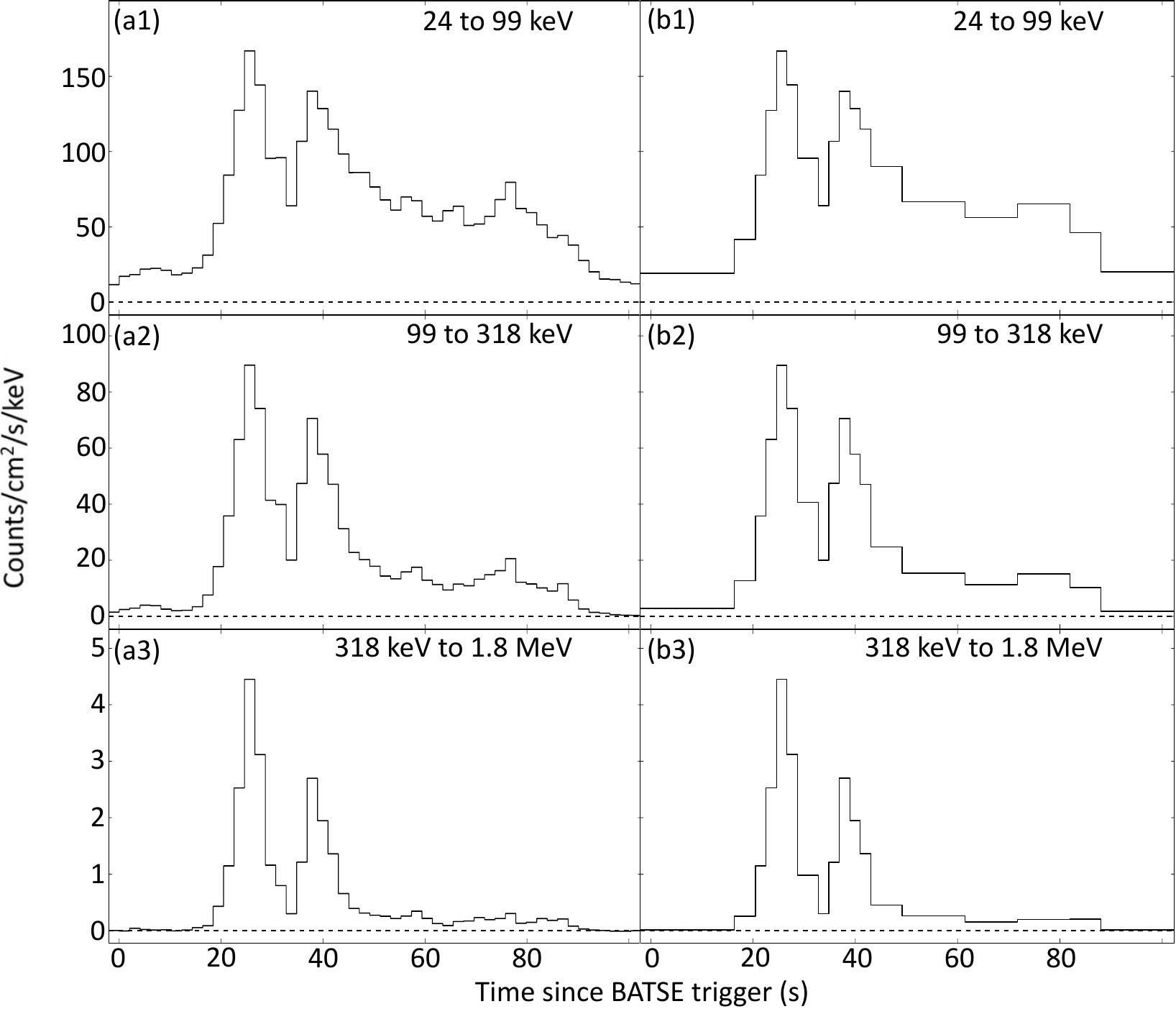}
\caption{\label{fig03}GRB 990123 count light curves as observed with BATSE LAD0 ($\sim$20 keV--2 MeV). (a1--3) 2.048 s time-resolved light curves. (b1--3) Light curves with the binning used for our time-resolved spectral analysis.}
\end{center}
\end{figure}

\section{Data Selection}
\label{sec:data}

We focus this first article of a series on a limited sample of three famous bright long GRBs detected with BATSE. These GRBs are of particular interest because they have often been and still are cited to support or discard theoretical interpretations for GRB prompt emission.

GRB~990123 was the first GRB to be simultaneously detected during its prompt phase in the $\gamma$-ray and optical bands~\citep[$\gamma$-ray reference here,][]{Akerlof:1999}. The prompt emission spectra of this burst in the keV-MeV energy range was considered as adequately fitted to a single Band function \citep{Briggs:1999}. A redshift estimate for the burst of 1.61 was reported by~\citet{Kelson:1999} and~\citet{Hjorth:1999}.

GRB~941017---which was also observed with the Energetic Gamma Ray Experiment (EGRET) on board {\it CGRO}---was the first burst of the pre-{\it Fermi} era for which a strong deviation at high energy from the Band function was reported \citep{Gonzalez:2003}. GRB~941017 was considered as adequately fitted to a combination of a Band function fading with time and an additional PL remaining constant and incompatible with the synchrotron models. Unfortunately no redshift was measured for this GRB.

Finally, we also analyzed GRB~970111 that was thought to be adequately fitted to a Band function but with positive values of $\alpha_\mathrm{Band}$ at early times. The photometric redshift of this burst was estimated to be 0.2$<$z$<$1.4 based on its probable association with host galaxies at this redshift range~\citep{Gorosabel:1998}. Unfortunately, many fainter galaxies may lie in the same region of the sky and the association with the detected objects is not secure, which jeopardizes the validity of this photometric redshift.

The light curves of GRBs 941017, 970111 and 990123 are presented in Figures \ref{fig01}, \ref{fig02} and \ref{fig03}, respectively.

Detailed descriptions of the BATSE instrument are given in several references such as \citet{Fishman:1989}, \citet{Pendleton:1995}, \citet{Preece:2000} and \citet{Kaneko:2006}; therefore, only a brief description sufficient to understand the data selection is given below.

BATSE consisted of 8 individual identical detector modules located at each corner of the {\it CGRO} spacecraft comprising the faces of a regular octahedron. Each module was built of two NaI(Tl) scintillation detectors coupled with photomultiplier tubes (PMTs): a Large Area Detector (LAD) optimized for sensitivity and directional response, and a Spectroscopy Detector (SD) optimized for energy coverage and resolution. While the LADs had a constant energy range extending from 20 keV up to 1.9\,MeV,  the SDs had an adjustable energy range between 10 keV and 100 MeV depending on the gain of the PMTs. Despite their significantly better spectral capabilities, the SDs had a collecting area 16 times smaller than the LADs reducing considerably their sensitivity. The data collected by each of the 16 BATSE detectors can be analyzed either individually or simultaneously in joint fits.

\begin{table*}[ht!]
\caption{\label{tab01}Model parameter values resulting from the time-integrated spectral analysis of GRBs~941017,~970111 and~990123 with their 1-$\sigma$ uncertainties as presented in Section~\ref{sec:tisa}.}
\begin{center}
{\tiny
\begin{tabular}{|c|c|c|c|c|c|c|}
\hline
\multicolumn{1}{|c|}{Models} &\multicolumn{3}{|c|}{Base Component} & \multicolumn{2}{|c|}{Additional Component} & \multicolumn{1}{|c|}{Cstat/dof} \\
\hline
     &\multicolumn{3}{|c|}{CPL or Band} & \multicolumn{1}{|c|}{BB} & \multicolumn{1}{|c|}{ PL} & \multicolumn{1}{|c|}{}\\
\hline
\multicolumn{1}{|c|}{Parameters} & \multicolumn{1}{|c|}{E$_{\rm peak}$} & \multicolumn{1}{|c|}{$\alpha$} & \multicolumn{1}{|c|}{$\beta$} & \multicolumn{1}{|c|}{kT} & \multicolumn{1}{|c|}{Index} & \\
\multicolumn{1}{|c|}{} & \multicolumn{1}{|c|}{(keV)} & \multicolumn{1}{|c|}{} & \multicolumn{1}{|c|}{} & \multicolumn{1}{|c|}{(keV)} & \multicolumn{1}{|c|}{} & \\
\hline
\multicolumn{7}{|c|}{ }   \\ 
 \multicolumn{7}{|c|}{GRB 941017 from T$_\mathrm{0}$-4.096 s to T$_\mathrm{0}$+118.784 s} \\
 \multicolumn{7}{|c|}{ }   \\ 
 Band & 330$\pm$5 & -0.78$\pm$0.01 & -2.35$\pm$0.03 & -- & -- & 205.8/9 \\
 C$_\mathrm{nTh}$+PL & 270$\pm$4 & +0.12$\pm$0.07 & -- & -- & -1.62$\pm$0.01 & 75.3/8 \\
 C$_\mathrm{nTh}$+PL & 272$\pm$9 & +0.01$\pm$0.25 & -2.29$\pm$0.15 & -- & -1.99$\pm$0.33 & 28.2/7 \\ 
 C$_\mathrm{nTh}$+C$_\mathrm{Th}$ & 784$\pm$44 & -1.33$\pm$0.02 & -- & 56.0$\pm$0.66 & -- & 25.1/8 \\
 C$_\mathrm{nTh}$+C$_\mathrm{Th}$ & 674$\pm$75 & -1.30$\pm$0.03 & -2.41$\pm$0.22 & 56.3$\pm$0.73 & -- & 23.6/7 \\
 C$_\mathrm{nTh}$+C$_\mathrm{Th}$+PL & 451$\pm$52 & -0.62$\pm$0.27 & -- & 51.8$\pm$2.3 & -1.70$\pm$0.03 & 18.3/6 \\
\multicolumn{7}{|c|}{ }   \\ 
\multicolumn{7}{|c|}{GRB 970111 from T$_\mathrm{0}$-2.304 s to T$_\mathrm{0}$+42.752 s} \\
\multicolumn{7}{|c|}{ }   \\ 
 CPL & 160$\pm$1 & -0.76$\pm$0.01 & -- & -- & -- & 98.7/10 \\
 Band & 159$\pm$1 & -0.76$\pm$0.02 & -4.55$\pm$0.62 & -- & -- & 96.7/9 \\
 C$_\mathrm{nTh}$+PL & 163$\pm$1 & -0.20$\pm$0.08 & -- & -- & -2.26$\pm$0.06 & 38.5/8 \\
 C$_\mathrm{nTh}$+C$_\mathrm{Th}$ & 151$\pm$3 & -1.21$\pm$0.06 & -- & 42.0$\pm$1.0 & -- & 24.3/8 \\
 C$_\mathrm{nTh}$+C$_\mathrm{Th}$+PL & 144$\pm$8 & -1.14$\pm$0.48 & -- & 43.2$\pm$2.1 & -1.72$\pm$0.80 & 23.0/6 \\
\multicolumn{7}{|c|}{ }   \\ 
\multicolumn{7}{|c|}{GRB 990123 from T$_\mathrm{0}$ to T$_\mathrm{0}$+100.352 s} \\
\multicolumn{7}{|c|}{ }   \\ 
 Band & 609$\pm$10 & -0.89$\pm$0.01 & -2.66$\pm$0.09 & -- & -- & 382.4/9 \\
 C$_\mathrm{nTh}$+PL & 432$\pm$8 & -0.10$\pm$0.05 & -- & -- & -1.58$\pm$0.01 & 123.3/8 \\
 C$_\mathrm{nTh}$+C$_\mathrm{Th}$ & 999$\pm$41 & -1.20$\pm$0.02 & -- & 81.7$\pm$1.1 & -- & 99.6/8 \\
 C$_\mathrm{nTh}$+C$_\mathrm{Th}$+PL & 614$\pm$60 & -0.40$\pm$0.20 & -- & 63.8$\pm$4.4 & -1.68$\pm$0.03 & 68.3/6 \\
\multicolumn{7}{|c|}{ }   \\ 
\hline
\end{tabular}
}
\end{center}
\end{table*}

In our analysis, we only used LAD data because the sensitivity of the SEDs was too low for the type of analysis presented here (i.e., time-resolved analysis using multiple spectral components). We used the LAD Continuous (CONT) data for which the energy range is divided into 16 energy channels with an accumulation time of 2.048s. For each of the three GRBs, we used the LAD detector with the smallest angle to the source (i.e., LAD4 for GRB 941017 and LAD0 for GRBs 970111 and 990123). The response matrices were generated using the best source locations for the three GRBs. 

\section{Analysis Methodology}
\label{sec:analysis}

We followed here the same procedure as described in detail in \citet{Guiriec:2011a,Guiriec:2013a,Guiriec:2015a,Guiriec:2015b}. We first performed a time-integrated analysis of the bursts to identify the main spectral components of the prompt emission. Then we analyzed the bursts on fine time scales to follow the evolution of the various components and to verify that the observed spectral features are not merely artifacts due, for instance, to strong spectral evolution.

For this analysis, we used the fitting package Rmfit and we determined the best parameters of the various tested models as well as their 1-$\sigma$ uncertainties by minimizing Castor C-Statistic (hereafter, Cstat). Cstat is a likelihood technique converging to a $\chi^\mathrm{2}$ for a specific data set when there is ``enough'' data.

As proposed in ~\citet{Guiriec:2011a,Guiriec:2013a,Guiriec:2015a,Guiriec:2015b} we fitted our C$_\mathrm{nTh}$+C$_\mathrm{Th}$+C$_\mathrm{nTh2}$ model to the three GRBs where: (i) C$_\mathrm{nTh}$ is a component with a non-thermal spectral shape that we approximated with either a CPL or Band function; (ii) C$_\mathrm{Th}$ is either $\varnothing$ or a thermal-like component that we approximated with a black  body (BB) spectrum in the current analysis; and (iii) C$_\mathrm{nTh2}$ is either $\varnothing$ or a second non-thermal component that we approximated with a PL. While we left all the parameters of C$_\mathrm{nTh}$+C$_\mathrm{Th}$+C$_\mathrm{nTh2}$ free in the time-integrated spectral analysis we froze $\alpha_\mathrm{nTh}$ to -0.7, $\beta_\mathrm{nTh}$ to $<$-5 (i.e., CPL) and $\alpha_\mathrm{PL}$ to -1.5 as proposed in \citet{Guiriec:2015a} in the time-resolved one (i.e., C$_\mathrm{nTh}$+C$_\mathrm{Th}$+PL$_\mathrm{5 params}$).


\section{Results}
\label{sec:results}
\subsection{Time-Integrated Spectral Analysis}
\label{sec:tisa}

\begin{figure}[ht!]
\begin{center}
\includegraphics[totalheight=0.35\textheight, clip]{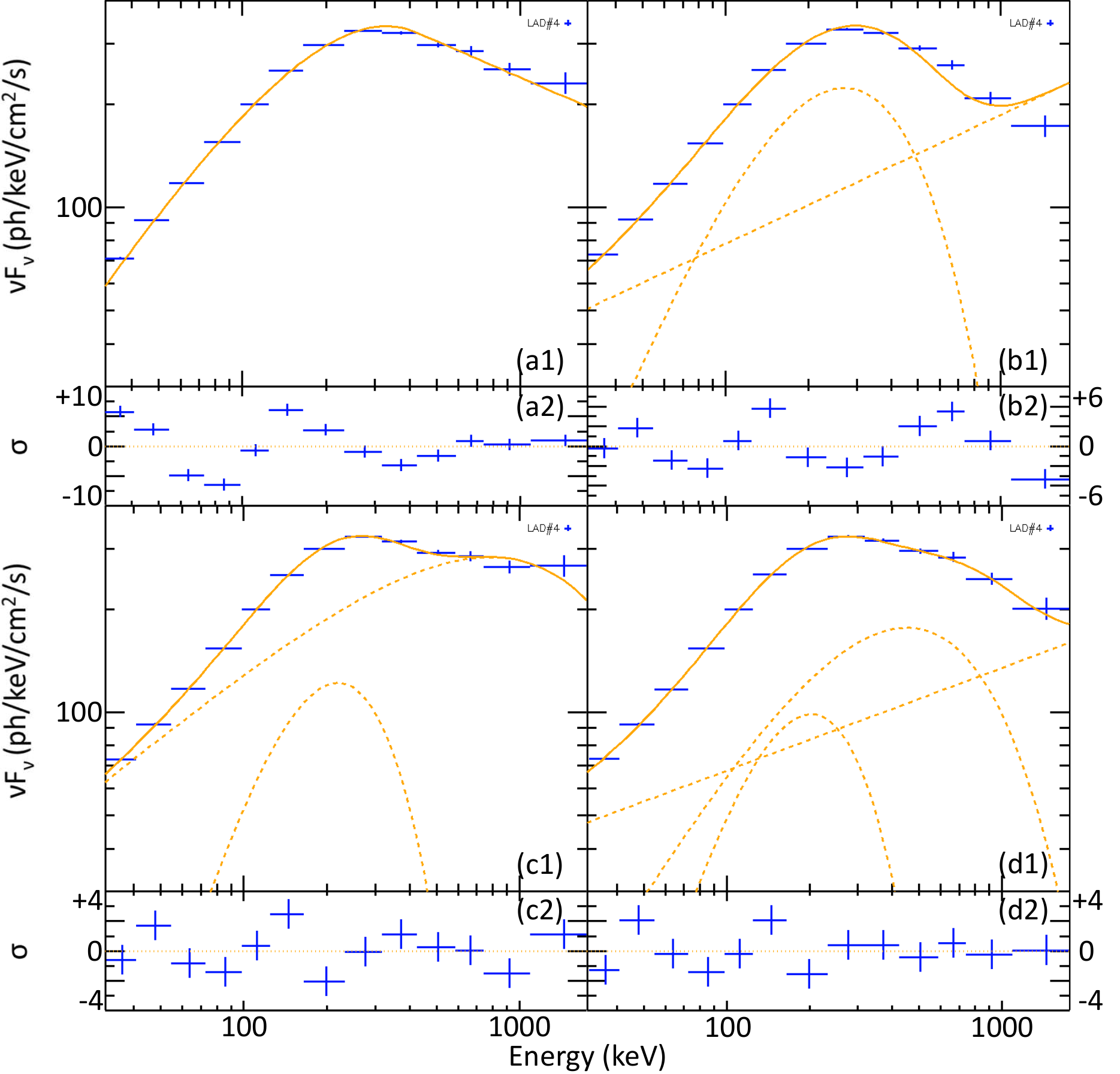}
\caption{\label{fig04}Time-integrated BATSE (LAD4) spectra of GRB~941017 (from T$_0$-4.096 s to T$_0$+118.784 s) when fitted to (a) a Band function alone, (b) C$_\mathrm{nTh}$+PL, (c) C$_\mathrm{nTh}$+C$_\mathrm{Th}$, and (d) C$_\mathrm{nTh}$+C$_\mathrm{Th}$+PL. The deconvolved $\nu$F$_\nu$ spectra are presented in panels (a1), (b1), (c1) and (d1)---the dashed lines correspond to the individual components of the fitted model and the solid ones to the total emission. Panels (a2), (b2), (c2) and (d2) correspond to the residuals of the fits.}
\end{center}
\end{figure}

\begin{figure}[ht!]
\begin{center}
\includegraphics[totalheight=0.35\textheight, clip]{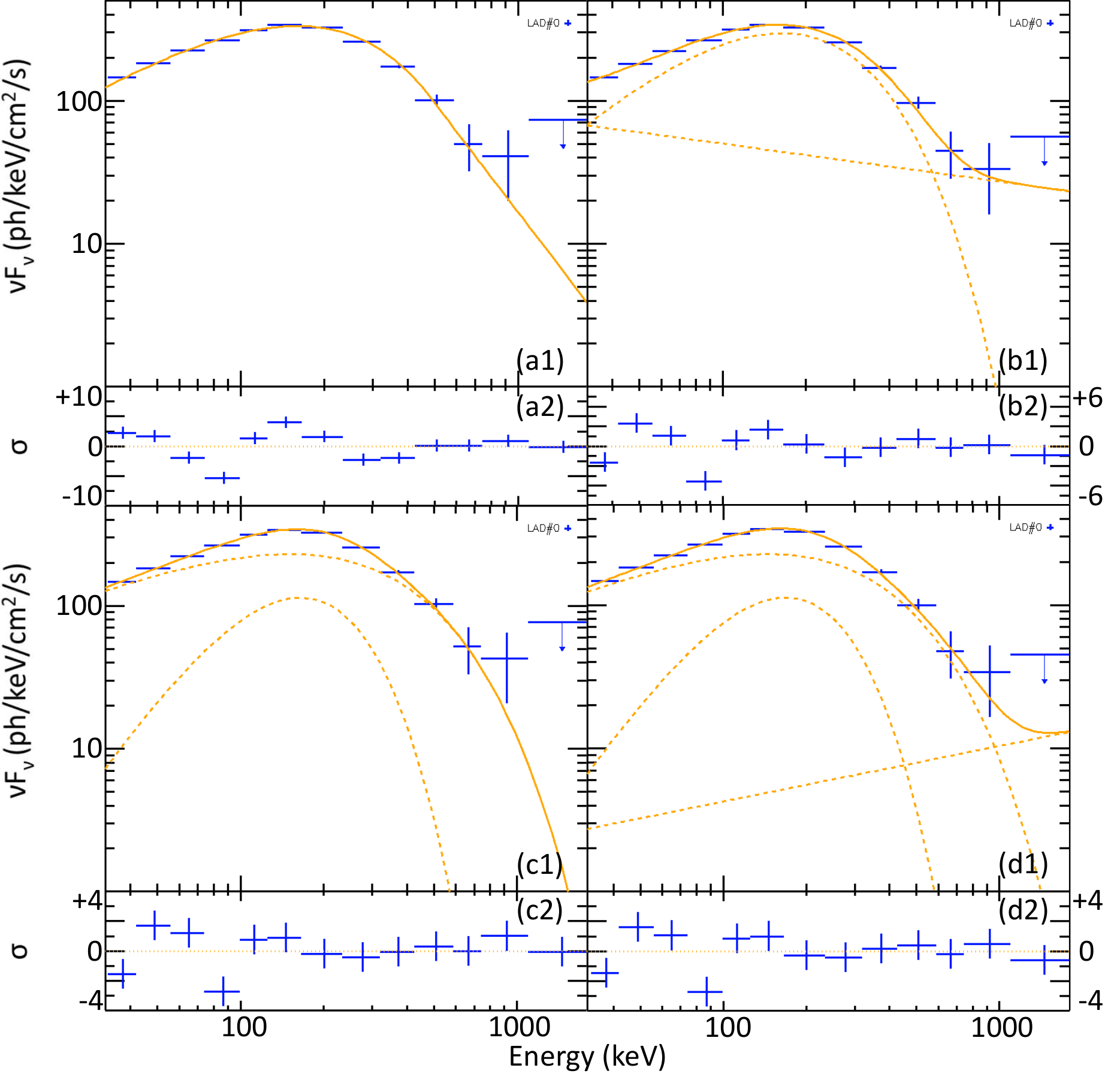}
\caption{\label{fig05}Time-integrated BATSE (LAD0) spectra of GRB~970111 (from T$_0$-2.304 s to T$_0$+42.752 s) when fitted to (a) a Band function alone, (b) C$_\mathrm{nTh}$+PL, (c) C$_\mathrm{nTh}$+C$_\mathrm{Th}$, and (d) C$_\mathrm{nTh}$+C$_\mathrm{Th}$+PL. The deconvolved $\nu$F$_\nu$ spectra are presented in panels (a1), (b1), (c1) and (d1)---the dashed lines correspond to the individual components of the fitted model and the solid ones to the total emission. Panels (a2), (b2), (c2) and (d2) correspond to the residuals of the fits.}
\end{center}
\end{figure}

\begin{figure}[ht!]
\begin{center}
\includegraphics[totalheight=0.35\textheight, clip]{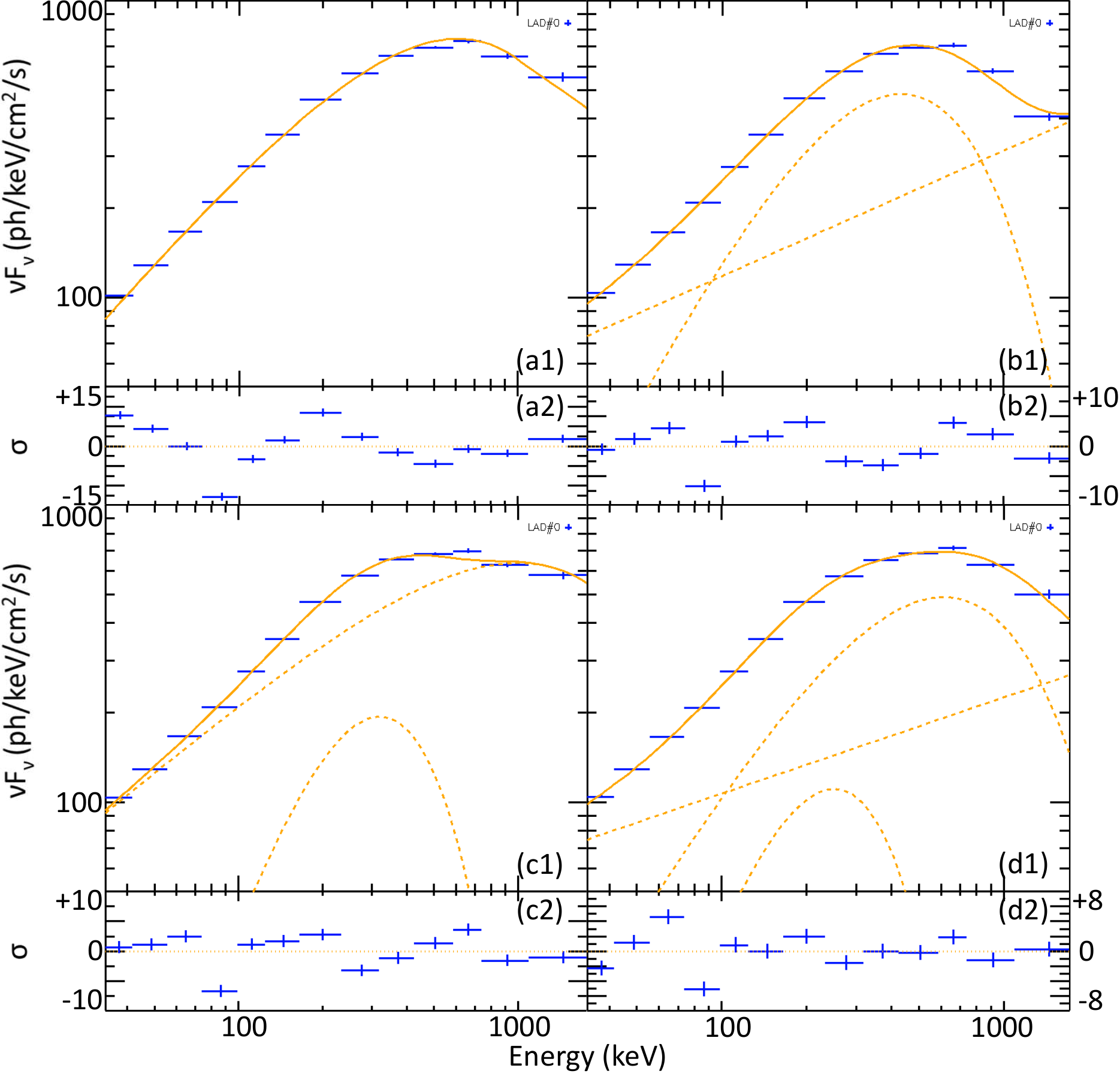}
\caption{\label{fig06}Time-integrated BATSE (LAD0) spectra of GRB~990123 (from T$_0$ to T$_0$+100.352 s) when fitted to (a) a Band function alone, (b) C$_\mathrm{nTh}$+PL, (c) C$_\mathrm{nTh}$+C$_\mathrm{Th}$, and (d) C$_\mathrm{nTh}$+C$_\mathrm{Th}$+PL. The deconvolved $\nu$F$_\nu$ spectra are presented in panels (a1), (b1), (c1) and (d1)---the dashed lines correspond to the individual components of the fitted model and the solid ones to the total emission. Panels (a2), (b2), (c2) and (d2) correspond to the residuals of the fits.}
\end{center}
\end{figure}

Table~\ref{tab01} summarizes the results of the time-integrated analysis for the three GRBs and the more relevant models are presented in Figures~\ref{fig04} to~\ref{fig06}. It is clear from the Cstat values that a Band function alone is not a good description of the time-integrated spectra; this is also supported by the strong wavy pattern observed in the residuals of the Band-only fits as shown in panels (a2) of Figures~\ref{fig04} to~\ref{fig06}.

C$_\mathrm{nTh}$+C$_\mathrm{Th}$ significantly improves the Band-only fits by 181, 72 and 283 units of Cstat for only one additional free parameter for GRBs~941017, 970111 and 990123, respectively. For the three GRBs, C$_\mathrm{nTh}$ is adequately approximated with a CPL in the C$_\mathrm{nTh}$+C$_\mathrm{Th}$ scenario. 

Although not as good as C$_\mathrm{nTh}$+C$_\mathrm{Th}$, C$_\mathrm{nTh}$+PL is also significantly better than Band-alone. It is not surprising because the additional PL usually overpowers C$_\mathrm{nTh}$ below few tens of keV and above few MeV as reported from {\it Fermi} data \citep{Abdo:2009:GRB090902B,Ackermann:2010:GRB090510,Ackermann:2011:GRB090926A,Guiriec:2010,Guiriec:2015a}. Indeed, since the energy range of BATSE only extends from 20 keV up to 2 MeV, the additional PL, if it exists, is most likely highly subdominant in this energy band compared to the other components. With a broader energy range, which extends from 8 keV up to 40 MeV, firm identifications of the additional PL are much easier with the {\it Fermi}/Gamma-ray Burst Monitor (GBM) as shown in \cite{Guiriec:2010,Guiriec:2015a}. Therefore, we do not expect to usually have an as strong Cstat improvement with the additional PL over the BATSE energy range as with C$_\mathrm{Th}$, but its possible existence cannot be discarded.


A Cstat improvement of 31 units for two additional free parameters is obtained when adding a PL to C$_\mathrm{nTh}$+C$_\mathrm{Th}$ for GRB 990123 (i.e., C$_\mathrm{nTh}$+C$_\mathrm{Th}$+PL) suggesting the simultaneous presence of the three components as presented in several {\it Fermi} GRBs in \citet{Guiriec:2015a}. For GRB 941017, only a limited Cstat improvement is obtained with C$_\mathrm{nTh}$+C$_\mathrm{Th}$+PL, but since the additional PL is supported by the joint analysis of BATSE and EGRET data in \citet{Gonzalez:2003}, we can, therefore, be confident about its existence. Claiming the possible presence of an additional PL component is much more speculative in the case of GRB 970111. Indeed, even if C$_\mathrm{nTh}$+C$_\mathrm{Th}$+PL can be fitted to the data of this GRB, the parameters of the additional PL are not well constrained and its flux is compatible with 0; however, we will see that the time-resolved analysis may support the presence of the additional PL albeit weaker than in the two other GRBs.

In addition to the discussion above, Figures \ref{fig04} to \ref{fig06} show a clear flattening of the residuals in the multi-component scenarios and the systematic wavy patterns observed in the residuals of the Band-only fits are replaced by much more random ones with C$_\mathrm{nTh}$+C$_\mathrm{Th}$+PL; therefore, the resulting spectral shape of this three component model appears as a valid option to describe the spectra of the three GRBs.

\begin{figure*}[ht!]
\begin{center}
\includegraphics[totalheight=0.80\textheight, clip]{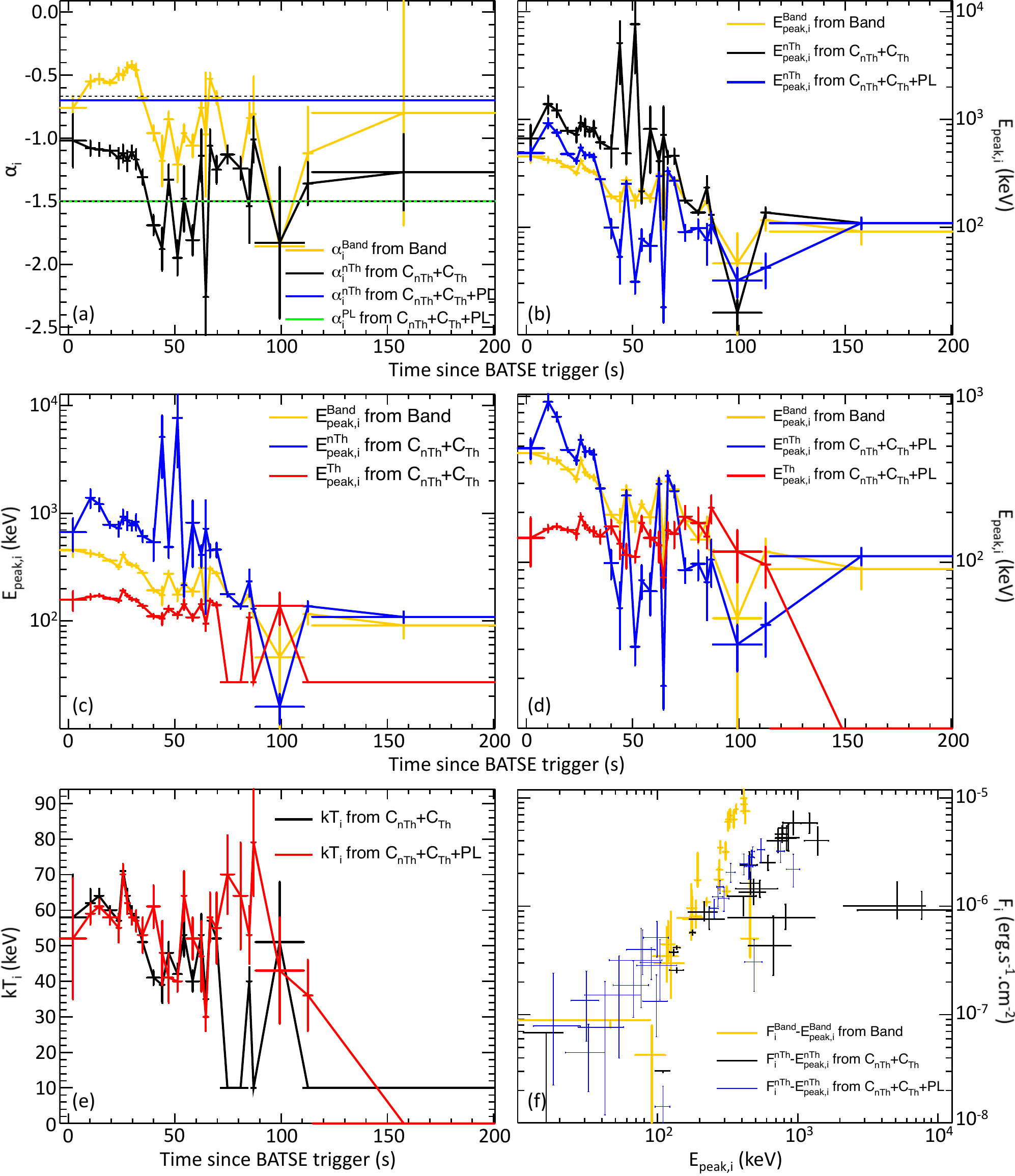}
\caption{\label{fig07}GRB~941017 -- Evolution of of the spectral parameters for Band, C$_\mathrm{nTh}$+C$_\mathrm{Th}$ and C$_\mathrm{nTh}$+C$_\mathrm{Th}$+PL. (a) Evolution of $\alpha_\mathrm{i}$. The horizontal dashed black lines at -2/3 and -3/2 correspond to the limits above which the valus of $\alpha_\mathrm{i}$ are incompatible with the synchrotron emission from electrons in the pure slow and fast cooling regime, respectively. (b) Evolution of E$_\mathrm{peak,i}$ for Band,  C$_\mathrm{nTh}$+C$_\mathrm{Th}$ and C$_\mathrm{nTh}$+C$_\mathrm{Th}$+PL. (c) Comparison of the evolution of E$_\mathrm{peak,i}^\mathrm{Band}$ from Band with E$_\mathrm{peak,i}^\mathrm{nTh}$ and E$_\mathrm{peak,i}^\mathrm{Th}$ from C$_\mathrm{nTh}$+C$_\mathrm{Th}$. (d) Comparison of the evolution of E$_\mathrm{peak,i}^\mathrm{Band}$ from Band with E$_\mathrm{peak,i}^\mathrm{nTh}$ and E$_\mathrm{peak,i}^\mathrm{Th}$ from C$_\mathrm{nTh}$+C$_\mathrm{Th}$+PL. (e) Comparison of the evolution of kT$_\mathrm{i}$ for C$_\mathrm{nTh}$+C$_\mathrm{Th}$ and C$_\mathrm{nTh}$+C$_\mathrm{Th}$+PL. (f) F$_\mathrm{i}^\mathrm{Band}$--E$_\mathrm{peak,i}^\mathrm{Band}$ and F$_\mathrm{i}^\mathrm{nTh}$--E$_\mathrm{peak,i}^\mathrm{nTh}$ for Band, C$_\mathrm{nTh}$+C$_\mathrm{Th}$ and C$_\mathrm{nTh}$+C$_\mathrm{Th}$+PL.}
\end{center}
\end{figure*}

\begin{figure*}[ht!]
\begin{center}
\includegraphics[totalheight=0.80\textheight, clip]{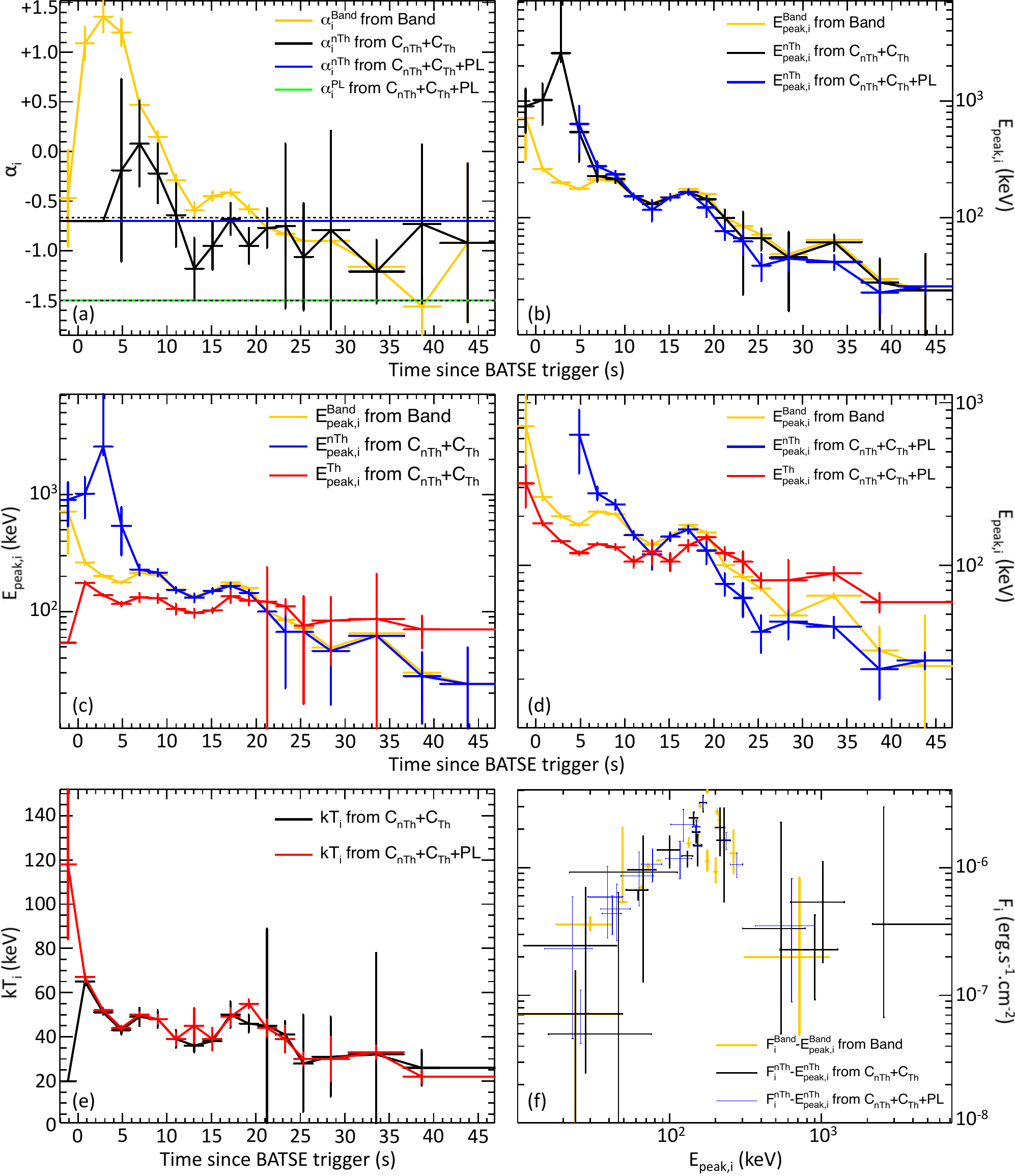}
\caption{\label{fig08}GRB~970111 -- Evolution of of the spectral parameters for Band, C$_\mathrm{nTh}$+C$_\mathrm{Th}$ and C$_\mathrm{nTh}$+C$_\mathrm{Th}$+PL. (a) Evolution of $\alpha_\mathrm{i}$. The horizontal dashed black lines at -2/3 and -3/2 correspond to the limits above which the values of $\alpha_\mathrm{i}$ are incompatible with the synchrotron emission from electrons in the pure slow and fast cooling regime, respectively. (b) Evolution of E$_\mathrm{peak,i}$ for Band,  C$_\mathrm{nTh}$+C$_\mathrm{Th}$ and C$_\mathrm{nTh}$+C$_\mathrm{Th}$+PL. (c) Comparison of the evolution of E$_\mathrm{peak,i}^\mathrm{Band}$ from Band with E$_\mathrm{peak,i}^\mathrm{nTh}$ and E$_\mathrm{peak,i}^\mathrm{Th}$ from C$_\mathrm{nTh}$+C$_\mathrm{Th}$. (d) Comparison of the evolution of E$_\mathrm{peak,i}^\mathrm{Band}$ from Band with E$_\mathrm{peak,i}^\mathrm{nTh}$ and E$_\mathrm{peak,i}^\mathrm{Th}$ from C$_\mathrm{nTh}$+C$_\mathrm{Th}$+PL. (e) Comparison of the evolution of kT$_\mathrm{i}$ for C$_\mathrm{nTh}$+C$_\mathrm{Th}$ and C$_\mathrm{nTh}$+C$_\mathrm{Th}$+PL. (f) F$_\mathrm{i}^\mathrm{Band}$--E$_\mathrm{peak,i}^\mathrm{Band}$ and F$_\mathrm{i}^\mathrm{nTh}$--E$_\mathrm{peak,i}^\mathrm{nTh}$ for Band, C$_\mathrm{nTh}$+C$_\mathrm{Th}$ and C$_\mathrm{nTh}$+C$_\mathrm{Th}$+PL.}
\end{center}
\end{figure*}

\begin{figure*}[ht!]
\begin{center}
\includegraphics[totalheight=0.80\textheight, clip]{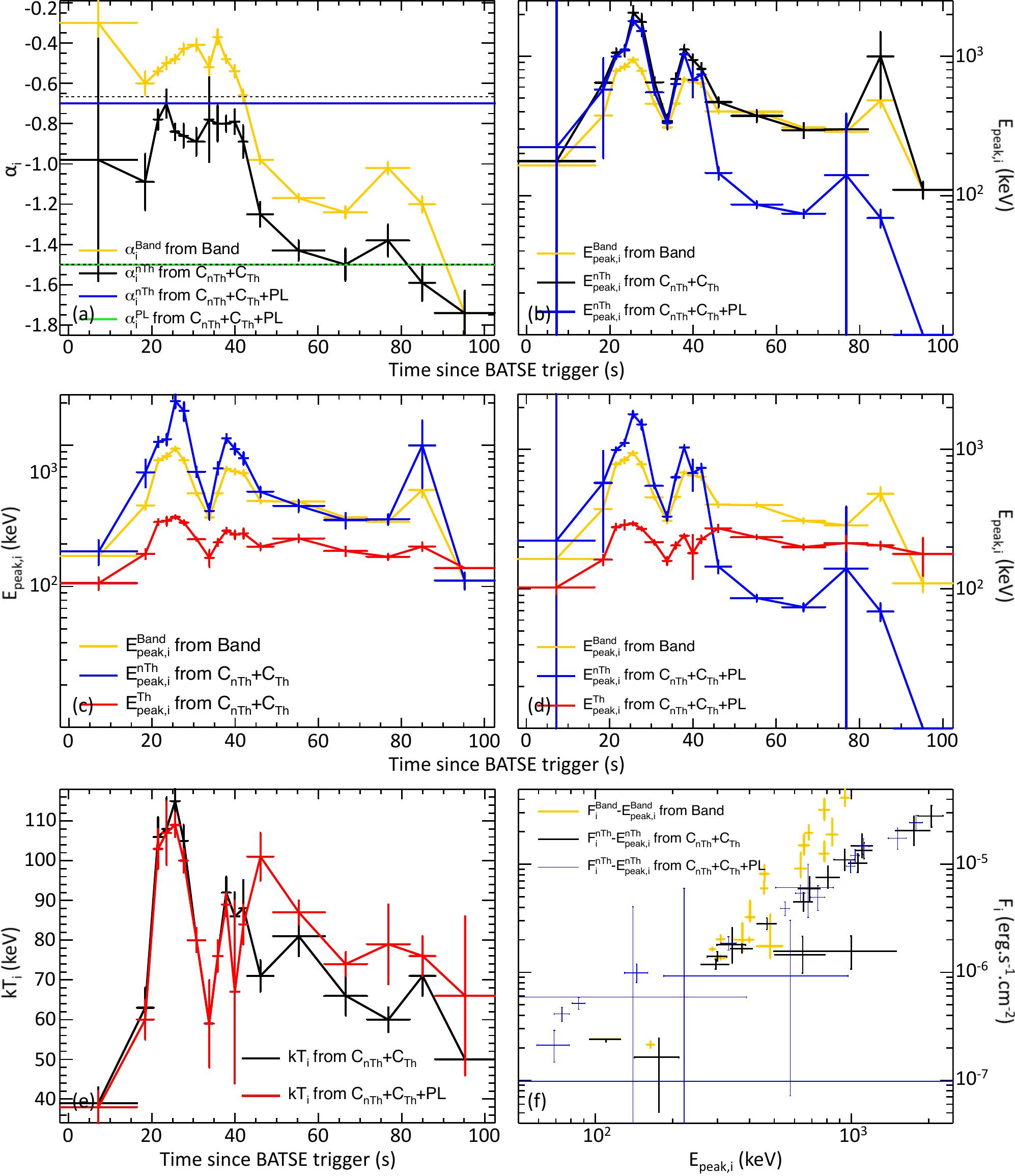}
\caption{\label{fig09}GRB~990123 -- Evolution of of the spectral parameters for Band, C$_\mathrm{nTh}$+C$_\mathrm{Th}$ and C$_\mathrm{nTh}$+C$_\mathrm{Th}$+PL. (a) Evolution of $\alpha_\mathrm{i}$. The horizontal dashed black lines at -2/3 and -3/2 correspond to the limits above which the values of $\alpha_\mathrm{i}$ are incompatible with the synchrotron emission from electrons in the pure slow and fast cooling regime, respectively. (b) Evolution of E$_\mathrm{peak,i}$ for Band,  C$_\mathrm{nTh}$+C$_\mathrm{Th}$ and C$_\mathrm{nTh}$+C$_\mathrm{Th}$+PL. (c) Comparison of the evolution of E$_\mathrm{peak,i}^\mathrm{Band}$ from Band with E$_\mathrm{peak,i}^\mathrm{nTh}$ and E$_\mathrm{peak,i}^\mathrm{Th}$ from C$_\mathrm{nTh}$+C$_\mathrm{Th}$. (d) Comparison of the evolution of E$_\mathrm{peak,i}^\mathrm{Band}$ from Band with E$_\mathrm{peak,i}^\mathrm{nTh}$ and E$_\mathrm{peak,i}^\mathrm{Th}$ from C$_\mathrm{nTh}$+C$_\mathrm{Th}$+PL. (e) Comparison of the evolution of kT$_\mathrm{i}$ for C$_\mathrm{nTh}$+C$_\mathrm{Th}$ and C$_\mathrm{nTh}$+C$_\mathrm{Th}$+PL. (f) F$_\mathrm{i}^\mathrm{Band}$--E$_\mathrm{peak,i}^\mathrm{Band}$ and F$_\mathrm{i}^\mathrm{nTh}$--E$_\mathrm{peak,i}^\mathrm{nTh}$ for Band, C$_\mathrm{nTh}$+C$_\mathrm{Th}$ and C$_\mathrm{nTh}$+C$_\mathrm{Th}$+PL.}
\end{center}
\end{figure*}

\subsection{Time-Resolved Spectral Analysis}
\label{sec:trsa}

In this Section we only discuss the most relevant models resulting from the time-integrated spectral analysis: Band, C$_\mathrm{nTh}$+C$_\mathrm{Th}$ and C$_\mathrm{nTh}$+C$_\mathrm{Th}$+PL. Here, C$_\mathrm{nTh}$ is approximated with a CPL for both C$_\mathrm{nTh}$+C$_\mathrm{Th}$ and C$_\mathrm{nTh}$+C$_\mathrm{Th}$+PL. With 7 free parameters, it is often challenging to fit C$_\mathrm{nTh}$+C$_\mathrm{Th}$+PL to the data in the fine time-intervals and to get meaningful constrains on the values of the spectral parameters; therefore, we fixed the spectral index values of C$_\mathrm{nTh}$ and of the additional PL to -0.7 and -1.5, respectively, as proposed in \citet{Guiriec:2015a}. Although these parameter estimates may not be the most accurate ones, they are good enough in the context of our analysis, which does not plan to study the detailed spectral shape of the various components, but their global shape and evolution. A more accurate estimate of these parameters based on a larger sample of GRBs will be the topic of a future article.

The results of the time-resolved spectral analysis are presented in Tables~\ref{tab04} to~\ref{tab06} as well as in Figures~\ref{fig15} to~\ref{fig20}.

\subsubsection{Band versus C$_\mathrm{nTh}$+C$_\mathrm{Th}$}

The $\nu$F$_\nu$ spectra resulting from the Band-only and C$_\mathrm{nTh}$+C$_\mathrm{Th}$ fits to the time-resolved data of GRBs~941017,~970111 and~990123 are overplotted in Figures~\ref{fig15},~\ref{fig16} and~\ref{fig17}, respectively.

For only one additional free parameter, the Cstat values obtained when fitting C$_\mathrm{nTh}$+C$_\mathrm{Th}$ to the data are overall much lower than these resulting from the fit to a Band function alone (see Tables~\ref{tab04} to \ref{tab06}.)

\begin{figure}[ht!]
\begin{center}
\includegraphics[totalheight=0.32\textheight, clip]{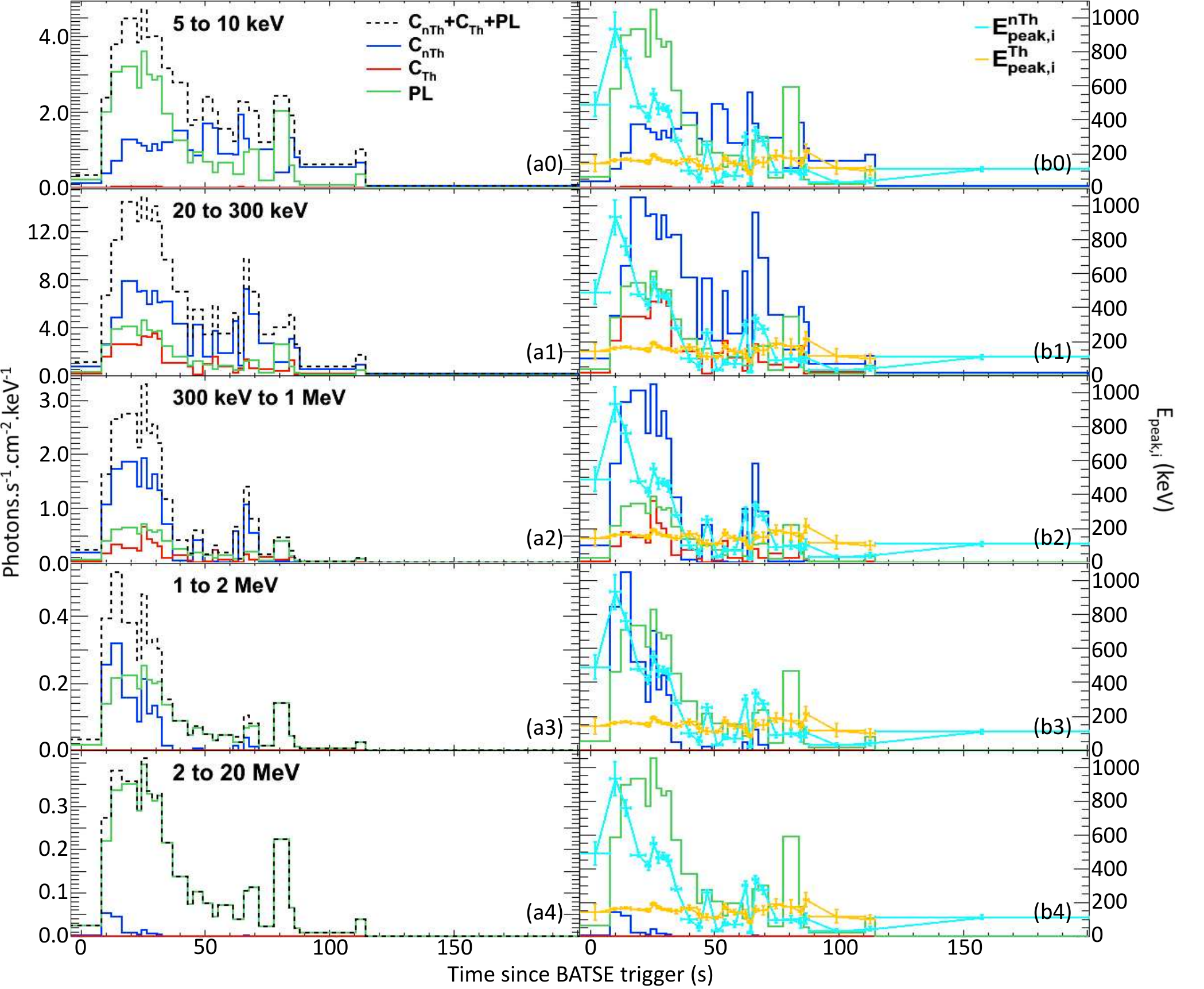}
\caption{\label{fig10}GRB 941017 -- Reconstructed photon light curves resulting from the spectral analysis using the C$_\mathrm{nTh}$+C$_\mathrm{Th}$+PL model. (a0--4) The reconstructed photon light curves of the non-thermal (C$_\mathrm{nTh}$), the thermal-like (C$_\mathrm{Th}$) and the PL components are displayed in blue, red and green, respectively. The black dashed lines correspond to the sum of the three components (C$_\mathrm{nTh}$+C$_\mathrm{Th}$+PL). The light curves (a1--3) are displayed in the same energy bands as the count light curves presented in Figure~\ref{fig01} and with the same time intervals as in Figure~\ref{fig01}b. (b0--4) The C$_\mathrm{nTh}$, C$_\mathrm{Th}$ and PL component are displayed in blue, red and green, respectively, together with the evolution of the $\nu$F$_\nu$ spectral peaks of C$_\mathrm{nTh}$, E$_\mathrm{peak}^\mathrm{nTh}$, in cyan and of C$_\mathrm{Th}$, E$_\mathrm{peak}^\mathrm{Th}$, in orange. The temperature of the thermal-like component C$_\mathrm{Th}$ is obtained by dividing E$_\mathrm{peak}^\mathrm{Th}$ by $\sim$2.5.}
\end{center}
\end{figure}
\begin{figure}[ht!]
\begin{center}
\includegraphics[totalheight=0.32\textheight, clip]{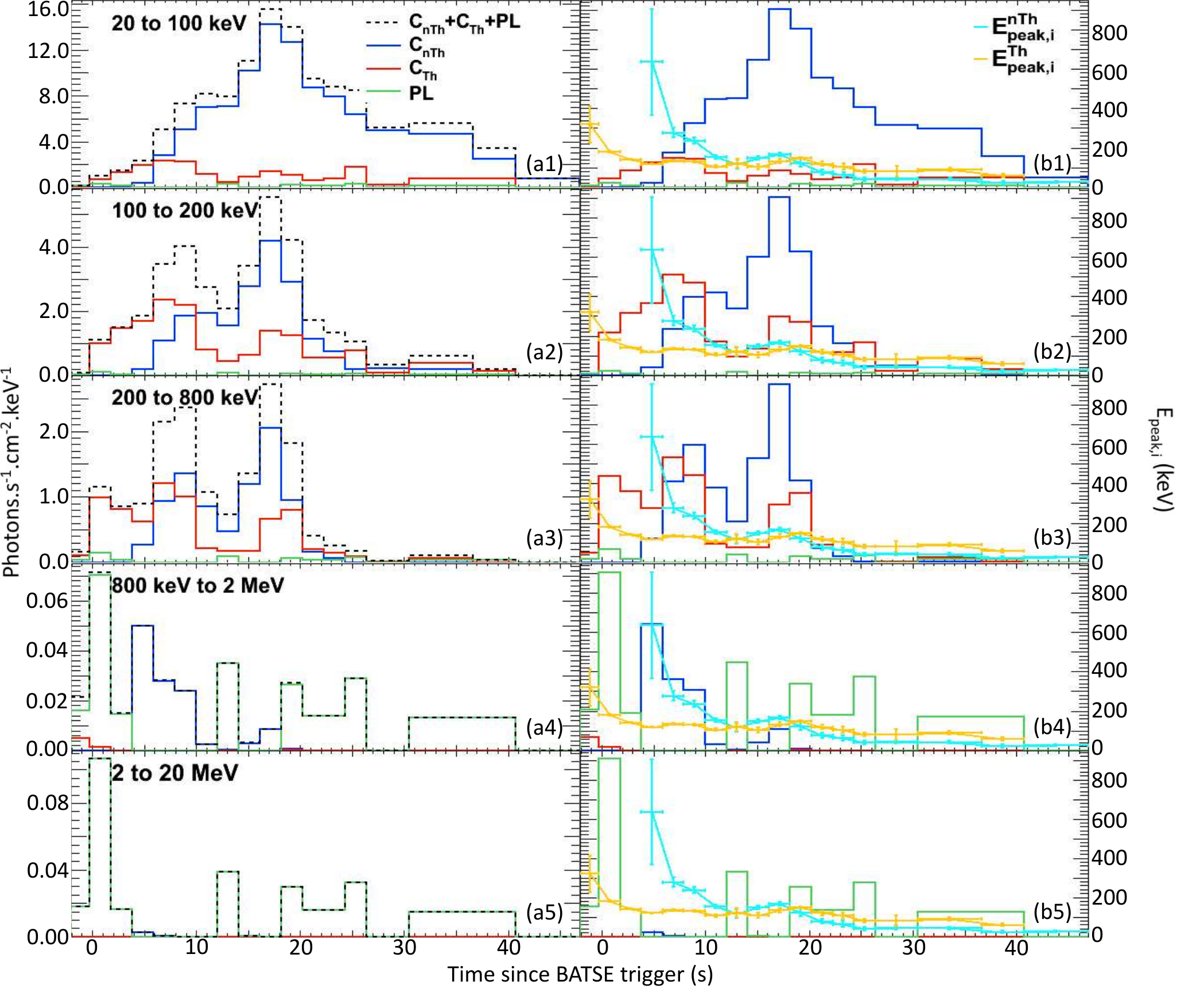}
\caption{\label{fig11}GRB 970111 -- Reconstructed photon light curves resulting from the spectral analysis using the C$_\mathrm{nTh}$+C$_\mathrm{Th}$+PL model. (a0--5) The reconstructed photon light curves of the non-thermal (C$_\mathrm{nTh}$), the thermal-like (C$_\mathrm{Th}$) and the PL components are displayed in blue, red and green, respectively. The black dashed lines correspond to the sum of the three components (C$_\mathrm{nTh}$+C$_\mathrm{Th}$+PL). The light curves (a1--4) are displayed in the same energy bands as the count light curves presented in Figure~\ref{fig02} and with the same time intervals as in Figure~\ref{fig02}b. (b0--5) The C$_\mathrm{nTh}$, C$_\mathrm{Th}$ and PL component are displayed in blue, red and green, respectively, together with the evolution of the $\nu$F$_\nu$ spectral peaks of C$_\mathrm{nTh}$, E$_\mathrm{peak}^\mathrm{nTh}$, in cyan and of C$_\mathrm{Th}$, E$_\mathrm{peak}^\mathrm{Th}$, in orange. The temperature of the thermal-like component C$_\mathrm{Th}$ is obtained by dividing E$_\mathrm{peak}^\mathrm{Th}$ by $\sim$2.5.}
\end{center}
\end{figure}

The similarities of BATSE and {\it Fermi} results are striking when comparing Band and C$_\mathrm{nTh}$+C$_\mathrm{Th}$. Indeed, $\alpha_\mathrm{i}^\mathrm{nTh}$ is systematically lower than $\alpha_\mathrm{i}^\mathrm{Band}$ (panels (a) of Figures~\ref{fig07} to \ref{fig09}) and E$_\mathrm{peak,i}^\mathrm{nTh}$ is systematically higher than E$_\mathrm{peak,i}^\mathrm{Band}$ (panels (c) of Figures~\ref{fig07} to \ref{fig09}) as reported in \citet{Guiriec:2011a,Guiriec:2013a,Guiriec:2015a,Guiriec:2015b}\footnote{The index i indicates that we are discussing time-resolved values. i is the time-interval index.}.

The $\alpha_\mathrm{i}^\mathrm{Band}$ values exhibit strong evolution and discontinuities. While usually incompatible with synchrotron scenario limits during the first few tens of seconds, they become consistent with synchrotron emission from electrons in the slow (i.e., $\alpha_\mathrm{i}^\mathrm{Band}$$<$-2/3) and even in the fast (i.e., $\alpha_\mathrm{i}^\mathrm{Band}$$<$-3/2) cooling regimes at late times. Similarly to the results presented in \citet{Guiriec:2013a,Guiriec:2015b}, the values of $\alpha_{i}^\mathrm{Band}$ are clearly positive and incompatible with non-thermal processes during the first $\sim$10 s of GRB~970111; this also corresponds to the time period during which C$_\mathrm{Th}$ overpowers C$_\mathrm{nTh}$ and therefore the spectral parameters of Band-alone are biased towards those of C$_\mathrm{Th}$ (see the several first time-resolved spectra of GRB~970111 in Figure~\ref{fig16}).

Conversely to $\alpha_\mathrm{i}^\mathrm{Band}$, the values of $\alpha_\mathrm{i}^\mathrm{nTh}$ are always compatible with at least the synchrotron slow cooling scenario (i.e., $\alpha_\mathrm{i}^\mathrm{nTh}$$<$-2/3). As we will see in the next Section, the discontinuity of $\alpha_\mathrm{i}^\mathrm{nTh}$ for GRB~990123---whose values drop from $\sim$-0.8 down to $<$-1.5 after $\sim$T$_\mathrm{0}$+40 s---is consistent with the presence of an intense additional PL at late times (see Figure~\ref{fig09}a).

Overall, the temperature of C$_\mathrm{Th}$ (i.e., kT$\sim$E$_\mathrm{peak,i}^\mathrm{Th}$/2.7; we choose to plot E$_\mathrm{peak,i}^\mathrm{Th}$ instead of kT for an easier comparison to E$_\mathrm{peak,i}^\mathrm{nTh}$ and E$_\mathrm{peak,i}^\mathrm{Band}$) varies much less than E$_\mathrm{peak,i}^\mathrm{Band}$ (see panels (c) of Figures~\ref{fig07} to~\ref{fig09}.) While the temperature kT of the C$_\mathrm{Th}$ varies between $\sim$40 and $\sim$120 keV for GRB~990123, its evolution is much more limited for GRBs~941017 and 970111 in which kT varies from $\sim$30 to $\sim$70 keV and from $\sim$20 to $\sim$65 keV, respectively (see panels (e) of Figures~\ref{fig07} to~\ref{fig09}.).

E$_\mathrm{peak,i}^\mathrm{Th}$ is usually lower than E$_\mathrm{peak,i}^\mathrm{nTh}$. As noted in \citet{Guiriec:2011a,Guiriec:2013a,Guiriec:2015a,Guiriec:2015b}, the discrepancies between E$_\mathrm{peak,i}^\mathrm{nTh}$ and E$_\mathrm{peak,i}^\mathrm{Band}$ are strongest at early times when the intensity of C$_\mathrm{Th}$ is the highest; at later times, when the C$_\mathrm{Th}$ contribution decreases, the values of E$_\mathrm{peak,i}^\mathrm{nTh}$ and E$_\mathrm{peak,i}^\mathrm{Band}$ become extremely similar. This is particularly striking in the case of GRB~970111 (see Figure~\ref{fig08}c): from $\sim$T$_\mathrm{0}$ to $\sim$T$_\mathrm{0}$+6 s, E$_\mathrm{peak,i}^\mathrm{Band}$ evolves like E$_\mathrm{peak,i}^\mathrm{Th}$ and not like E$_\mathrm{peak,i}^\mathrm{nTh}$, while after $\sim$T$_\mathrm{0}$+6 s E$_\mathrm{peak,i}^\mathrm{Band}$ and E$_\mathrm{peak,i}^\mathrm{nTh}$ are perfectly identical. It is also interesting to note that the values of E$_\mathrm{peak,i}^\mathrm{nTh}$ seem to fall below those of E$_\mathrm{peak,i}^\mathrm{Th}$ after $\sim$T$_\mathrm{0}$+20 s as also reported in \citep{Guiriec:2015a} for {\it Fermi} GRBs; this trend will be clearer in the C$_\mathrm{nTh}$+C$_\mathrm{Th}$+PL analysis presented in the next Section.

\begin{figure}[ht!]
\begin{center}
\includegraphics[totalheight=0.32\textheight, clip]{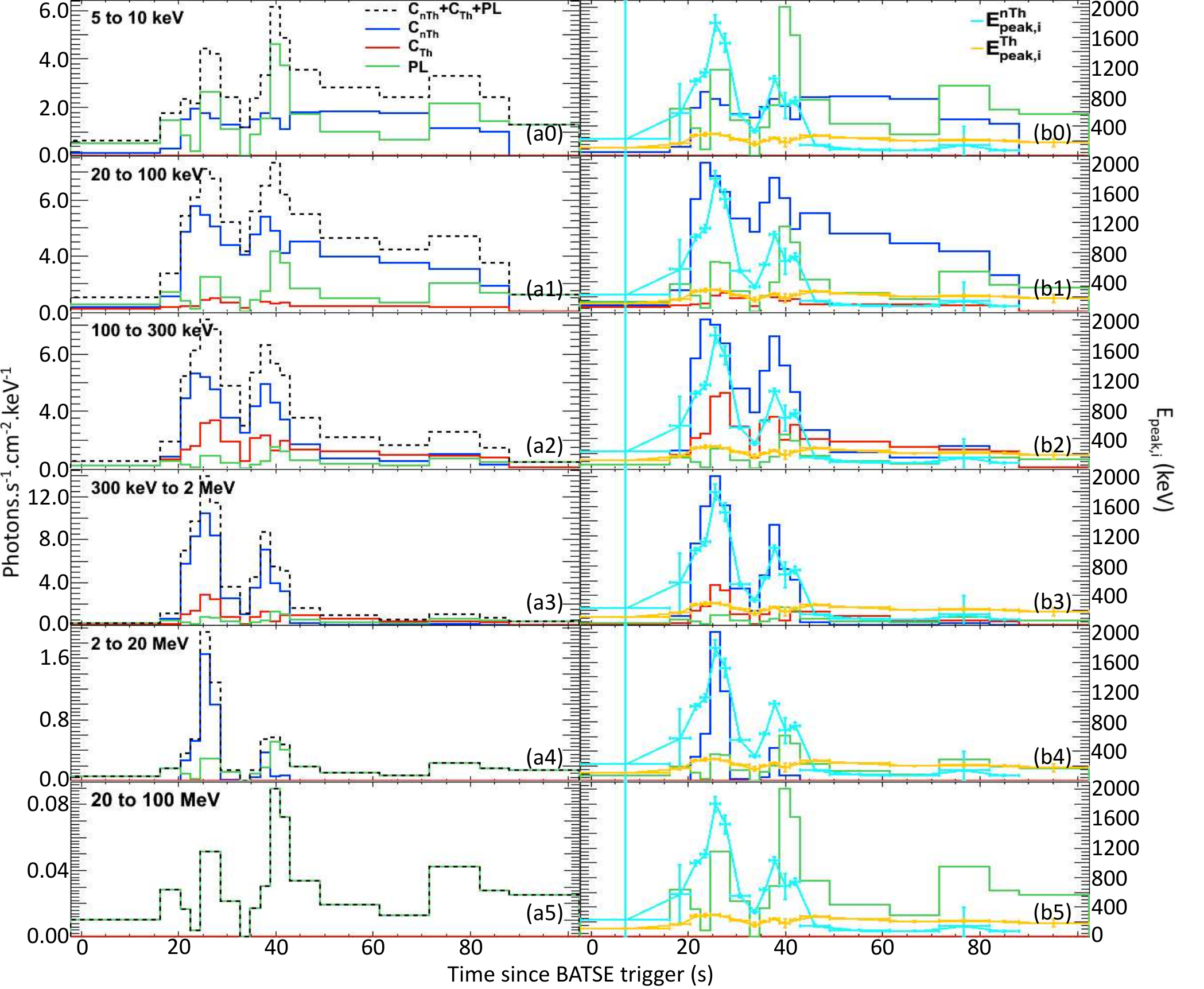}
\caption{\label{fig12}GRB 990123 -- Reconstructed photon light curves resulting from the spectral analysis using the C$_\mathrm{nTh}$+C$_\mathrm{Th}$+PL model. (a0--5) The reconstructed photon light curves of the non-thermal (C$_\mathrm{nTh}$), the thermal-like (C$_\mathrm{Th}$) and the PL components are displayed in blue, red and green, respectively. The black dashed lines correspond to the sum of the three components (C$_\mathrm{nTh}$+C$_\mathrm{Th}$+PL). The light curves (a1--3) are displayed in the same energy bands as the count light curves presented in Figure~\ref{fig03} and with the same time intervals as in Figure~\ref{fig03}b. (b0--5) The C$_\mathrm{nTh}$, C$_\mathrm{Th}$ and PL component are displayed in blue, red and green, respectively, together with the evolution of the $\nu$F$_\nu$ spectral peaks of C$_\mathrm{nTh}$, E$_\mathrm{peak}^\mathrm{nTh}$, in cyan and of C$_\mathrm{Th}$, E$_\mathrm{peak}^\mathrm{Th}$, in orange. The temperature of the thermal-like component C$_\mathrm{Th}$ is obtained by dividing E$_\mathrm{peak}^\mathrm{Th}$ by $\sim$2.5.}
\end{center}
\end{figure}

\subsubsection{C$_\mathrm{nTh}$+C$_\mathrm{Th}$ versus C$_\mathrm{nTh}$+C$_\mathrm{Th}$+PL}

The $\nu$F$_\nu$ spectra resulting from the C$_\mathrm{nTh}$+C$_\mathrm{Th}$+PL fits to the time-resolved data of GRBs~941017,~970111 and~990123 are displayed in Figures~\ref{fig18},~\ref{fig19} and~\ref{fig20}, respectively---together with the Band-only fit results.

In \citet{Guiriec:2015a}, we showed that the distributions of $\alpha_\mathrm{i}^\mathrm{nTh}$ and $\alpha_\mathrm{i}^\mathrm{PL}$ peaked at $\sim$-0.7 and at $\sim$-1.5, respectively, for a sample of {\it Fermi} GRBs fitted to C$_\mathrm{nTh}$+C$_\mathrm{Th}$+PL. When fixing $\alpha_\mathrm{i}^\mathrm{nTh}$ and $\alpha_\mathrm{i}^\mathrm{PL}$ of C$_\mathrm{nTh}$+C$_\mathrm{Th}$+PL to these typical values for all the time intervals of our current analysis of BATSE data, C$_\mathrm{nTh}$+C$_\mathrm{Th}$ and C$_\mathrm{nTh}$+C$_\mathrm{Th}$+PL result in similar Cstat values, overall, for the same number of free parameters (see Tables~\ref{tab04} to \ref{tab06}).

When C$_\mathrm{nTh}$+C$_\mathrm{Th}$ is fitted to the data of the three GRBs without the additional PL, the values of $\alpha_\mathrm{i}^\mathrm{nTh}$ cluster around -1 that is an average of $\sim$-0.7 and $\sim$-1.5 (see panels (a) of Figures~\ref{fig07} to~\ref{fig09}); this result is similar to the one reported in \citet{Guiriec:2015a} that showed the impact of the extension at low energy of the additional PL on the values of $\alpha$. For GRB 970111, which exhibits a weaker additional PL component (see Section~\ref{sec:tisa}), the $\alpha_\mathrm{i}^\mathrm{nTh}$ values remains closer to $\sim$-0.7 with C$_\mathrm{nTh}$+C$_\mathrm{Th}$.

As also reported in \citet{Guiriec:2015a}, the values of E$_\mathrm{peak,i}^\mathrm{nTh}$ obtained when fitting C$_\mathrm{nTh}$+C$_\mathrm{Th}$+PL to the data are usually lower than those resulting from the C$_\mathrm{nTh}$+C$_\mathrm{Th}$ fits (see panels (b) of Figures \ref{fig07} to \ref{fig09}); this is particularly striking for GRB 941017 as well as for GRB 990123 after $\sim$T$_\mathrm{0}$+45 s. This can easily be explained if we consider that the contribution of the additional PL at high energy (i.e., $>$$\sim$1 MeV) is included in C$_\mathrm{nTh}$ when fitting C$_\mathrm{nTh}$+C$_\mathrm{Th}$ to the data and makes, therefore, the spectra of C$_\mathrm{nTh}$ harder and their E$_\mathrm{peak,i}^\mathrm{nTh}$ values higher.

We mentioned in Section~\ref{sec:tisa} the difficulty to clearly identify the additional PL in the time-integrated spectrum of GRB 970111. It is interesting to note that the C$_\mathrm{Th}$+PL fit---with $\alpha_\mathrm{i}^\mathrm{PL}$=-1.5 and without any C$_\mathrm{nTh}$ component---to the data of GRB 970111 provides an adequate description of the spectra during the first few seconds of the burst. This has two important consequences: first, the use of C$_\mathrm{Th}$+PL at early time results in an initial monotonic cooling of the thermal-like component, C$_\mathrm{Th}$, while it was not the case for the first time interval when fitting C$_\mathrm{nTh}$+C$_\mathrm{Th}$ (see Figure~\ref{fig08}e); second, this would indicate that similarly to the cases of GRBs 080916C and 090926A published in \citet{Guiriec:2015a}, the additional PL would be initially strong in GRB 970111 while C$_\mathrm{nTh}$ is either weak or absent. The additional PL would then be present from the very beginning or even before the other components conversely to previous reports where the additional PL was considered to start with a delay compared to the other main non-thermal component \citep[see for instance][]{Ackermann:2010:GRB090510,Ackermann:2011:GRB090926A}.

\subsection{The C$_\mathrm{nTh}$+C$_\mathrm{Th}$+PL Scenario}
\label{sec:CnTh+CTh+PL}

In Section~\ref{sec:tisa} we discussed the presence of three spectral components in the time-integrated spectra of GRBs 941017, 970111 and 990123, and we followed their evolution with time in Section~\ref{sec:trsa}. We compared the differences in the spectral parameters when fitting the simplest model and the more complex ones, and we found strong similarities with the results reported in \citet{Guiriec:2011a,Guiriec:2013a,Guiriec:2015a,Guiriec:2015b} from {\it Fermi} data. Since the time-resolved analysis of BATSE data is consistent with the new paradigm proposed in \citet{Guiriec:2015a}, we focussed on the C$_\mathrm{nTh}$+C$_\mathrm{Th}$+PL scenario later on.

\begin{figure*}[ht!]
\begin{center}
\includegraphics[totalheight=0.39\textheight, clip]{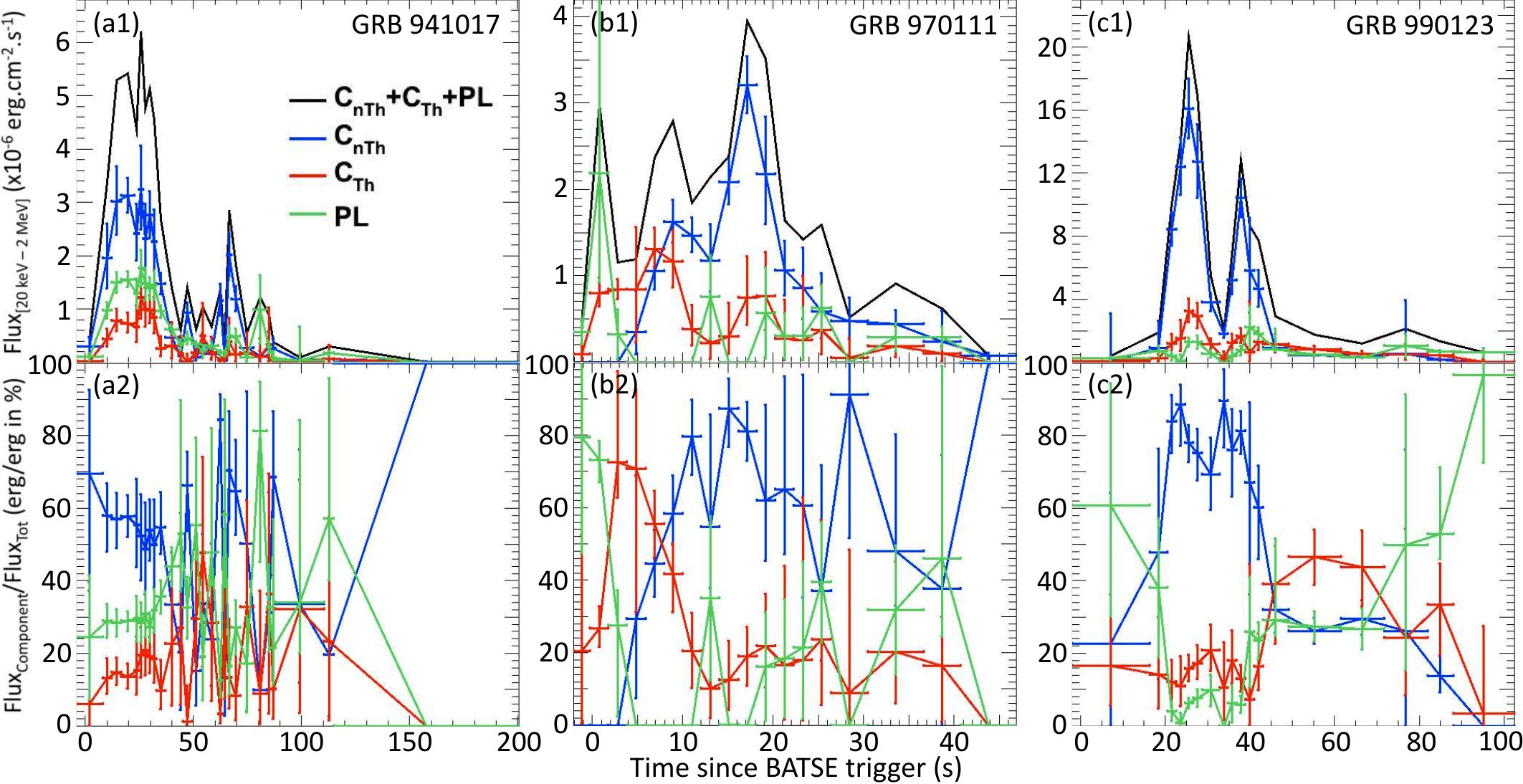}
\caption{\label{fig13}Energy flux evolution between 20 keV and 2 MeV for GRB~941017 (a1), GRB~970111 (b1) and GRB~990123 (c1) in the context of the C$_\mathrm{nTh}$+C$_\mathrm{Th}$+PL model. Panels (a2), (b2) and (c2) of each figure show the contribution of each component to the total energy flux of GRBs~941017,~970111 and~990123, respectively. For clarity, no uncertainty is displayed for the total energy flux (black line). Because of its narrow spectral shape (i.e., Planck function), the contribution of C$_\mathrm{Th}$ to the total 20 keV--2 MeV energy flux is very limited; however, it has an important relative contribution to the total emission at the C$_\mathrm{Th}$ spectral peak around 100--200 keV (see Figure~\ref{fig18} to~\ref{fig20} of Appendix~\ref{section:Time-resolved-analysis}).}
\end{center}
\end{figure*}

Figures~\ref{fig10} to \ref{fig12} show the reconstructed photon light curves in various energy bands resulting from the time-resolved spectral analysis using C$_\mathrm{nTh}$+C$_\mathrm{Th}$+PL for GRBs 941017, 970111 and 990123. The energy light curves between 20 keV and 2 MeV as well as the relative contribution of each component to the total energy are presented in Figure~\ref{fig13}.

\begin{center}
The C$_\mathrm{nTh}$ Component
\end{center}

Overall, C$_\mathrm{nTh}$ overpowers the other components in the BATSE energy range from 20 keV to 2 MeV for the three GRBs (see Figures~\ref{fig10} to \ref{fig12}). Its contribution is usually higher than 50\% of the total energy released between 20 keV and 2 MeV during the most intense part of the prompt emission (see Figure~\ref{fig13}). The hardness of C$_\mathrm{nTh}$ strongly evolves with time during each burst; indeed, E$_\mathrm{peak,i}^\mathrm{nTh}$ can vary from tens of keV up to several MeV in only a few seconds (see panels (d) of Figures~\ref{fig07} to \ref{fig10}).

The light curves of the C$_\mathrm{nTh}$ component usually exhibit multiple intensity peaks whose hardness is correlated to the flux; this is clearly evident with the E$_\mathrm{peak,i}^\mathrm{nTh}$ values tracking the photon fluxes as shown in panels (b) of Figures~\ref{fig10} to \ref{fig12}. The correlation between the energy flux of C$_\mathrm{nTh}$, F$_\mathrm{i}^\mathrm{nTh}$, and E$_\mathrm{peak,i}^\mathrm{nTh}$ will be studied in detail in Section~\ref{sec:F-Ep}.

\begin{center}
The C$_\mathrm{Th}$ Component
\end{center}

The contribution of the C$_\mathrm{Th}$ component to the total energy between 20 keV and 2 MeV is globally $\leq$20\% of the total emission during most of the burst duration, which is consistent with the results reported from {\it Fermi} data in \citet{Guiriec:2013a,Guiriec:2015a,Guiriec:2015b}. C$_\mathrm{Th}$ is usually more intense at early times. The temperature of C$_\mathrm{Th}$---approximated with E$_\mathrm{peak,i}^\mathrm{Th}$ for an easier comparison with E$_\mathrm{peak,i}^\mathrm{nTh}$---does not vary over the same amplitude as E$_\mathrm{peak,i}^\mathrm{nTh}$ (see panels (b) of Figures~\ref{fig10} to \ref{fig12}.); indeed, kT$_\mathrm{i}$ varies between $\sim$30 and $\sim$80 keV, between $\sim$20 and $\sim$140 keV and between $\sim$40 and $\sim$110 keV for GRBs 941017, 970111 and 990123, respectively (see panels (e) of Figures~\ref{fig07} to \ref{fig09}). Although a limited correlation may exist between the flux of C$_\mathrm{Th}$ and its temperature, kT$_\mathrm{i}$ remains more or less constant with time for GRBs 941017 and 990123; this result is similar to the one reported from the observations of GRBs 080916C and 090926A in \citet{Guiriec:2015a}.

The behavior of GRB~970111 is peculiar compared to the two other GRBs of this study but very similar to {\it Fermi} GRB~131014A published in \citet{Guiriec:2015b}. Indeed, GRB~970111 exhibits a initial nearly purely thermal episode\footnote{This is is consistent with~\citet{Ryde:2004} that reported a pure BB spectrum at early times for GRB~970111.}---despite the presence of the additional PL at early times---with a monotonic cooling from $\sim$120 keV down to $\sim$45 keV during the first five seconds of the burst (see Figures~\ref{fig08}e and~\ref{fig11}b). After $\sim$T$_\mathrm{0}$+5 s, kT$_\mathrm{i}$ continues to cool slowly from $\sim$45 down to $\sim$25 keV during the remaining $\sim$40 s of the studied emission period although there is very slight reheating consistent with the light curve peaks of the thermal-like component.

Despite some similarities in the light curve structures, there is no evidence for a perfect correlation either between the flux variations of C$_\mathrm{nTh}$ and C$_\mathrm{Th}$ nor between the variations of E$_\mathrm{peak,i}^\mathrm{nTh}$ and E$_\mathrm{peak,i}^\mathrm{Th}$ (see panels (d) of Figures \ref{fig07} to \ref{fig09} and panels (b) of Figures \ref{fig10} to \ref{fig12}). It is particularly striking that while E$_\mathrm{peak,i}^\mathrm{nTh}$ is usually much higher than E$_\mathrm{peak,i}^\mathrm{Th}$ at early times for all three GRBs, its values drop below E$_\mathrm{peak,i}^\mathrm{Th}$ at late times.

\begin{center}
The Additional PL Component
\end{center}

The overall contribution of the additional PL to the total energy between 20 keV and 2 MeV is roughly few tens of percents although it can be $>$70\% at early and late times (see Figure~\ref{fig13}).

It has been reported many times from Fermi data that the additional PL usually extends from a few keV up to tens to hundreds of MeV and maybe even up to GeVs; the additional PL usually overpowers C$_\mathrm{nTh}$ below few tens of keV and above several MeVs \citep{Abdo:2009:GRB090902B,Ackermann:2010:GRB090510,Ackermann:2011:GRB090926A,Guiriec:2011a,Guiriec:2015a}. Because of the limited energy range of BATSE, the additional PL is mostly subdominant compared to C$_\mathrm{nTh}$ over the whole observed energy band, and it only starts to be the dominant component at the very high-end of the spectrum (see Figures~\ref{fig18} to \ref{fig20}). Although C$_\mathrm{nTh}$ clearly outshines the additional PL below 20 keV in GRB 970111 (see Figure \ref{fig11}), there is evidence for an extension of the additional PL at low and high energies with a flux higher than in the other components in GRBs 941017 and 990123 (see Figures \ref{fig10} and \ref{fig12}). For a better comparison to the {\it Fermi} data, we extrapolated the C$_\mathrm{nTh}$+C$_\mathrm{Th}$+PL model derived from the 20 keV to 2 MeV BATSE data down to 5 keV---to mimic the low end of the energy range of the GBM NaI detectors---and beyond 2 MeV---to mimic the high energy coverage of GBM BGO detectors up to 40 MeV as well as the 20 to 100 MeV energy range of the {\it Fermi}/Large Area Telescope (LAT). If {\it Fermi} had observed GRBs 941017 and GRB990123---without considering the instrumental sensitivities nor possible $<$100 MeV breaks in the additional PL as suggested in \citet{Guiriec:2015a}---, it would have detected an additional PL outshining the other components at low and high energies similarly to the other bursts with additional PL detected by {\it Fermi}:

\begin{itemize}
\item If observed with {\it Fermi}, the additional PL in GRB 941017 would correspond to a broad pulse in the light curves with a fast rise and exponential decay type shape (see Figure~\ref{fig10}); despite the similarities in the timing of the most intense part of the additional PL with the other components, their evolution with time are quite different. The additional PL emission would be the most intense between $\sim$T$_\mathrm{0}$+10 s and $\sim$T$_\mathrm{0}$+30 s, then the emission would quickly decay but it would remain present until at least T$_\mathrm{0}$+100 s; this is in perfect agreement with the $>$ 1 MeV light curves of the data from the Total Absorption Shower Counter (TASC), which is the calorimeter of the Energetic Gamma-Ray Experiment Telescope (EGRET) on board CGRO, presented in Figure~1 in \citet{Gonzalez:2003}. The broad shape of the additional PL light curve is similar to the one reported for {\it Fermi} GRB 090902B in \citet{Abdo:2009:GRB090902B}.
\item The case of GRB~990123 is particularly interesting with respect to the additional PL. In \citet{Guiriec:2015a} we reported that the additional PLs identified in {\it Fermi} GRBs 080916C and 090926A, were, overall, low intensity components that sometimes strongly outshine the other ones during short periods of time of the order of a few seconds---short duration excesses associated with the additional PL were also reported in \citet{Ackermann:2011:GRB090926A} and \citet{Gonzalez:2012} for {\it Fermi} GRB 090926A and BATSE GRB 980923. The same behavior is observed in GRB~990123 (see Figure~\ref{fig12}). Indeed, if {\it Fermi} had observed this GRB, it would have detected a strong intensity peak at low energies below $\sim$10 keV around T$_\mathrm{0}$+40 s and lasting $\sim$5 s perfectly correlated in time with an intense peak in the light curve above a few MeV. As shown in Figure~\ref{fig12}, this intensity peak in the light curves would be mostly related to the additional PL. However, the contribution of the additional PL would be mostly hidden by the intense C$_\mathrm{nTh}$ component between a few tens of keV and several MeV as in GRB~080916C~\citep{Guiriec:2015a}.
\end{itemize}

\begin{figure*}[ht!]
\begin{center}
\includegraphics[totalheight=0.55\textheight, clip]{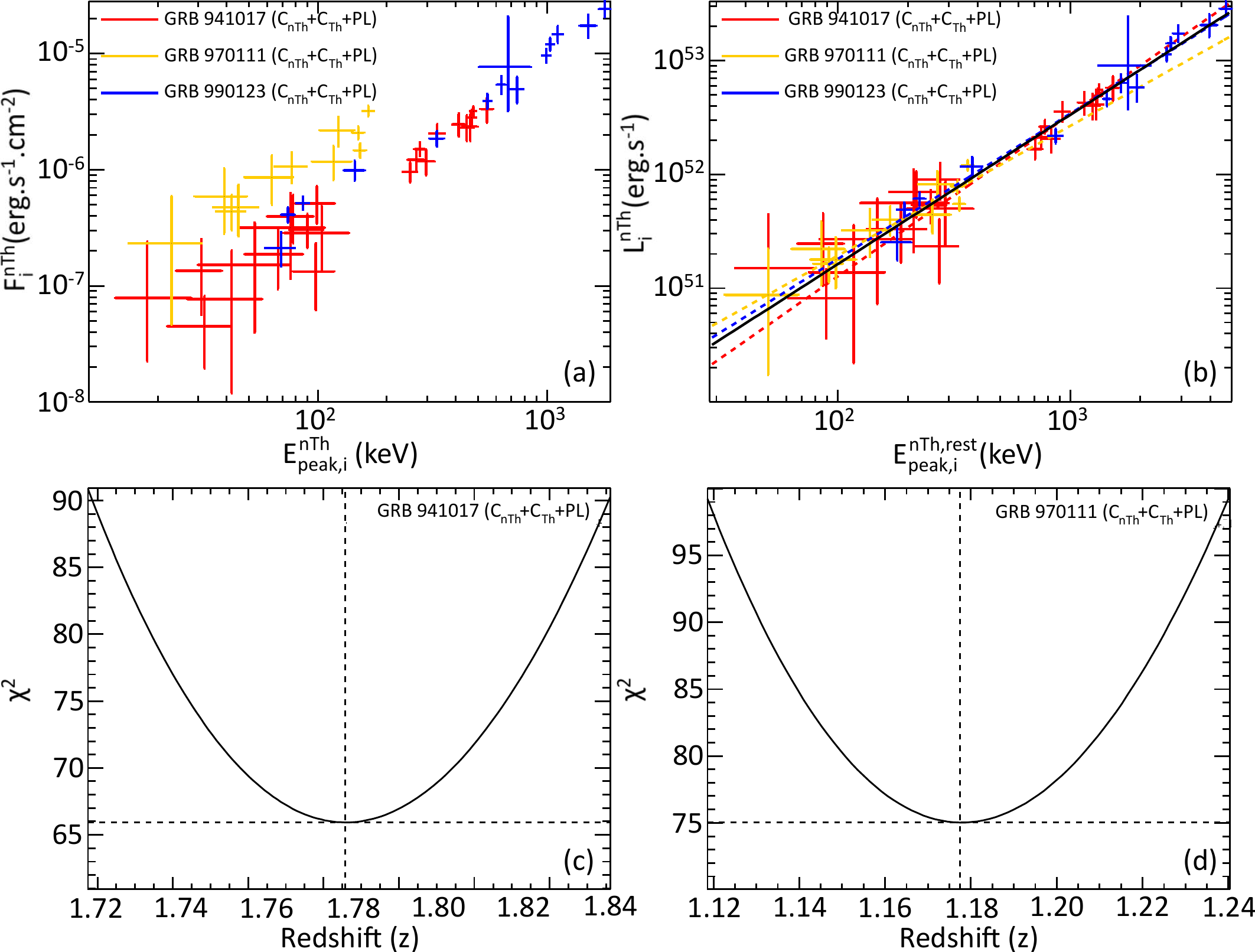}
\caption{\label{fig14}(a) Energy flux of C$_\mathrm{nTh}$ between 20 keV and 2 MeV, F$_\mathrm{i}^\mathrm{nTh}$, as a function of the $\nu$F$_\nu$ spectral peak of C$_\mathrm{nTh}$, E$_\mathrm{peak,i}^\mathrm{nTh}$, resulting from the C$_\mathrm{nTh}$+C$_\mathrm{Th}$+PL fits to the fine time intervals of GRBs~941017, 970111 and 990123 (i.e., F$_\mathrm{i}^\mathrm{nTh}$--E$_\mathrm{peak,i}^\mathrm{nTh}$ relation). (b) Luminosity of C$_\mathrm{nTh}$, L$_\mathrm{i}^\mathrm{nTh}$, as a function of the $\nu$F$_\nu$ spectral peak of C$_\mathrm{nTh}$ in the rest frame, E$_\mathrm{peak,i}^\mathrm{nTh,rest}$, resulting from the C$_\mathrm{nTh}$+C$_\mathrm{Th}$+PL fits to the fine time intervals of GRBs~941017, 970111 and 990123 (i.e., L$_\mathrm{i}^\mathrm{nTh}$--E$_\mathrm{peak,i}^\mathrm{nTh,rest}$ relation). The measured redshift  z=1.61 is used for GRB~990123, while the redshift estimates z=1.79 and z=1.18---using GRB~990123 as a reference for the redshift estimates---are used for GRBs~941017 and~970111, respectively. The color dashed lines correspond to the PL fits to the L$_\mathrm{i}^\mathrm{nTh}$--E$_\mathrm{peak,i}^\mathrm{nTh,rest}$ relations of each GRB, and the solid black line corresponds to the PL fit to the data of the three GRBs simultaneously. (c \& d) $\chi^\mathrm{2}$ profiles resulting from the redshift estimates of GRBs~941017 (z=1.79$\pm$0.07) and~970111 (z=1.18$\pm$0.06) using GRB~990123---with z=1.6---as a reference.}
\end{center}
\end{figure*}

\newpage
\subsection{F$_\mathrm{i}^\mathrm{nTh}$--E$_\mathrm{peak,i}^\mathrm{nTh}$ and L$_\mathrm{i}^\mathrm{nTh}$--E$_\mathrm{peak,i}^\mathrm{nTh,rest}$ relations, and distance estimates}
\label{sec:F-Ep}

\begin{center}
The F$_\mathrm{i}^\mathrm{nTh}$--E$_\mathrm{peak,i}^\mathrm{nTh}$ Relation
\end{center}

When fitting the $\gamma$-ray prompt emission of both short and long GRBs to a Band function, a correlation is often observed between the evolution of the $\gamma$-ray energy flux, F$_\mathrm{i}^\mathrm{Band}$, and the evolution of the corresponding E$_\mathrm{peak,i}^\mathrm{Band}$ within each burst~\citep[i.e., F$_\mathrm{i}^\mathrm{Band}$--E$_\mathrm{peak,i}^\mathrm{Band}$ relation -- see for instance][]{Golenetskii:1983,Borgonovo:2001,Liang:2004,Guiriec:2010,Guiriec:2015a,Ghirlanda:2011a,Ghirlanda:2011b,Lu:2012}. Although this correlation globally follows a PL, there is often large scatter in the data of each burst and the PL indices can be dramatically different from burst to burst. More importantly, there are many bursts that do not exhibit this correlation.

In~\citet{Guiriec:2013a,Guiriec:2015a,Guiriec:2015b} we introduced a new relation between the $\gamma$-ray energy flux and the spectral hardness. Indeed, by isolating the non-thermal component of the spectra from the thermal-like one---in fitting C$_\mathrm{nTh}$+C$_\mathrm{Th}$ or C$_\mathrm{nTh}$+C$_\mathrm{Th}$+PL to the data---we showed that a very strong correlation, intrinsic to the non-thermal component only, appears between the evolution of its energy flux, F$_\mathrm{i}^\mathrm{nTh}$, and the evolution of its $\nu$F$_\nu$ $\gamma$-ray spectral peak, E$_\mathrm{peak,i}^\mathrm{nTh}$ (i.e., F$_\mathrm{i}^\mathrm{nTh}$-- E$_\mathrm{peak,i}^\mathrm{nTh}$ relation). Conversely to the F$_\mathrm{i}^\mathrm{Band}$--E$_\mathrm{peak,i}^\mathrm{Band}$ relations, the F$_\mathrm{i}^\mathrm{nTh}$-- E$_\mathrm{peak,i}^\mathrm{nTh}$ relations have extremely similar indices for all GRBs when fitted to PLs.

Moreover, in the central engine frame, the F$_\mathrm{i}^\mathrm{nTh}$-- E$_\mathrm{peak,i}^\mathrm{nTh}$ relations translate into a unique L$_\mathrm{i}^\mathrm{nTh}$--E$_\mathrm{peak,i}^\mathrm{nTh,rest}$ relation---with the same index as the F$_\mathrm{i}^\mathrm{nTh}$--E$_\mathrm{peak,i}^\mathrm{nTh}$ relation---that seems to be universal for all short and long GRBs~\citep{Guiriec:2013a,Guiriec:2015a,Guiriec:2015b}.

The panels (f) of Figures~\ref{fig07} to~\ref{fig09} show the F$_\mathrm{i}^\mathrm{Band}$--E$_\mathrm{peak,i}^\mathrm{Band}$ and F$_\mathrm{i}^\mathrm{nTh}$-- E$_\mathrm{peak,i}^\mathrm{nTh}$ relations for the Band, C$_\mathrm{nTh}$+C$_\mathrm{Th}$ and C$_\mathrm{nTh}$+C$_\mathrm{Th}$+PL models. As already reported in~\citet{Guiriec:2013a,Guiriec:2015a,Guiriec:2015b} for other GRBs,  the energy flux and E$_\mathrm{peak}$ values are typically anti-correlated at very early times in the burst, which also usually corresponds to the highest values of E$_\mathrm{peak}$. The indices of the PLs fitted to these relations---after excluding the anti-correlated early phase---are reported in Table~\ref{tab02}. Apart from GRB~970111, the indices of the F$_\mathrm{i}^\mathrm{Band}$--E$_\mathrm{peak,i}^\mathrm{Band}$ relations are inconsistent with the F$_\mathrm{i}^\mathrm{nTh}$-- E$_\mathrm{peak,i}^\mathrm{nTh}$ ones; the indices of the F$_\mathrm{i}^\mathrm{Band}$--E$_\mathrm{peak,i}^\mathrm{Band}$ relations are also dramatically different from one burst to the other (i.e., from +1.4 to +3.2). Conversely, the indices of the F$_\mathrm{i}^\mathrm{nTh}$-- E$_\mathrm{peak,i}^\mathrm{nTh}$ relations are all clustered with the C$_\mathrm{nTh}$+C$_\mathrm{Th}$+PL model (between +1.3 and +1.5 -- see Figure~\ref{fig14}a); these values are perfectly consistent with the indices of +1.33$\pm$0.06, +1.38$\pm$0.04 and +1.43$\pm$0.03 reported in~\citet{Guiriec:2013a,Guiriec:2015a,Guiriec:2015b}, respectively.

\begin{table}[h!]
\caption{\label{tab02}Power-law indices of the F$_\mathrm{i}^\mathrm{Band}$--E$_\mathrm{peak,i}^\mathrm{Band}$ or F$_\mathrm{i}^\mathrm{nTh}$--E$_\mathrm{peak,i}^\mathrm{nTh}$ relations obtained when fitting Band-only or C$_\mathrm{nTh}$+C$_\mathrm{Th}$ or C$_\mathrm{nTh}$+C$_\mathrm{Th}$+PL to the time-resolved spectra of GRBs~941017,~970111 and~990123 (see Section~\ref{sec:F-Ep}).}
\begin{center}
{\tiny
\begin{tabular}{|c|c|c|c|}
\hline
\multicolumn{1}{|c|}{} &\multicolumn{3}{c|}{F$_\mathrm{i}^\mathrm{Band}$--E$_\mathrm{peak,i}^\mathrm{Band}$ or F$_\mathrm{i}^\mathrm{nTh}$--E$_\mathrm{peak,i}^\mathrm{nTh}$ relation PL-indices for} \\
\multicolumn{1}{|c|}{} &\multicolumn{1}{c|}{Band} &\multicolumn{1}{c|}{C$_\mathrm{nTh}$+C$_\mathrm{Th}$} &\multicolumn{1}{c|}{C$_\mathrm{nTh}$+C$_\mathrm{Th}$+PL} \\
\hline
GRB~941017 & +3.07$\pm$0.12 & +1.35$\pm$0.06 & +1.40$\pm$0.10 \\
GRB~970111 & +1.43$\pm$0.04 & +1.61$\pm$0.32 & +1.33$\pm$0.23 \\
GRB~990123 & +2.68$\pm$0.09 & +1.72$\pm$0.09 & +1.28$\pm$0.05 \\
\hline
\end{tabular}
}
\end{center}
\end{table}

\newpage
\begin{center}
Redshift Estimates Using the L$_\mathrm{i}^\mathrm{nTh}$--E$_\mathrm{peak,i}^\mathrm{nTh,rest}$ Relation
\end{center}

In our three BATSE-burst sample, GRB~990123 has an accurate redshift measurement of z$\sim$1.61. There are no redshift measurements available for GRB~941017 and GRB~970111, although host galaxies possibly associated to GRB~970111 resulted in a initial redshift range of 0.2$\leq$z$\leq$1.4~\citep{Gorosabel:1998}. Assuming that the L$_\mathrm{i}^\mathrm{nTh}$-- E$_\mathrm{peak,i}^\mathrm{nTh,rest}$ relation is universal---as suggested in~\citet{Guiriec:2013a,Guiriec:2015a}---, we estimated the distance of GRBs~941017 and 970111 using GRB~990123 as the reference; the L$_\mathrm{i}^\mathrm{nTh}$-- E$_\mathrm{peak,i}^\mathrm{nTh,rest}$ relation for GRB~990123 is displayed in Figure~\ref{fig14}b (dashed blue line). We varied the redshift to determine the values that minimize the distances between the dashed blue line and the data of GRBs~941017 and 970111. To do so, we performed a linear fit in the log--log space to account for the uncertainties on both the L$_\mathrm{i}^\mathrm{nTh}$ and E$_\mathrm{peak,i}^\mathrm{nTh,rest}$ quantities; the resulting $\chi^\mathrm{2}$ profiles are displayed in panels (c) and (d) of Figure~\ref{fig14} for GRBs~941017 and 970111, respectively. The redshift estimates using this technique are reported in Table~\ref{tab03}. Interestingly, the distance estimate for GRB~970111 (i.e., z=1.18$\pm$0.06) is compatible with the redshifts of the galaxies (i.e., 0.2$\leq$z$\leq$1.4) identified as possible hosts for this burst in~\citet{Gorosabel:1998}; however, this result should not be interpreted as a proof of the technique nor as a proof that the host galaxy is one of the objects identified between z=0.2 and z=1.4 since it may be a coincidence. The redshift estimate for GRB~941017 (i.e., z=1.79$\pm$0.07) is a typical value for a long GRB \citep[i.e., $<$z$>$$\sim$2.16 --][]{Jakobsson:2012}.

Figure~\ref{fig14}b shows the L$_\mathrm{i}^\mathrm{nTh}$-- E$_\mathrm{peak,i}^\mathrm{nTh,rest}$ relations for the three bursts using the redshift measurement for GRB~990123 and the redshift estimates for the two other bursts. The PL fits to each burst are displayed in dashed color lines. The solid black line is the fit to the whole data together, and it results in a PL index of +1.31$\pm$0.03, which is in perfect agreement with the results published in~\citet{Guiriec:2013a,Guiriec:2015a,Guiriec:2015b} using {\it Fermi} data.
\begin{table}[h]
\caption{\label{tab03}Measured and estimated redshifts (z) for GRBs~941017,~97011 and~990123 using GRB~990123 with z=1.61 as a reference (see Section~\ref{sec:F-Ep}).}
\begin{center}
\begin{tabular}{|c|c|c|}
\hline
 & Measured z & z estimates\\ 
\hline
GRB~941017 & -- & 1.79$\pm$0.07 \\
GRB~970111 & 0.2$\leq$z$\leq$1.4\footnote{This photometric redshift estimate, based on possible host-galaxy identifications reported in~\citet{Gorosabel:1998}, is not secure because fainter objects are present in this region of the sky and one of them may be the actual host galaxy of GRB~970111.} & 1.18$\pm$0.06 \\
GRB~990123 & $\sim$1.61 & -- \\
\hline
\end{tabular}
\end{center}
\end{table}

\vspace{-0.2cm}
\section{Conclusion}

In this article we tested the new paradigm for GRB prompt emission proposed in~\citet{Guiriec:2011a,Guiriec:2013a,Guiriec:2015a,Guiriec:2015b} to three bright and famous BATSE GRBs; according to this new model, GRB prompt emission is composed of three main components: (i) a thermal-like component, C$_\mathrm{Th}$, that we interpret as emission from the jet photosphere, (ii) a non-thermal component, C$_\mathrm{nTh}$, interpreted as synchrotron emission from charged particles propagating and accelerated within the GRB jet, and (iii) a second non-thermal component adequately fitted to a PL with or without cutoff, which extends from below few tens of keV up to tens or hundreds of MeV, and most likely of inverse Compton origin. The three components are not systematically present or detectable in all GRBs, especially the second non-thermal component, which is clearly identified only in a limited number of GRBs.

This new model is perfectly consistent with the prompt emission data of GRBs 941017, 970111 and 990123 recorded with BATSE. We identified the signature of the three components in the time-integrated spectra of the three bursts and followed their evolution through fine time-resolved analysis. The results are similar to those reported from {\it Fermi} GRBs:

\begin{itemize}
\item C$_\mathrm{nTh}$ is the most intense component from 20 keV to 2 MeV and accounts for $>$50--60\% of the total collected energy in this energy range. The $\nu$F$_\nu$ peak energy of C$_\mathrm{nTh}$, E$_\mathrm{peak,i}^\mathrm{nTh}$, exhibits strong variations perfectly correlated with the energy flux variations of this component, F$_\mathrm{i}^\mathrm{nTh}$ (i.e., F$_\mathrm{i}^\mathrm{nTh}$--E$_\mathrm{peak,i}^\mathrm{nTh}$ relation).
\item The C$_\mathrm{Th}$ contribution to the total energy between 20 keV and 2 MeV is usually $\leq$20\%. C$_\mathrm{Th}$ is present from the beginning of the burst and its intensity globally decreases with time. The temperature of C$_\mathrm{Th}$ is either globally constant during the whole burst duration (i.e., GRBs~941017 and~990123) or it globally decreases with time with a strong cooling phase during the first seconds (i.e., GRB~970111). In GRB~970111, C$_\mathrm{Th}$ contributes for $\geq$70\% of the total energy during an initial nearly purely thermal episode lasting for a few seconds similarly to {\it Fermi} GRB~131014A~\citep{Guiriec:2015b}.
\item The flux of the additional PL is very significant for GRBs~941017 and~990123, but only marginal for GRB~970111. This component is present from the very beginning to the very end of the burst, and contributes to $\leq$30\% of the total energy between 20 keV and 2 MeV. Extrapolation of the C$_\mathrm{nTh}$+C$_\mathrm{Th}$+PL model over a broader energy range covered by {\it Fermi} (i.e., 5 keV to 100 MeV) shows that the PL would overpower the other components below a few tens of keV and above a few MeV; this is striking for GRB~941017. In addition, for GRB~990123, intense and short duration light-curve structures associated to the additional PL would be present in the low and high energy bands ($\leq$10 keV and $\geq$20 MeV, respectively) and perfectly correlated in time such as reported for {\it Fermi} GRBs~\citep{Guiriec:2015a}.
\end{itemize}

The relative contribution of each component to the total energy strongly evolves with time showing that there is no perfect correlation between the three components. There is also no clear correlation between E$_\mathrm{peak,i}^\mathrm{Th}$ and E$_\mathrm{peak,i}^\mathrm{nTh}$.

The PL indices of the F$_\mathrm{i}^\mathrm{nTh}$--E$_\mathrm{peak,i}^\mathrm{nTh}$ relations are very similar for the three GRBs (i.e., $\sim$1.3--1.4) and they are also similar to the values reported from {\it Fermi} GRBs in~\citet{Guiriec:2013a,Guiriec:2015a,Guiriec:2015b}.

Assuming that the L$_\mathrm{i}^\mathrm{nTh}$--E$_\mathrm{peak,i}^\mathrm{nTh,rest}$ relation is universal as suggested in~\citet{Guiriec:2013a,Guiriec:2015a}, and using GRB~990123---which has a measured redshift z$\sim$1.61---as reference, we estimated the redshift of GRBs~970111 and 941017 to be z=1.18$\pm$0.06 and z=1.79$\pm$0.07, respectively. The redshift estimate for GRB~970111 is compatible with the redshifts of the galaxies identified as possible hosts in~\citep[0.2$\leq$z$\leq$1.4 --][]{Gorosabel:1998}; however, the actual host may be a fainter galaxy not yet identified. Despite the lack of constraints for the distance of GRB~941017, the redshift estimate for this burst has a value perfectly typical for long GRBs.

\section{Acknowledgments}

S.G. was supported by the NASA Postdoctoral Program (NPP) at the NASA/Goddard Space Flight Center, administered by Oak Ridge Associated Universities through a contract with NASA as well as by the NASA grants NNH11ZDA001N and NNH13ZDA001N awarded to S.G. during the cycles 5 and 7 of the NASA Fermi Guest Investigator Program. M.M.G was supported by DGAPA UNAM grant number IG100414-3.

\newpage

\begin{appendix}

\section{Spectral Analysis Results}
\label{section:Spectral-analysis-results}

The values of the spectral parameters resulting from the fine-time analysis of GRBs~941017,~970111 and~990123 as presented in Sections~\ref{sec:trsa} and~\ref{sec:CnTh+CTh+PL} are reported in Tables~\ref{tab04},~\ref{tab05} \&~\ref{tab06}, respectively.

\section{Fine Time-Resolved Analysis}
\label{section:Time-resolved-analysis}

Figure~\ref{fig15} and~\ref{fig16} show the results of the fine time-resolved analysis of GRB~941017 with a Band function alone and the C$_\mathrm{nTh}$+C$_\mathrm{Th}$ or the C$_\mathrm{nTh}$+C$_\mathrm{Th}$+PL models, respectively, as presented in Section~\ref{sec:trsa}.

Figure~\ref{fig17} and~\ref{fig18} show the results of the fine time-resolved analysis of GRB~970111 with a Band function alone and the C$_\mathrm{nTh}$+C$_\mathrm{Th}$ or the C$_\mathrm{nTh}$+C$_\mathrm{Th}$+PL models as presented in Section~\ref{sec:trsa}.

Figure~\ref{fig19} and~\ref{fig20} show the results of the fine time-resolved analysis of GRB~990123 with a Band function alone and the C$_\mathrm{nTh}$+C$_\mathrm{Th}$ or the C$_\mathrm{nTh}$+C$_\mathrm{Th}$+PL models as presented in Section~\ref{sec:trsa}.

\end{appendix}

\newpage

\setcounter{figure}{0}
\renewcommand{\thefigure}{A\arabic{figure}}

\newpage

\begin{figure*}
\begin{center}
\includegraphics[totalheight=0.185\textheight, clip]{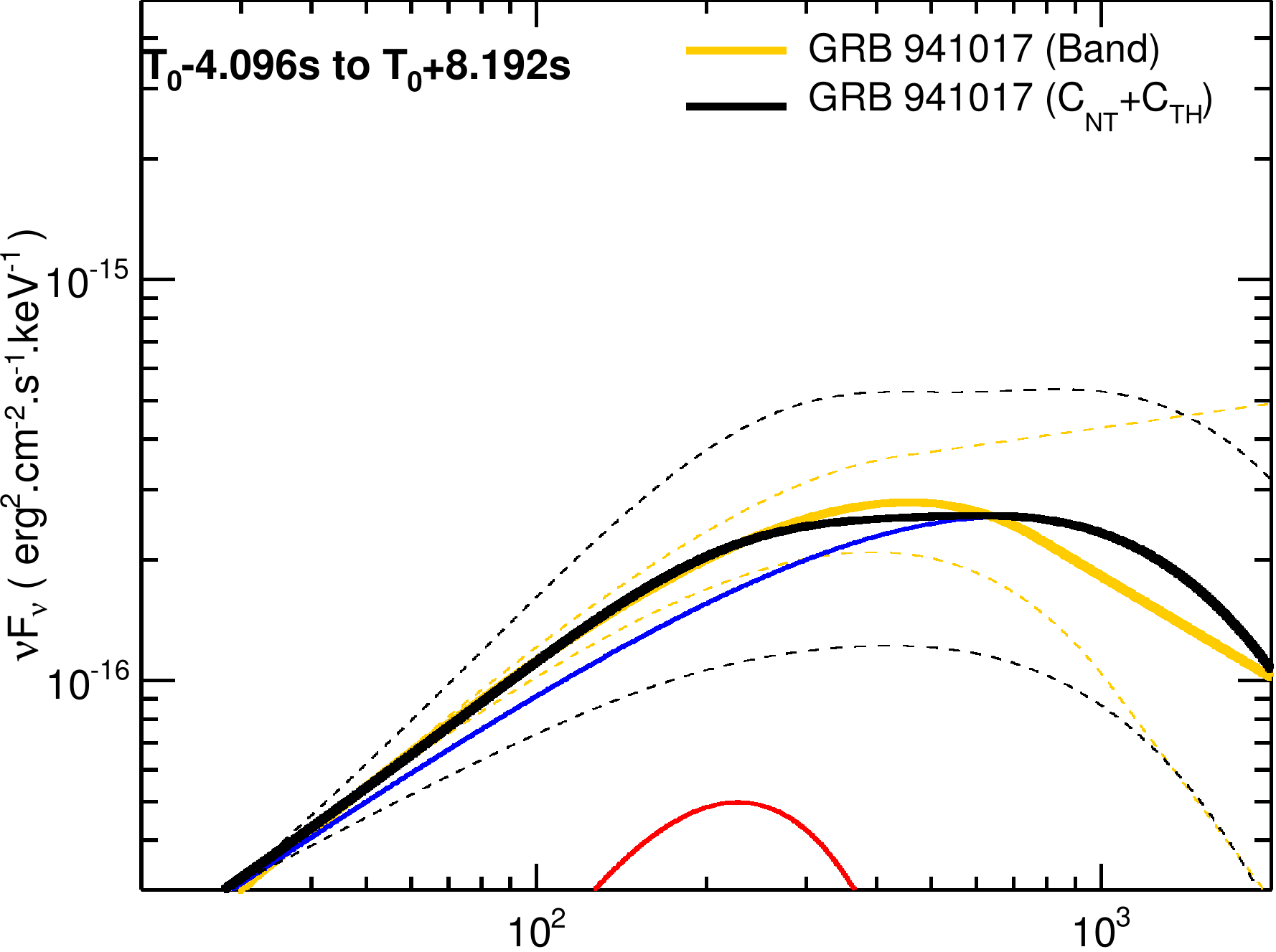}
\includegraphics[totalheight=0.185\textheight, clip]{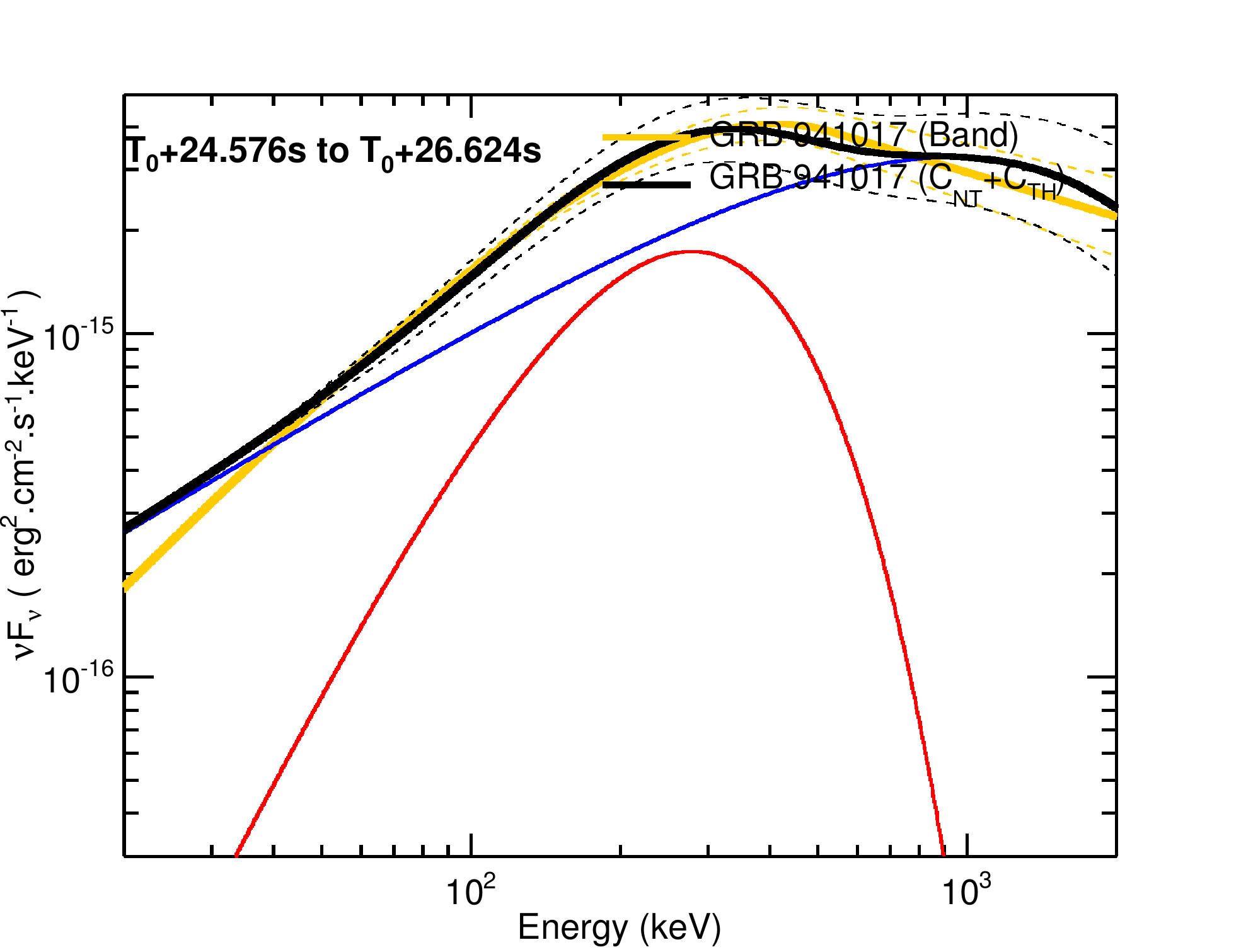}
\includegraphics[totalheight=0.185\textheight, clip]{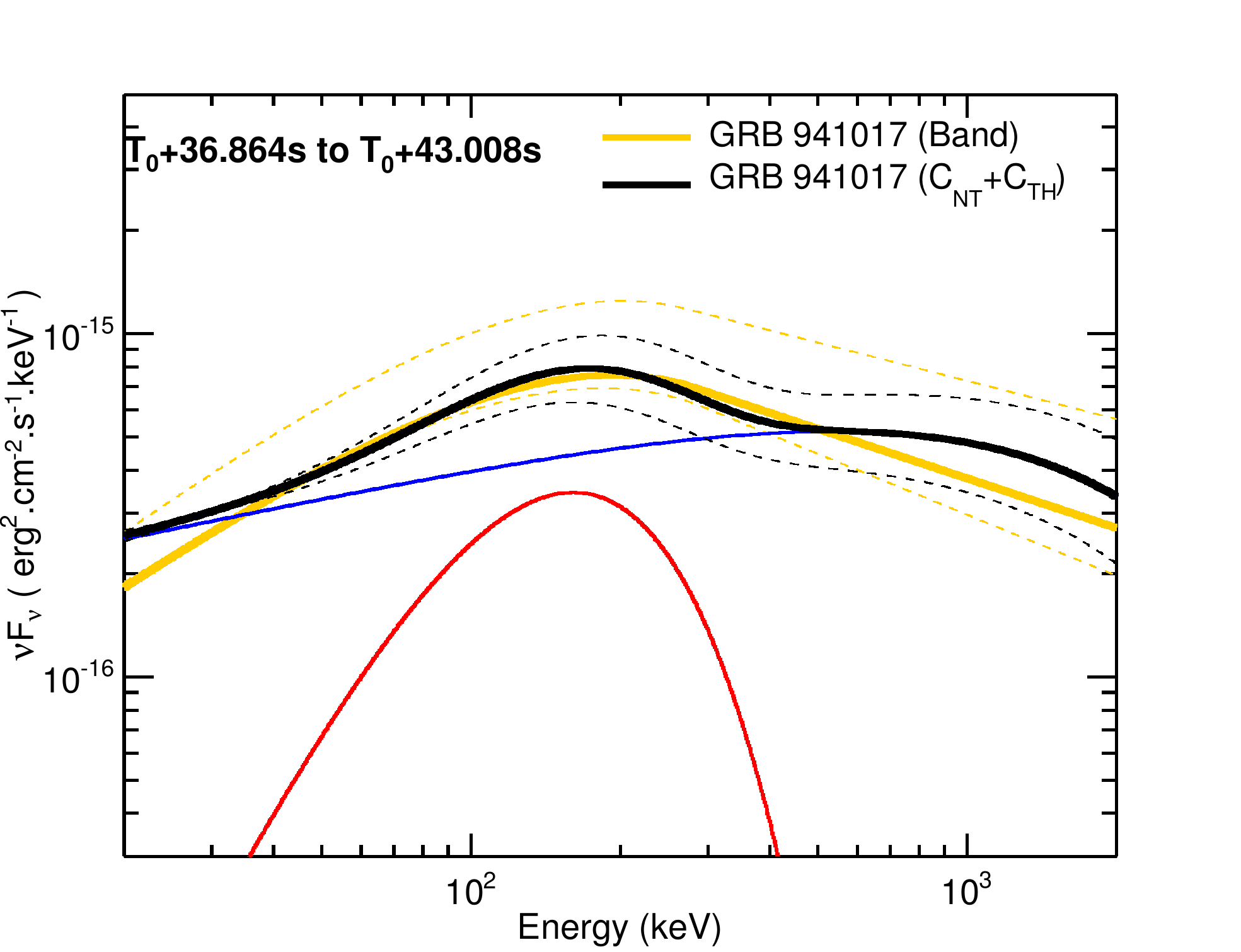}

\includegraphics[totalheight=0.185\textheight, clip]{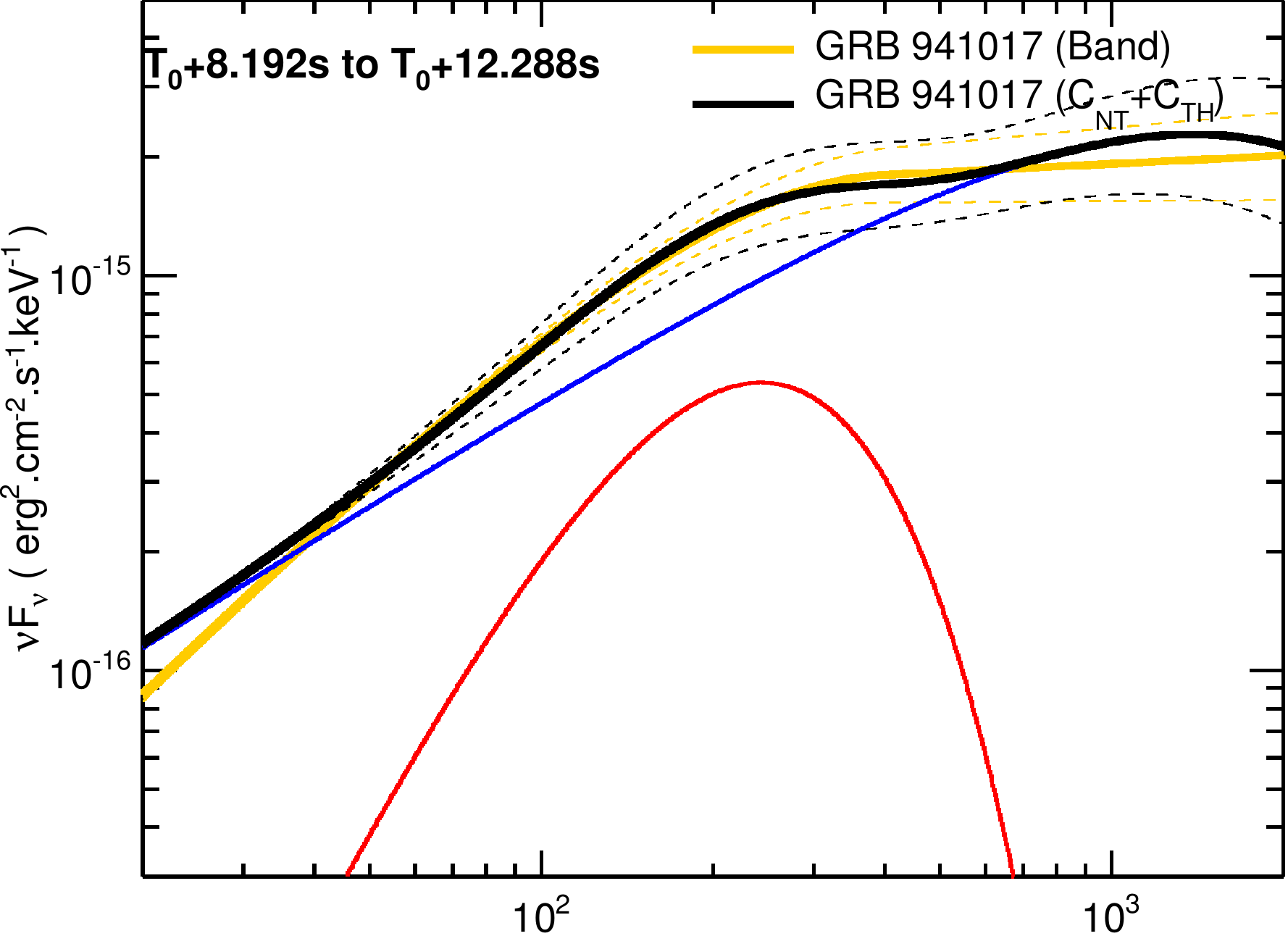}
\includegraphics[totalheight=0.185\textheight, clip]{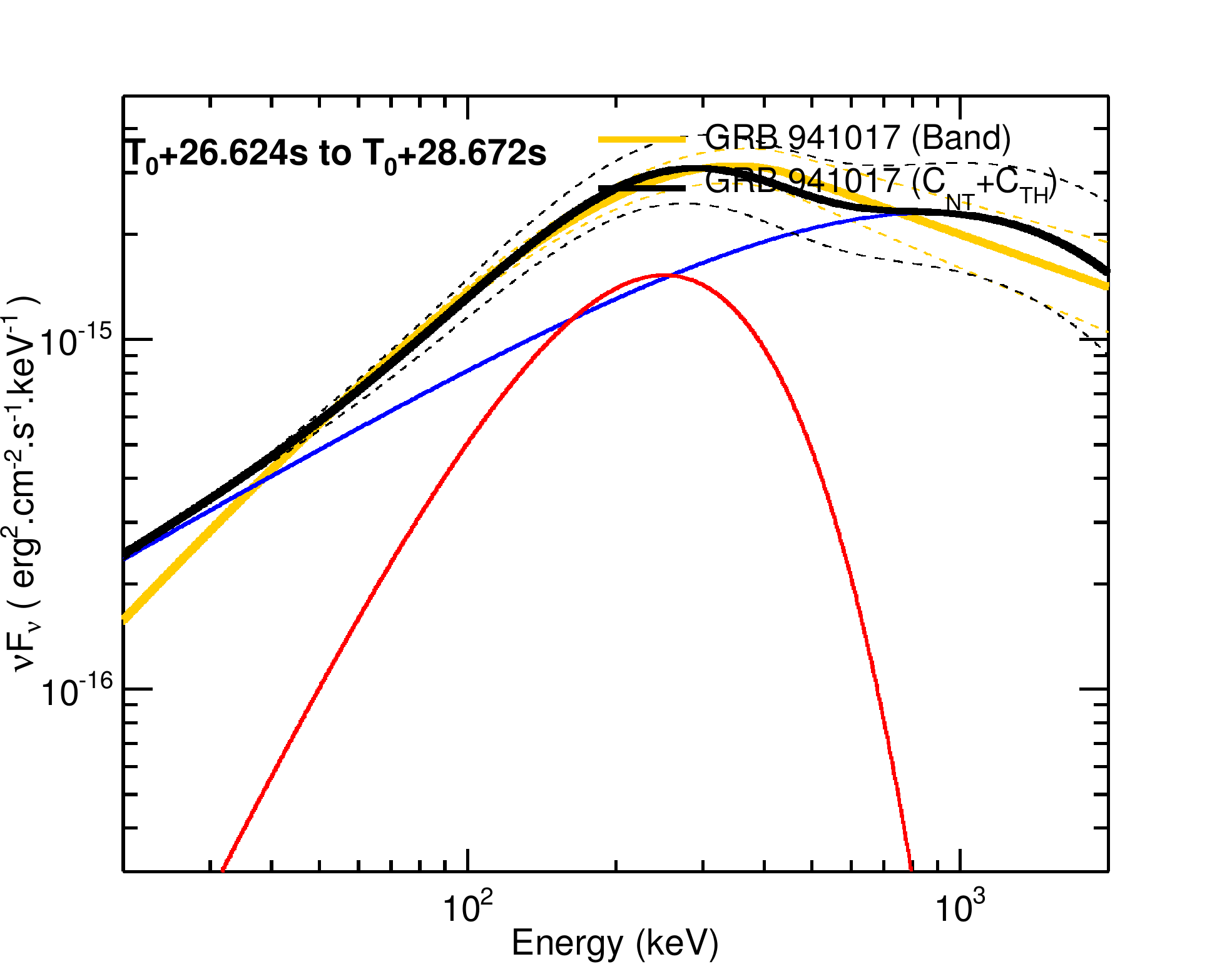}
\includegraphics[totalheight=0.185\textheight, clip]{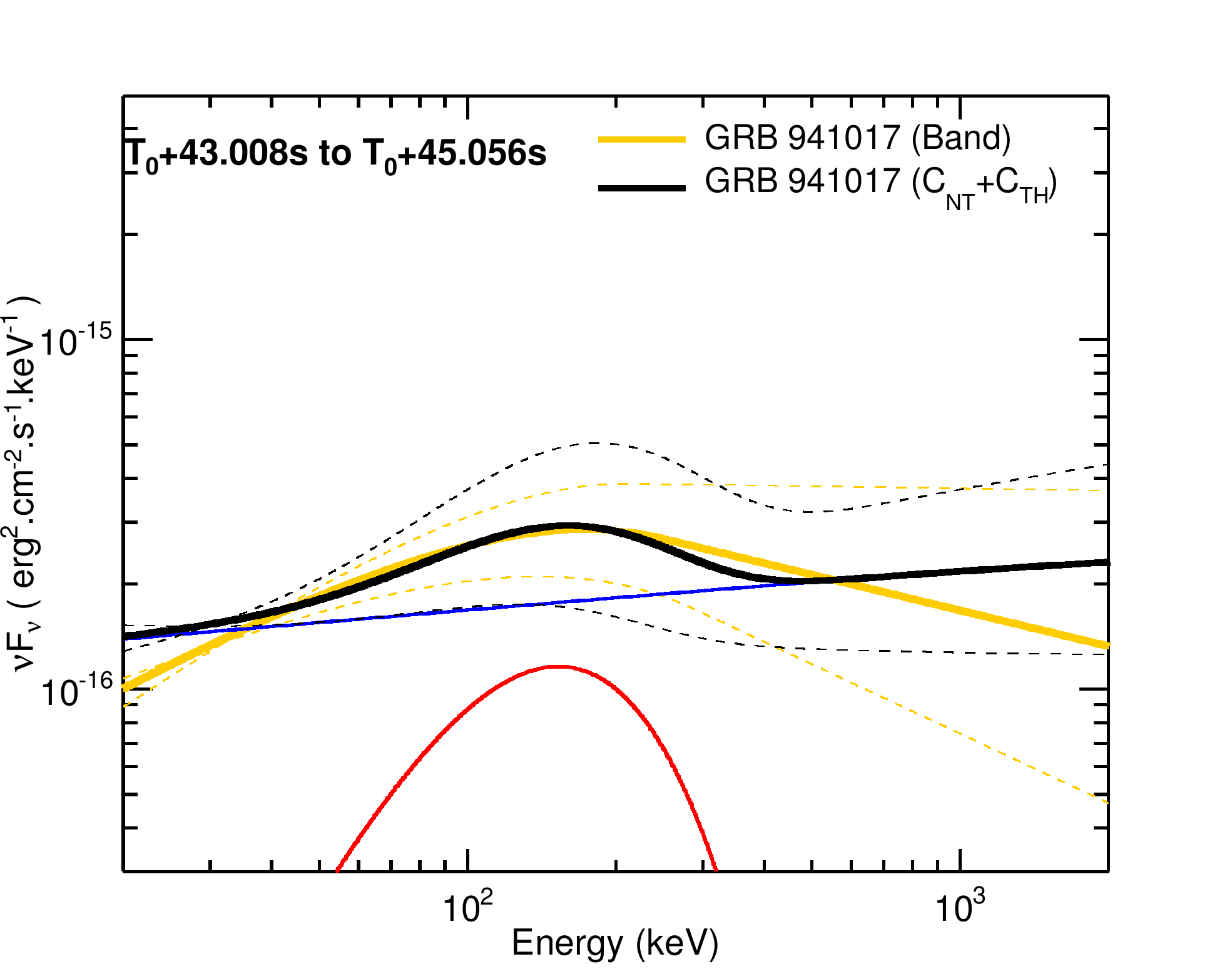}

\includegraphics[totalheight=0.185\textheight, clip]{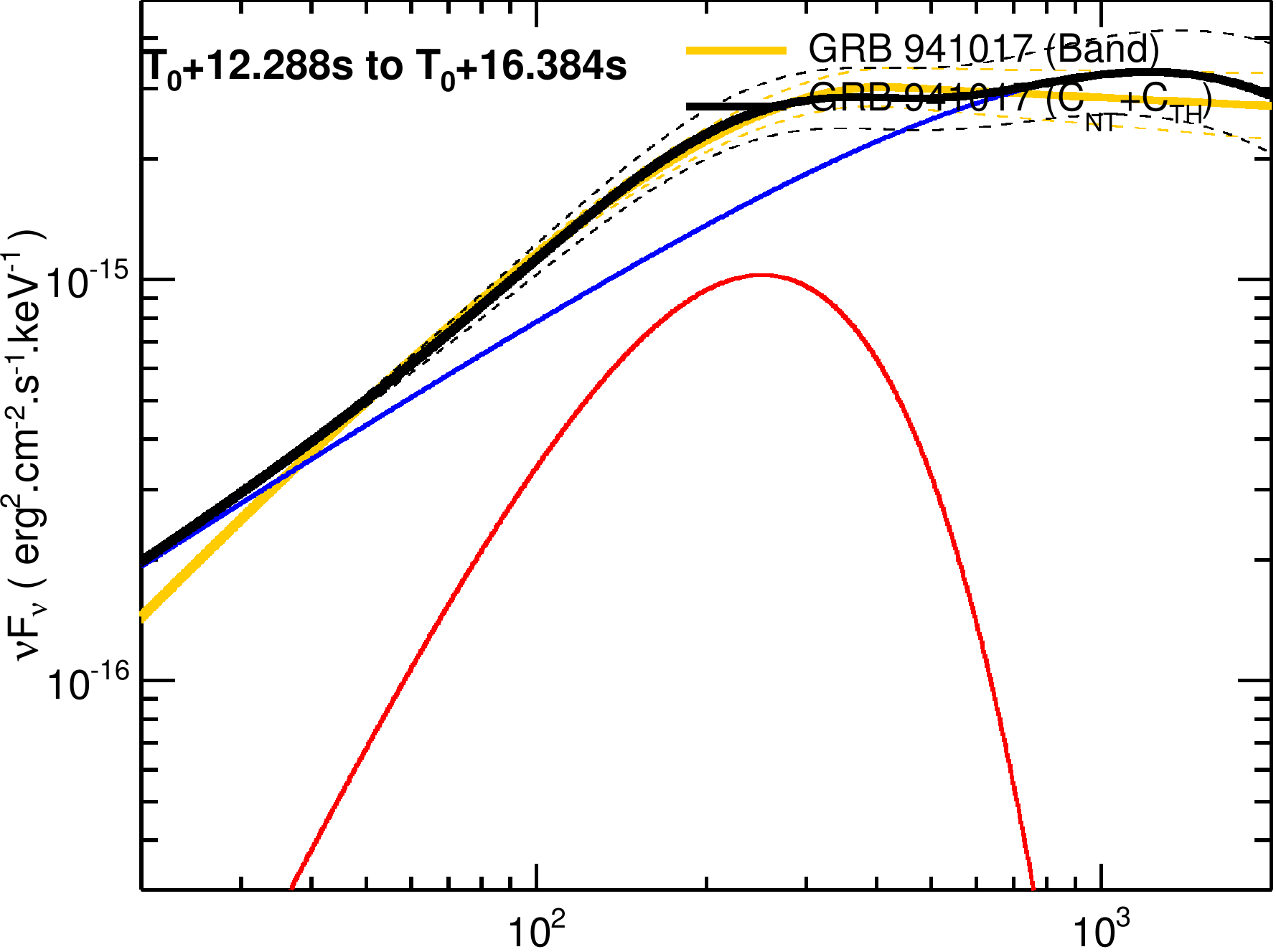}
\includegraphics[totalheight=0.185\textheight, clip]{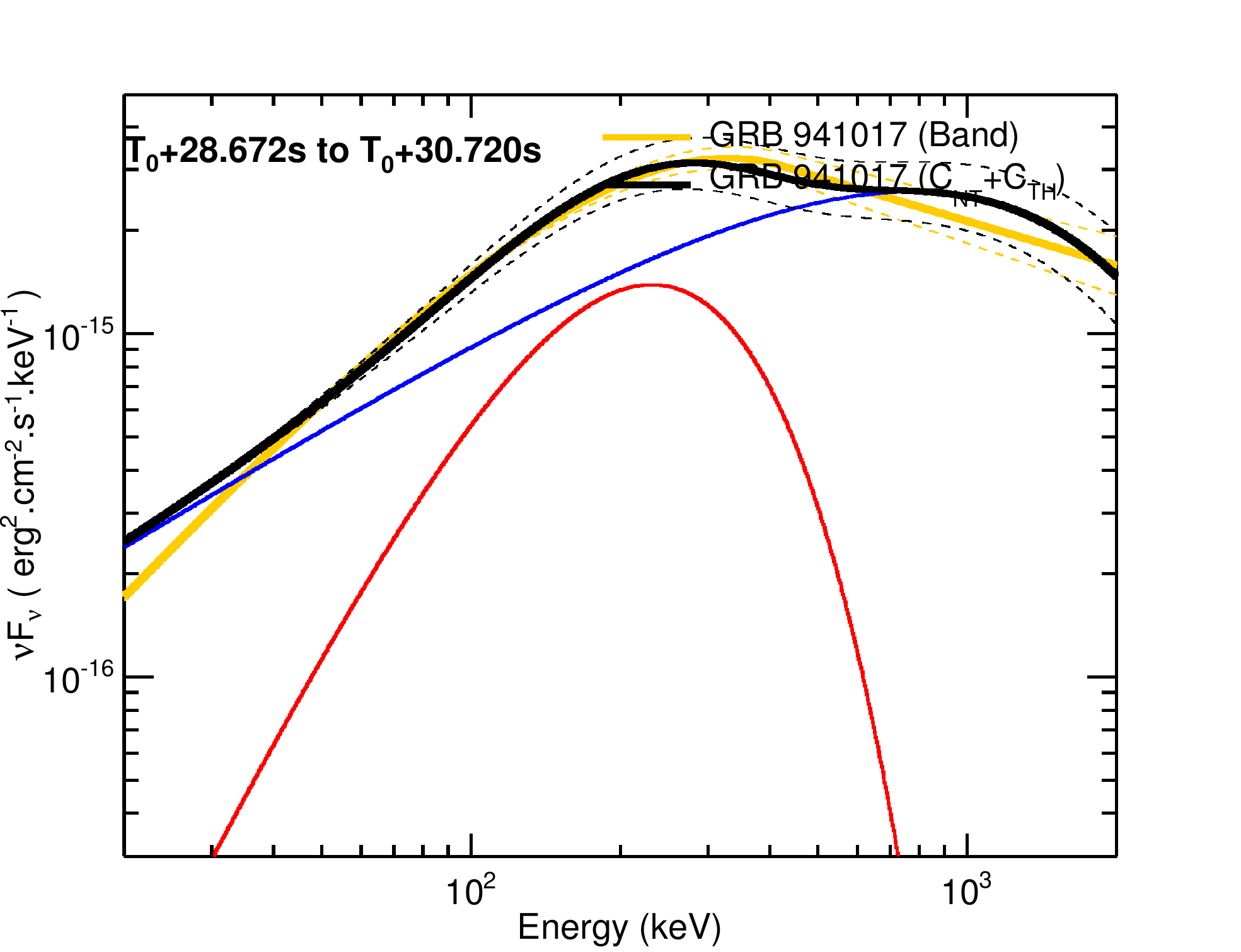}
\includegraphics[totalheight=0.185\textheight, clip]{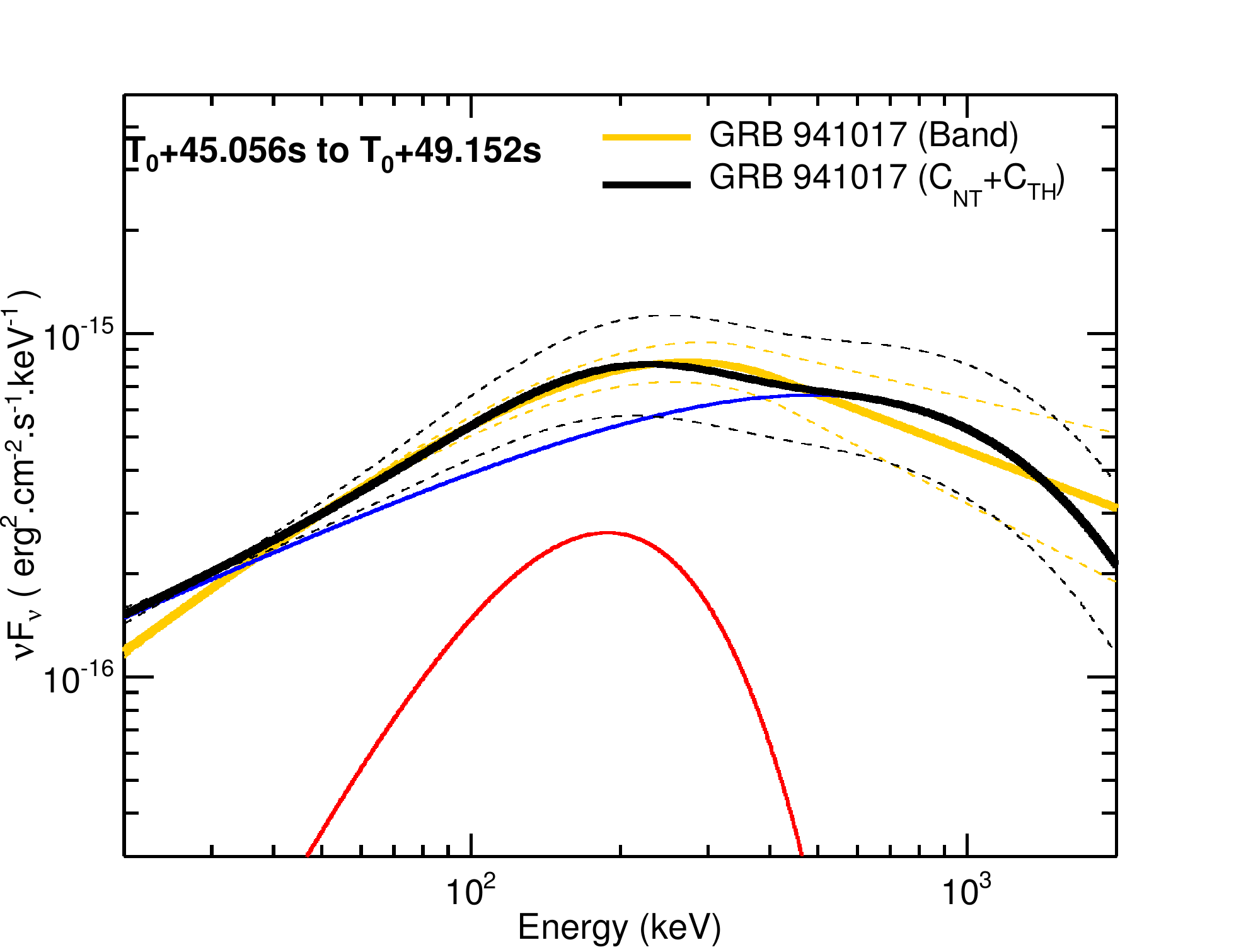}

\includegraphics[totalheight=0.185\textheight, clip]{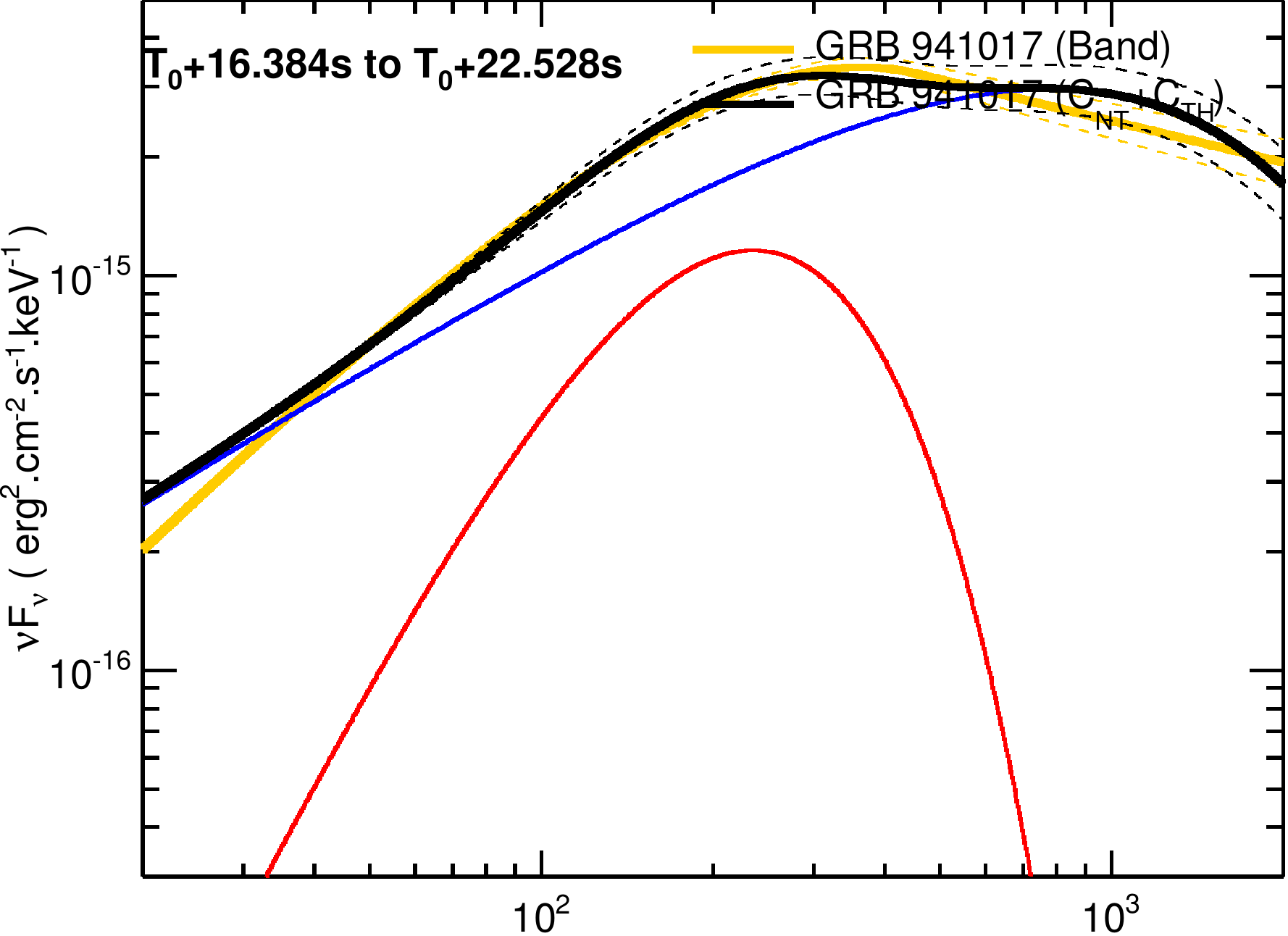}
\includegraphics[totalheight=0.185\textheight, clip]{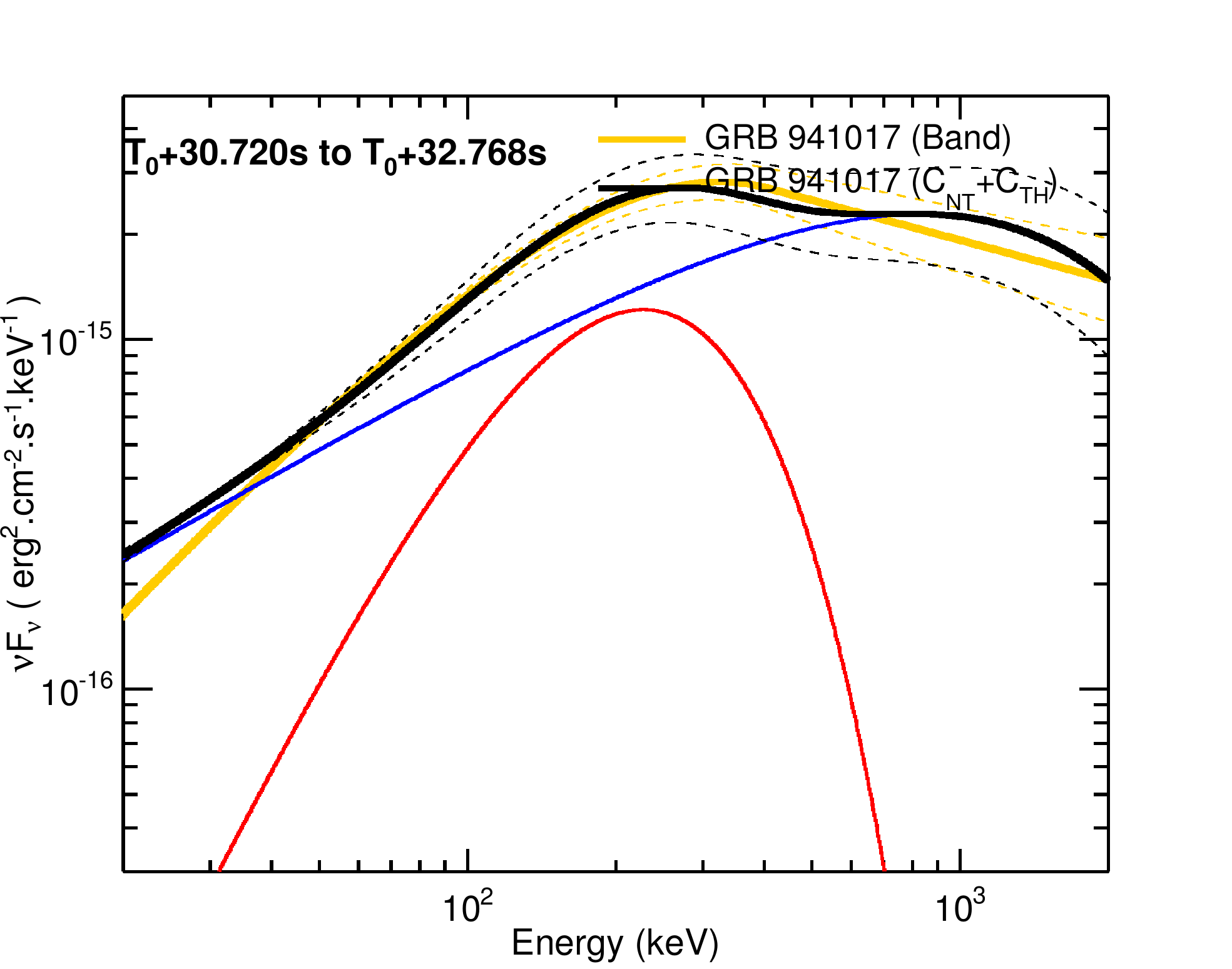}
\includegraphics[totalheight=0.185\textheight, clip]{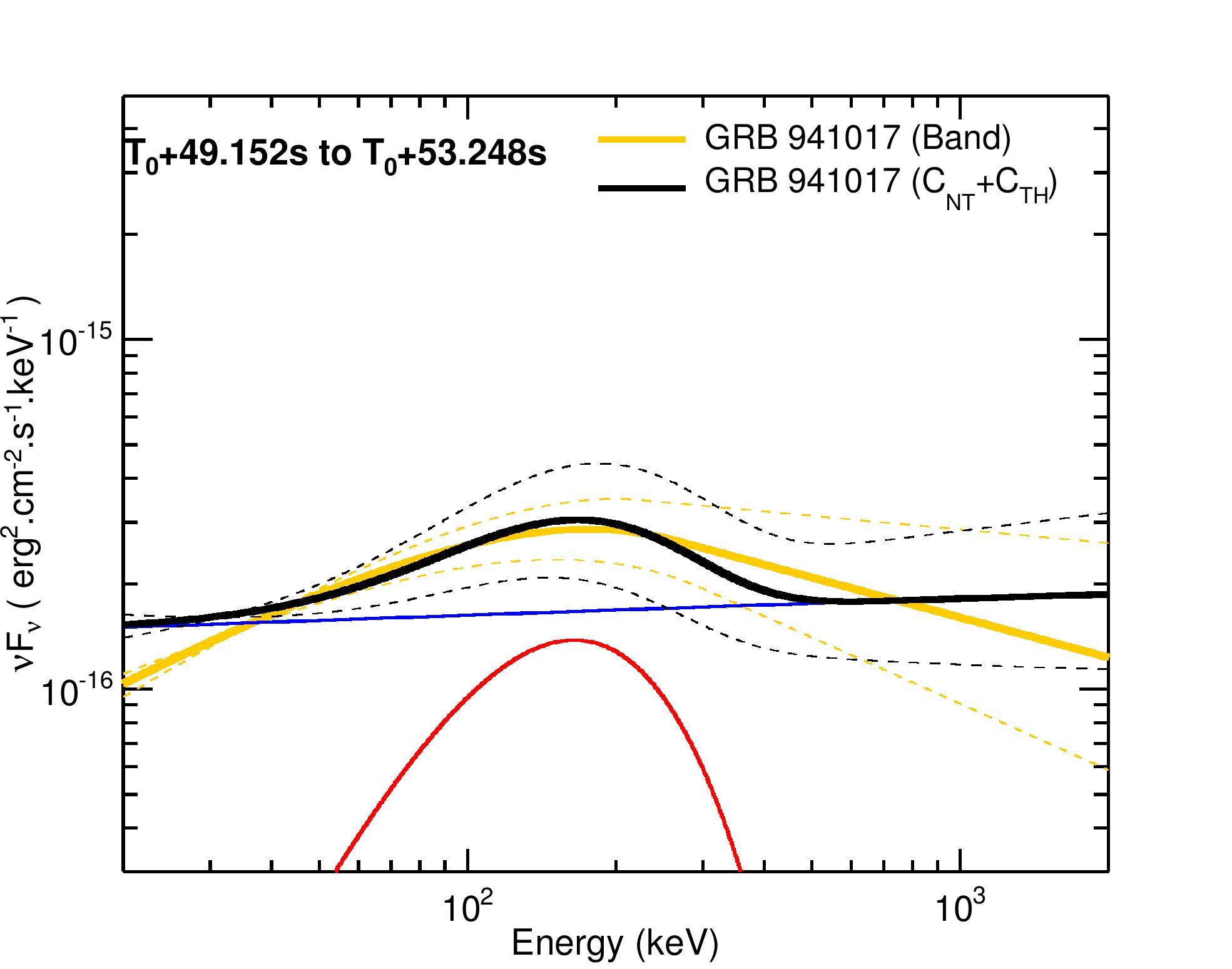}

\includegraphics[totalheight=0.195\textheight, clip]{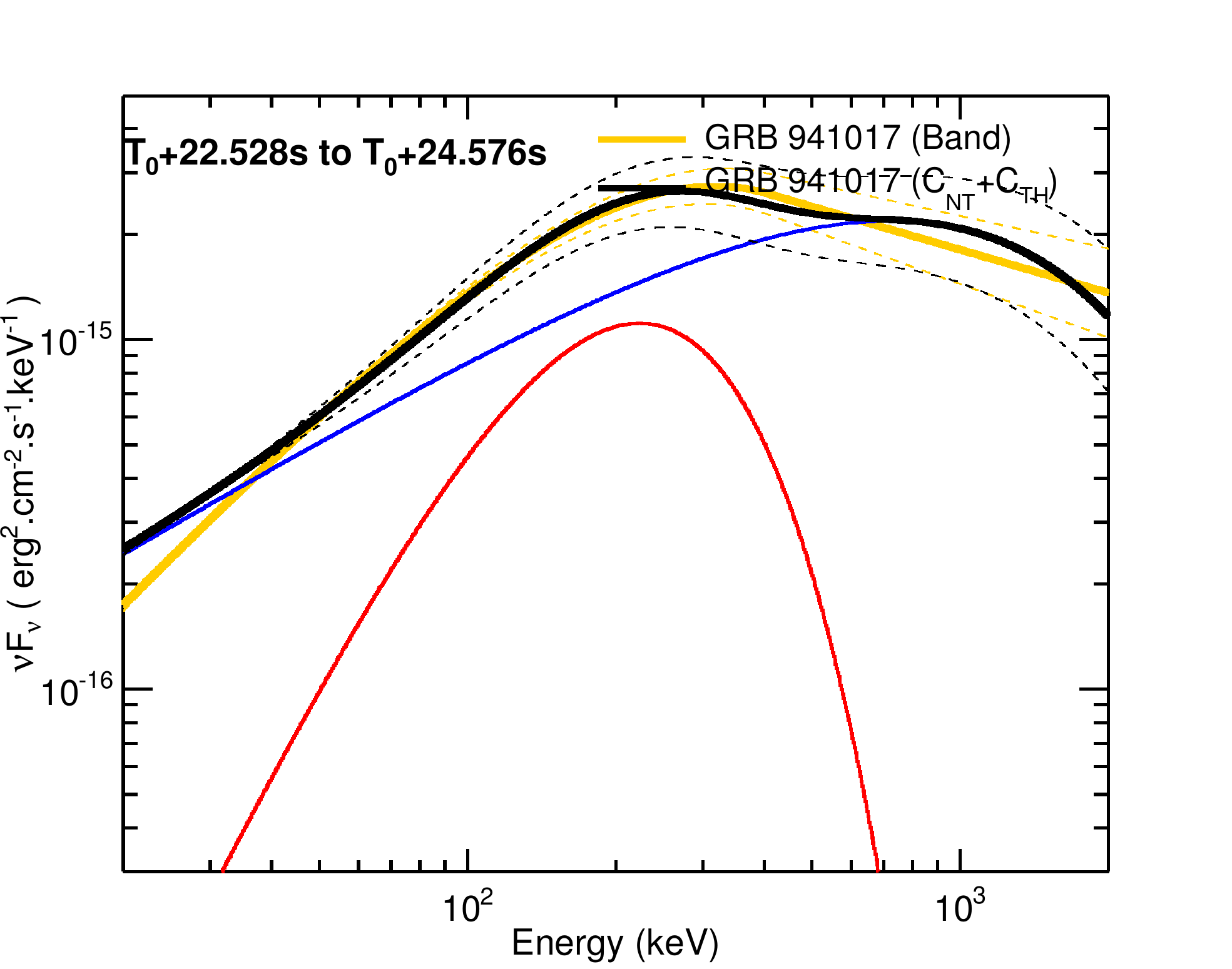}
\includegraphics[totalheight=0.195\textheight, clip]{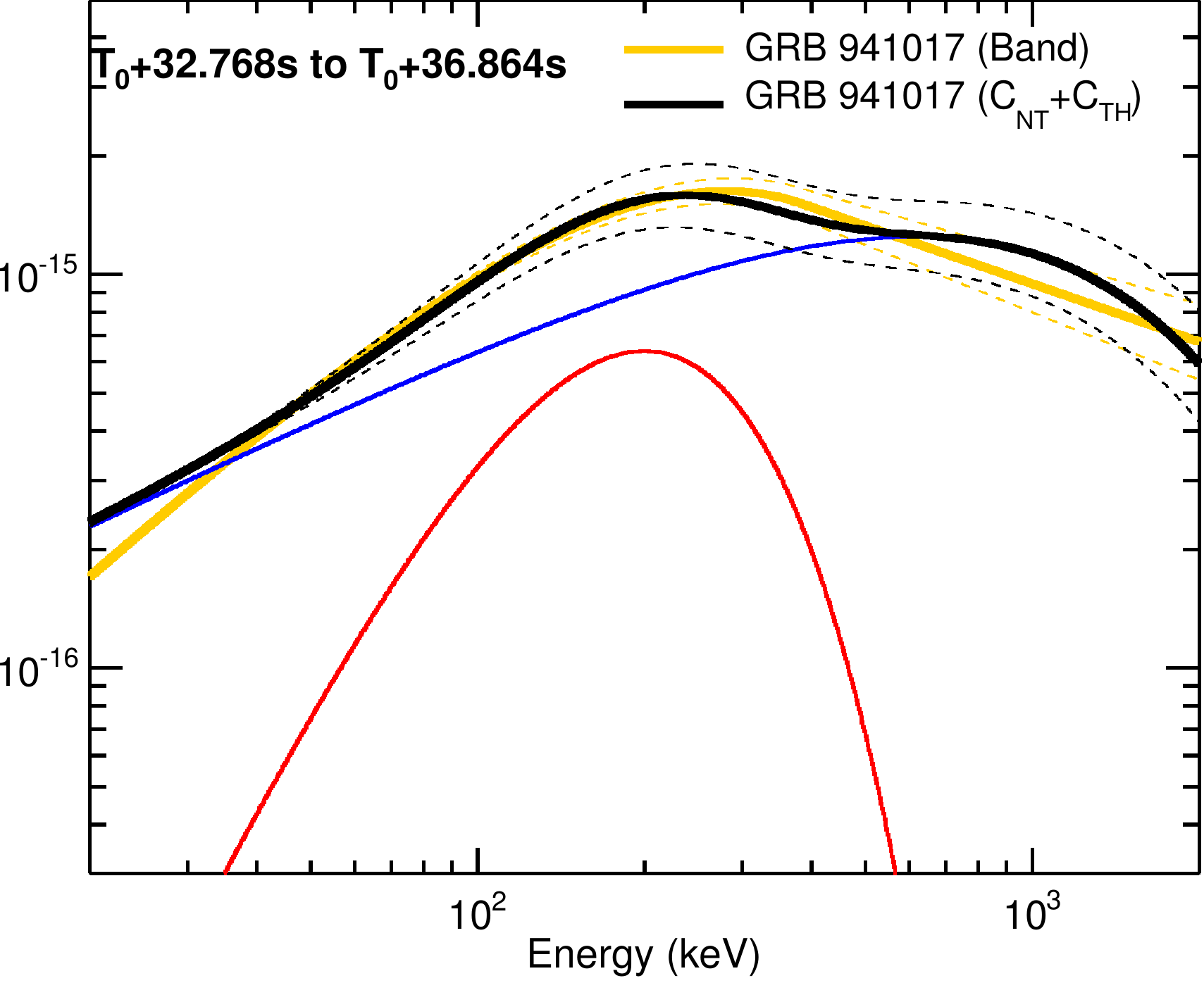}
\includegraphics[totalheight=0.195\textheight, clip]{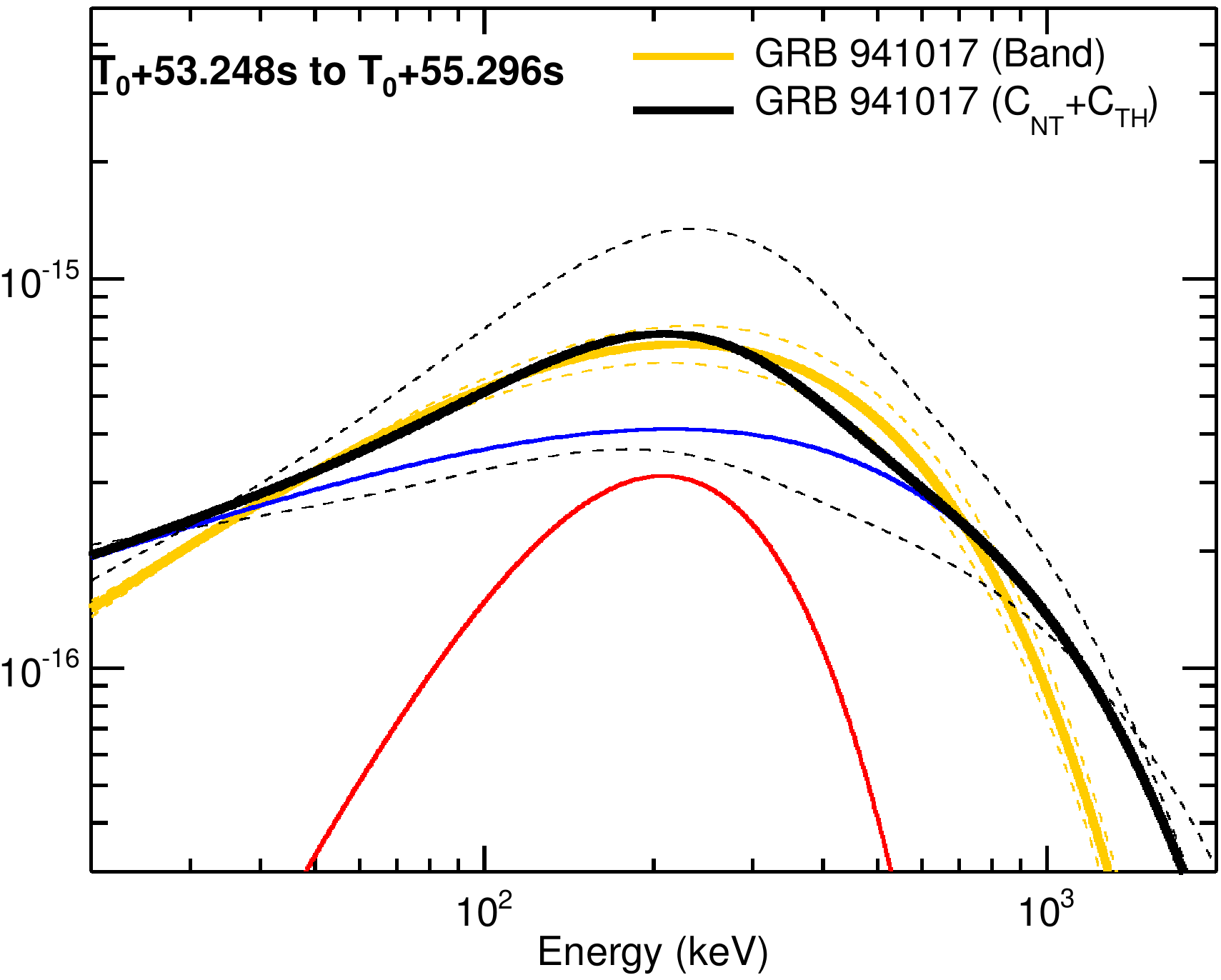}

\end{center}
\end{figure*}

\newpage

\begin{figure*}
\begin{center}
\includegraphics[totalheight=0.185\textheight, clip]{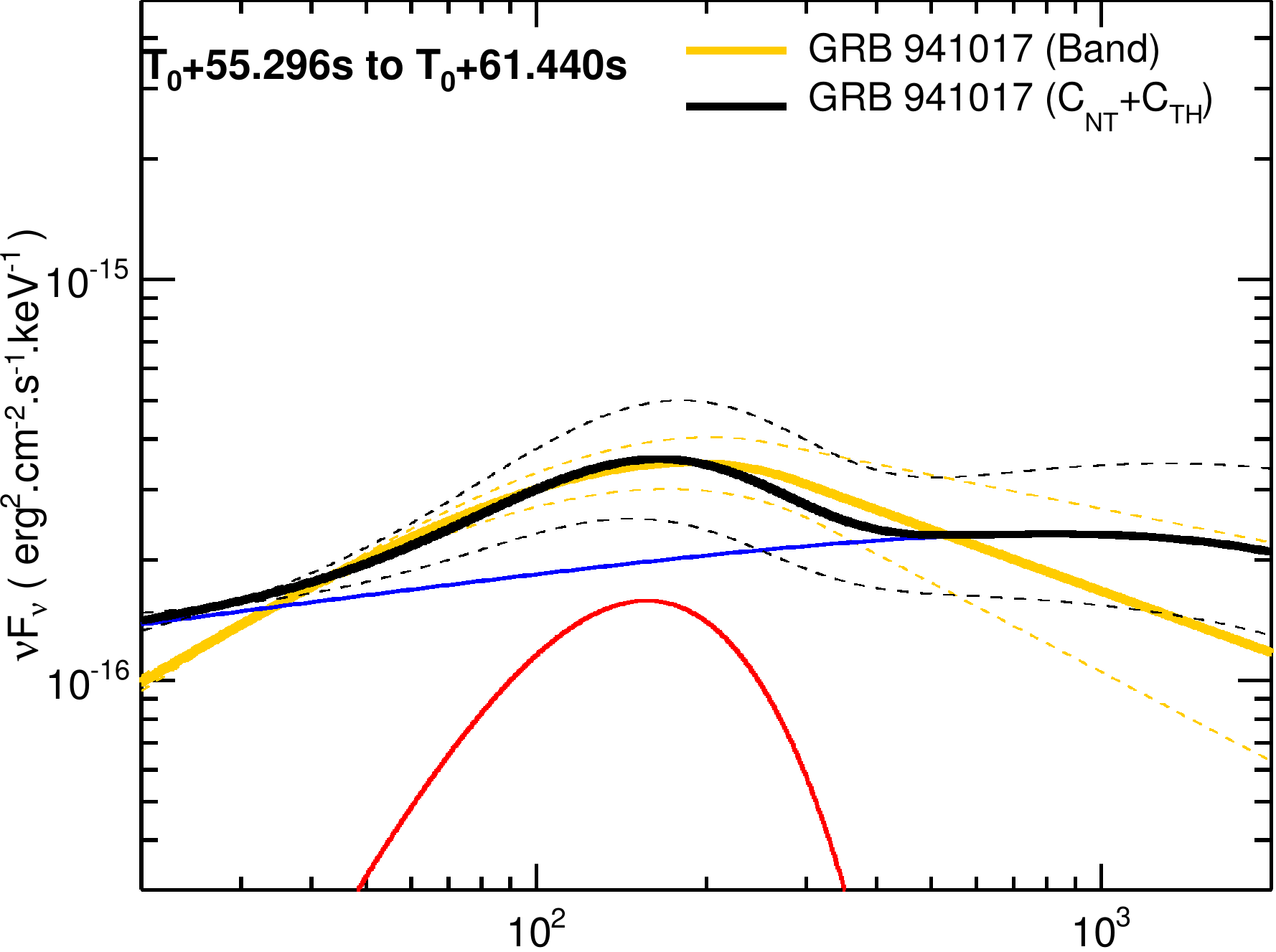}
\includegraphics[totalheight=0.185\textheight, clip]{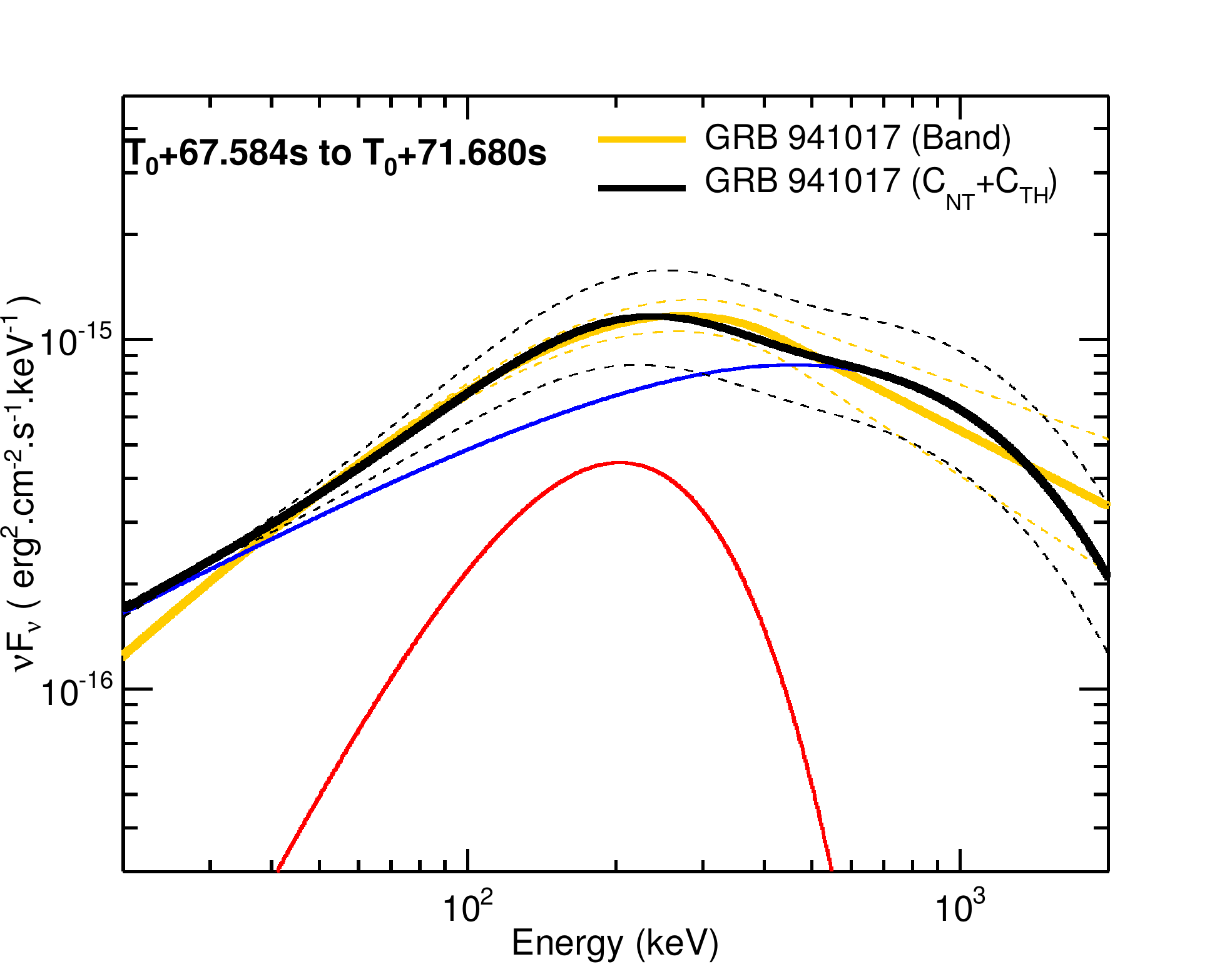}
\includegraphics[totalheight=0.185\textheight, clip]{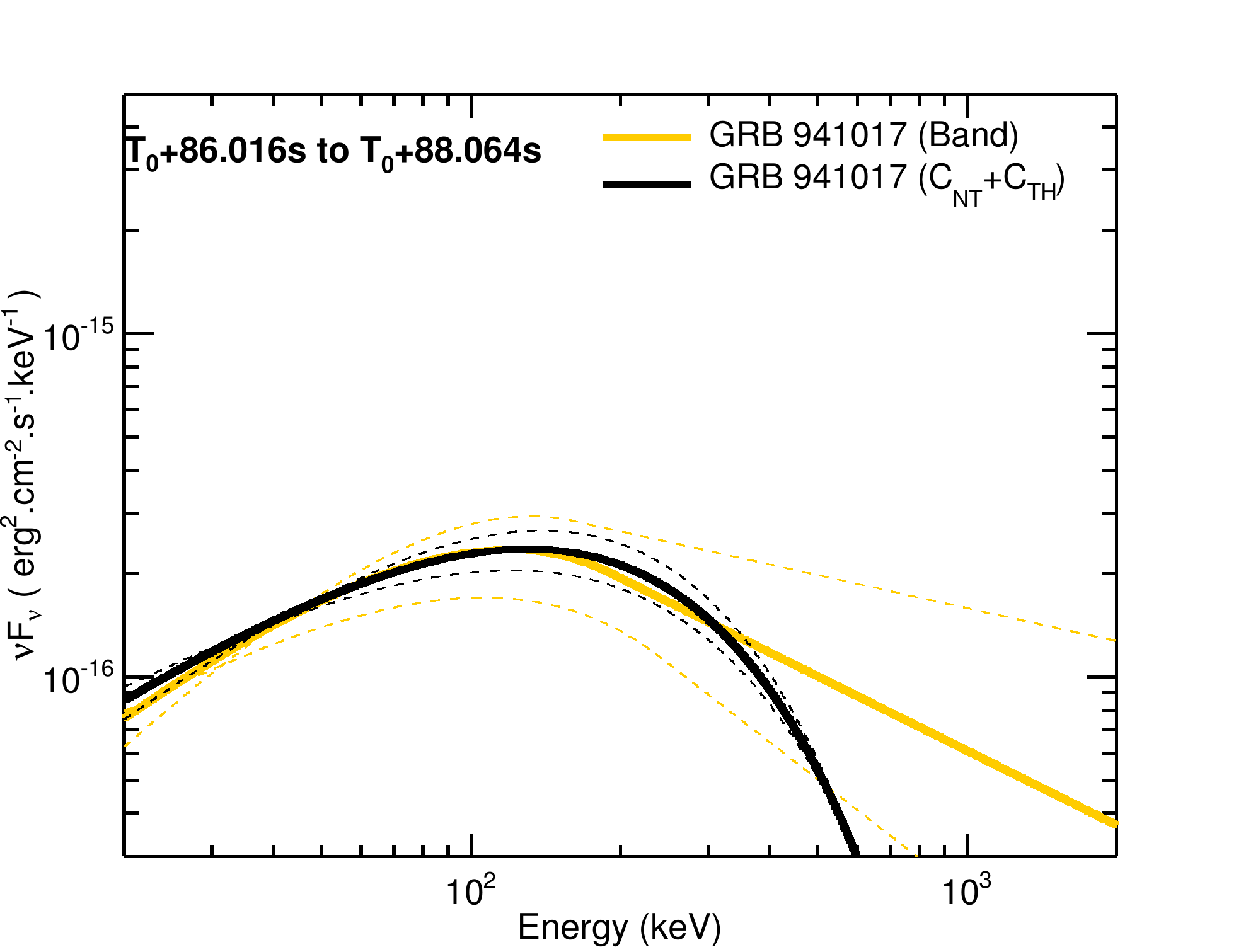}

\includegraphics[totalheight=0.185\textheight, clip]{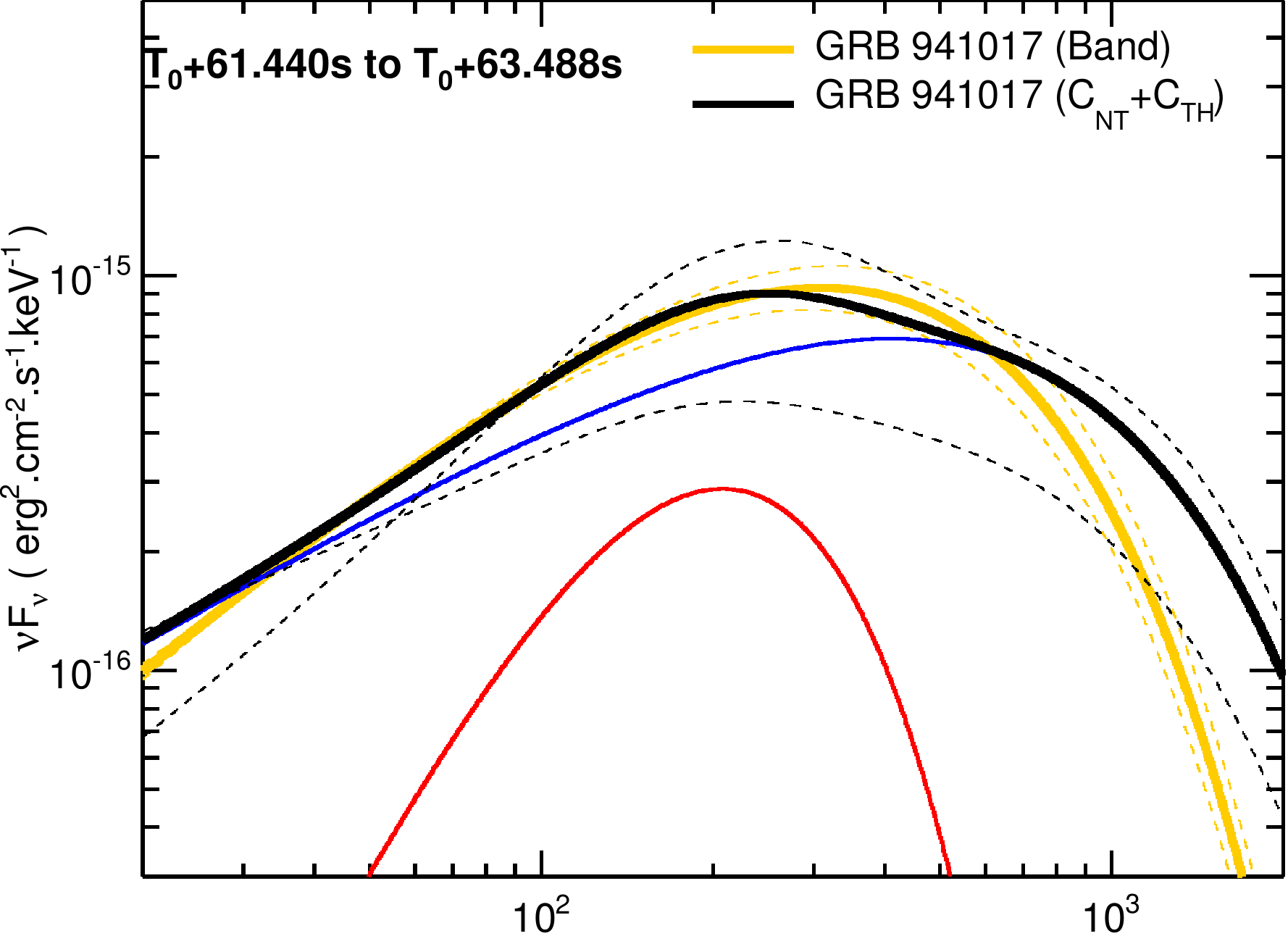}
\includegraphics[totalheight=0.185\textheight, clip]{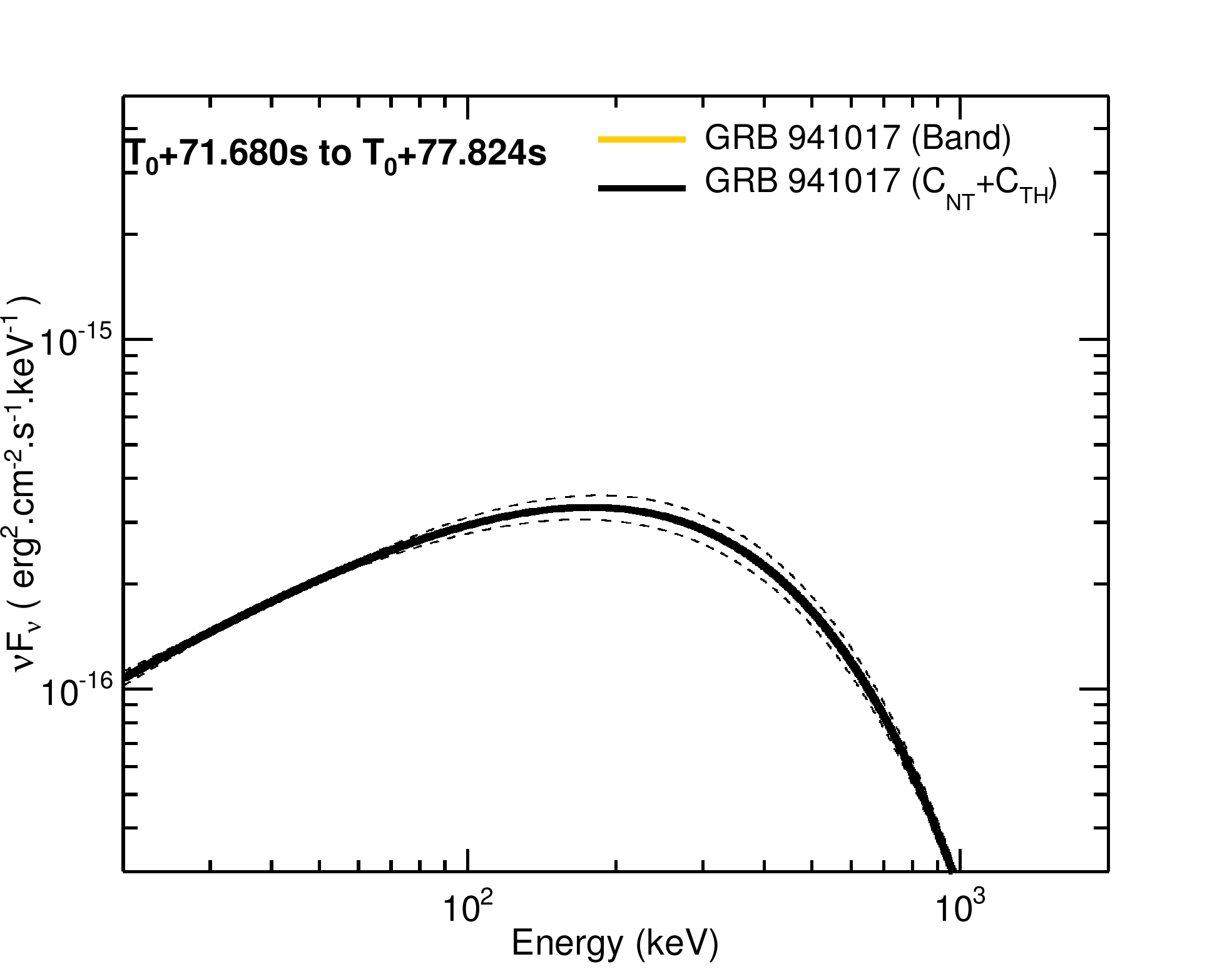}
\includegraphics[totalheight=0.185\textheight, clip]{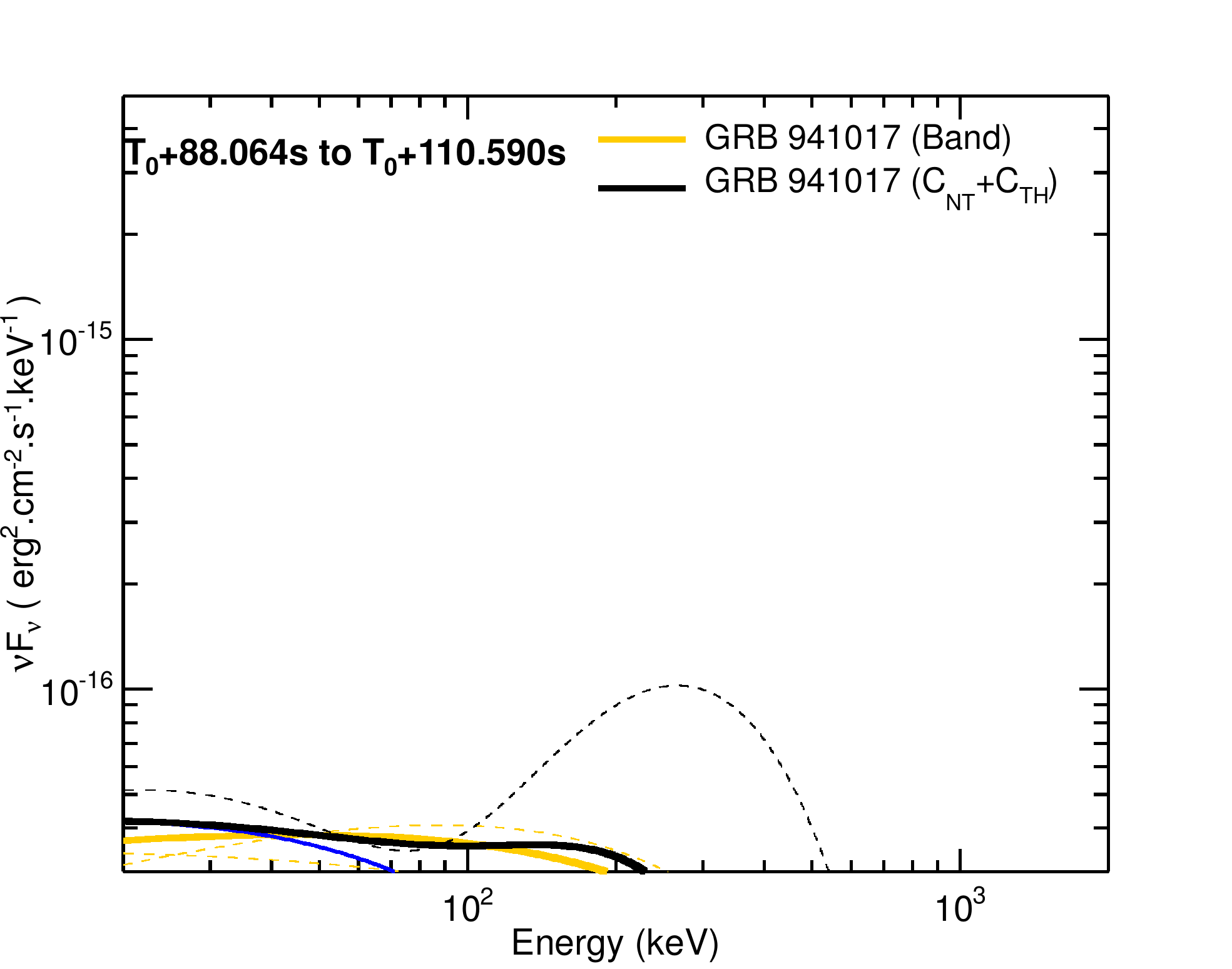}

\includegraphics[totalheight=0.185\textheight, clip]{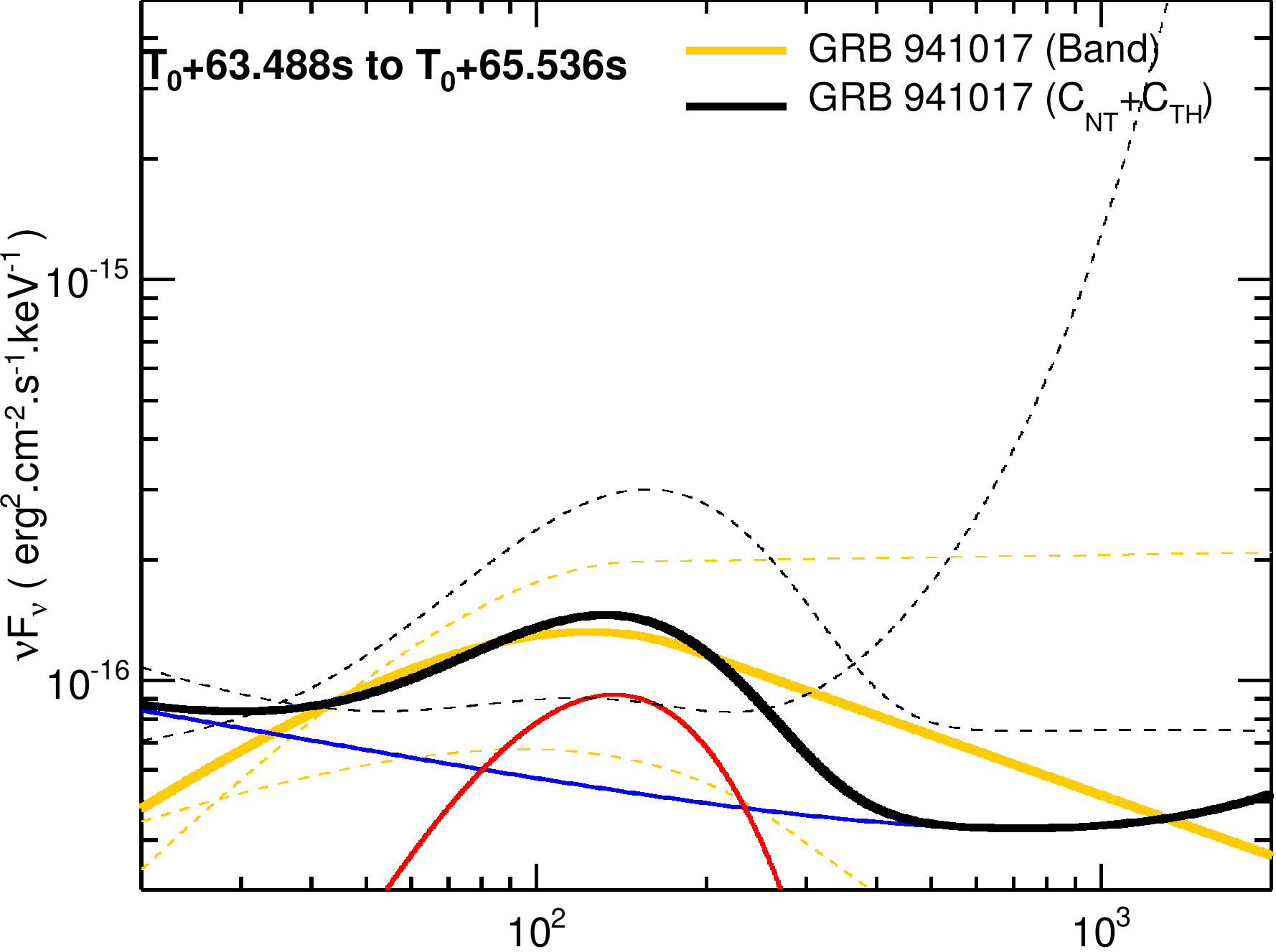}
\includegraphics[totalheight=0.185\textheight, clip]{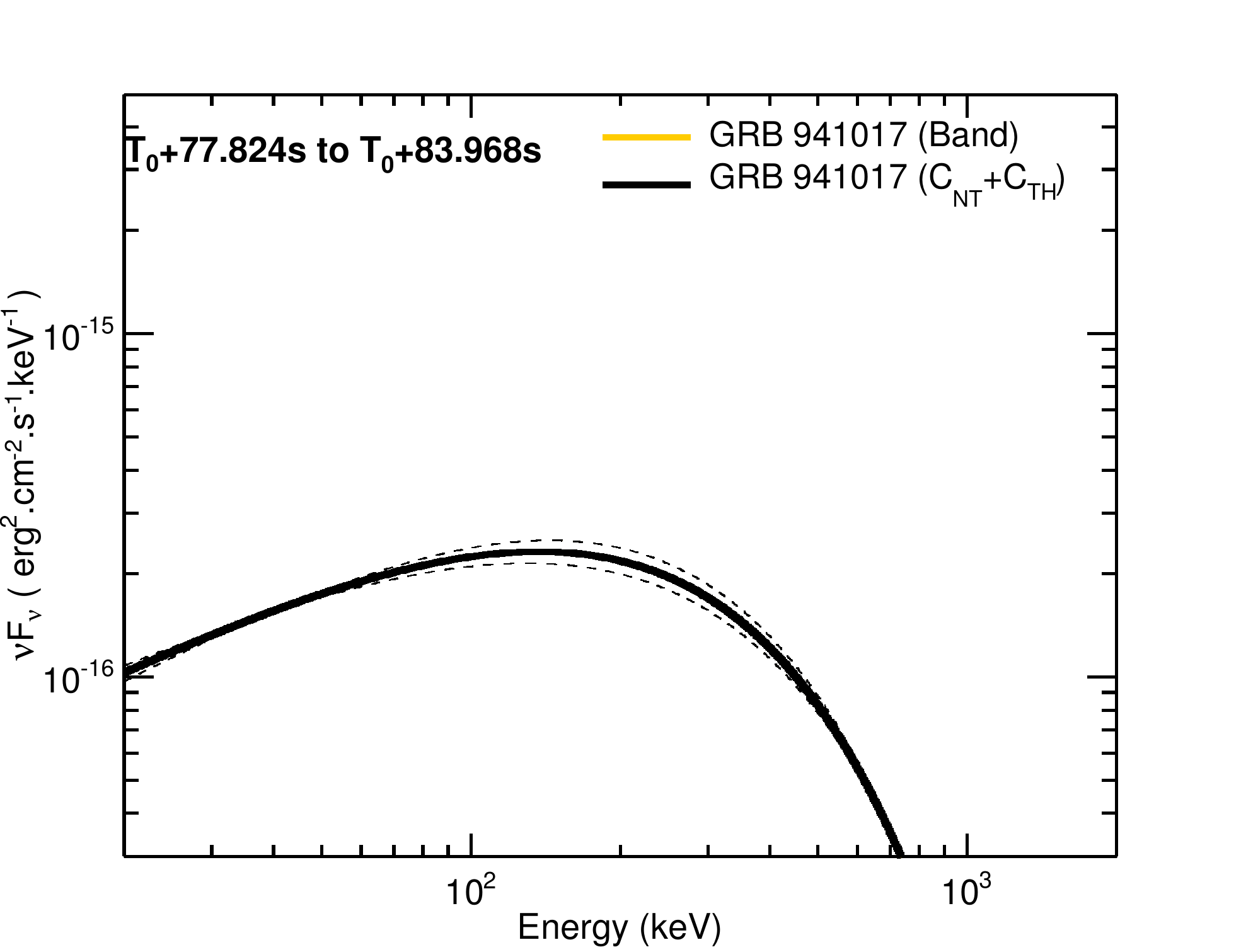}
\includegraphics[totalheight=0.185\textheight, clip]{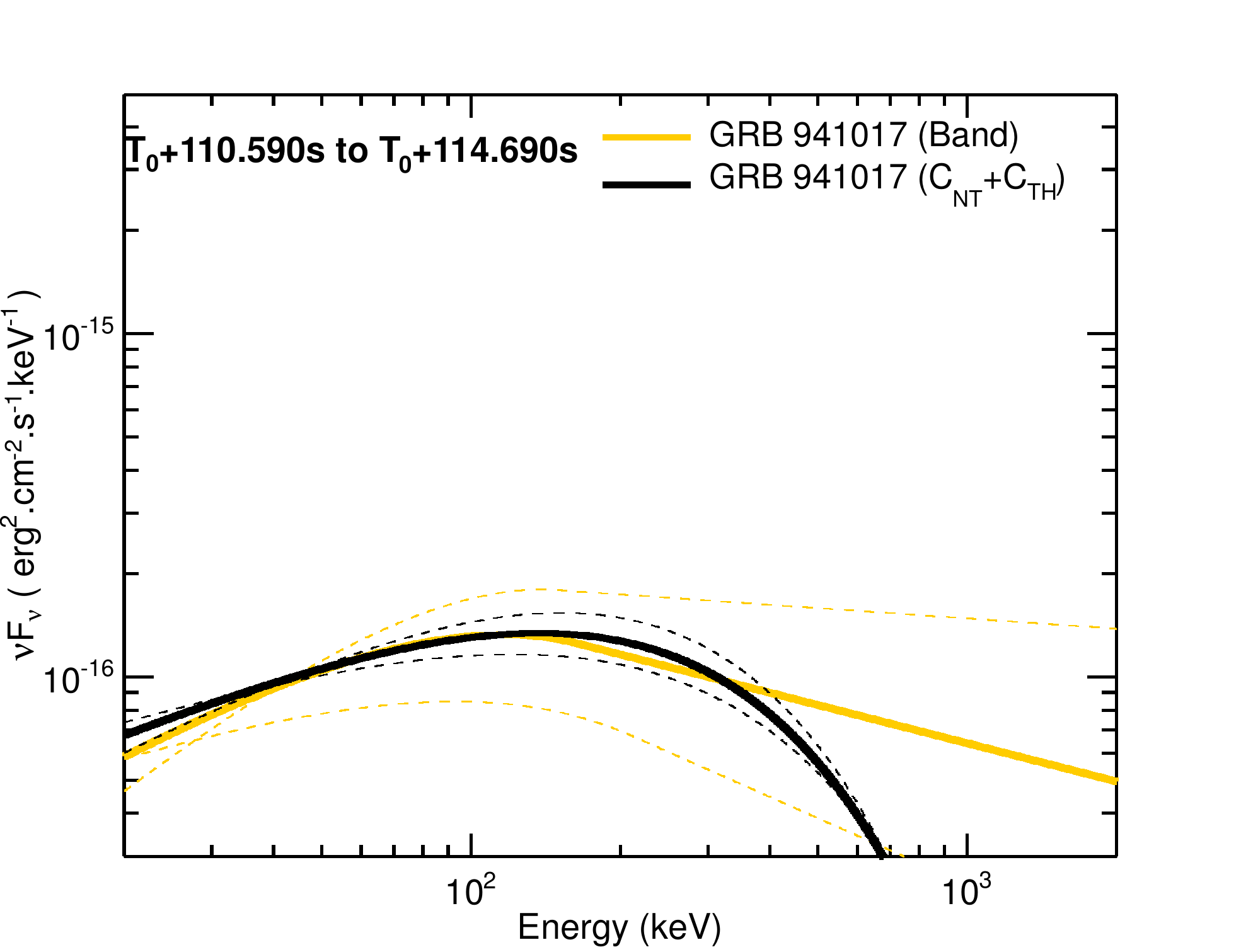}

\includegraphics[totalheight=0.195\textheight, clip]{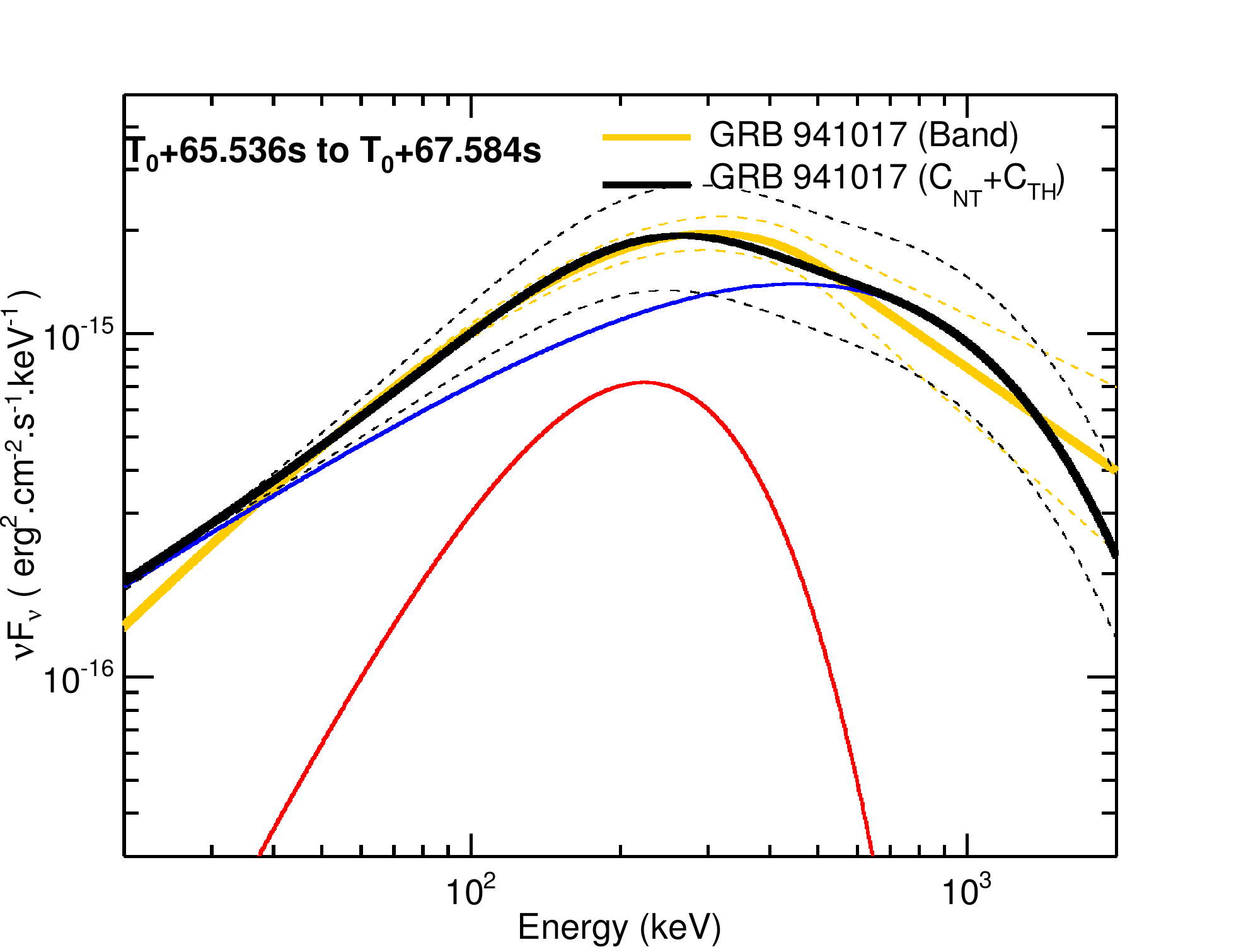}
\includegraphics[totalheight=0.195\textheight, clip]{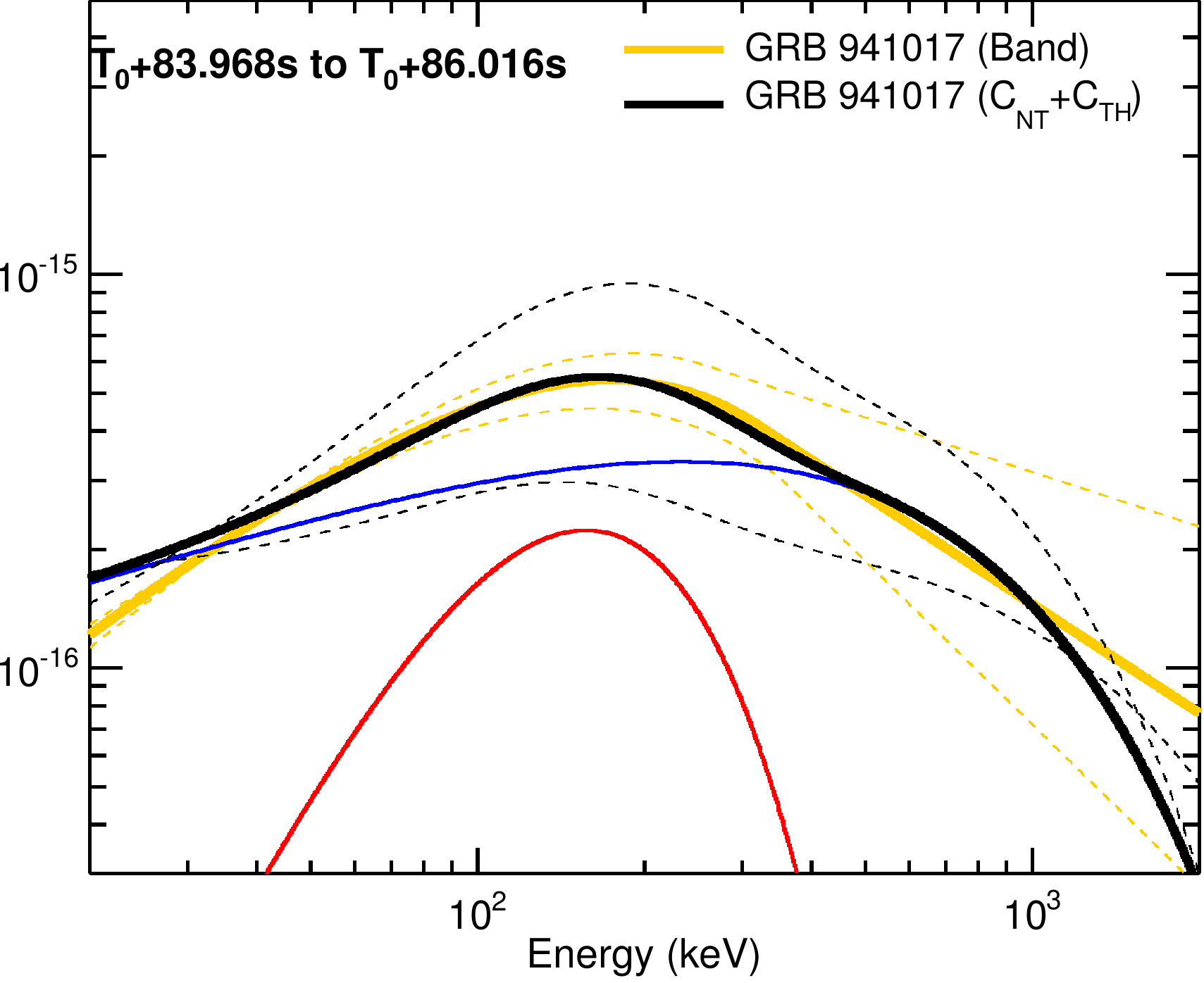}
\includegraphics[totalheight=0.195\textheight, clip]{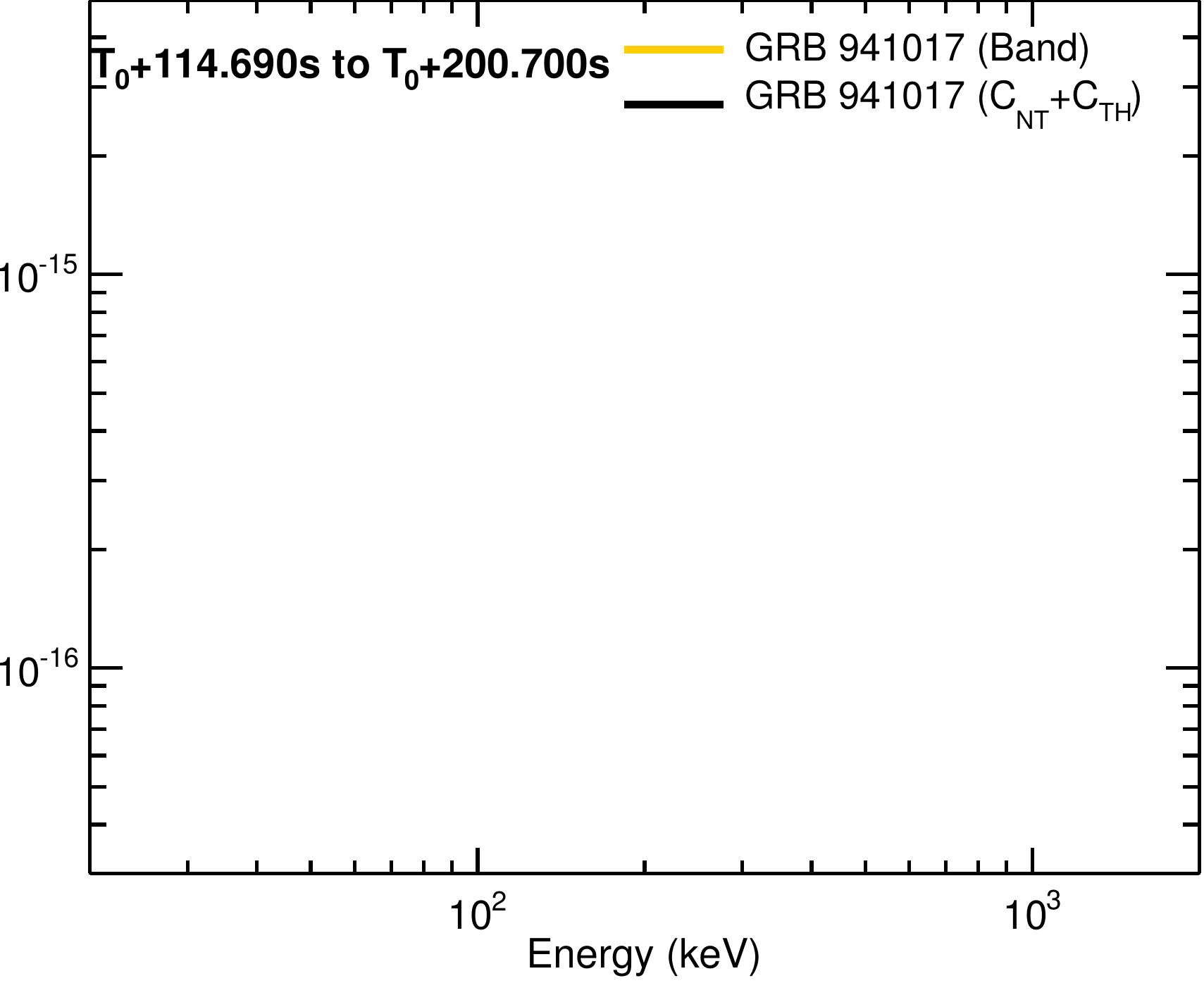}

\caption{\label{fig15}GRB~941017 : $\nu$F$_\nu$ spectra resulting from the fine-time analysis presented in Section~\ref{sec:trsa}. The solid yellow and black lines correspond to the best Band-only and C$_\mathrm{nTh}$+C$_\mathrm{Th}$ fits, respectively. The dashed yellow and black lines correspond to the 1--$\sigma$ confidence regions of the Band-only and C$_\mathrm{nTh}$+C$_\mathrm{Th}$ fits, respectively. The solid blue and red lines correspond to C$_\mathrm{nTh}$ and C$_\mathrm{Th}$ resulting from the best C$_\mathrm{nTh}$+C$_\mathrm{Th}$ fits (i.e., solid black line) to the data, respectively.}
\end{center}
\end{figure*}

\newpage

\begin{figure*}
\begin{center}
\includegraphics[totalheight=0.185\textheight, clip]{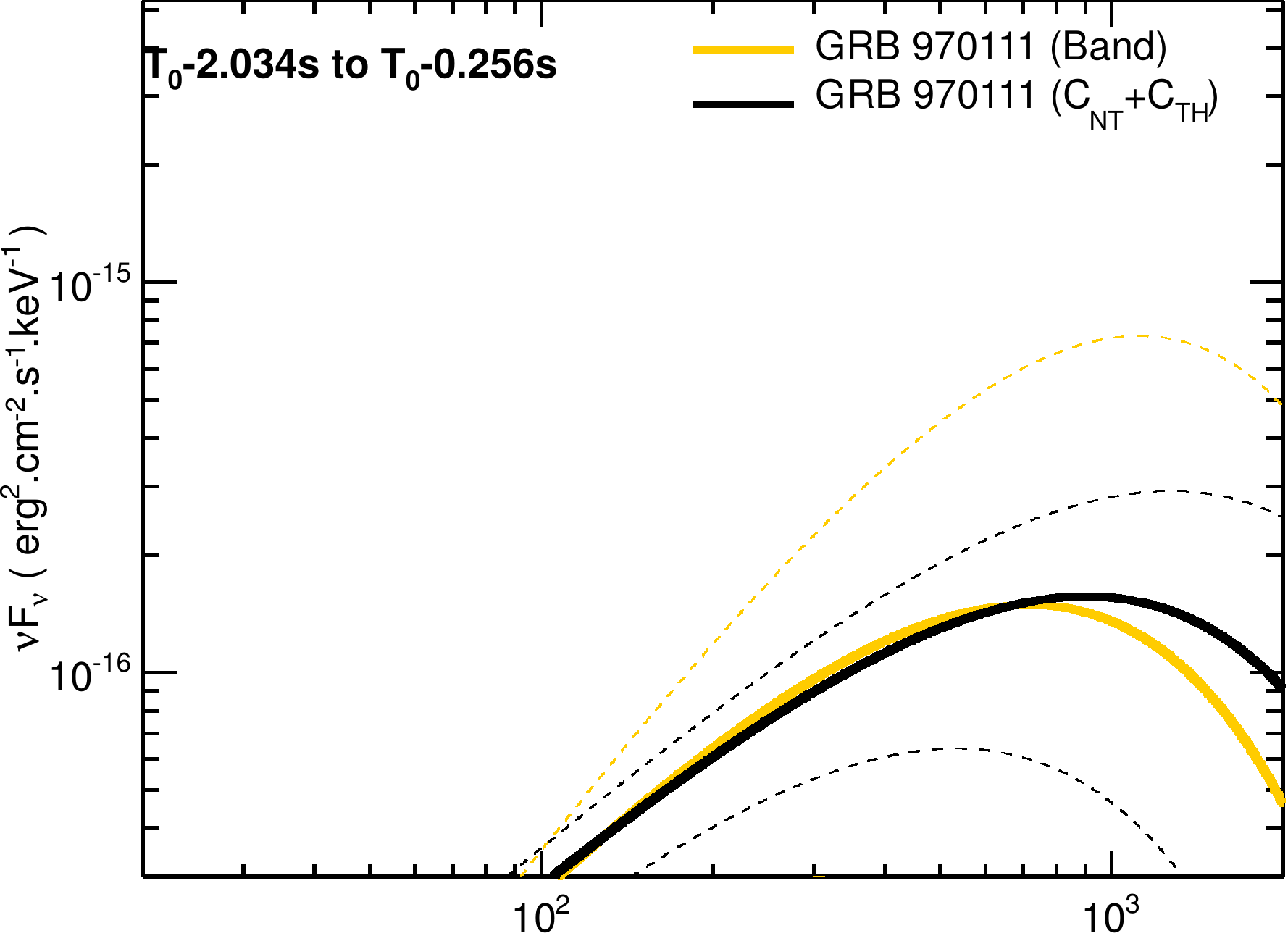}
\includegraphics[totalheight=0.185\textheight, clip]{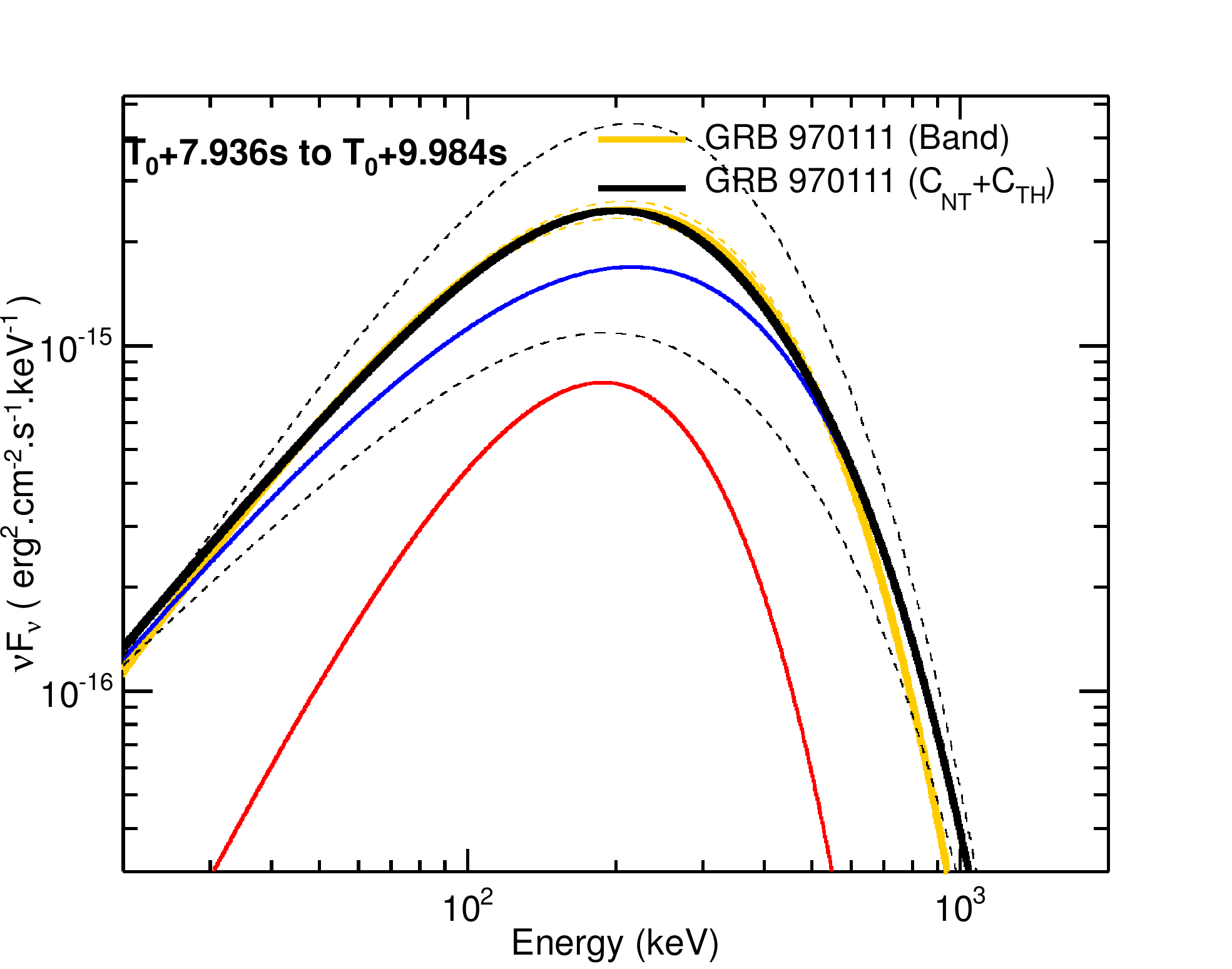}
\includegraphics[totalheight=0.185\textheight, clip]{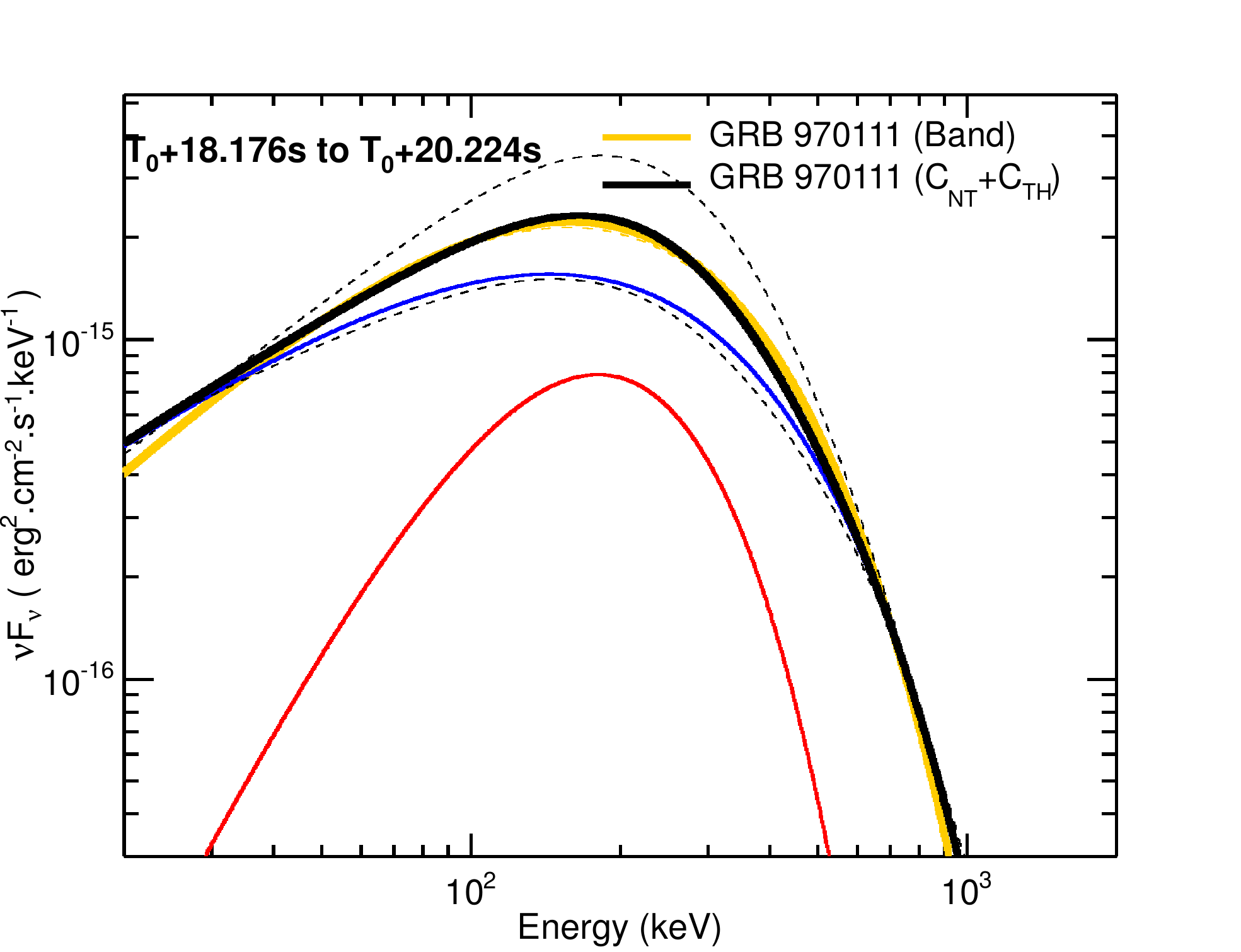}

\includegraphics[totalheight=0.185\textheight, clip]{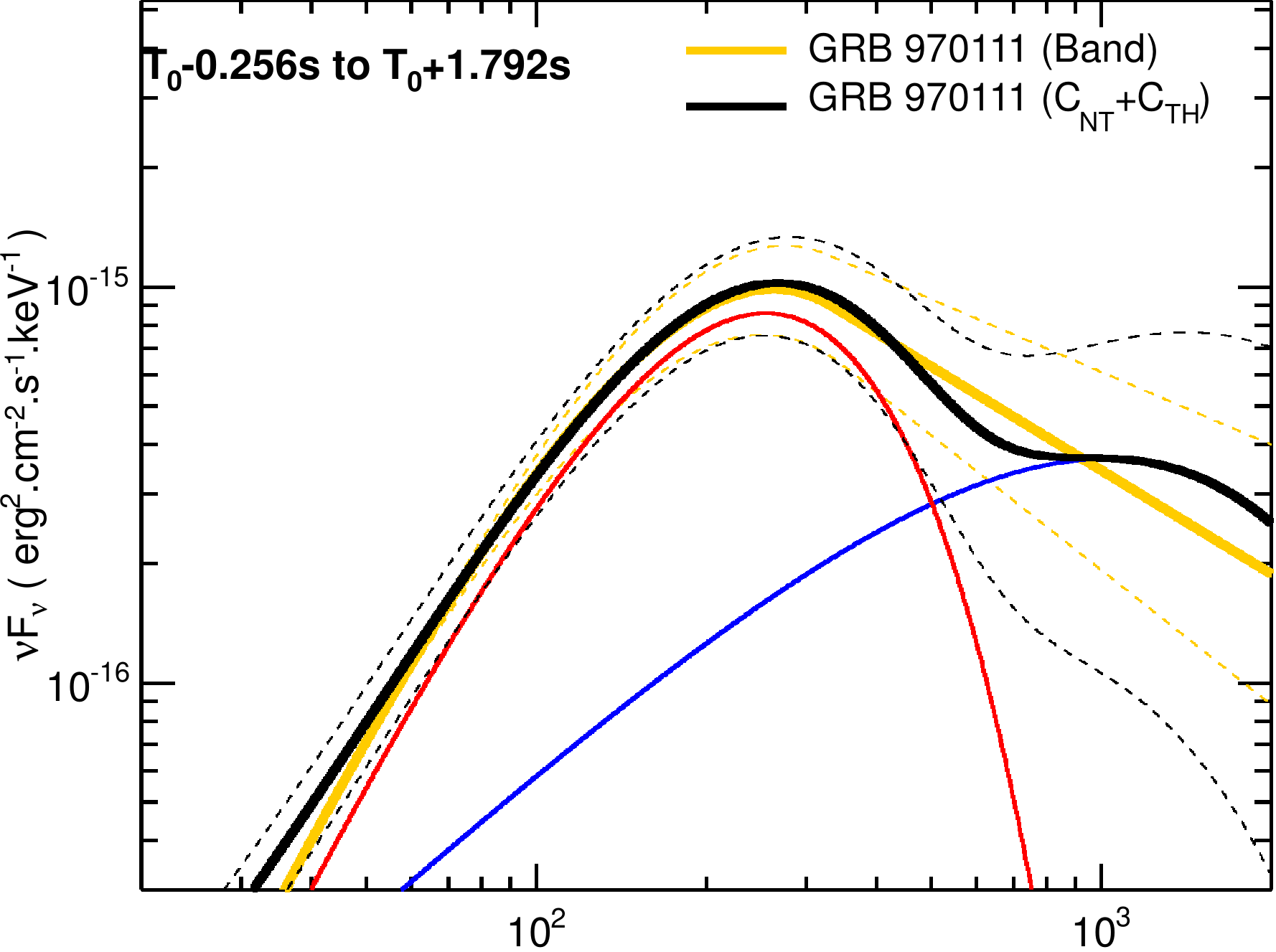}
\includegraphics[totalheight=0.185\textheight, clip]{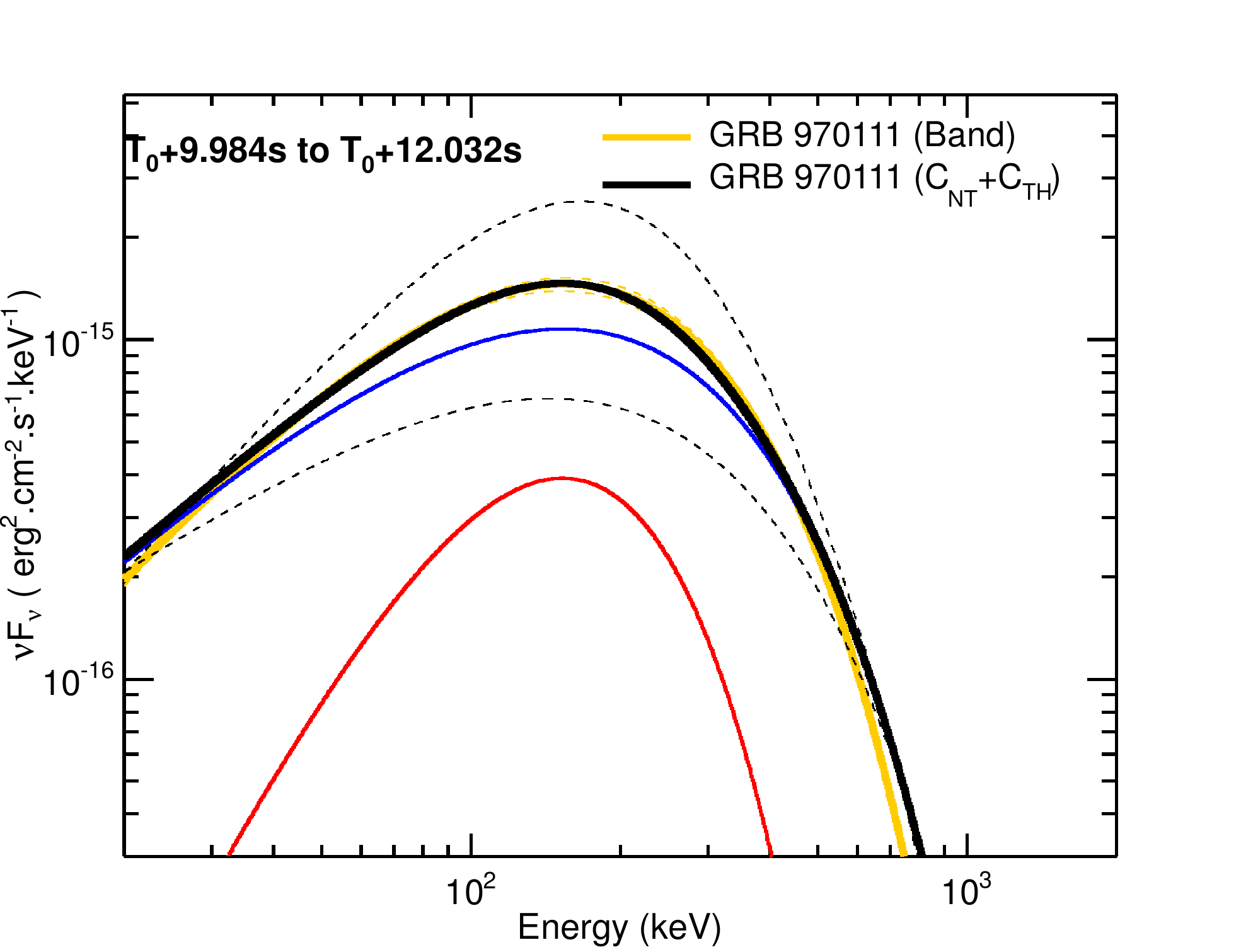}
\includegraphics[totalheight=0.185\textheight, clip]{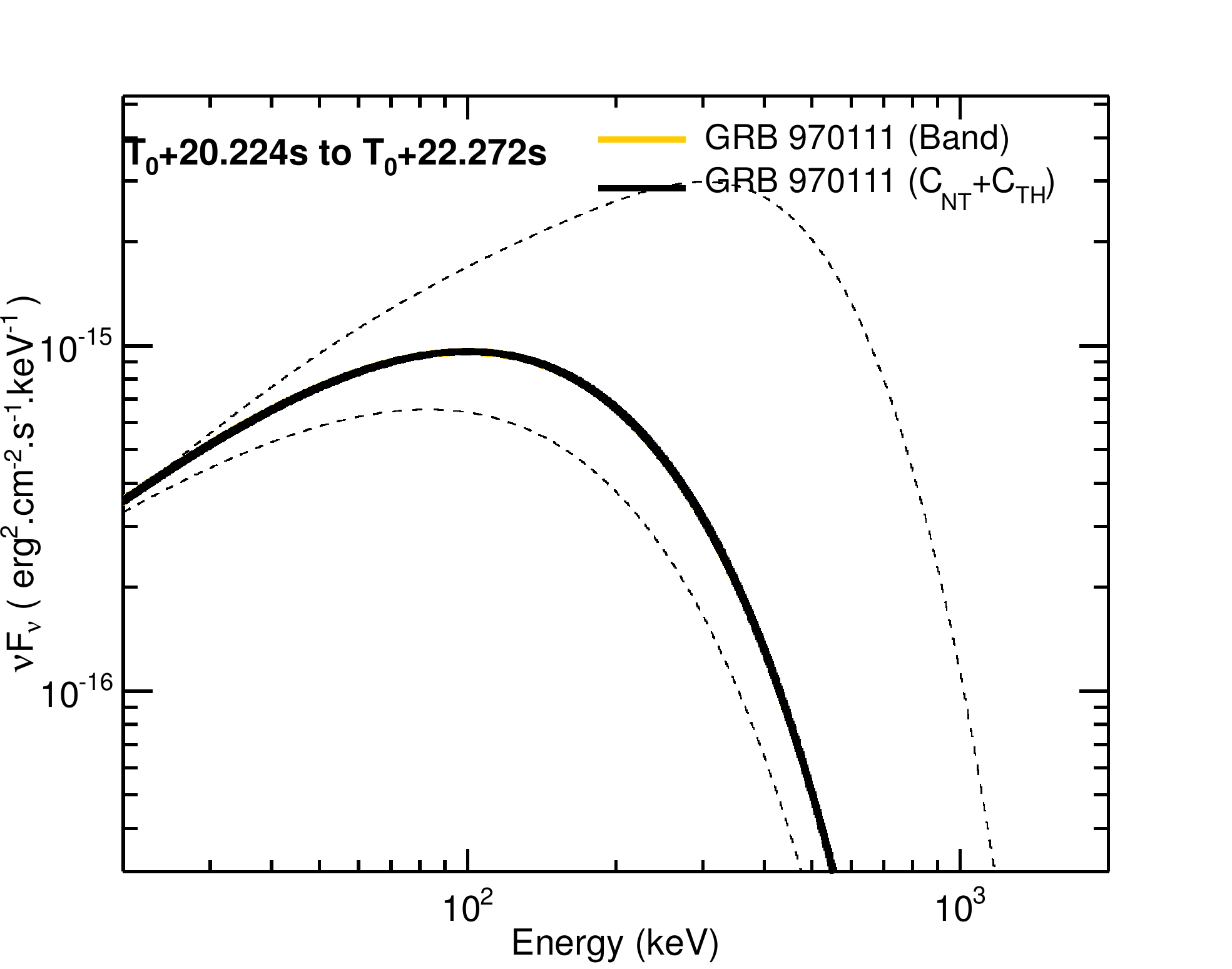}

\includegraphics[totalheight=0.185\textheight, clip]{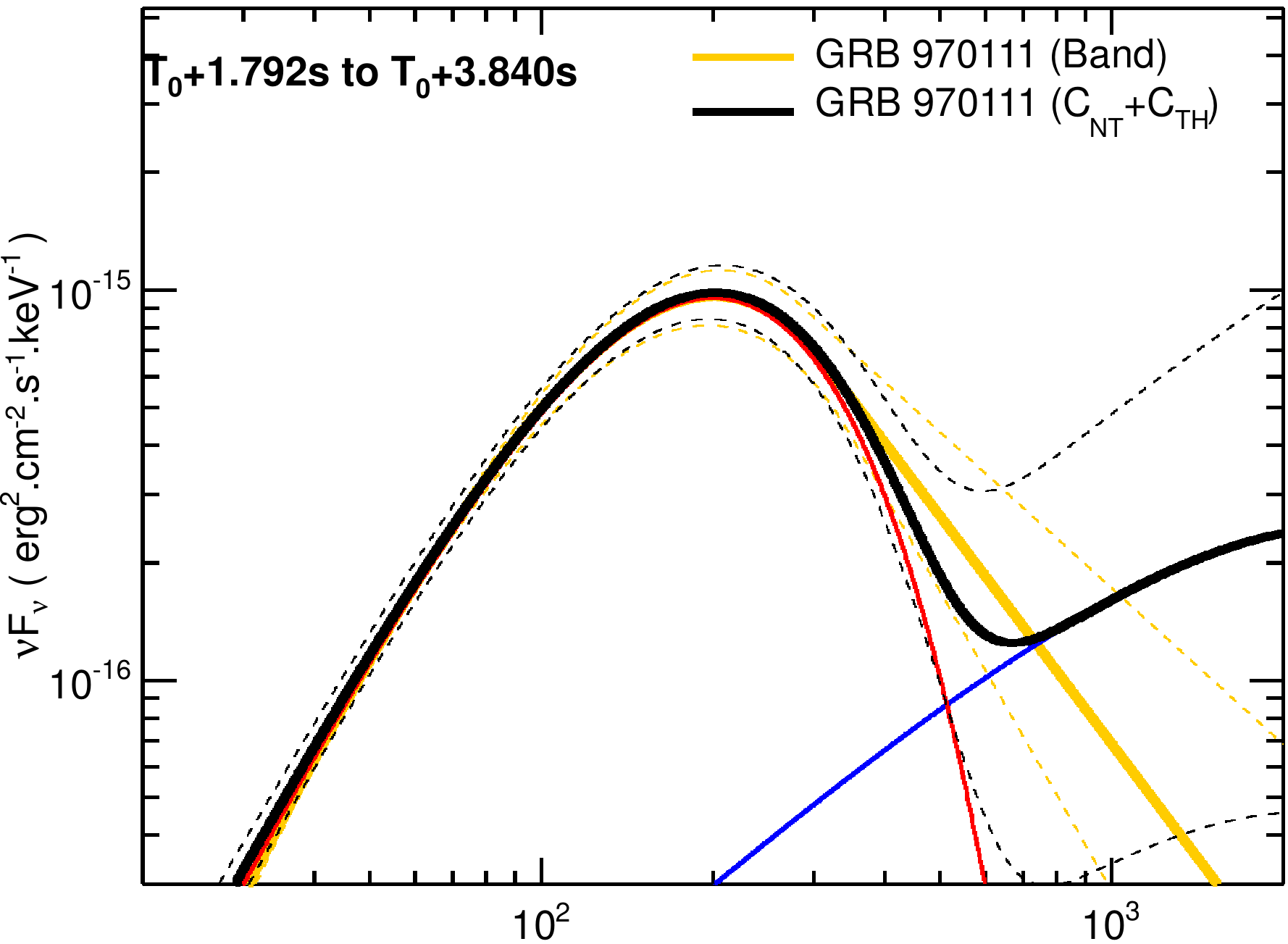}
\includegraphics[totalheight=0.185\textheight, clip]{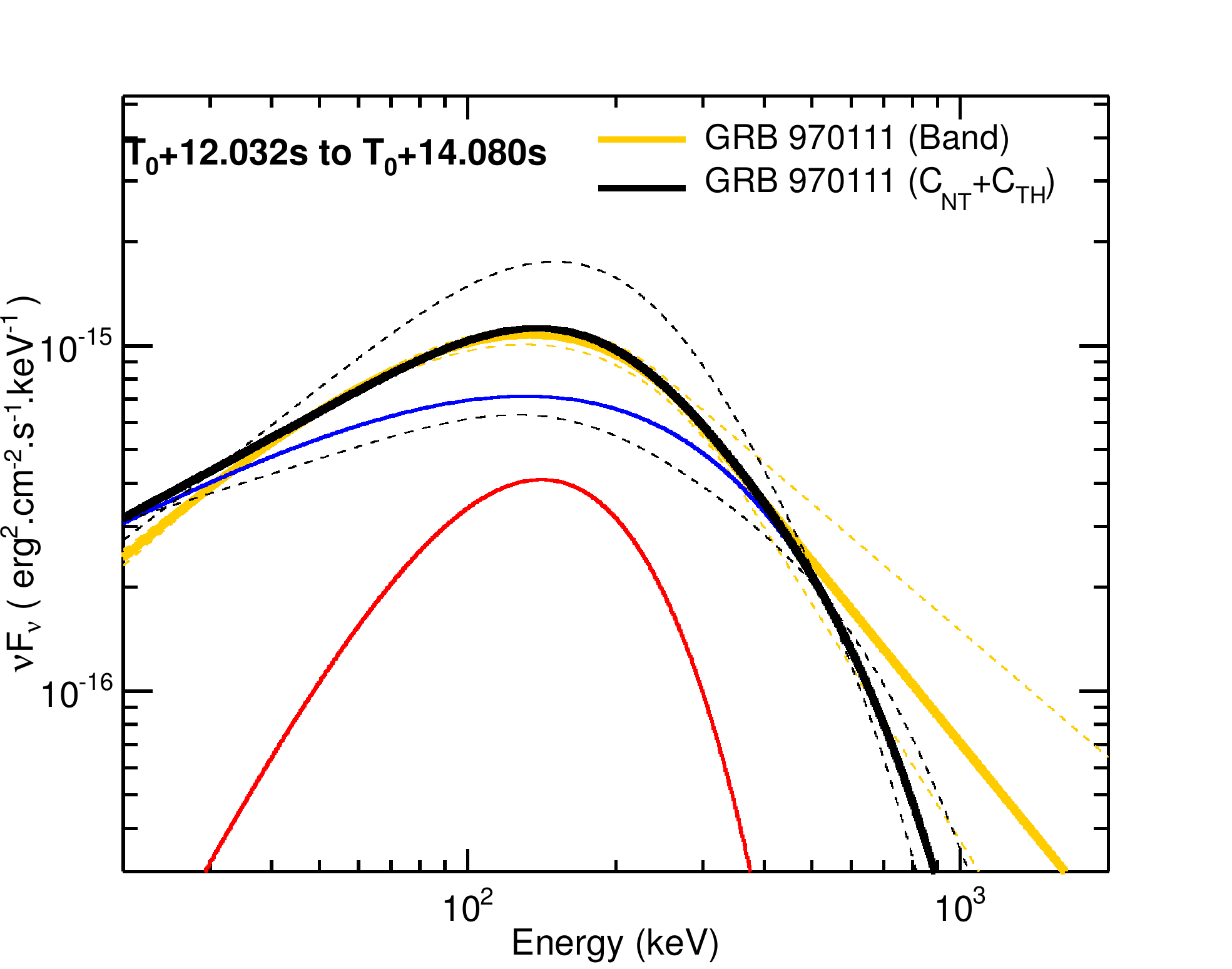}
\includegraphics[totalheight=0.185\textheight, clip]{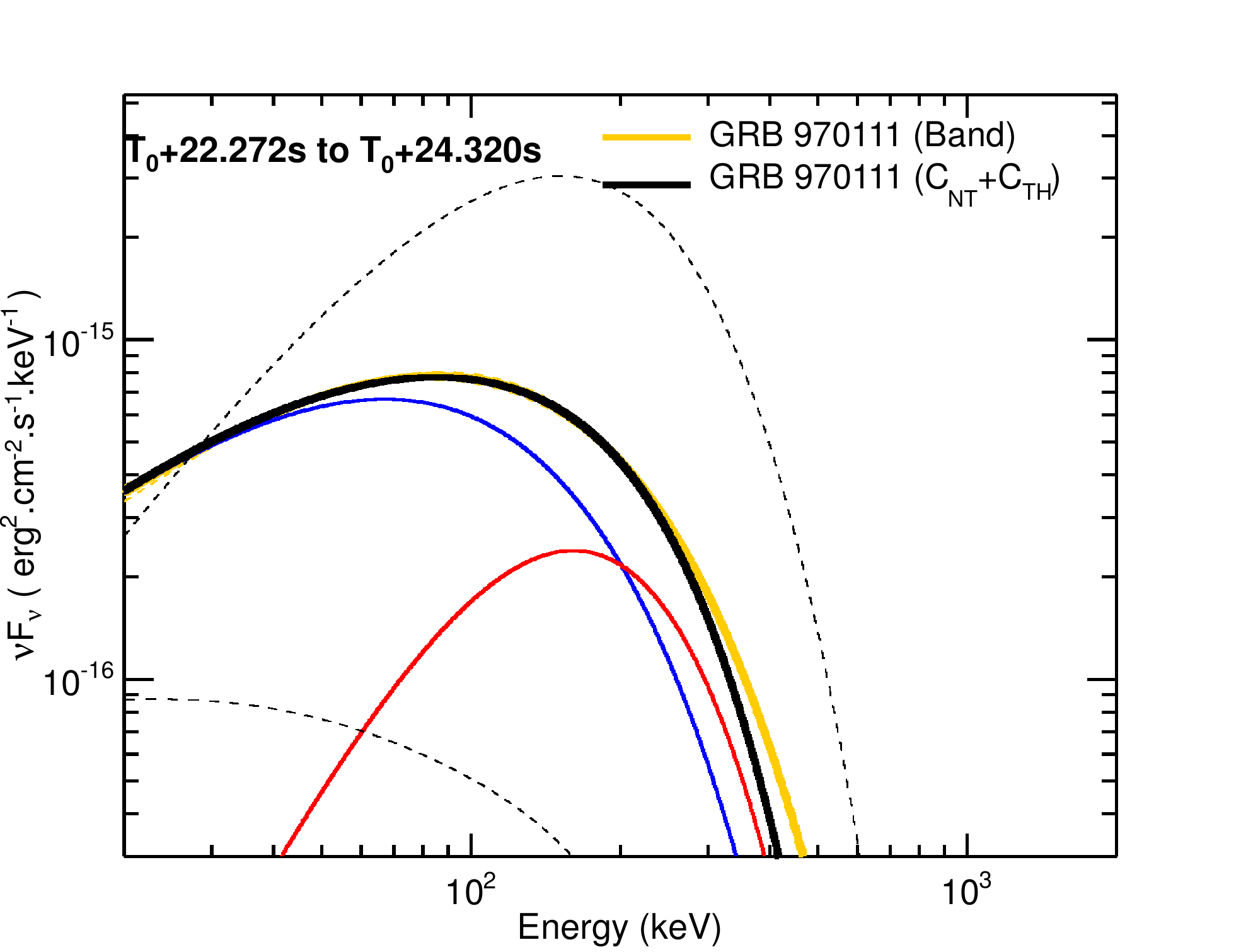}

\includegraphics[totalheight=0.185\textheight, clip]{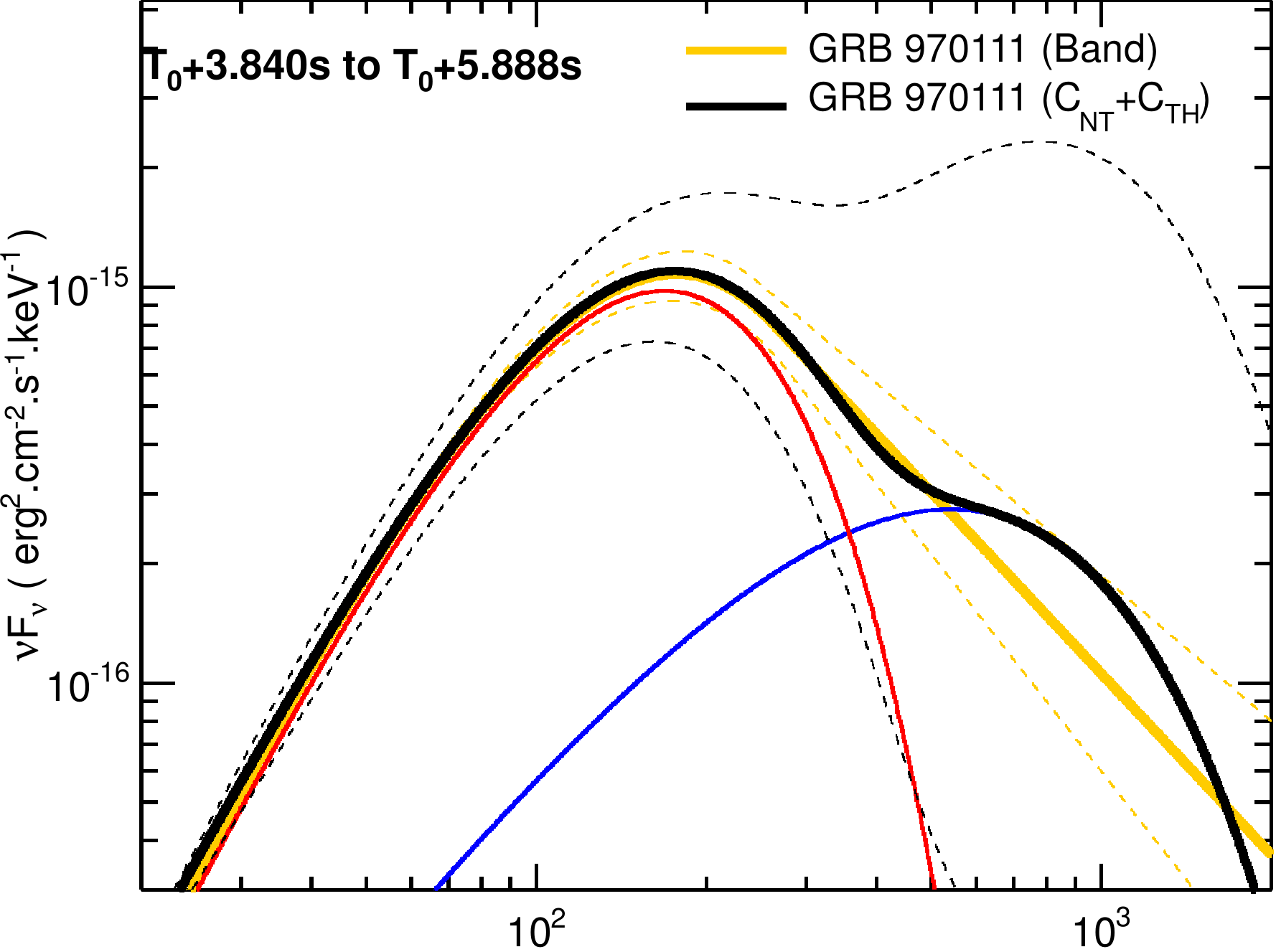}
\includegraphics[totalheight=0.185\textheight, clip]{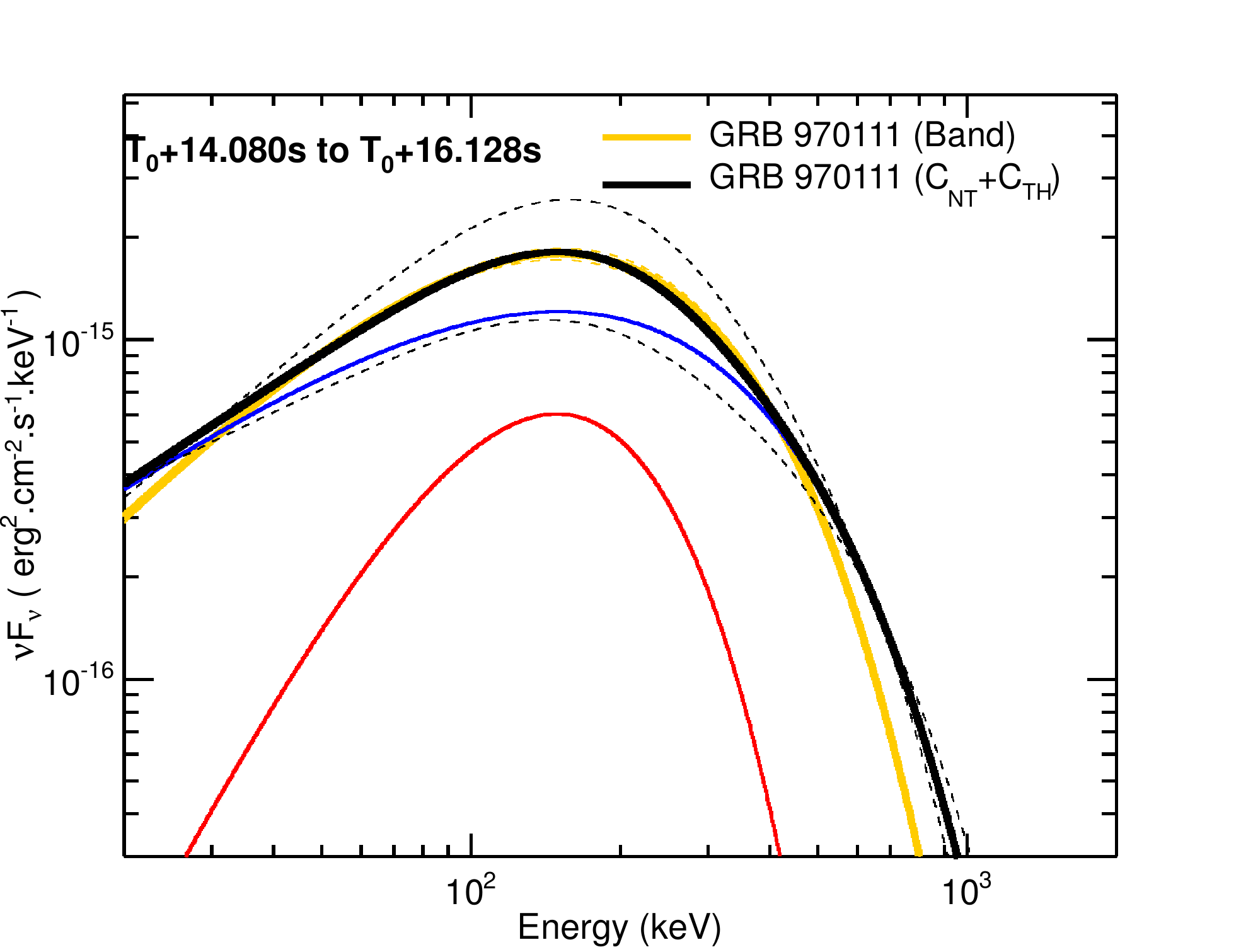}
\includegraphics[totalheight=0.185\textheight, clip]{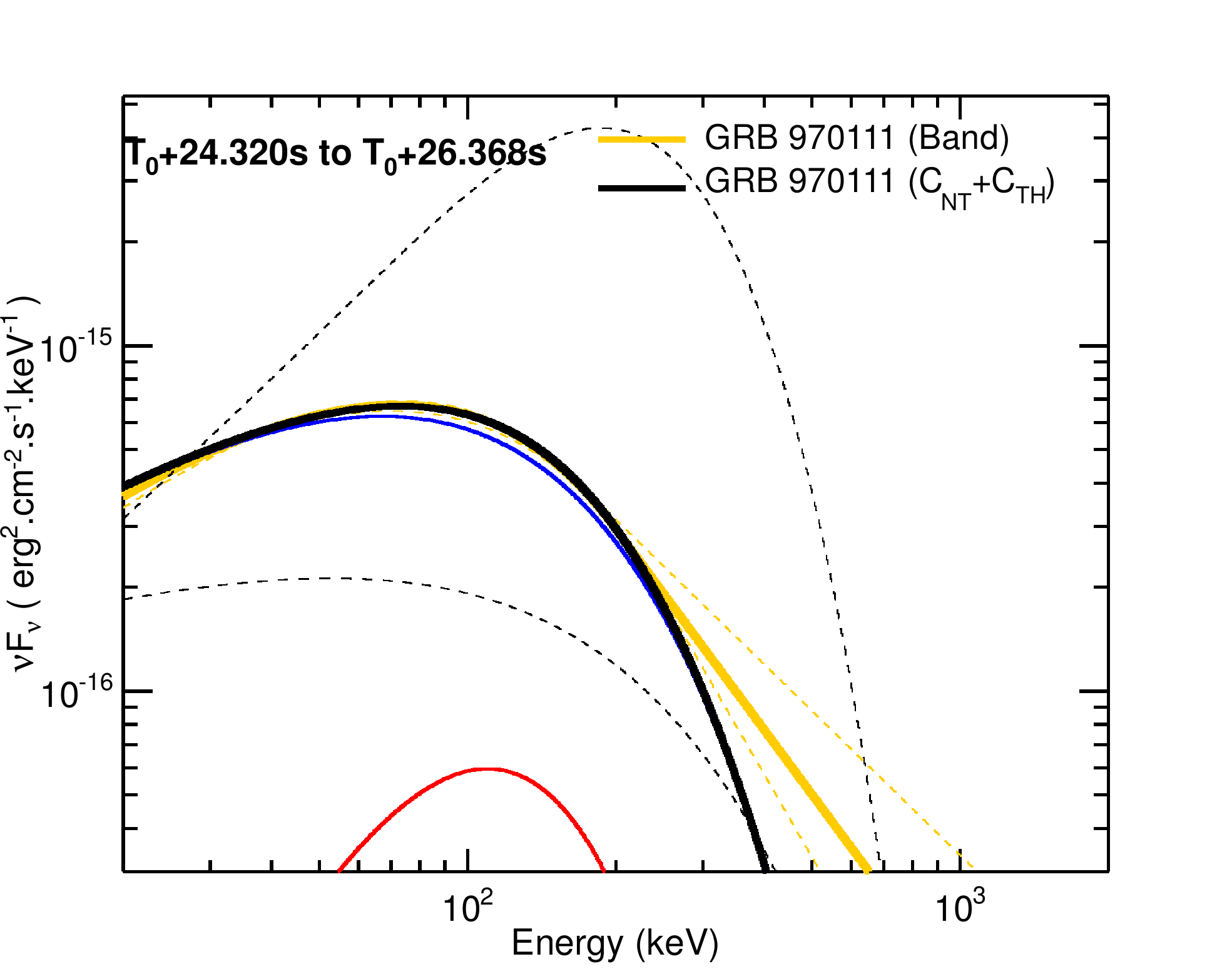}

\includegraphics[totalheight=0.195\textheight, clip]{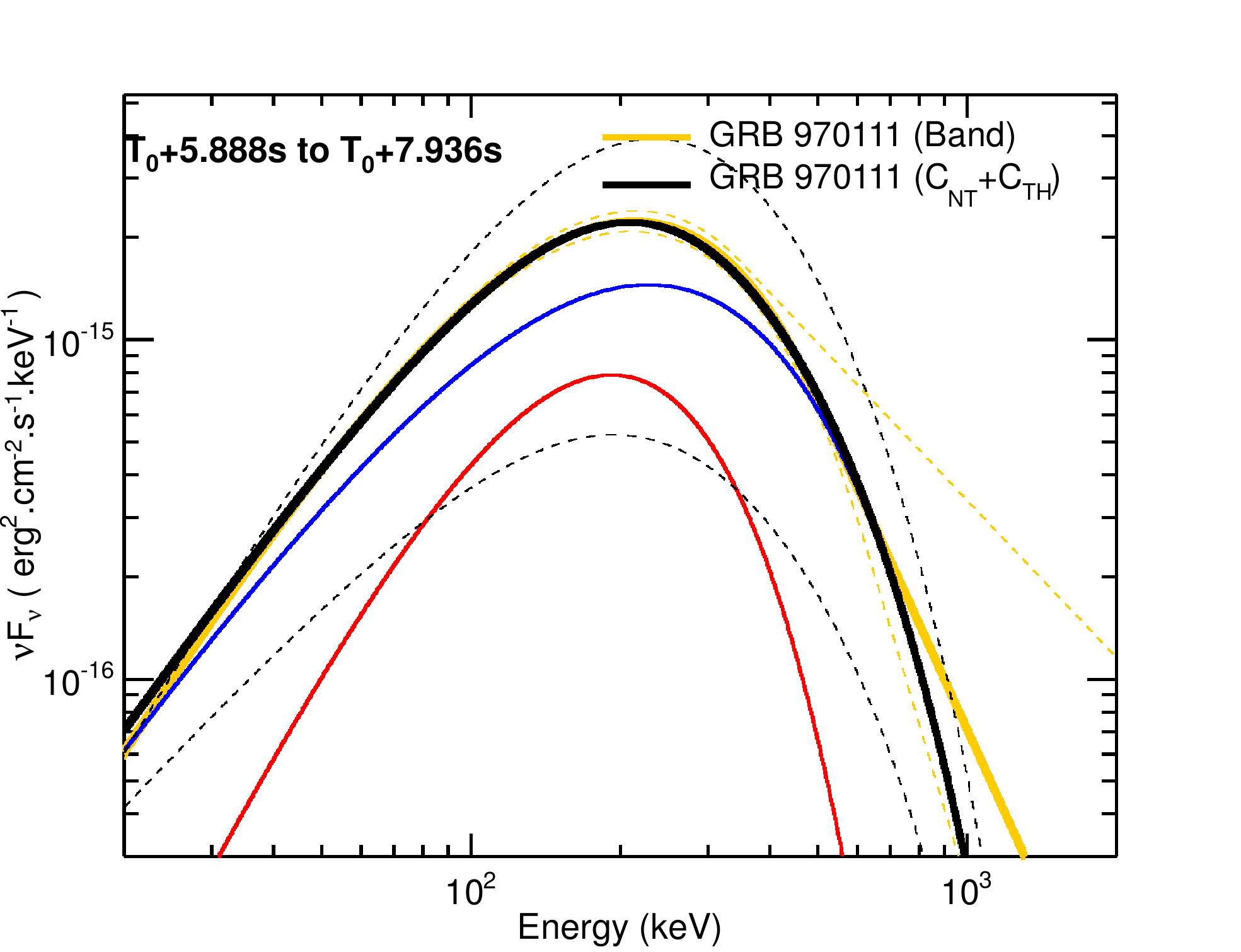}
\includegraphics[totalheight=0.195\textheight, clip]{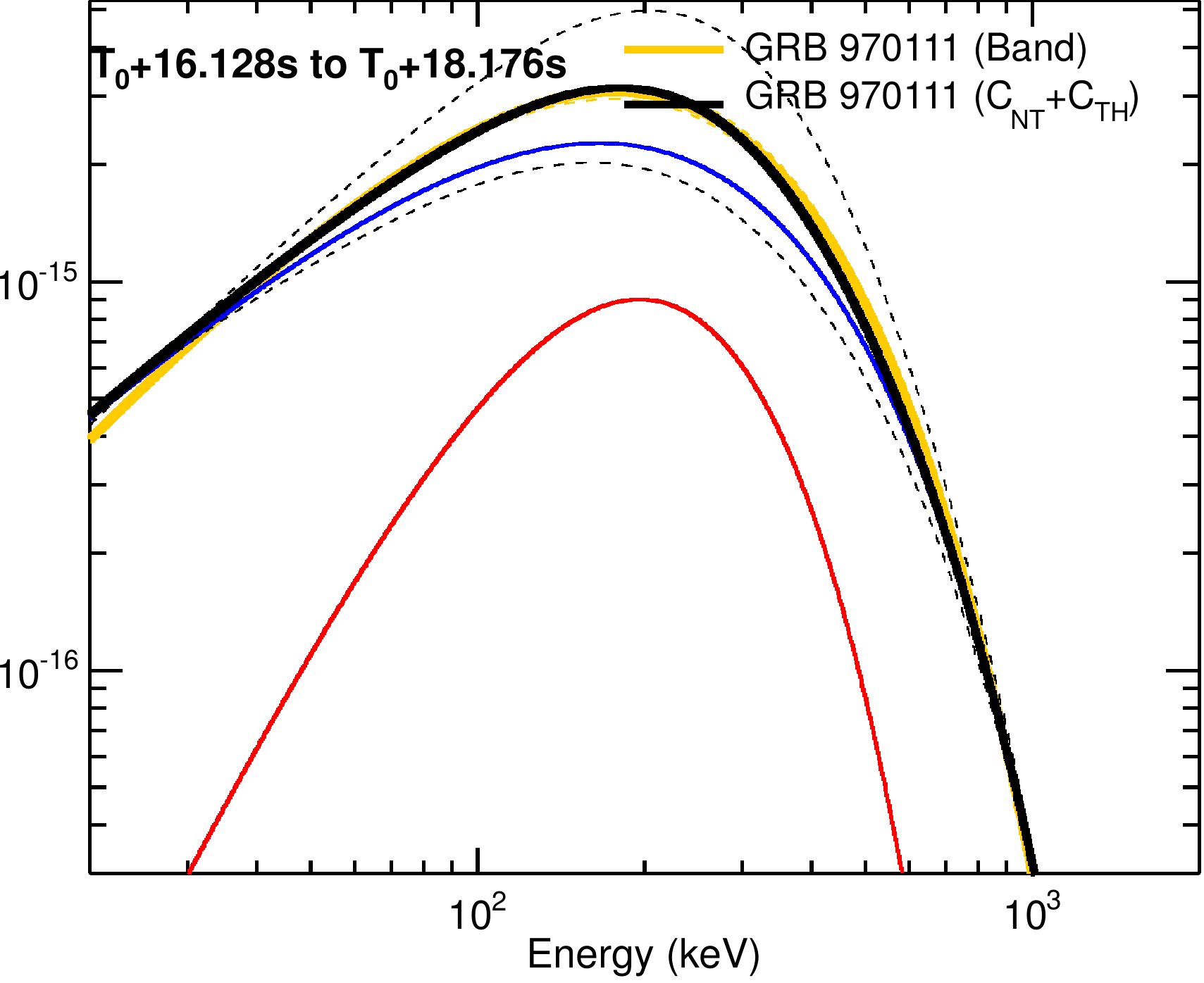}
\includegraphics[totalheight=0.195\textheight, clip]{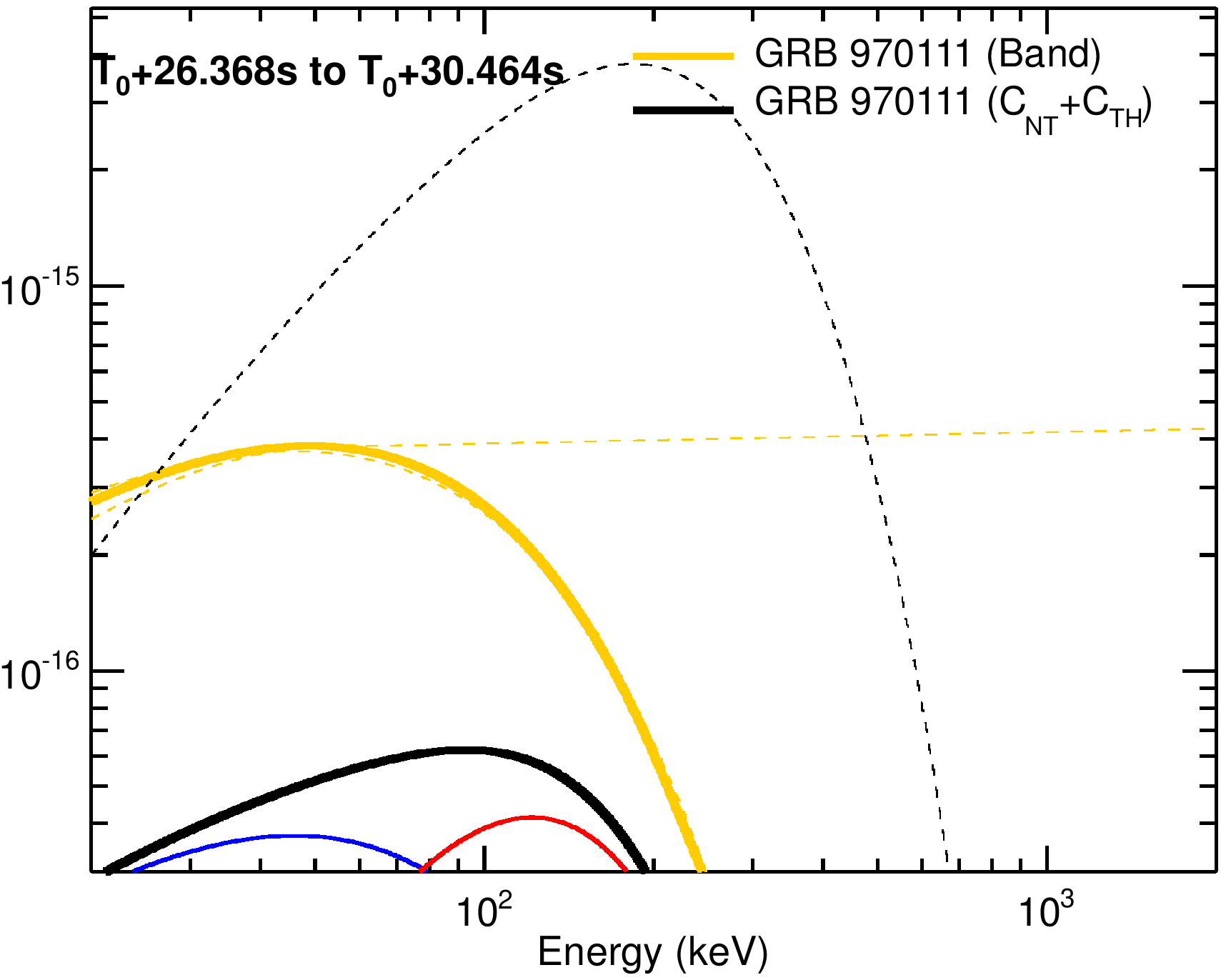}

\end{center}
\end{figure*}

\newpage

\begin{figure*}
\begin{center}
\includegraphics[totalheight=0.185\textheight, clip]{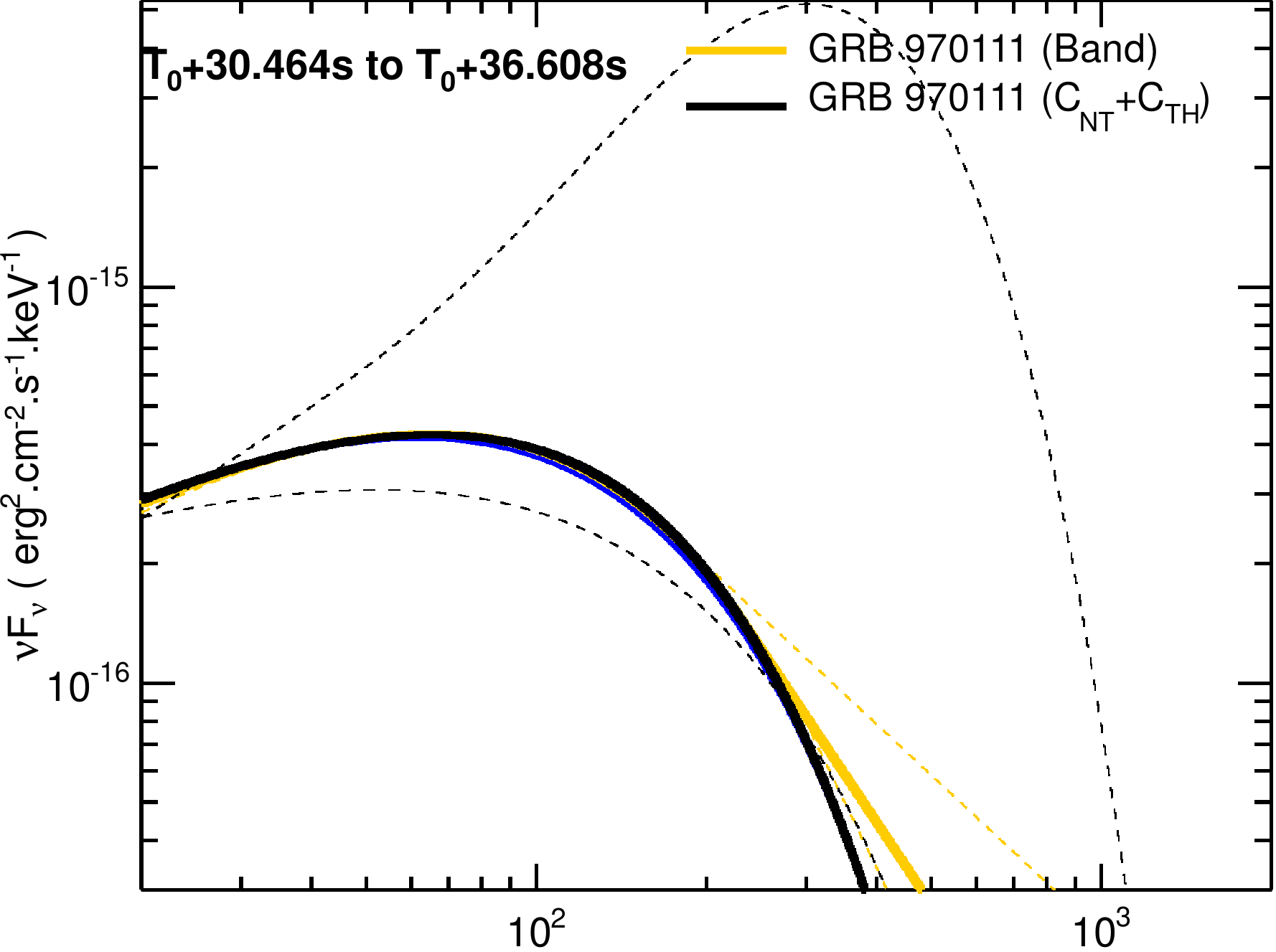}
\includegraphics[totalheight=0.185\textheight, clip]{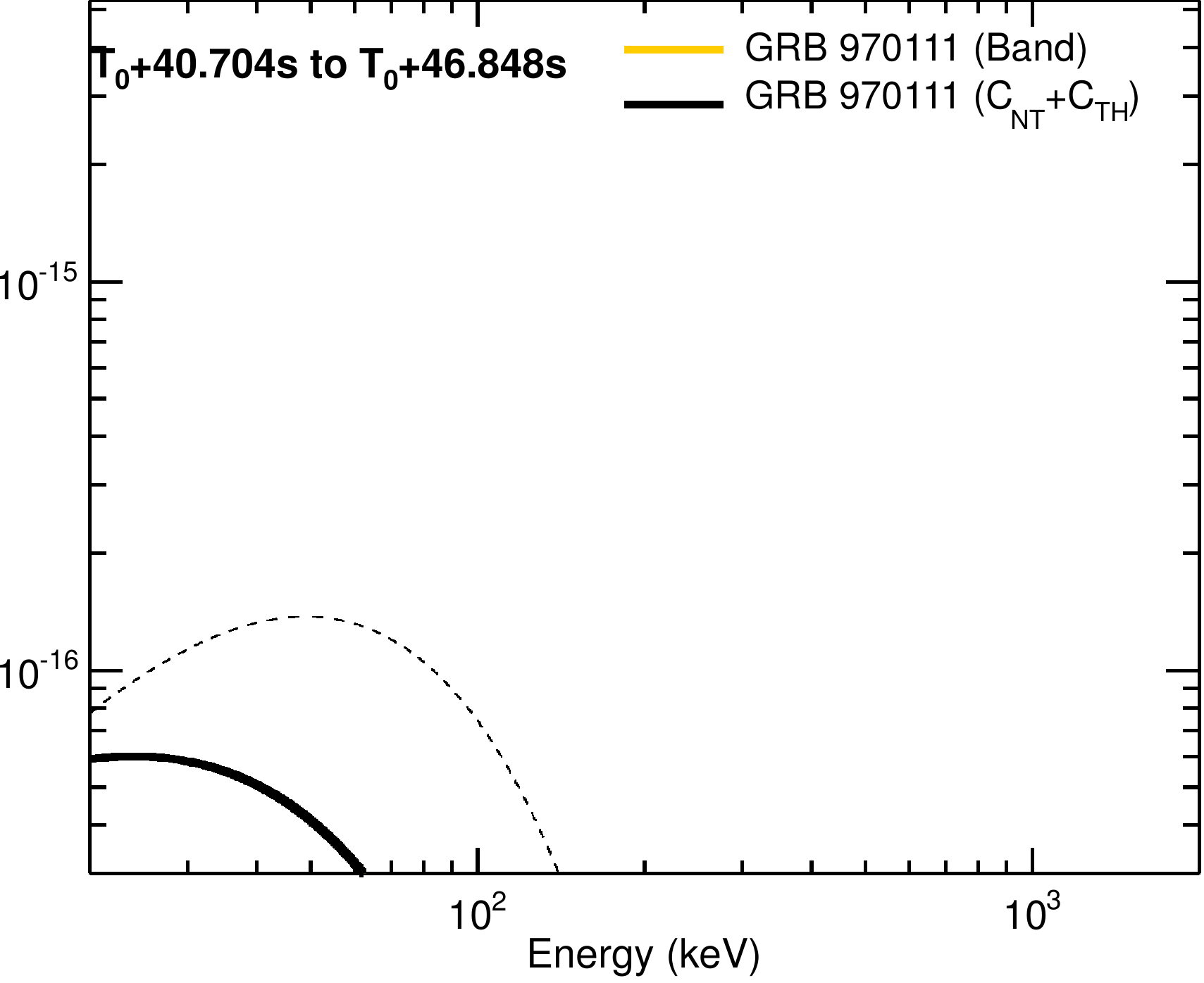}

\hspace{-6.cm}
\includegraphics[totalheight=0.195\textheight, clip]{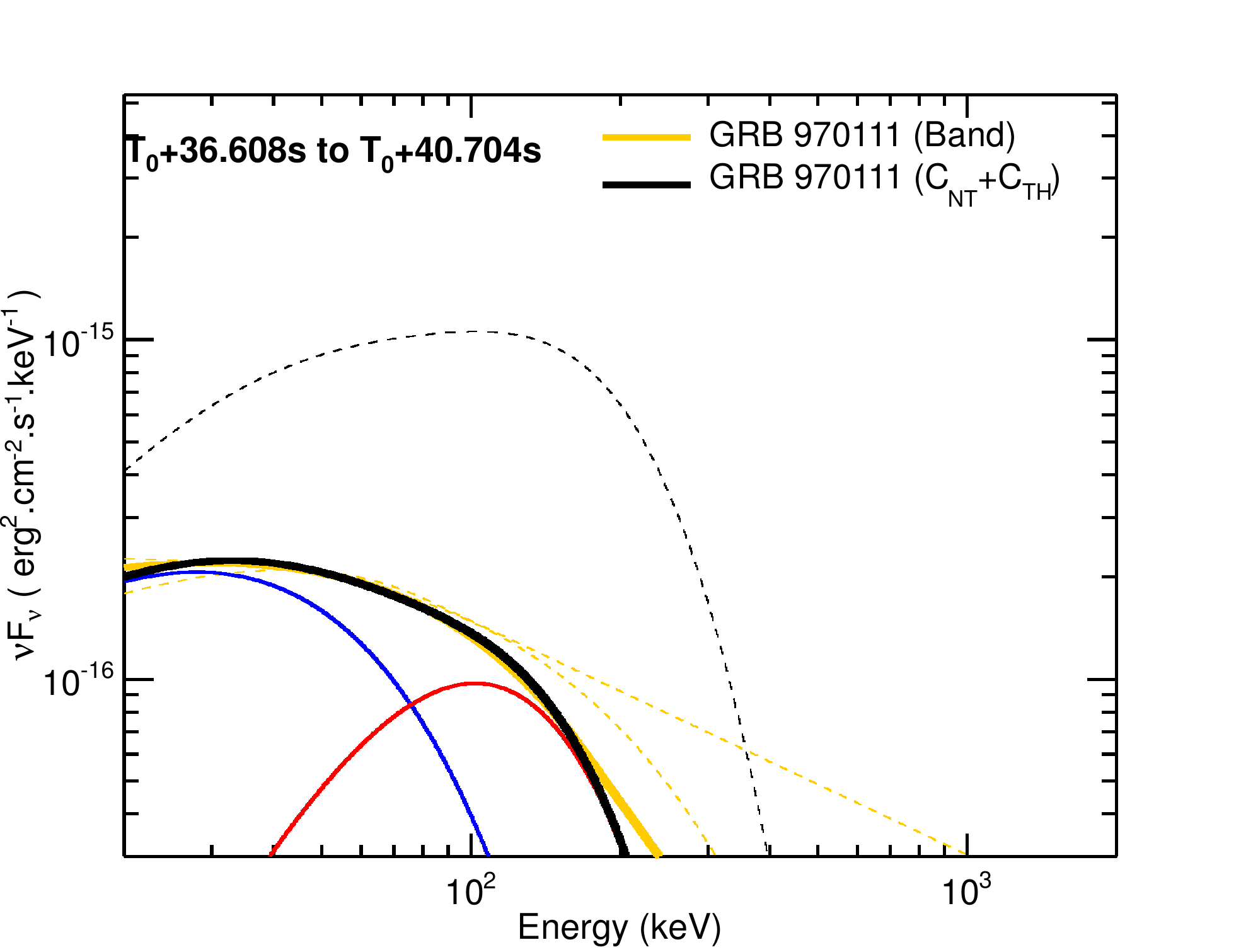}

\caption{\label{fig16}GRB~970111 : $\nu$F$_\nu$ spectra resulting from the fine-time analysis presented in Section~\ref{sec:trsa}. The solid yellow and black lines correspond to the best Band-only and C$_\mathrm{nTh}$+C$_\mathrm{Th}$ fits, respectively. The dashed yellow and black lines correspond to the 1--$\sigma$ confidence regions of the Band-only and C$_\mathrm{nTh}$+C$_\mathrm{Th}$ fits, respectively. The solid blue and red lines correspond to C$_\mathrm{nTh}$ and C$_\mathrm{Th}$ resulting from the best C$_\mathrm{nTh}$+C$_\mathrm{Th}$ fits (i.e., solid black line) to the data, respectively.}
\end{center}
\end{figure*}

\newpage

\begin{figure*}
\begin{center}
\includegraphics[totalheight=0.185\textheight, clip]{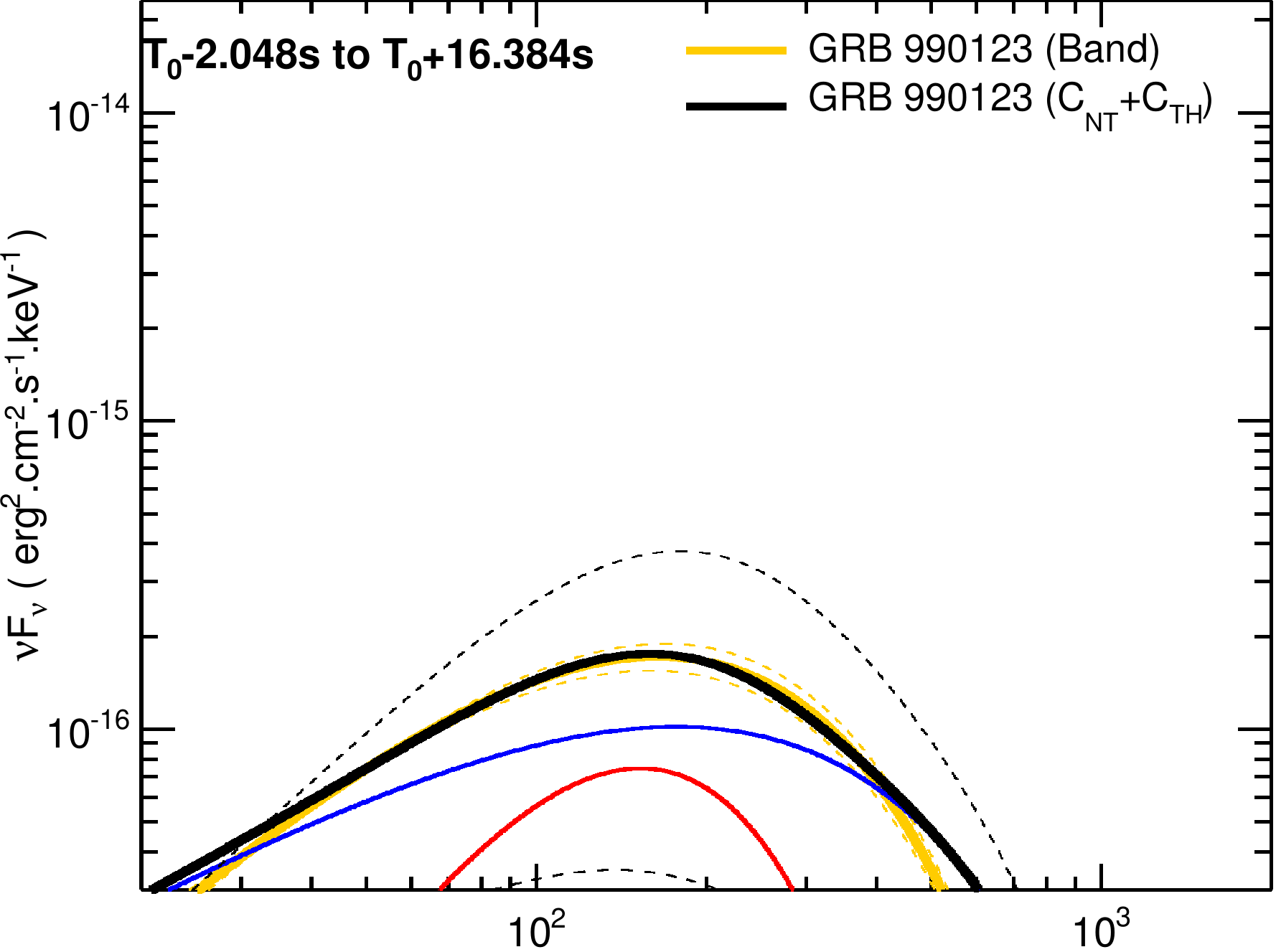}
\includegraphics[totalheight=0.185\textheight, clip]{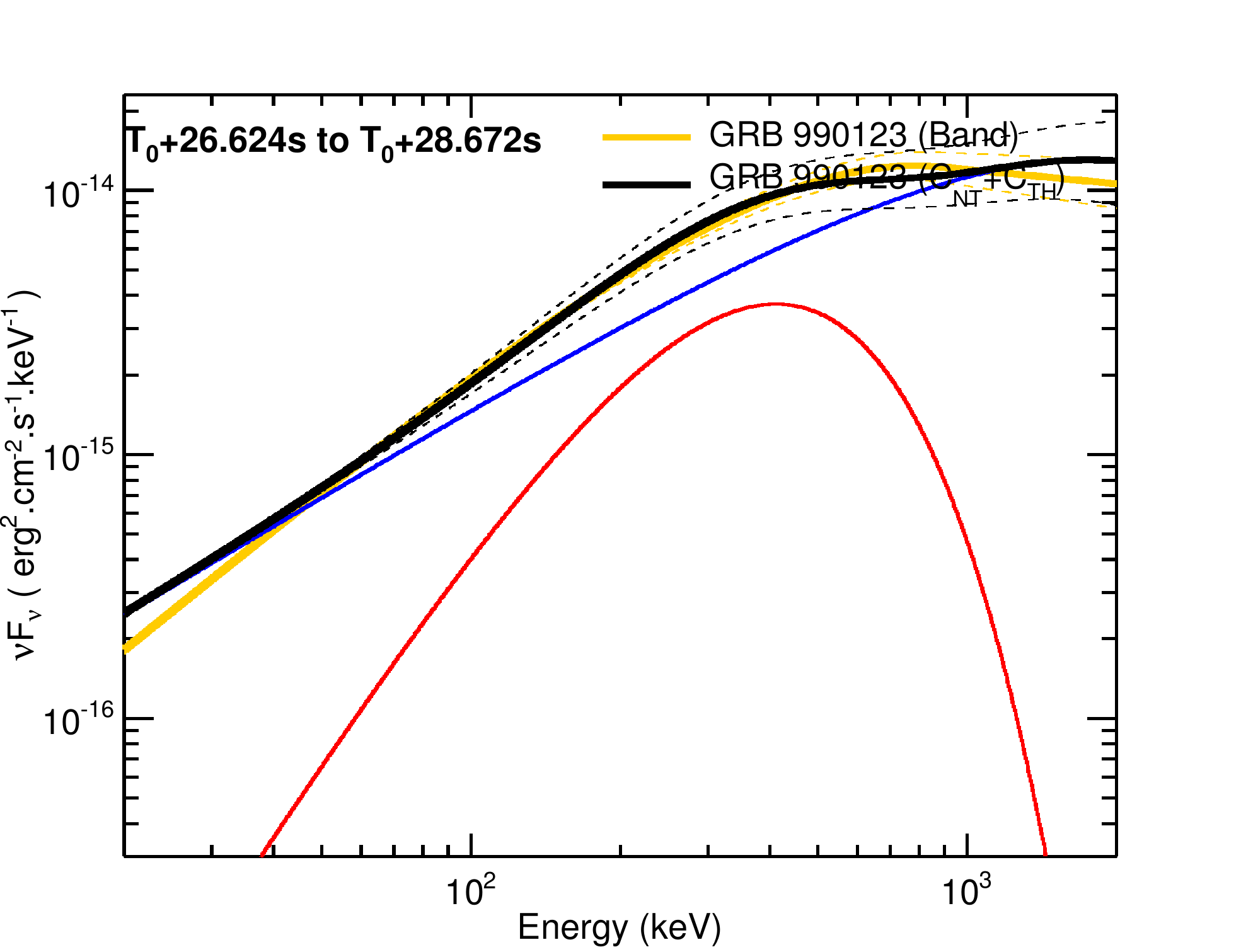}
\includegraphics[totalheight=0.185\textheight, clip]{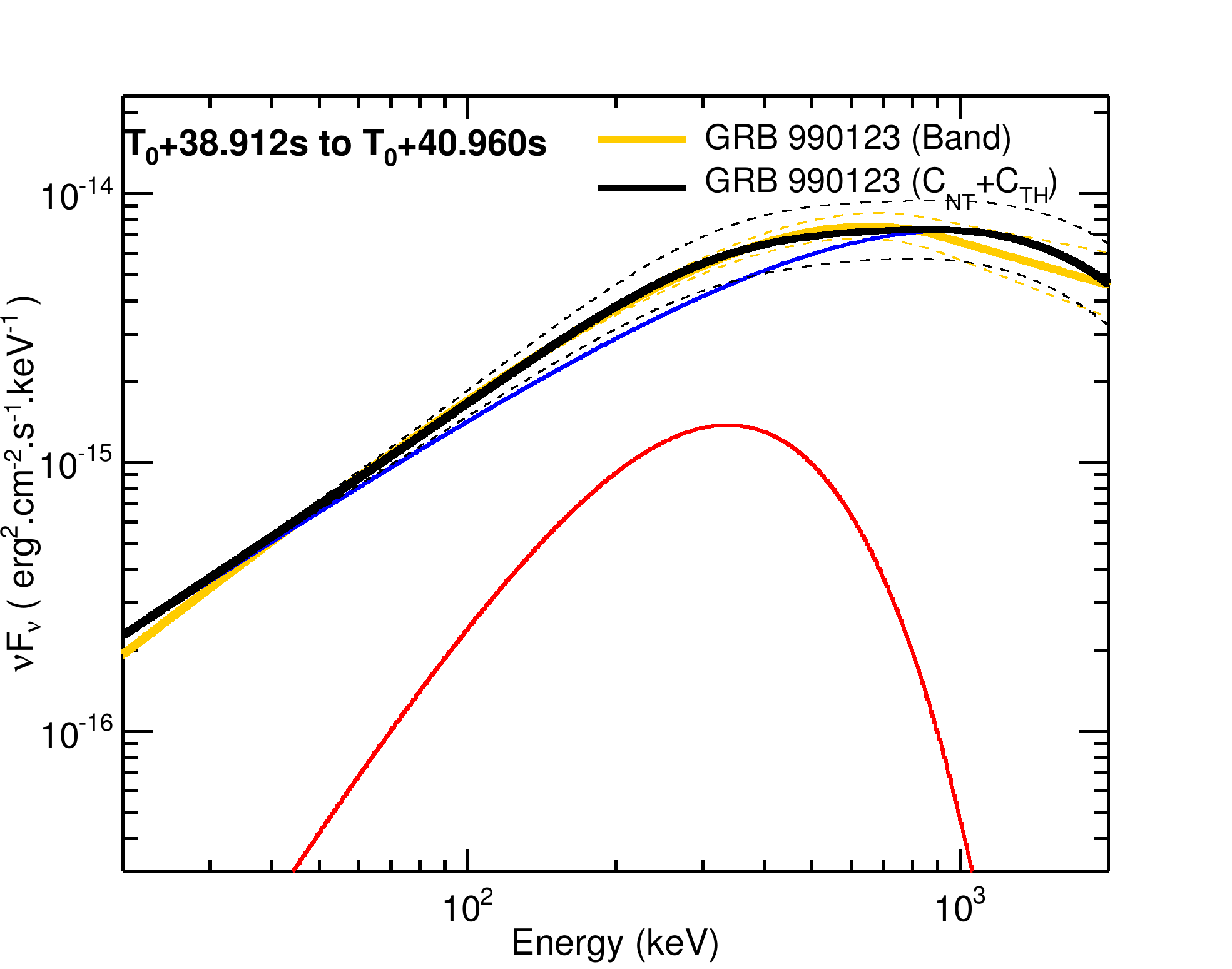}

\includegraphics[totalheight=0.185\textheight, clip]{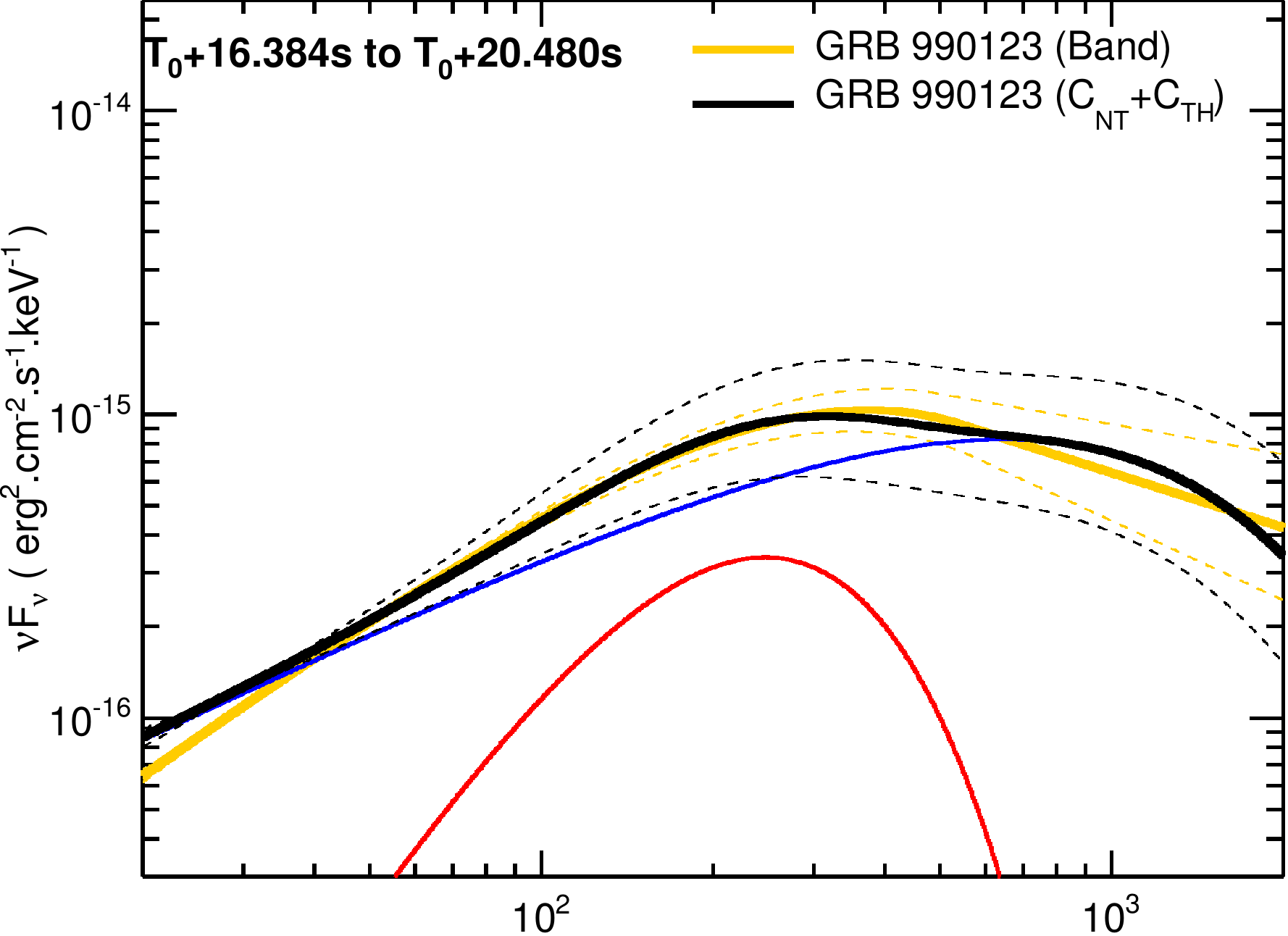}
\includegraphics[totalheight=0.185\textheight, clip]{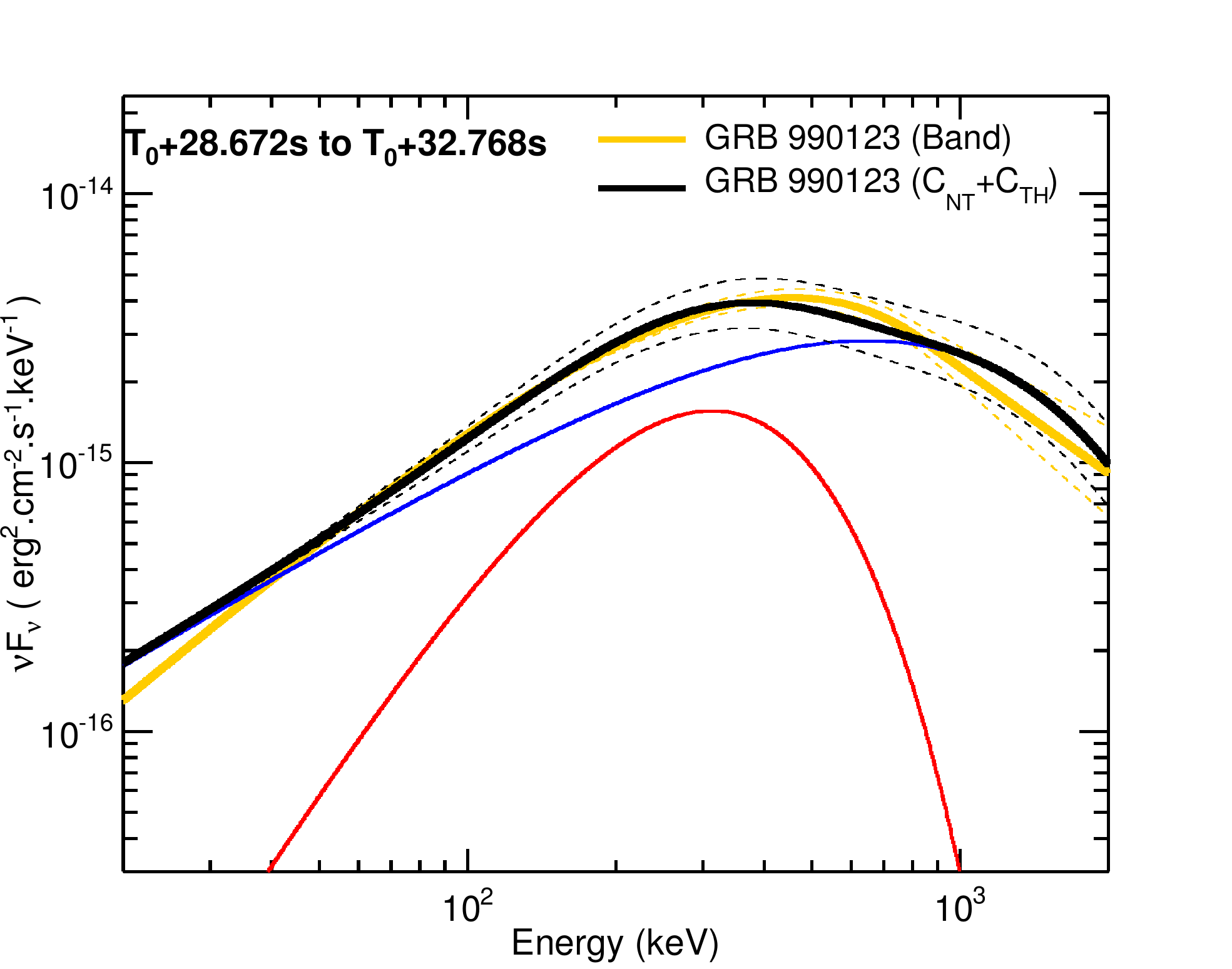}
\includegraphics[totalheight=0.185\textheight, clip]{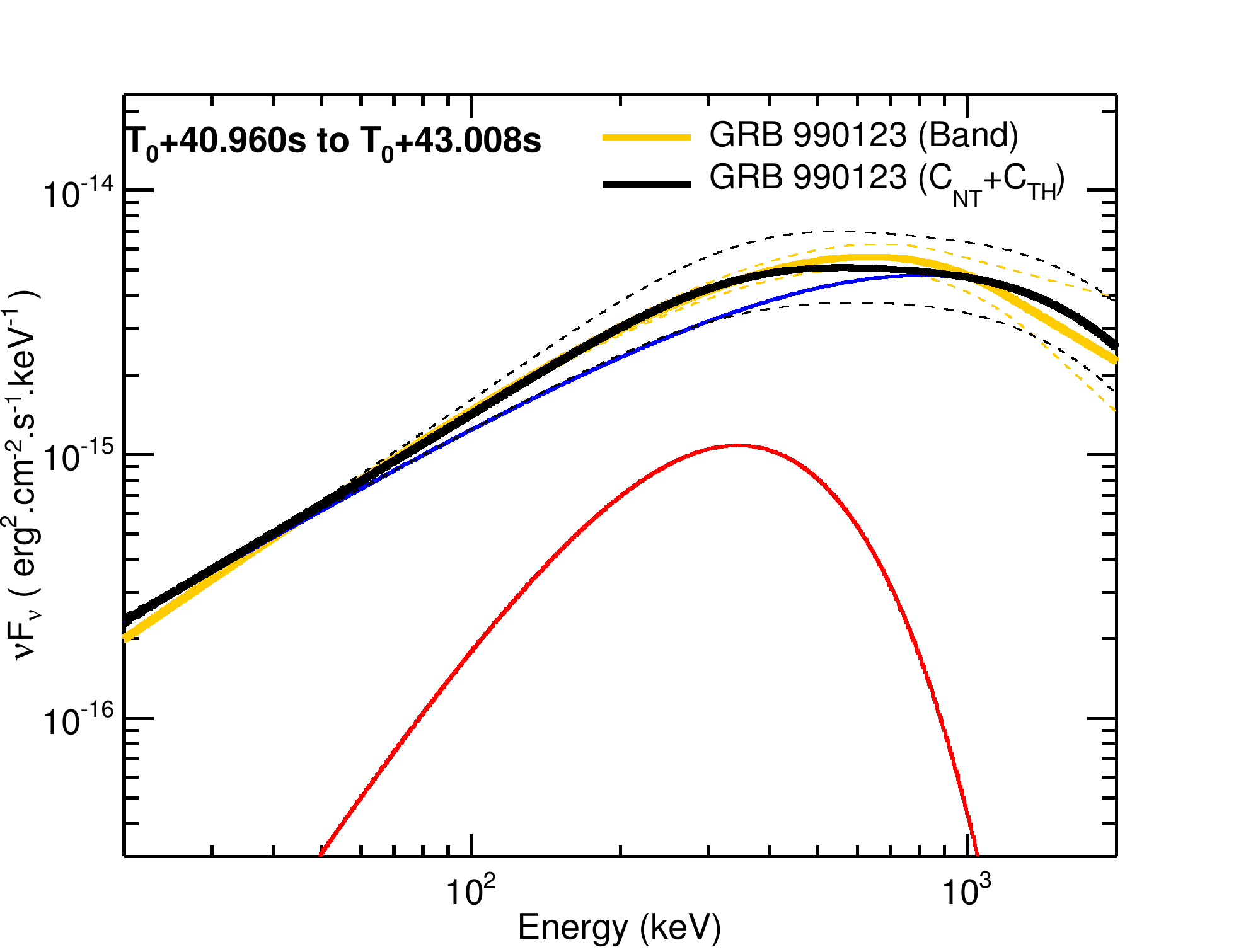}

\includegraphics[totalheight=0.185\textheight, clip]{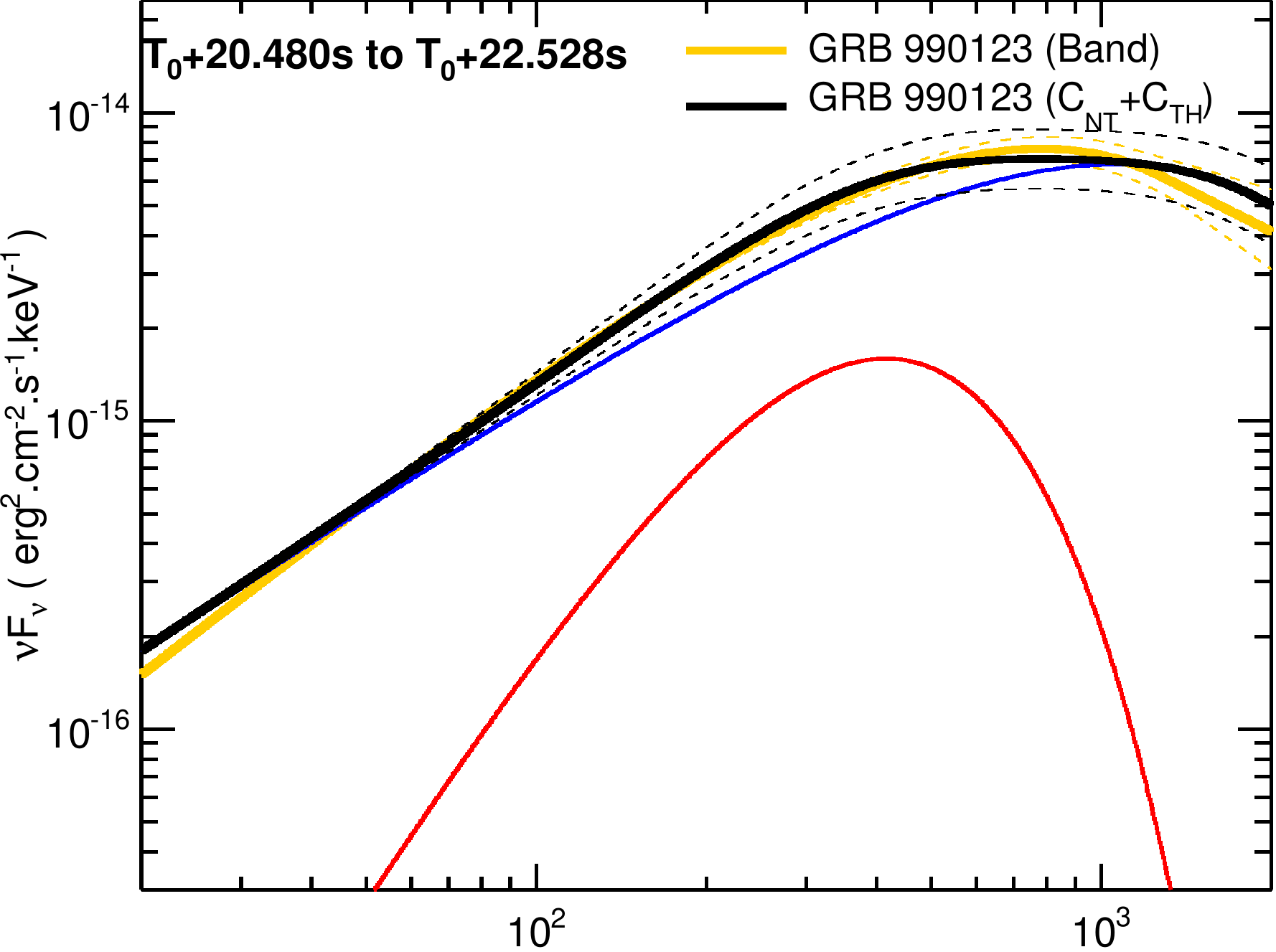}
\includegraphics[totalheight=0.185\textheight, clip]{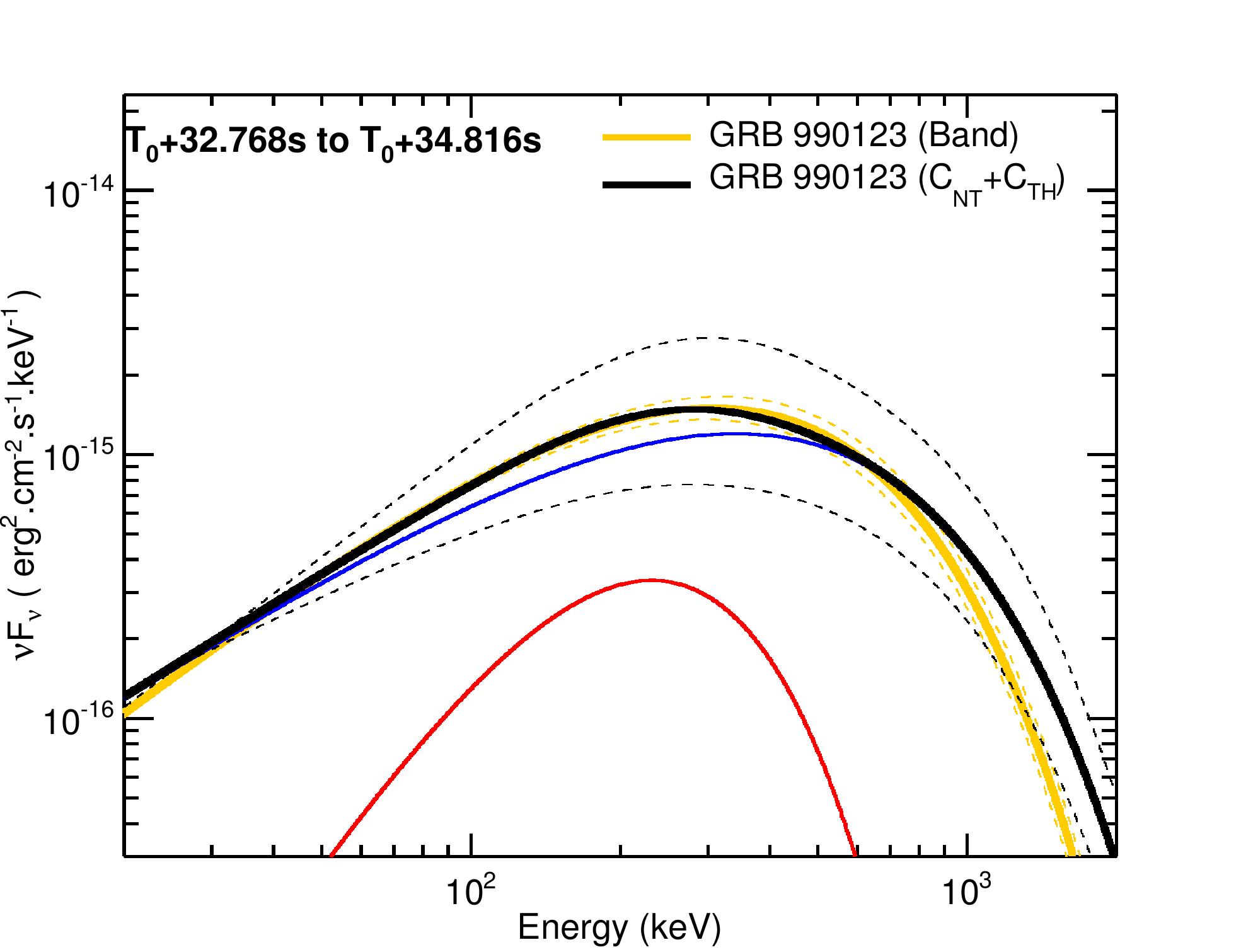}
\includegraphics[totalheight=0.185\textheight, clip]{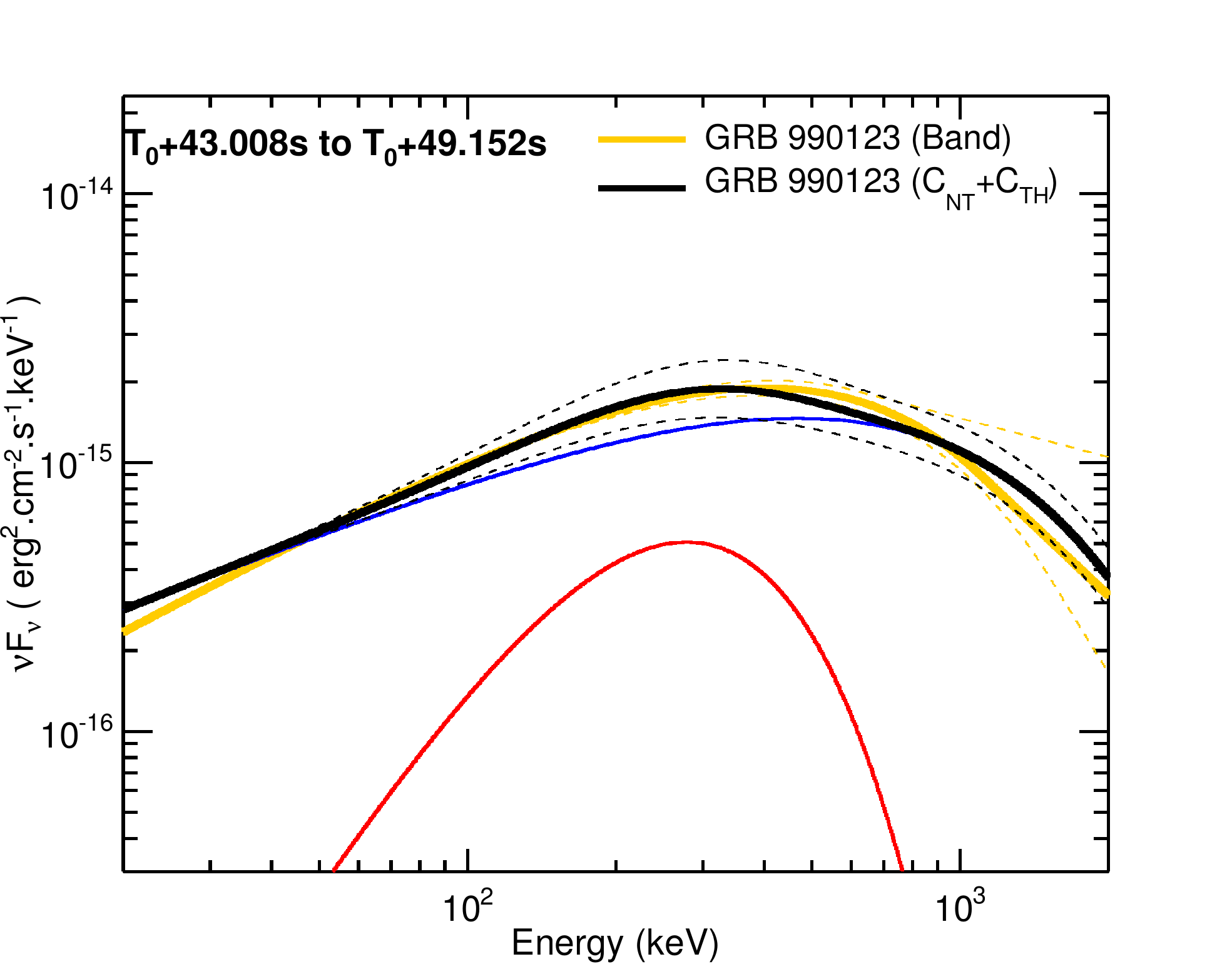}

\includegraphics[totalheight=0.185\textheight, clip]{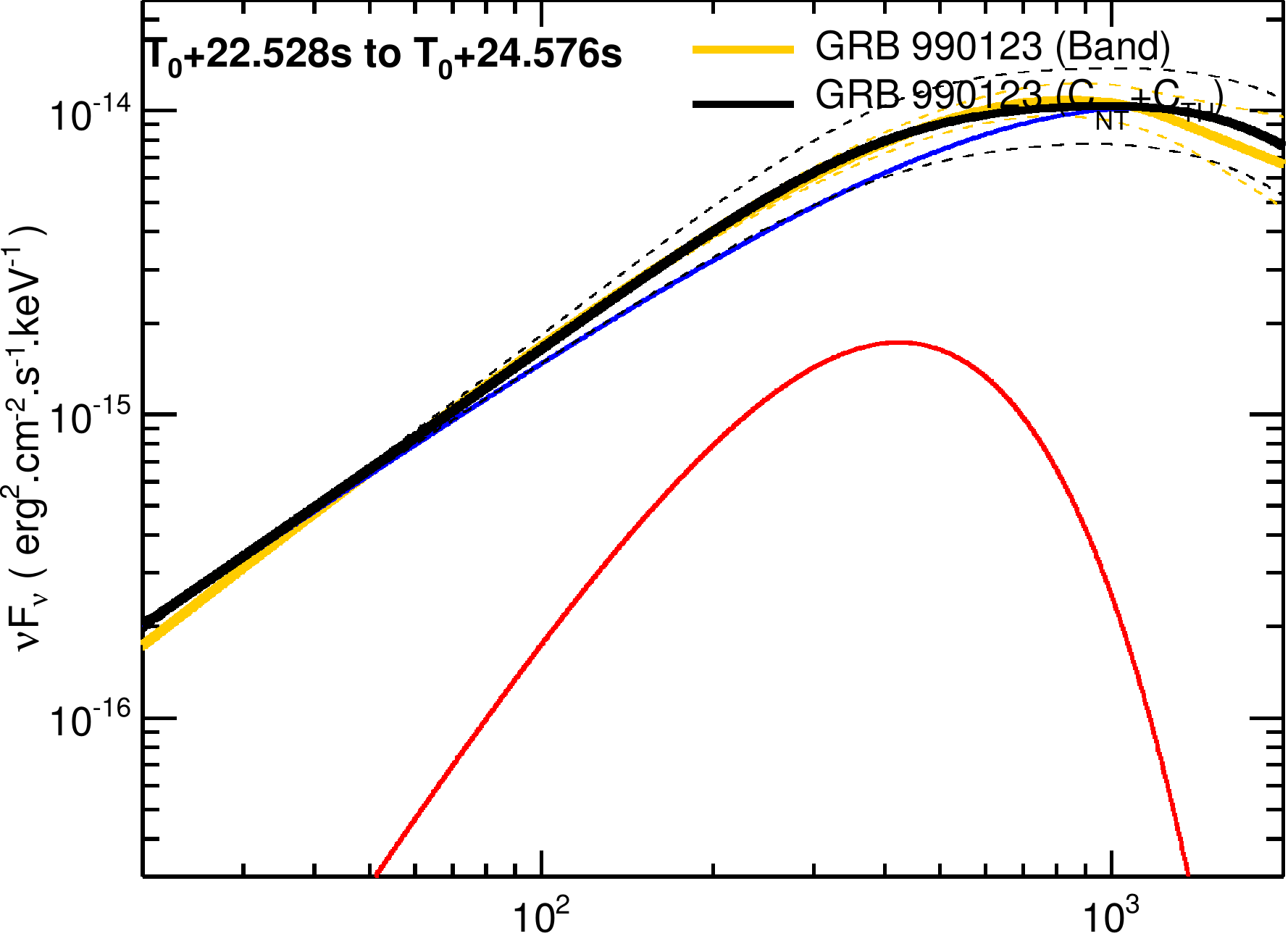}
\includegraphics[totalheight=0.185\textheight, clip]{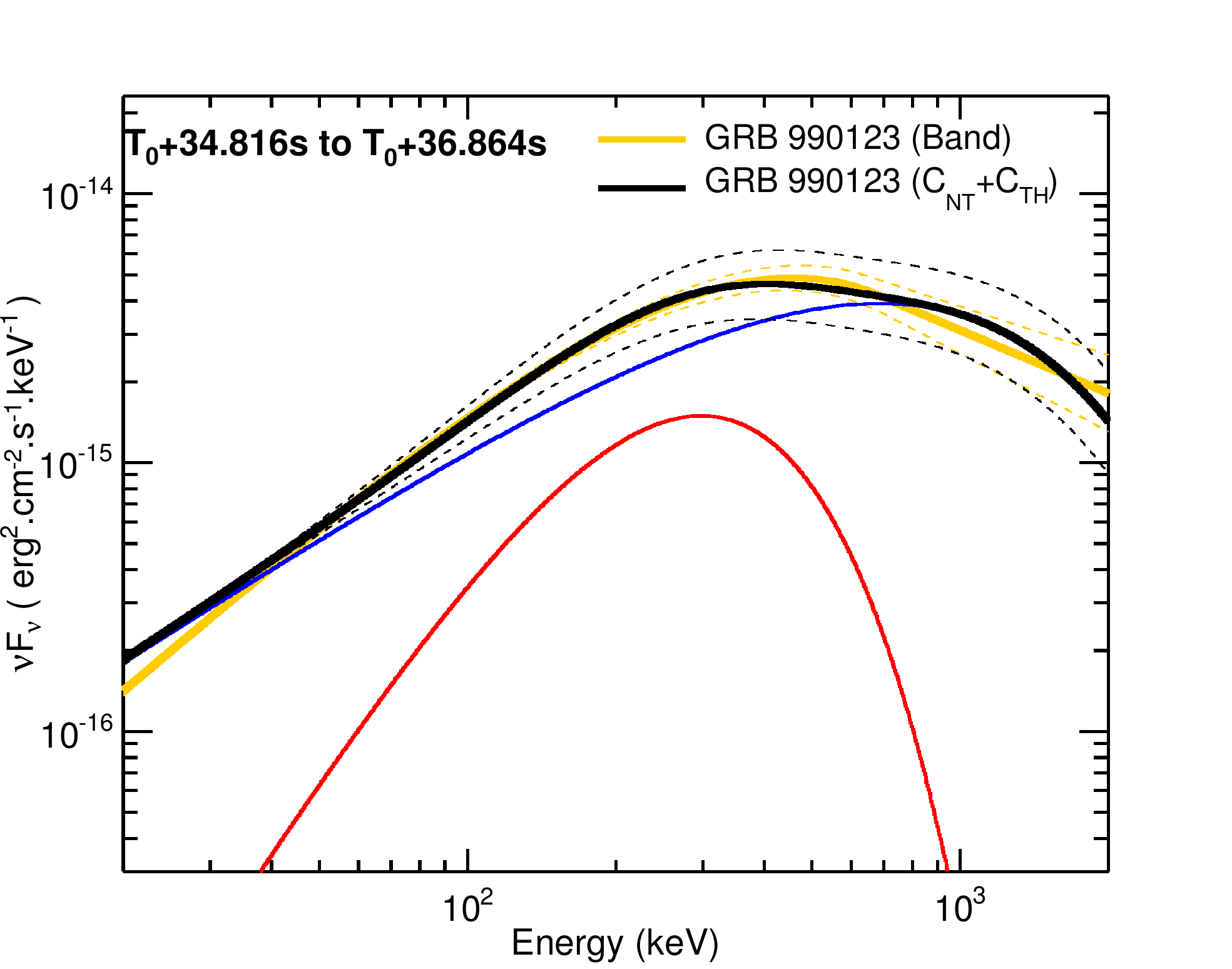}
\includegraphics[totalheight=0.185\textheight, clip]{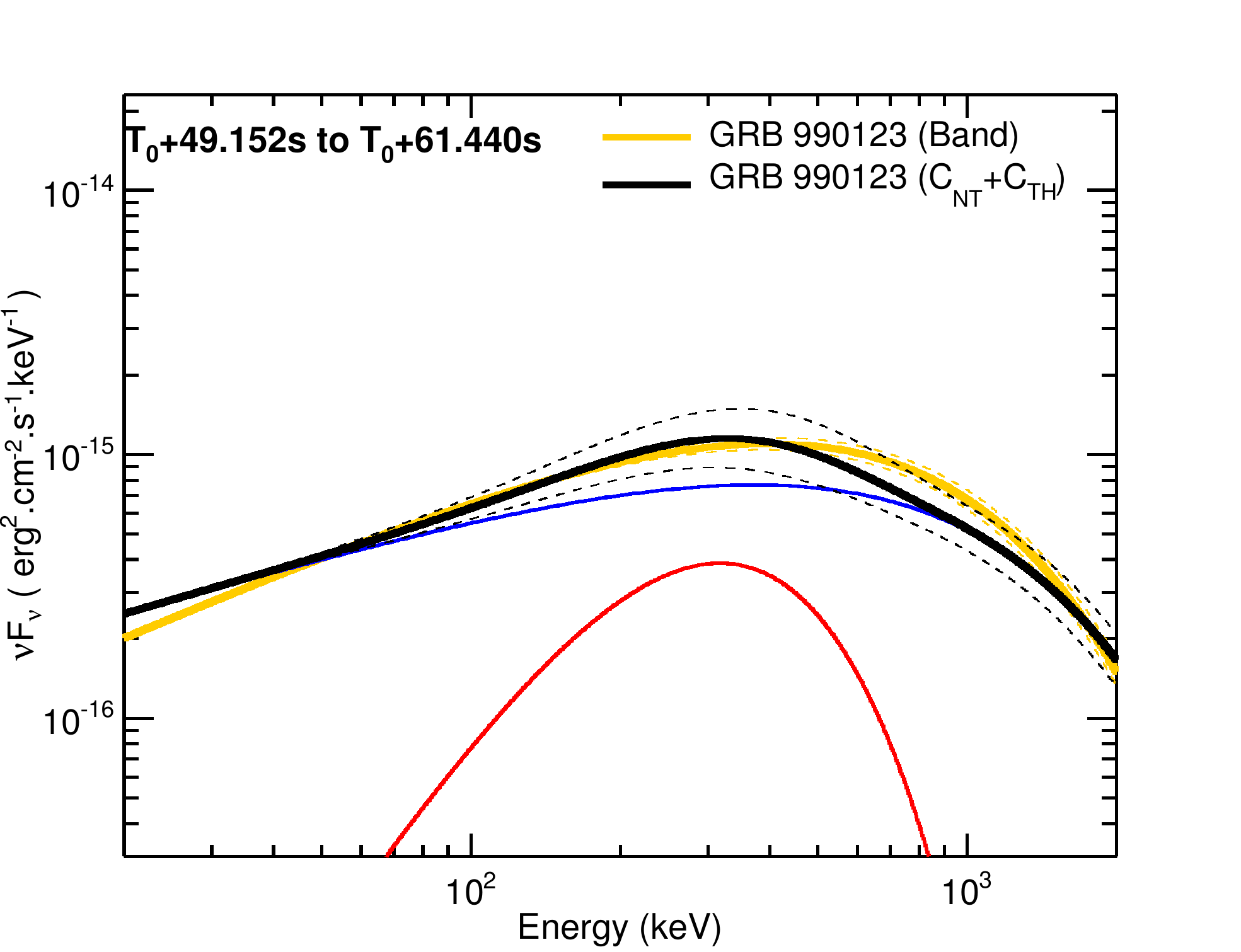}

\includegraphics[totalheight=0.195\textheight, clip]{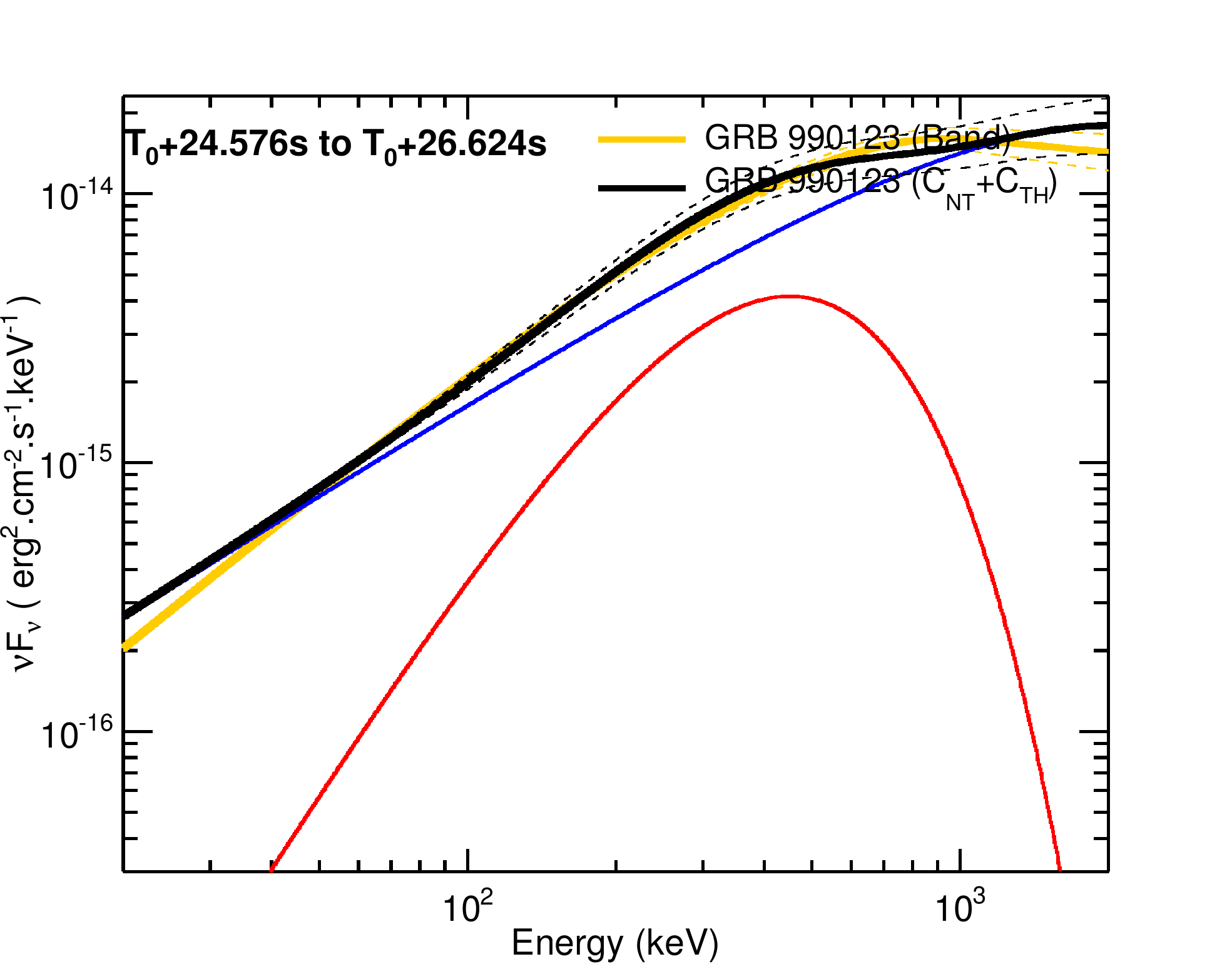}
\includegraphics[totalheight=0.195\textheight, clip]{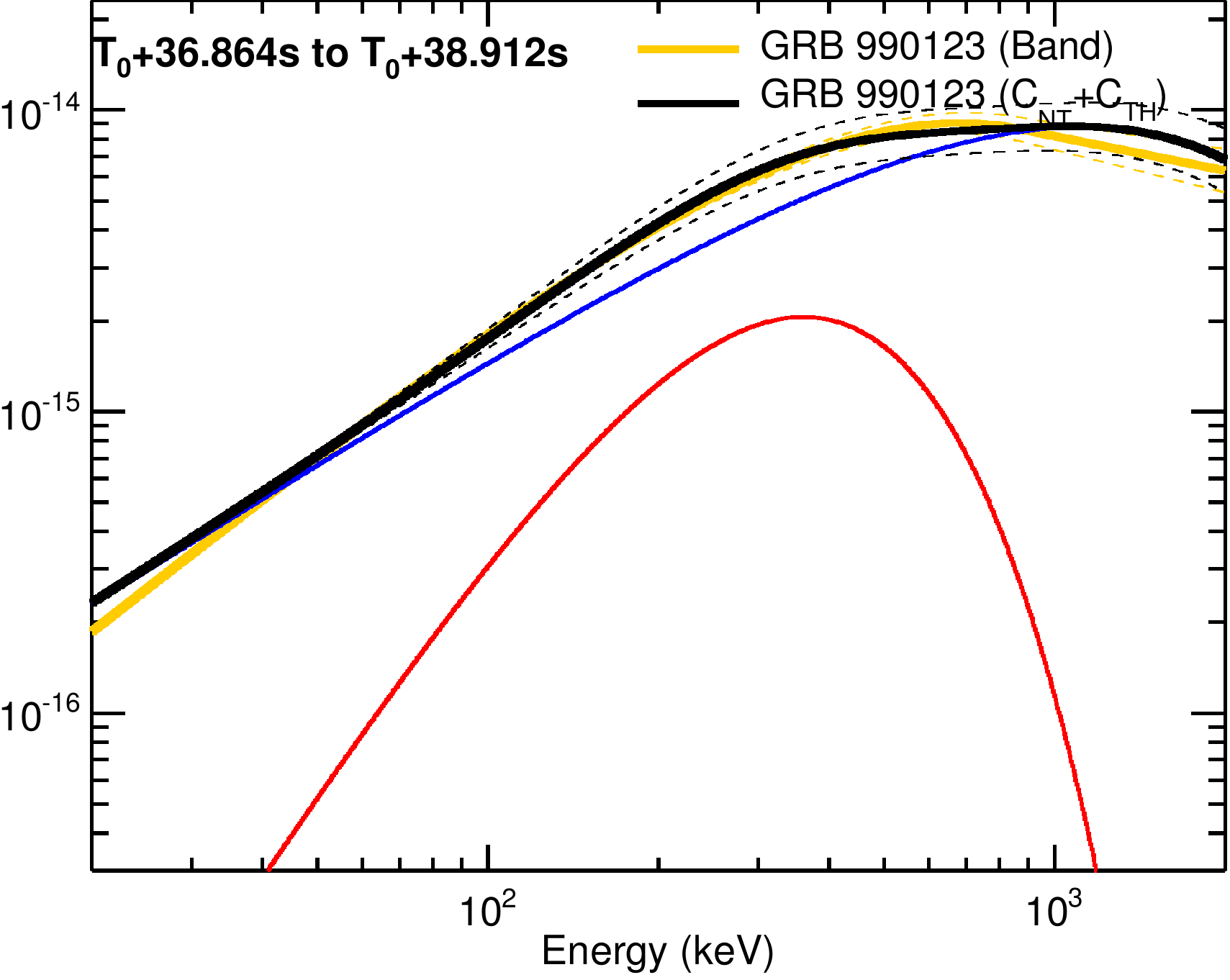}
\includegraphics[totalheight=0.195\textheight, clip]{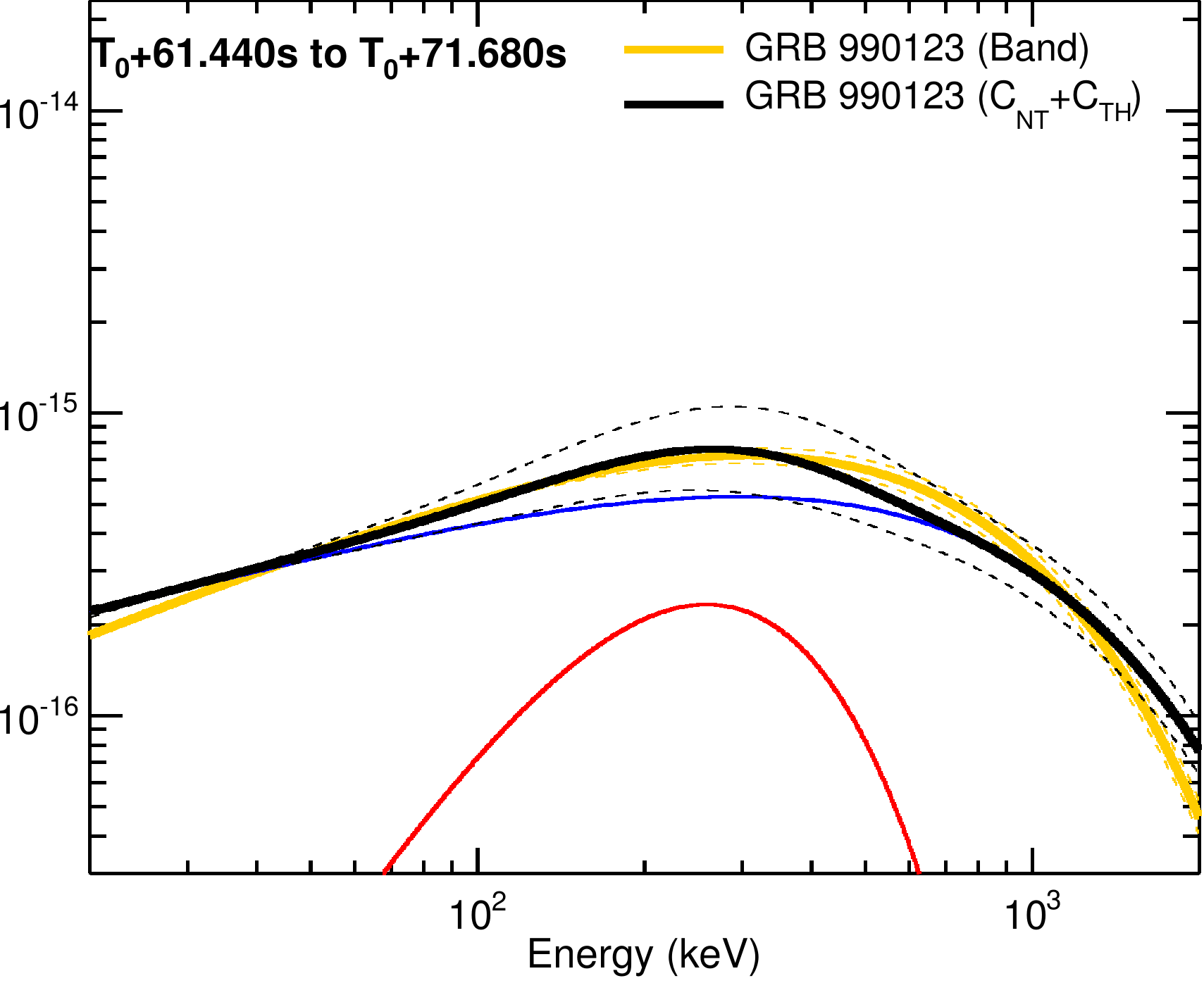}

\end{center}
\end{figure*}

\newpage

\begin{figure*}
\begin{center}
\includegraphics[totalheight=0.185\textheight, clip]{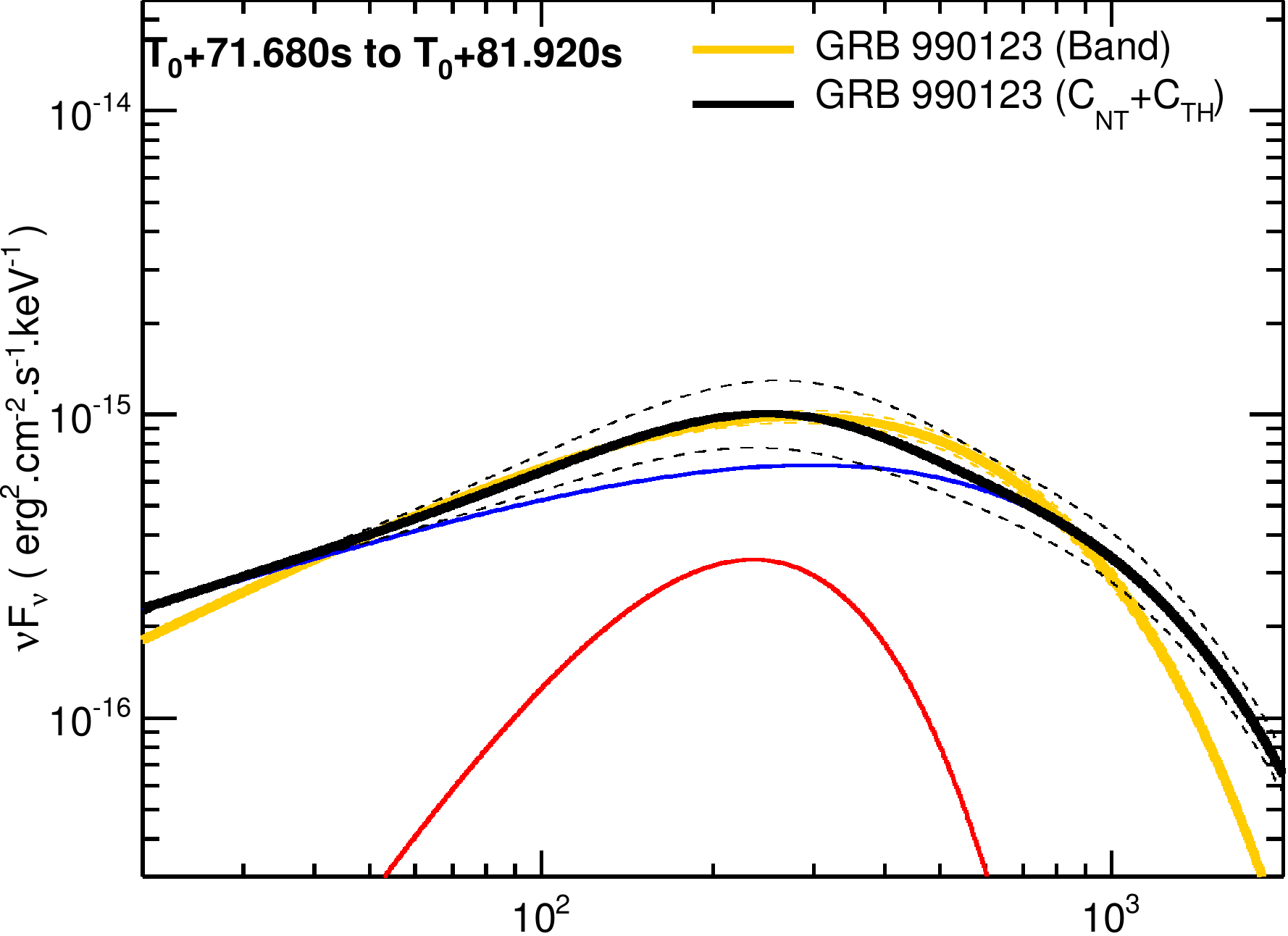}
\includegraphics[totalheight=0.185\textheight, clip]{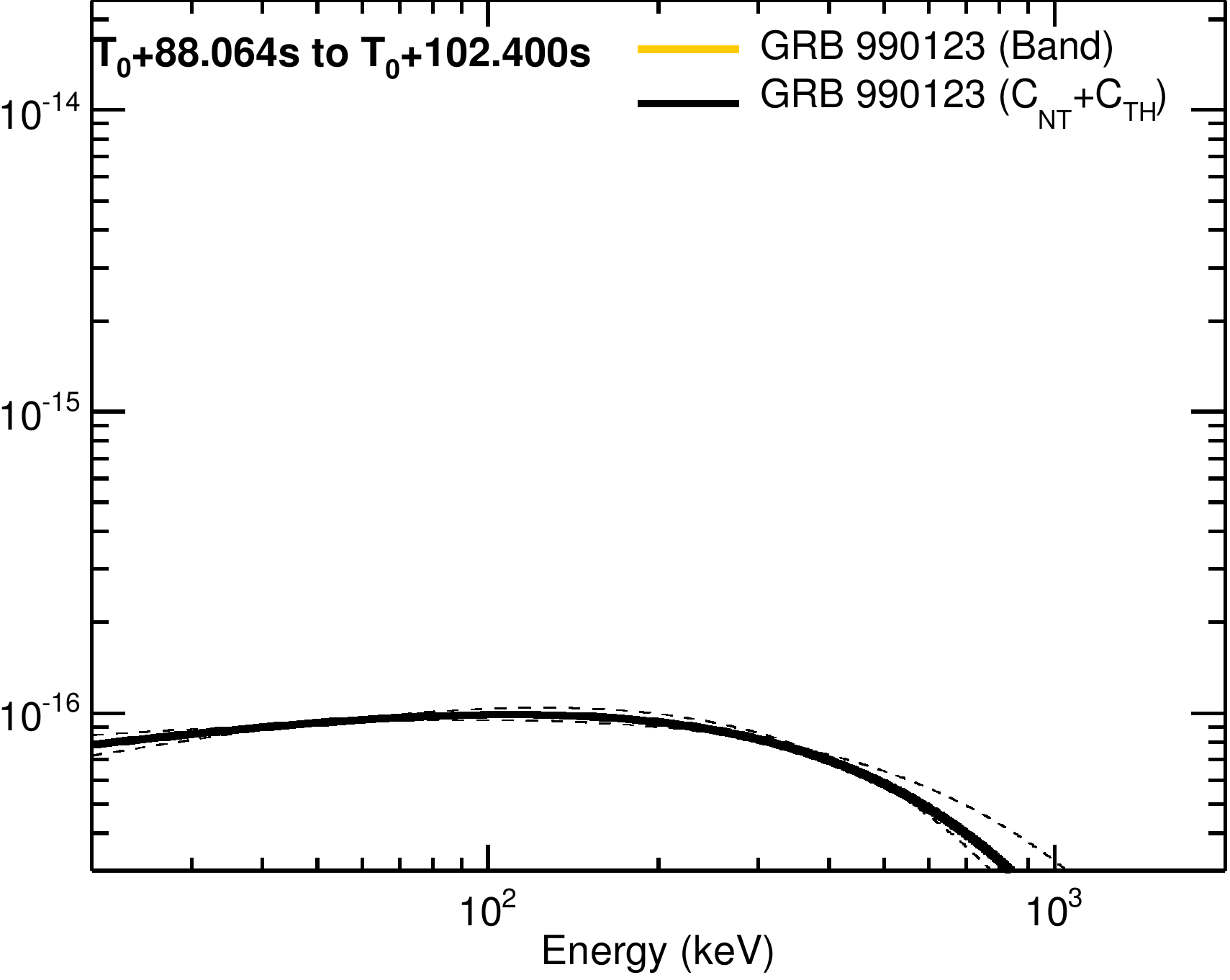}

\hspace{-6.cm}
\includegraphics[totalheight=0.195\textheight, clip]{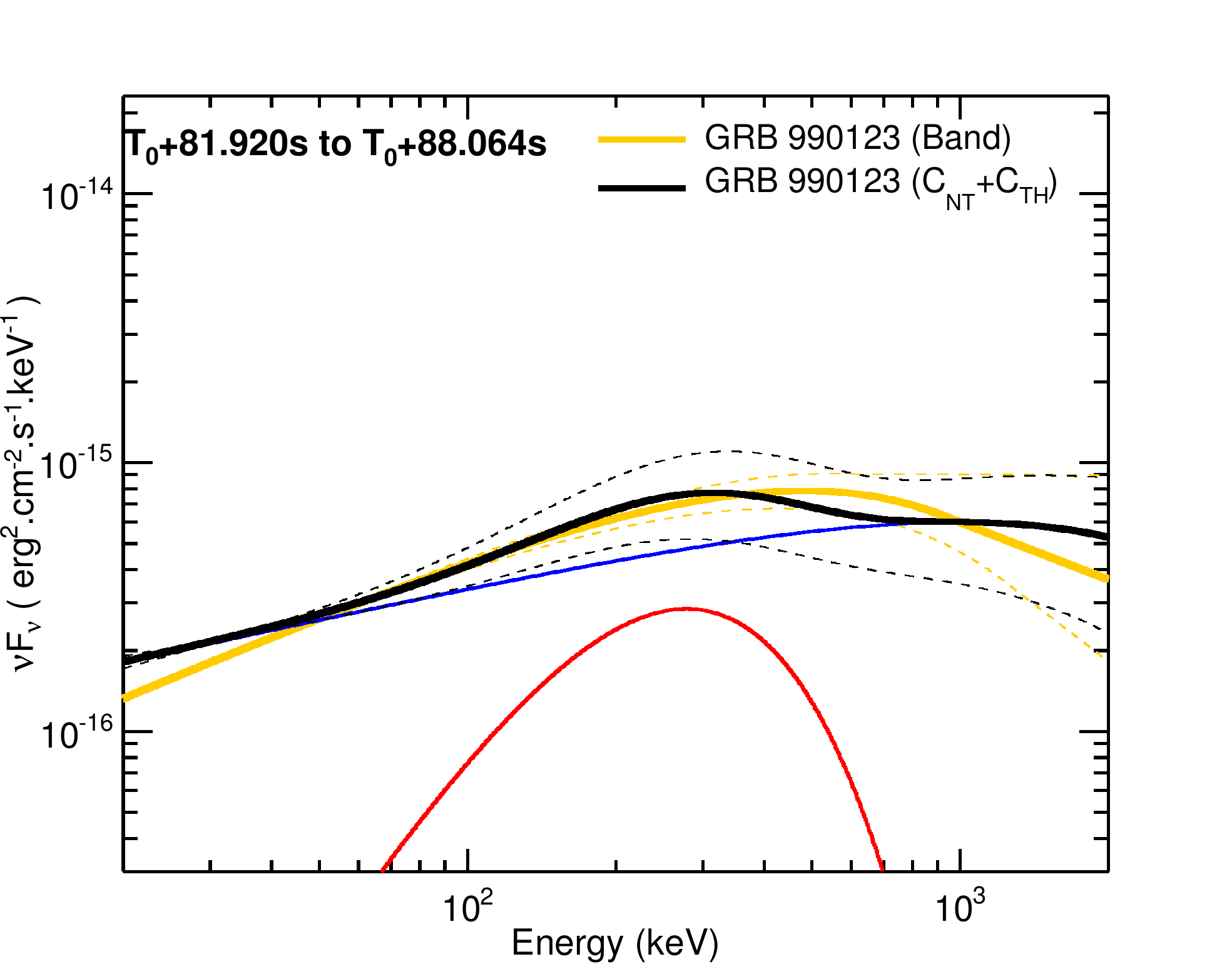}

\caption{\label{fig17}GRB~990123 : $\nu$F$_\nu$ spectra resulting from the fine-time analysis presented in Section~\ref{sec:trsa}. The solid yellow and black lines correspond to the best Band-only and C$_\mathrm{nTh}$+C$_\mathrm{Th}$ fits, respectively. The dashed yellow and black lines correspond to the 1--$\sigma$ confidence regions of the Band-only and C$_\mathrm{nTh}$+C$_\mathrm{Th}$ fits, respectively. The solid blue and red lines correspond to C$_\mathrm{nTh}$ and C$_\mathrm{Th}$ resulting from the best C$_\mathrm{nTh}$+C$_\mathrm{Th}$ fits (i.e., solid black line) to the data, respectively.}
\end{center}
\end{figure*}

\newpage

\begin{figure*}
\begin{center}
\includegraphics[totalheight=0.185\textheight, clip]{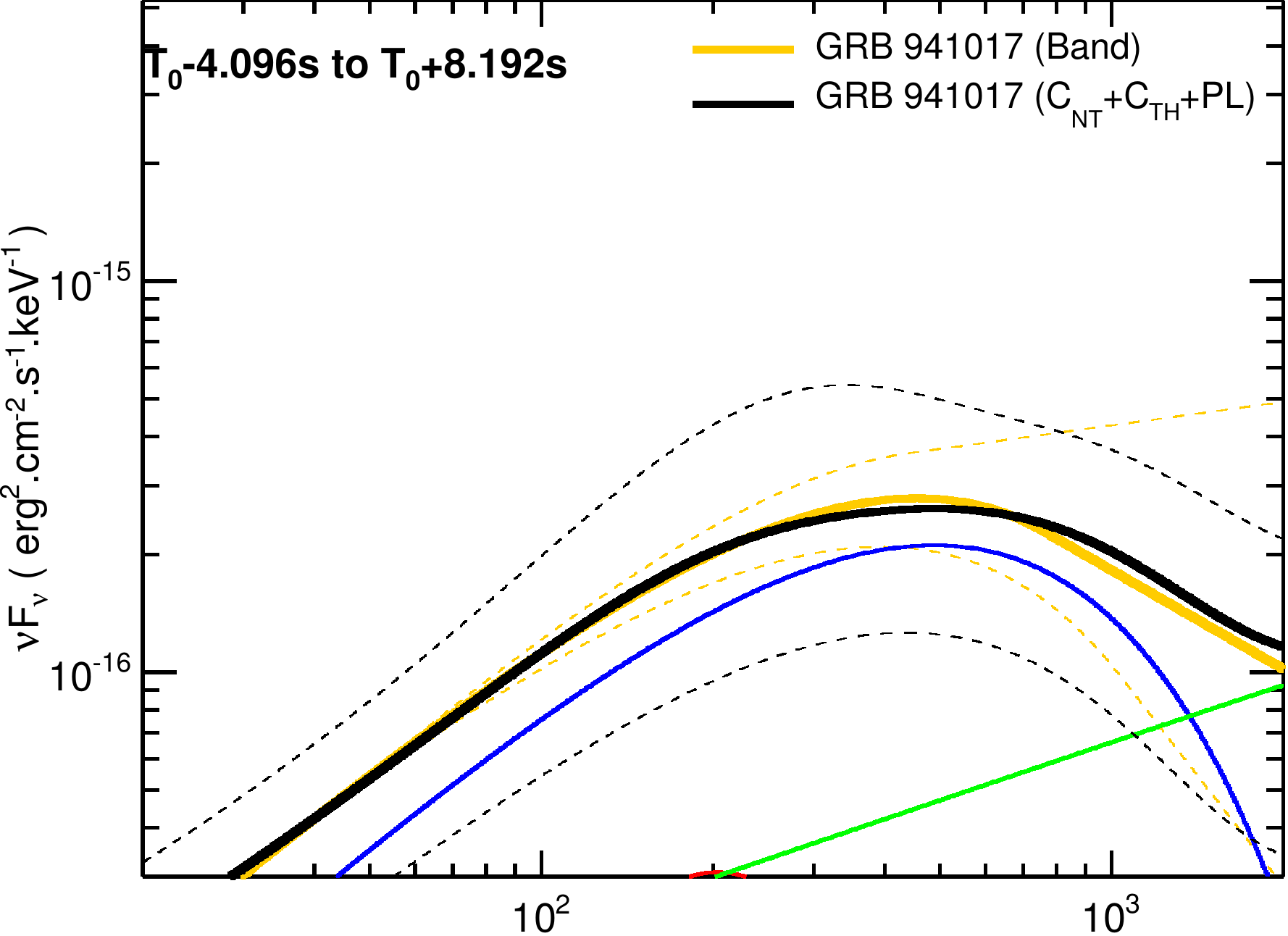}
\includegraphics[totalheight=0.185\textheight, clip]{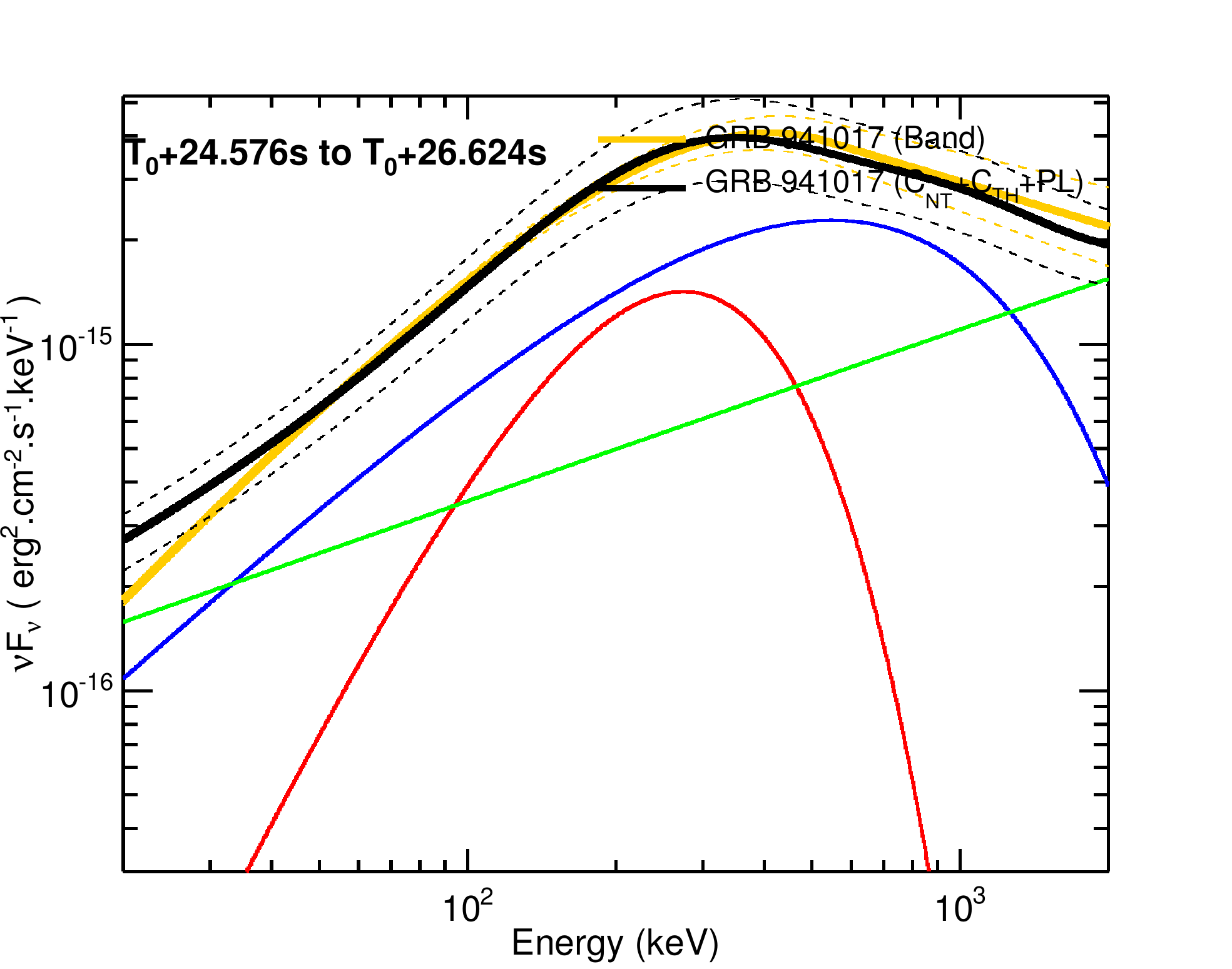}
\includegraphics[totalheight=0.185\textheight, clip]{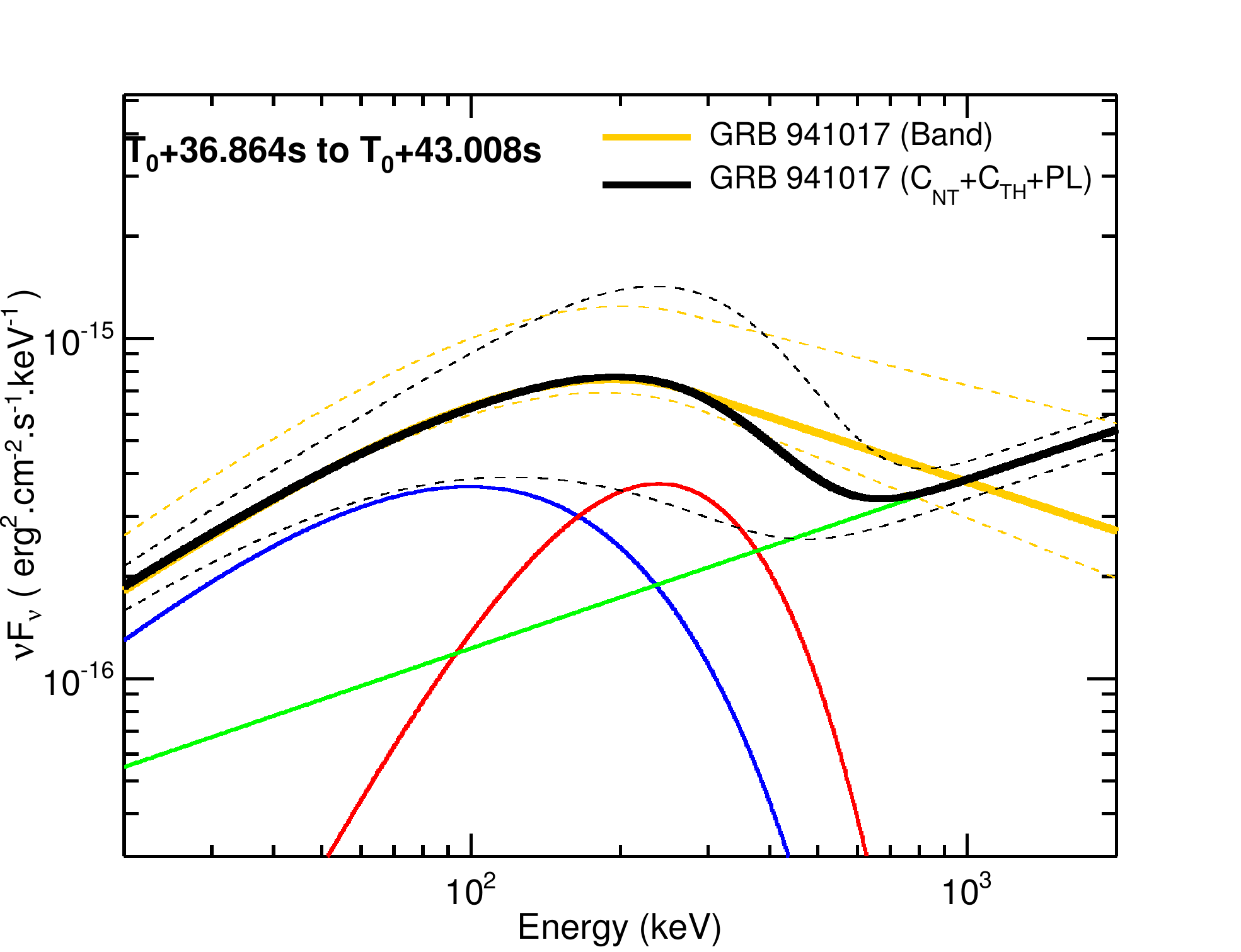}

\includegraphics[totalheight=0.185\textheight, clip]{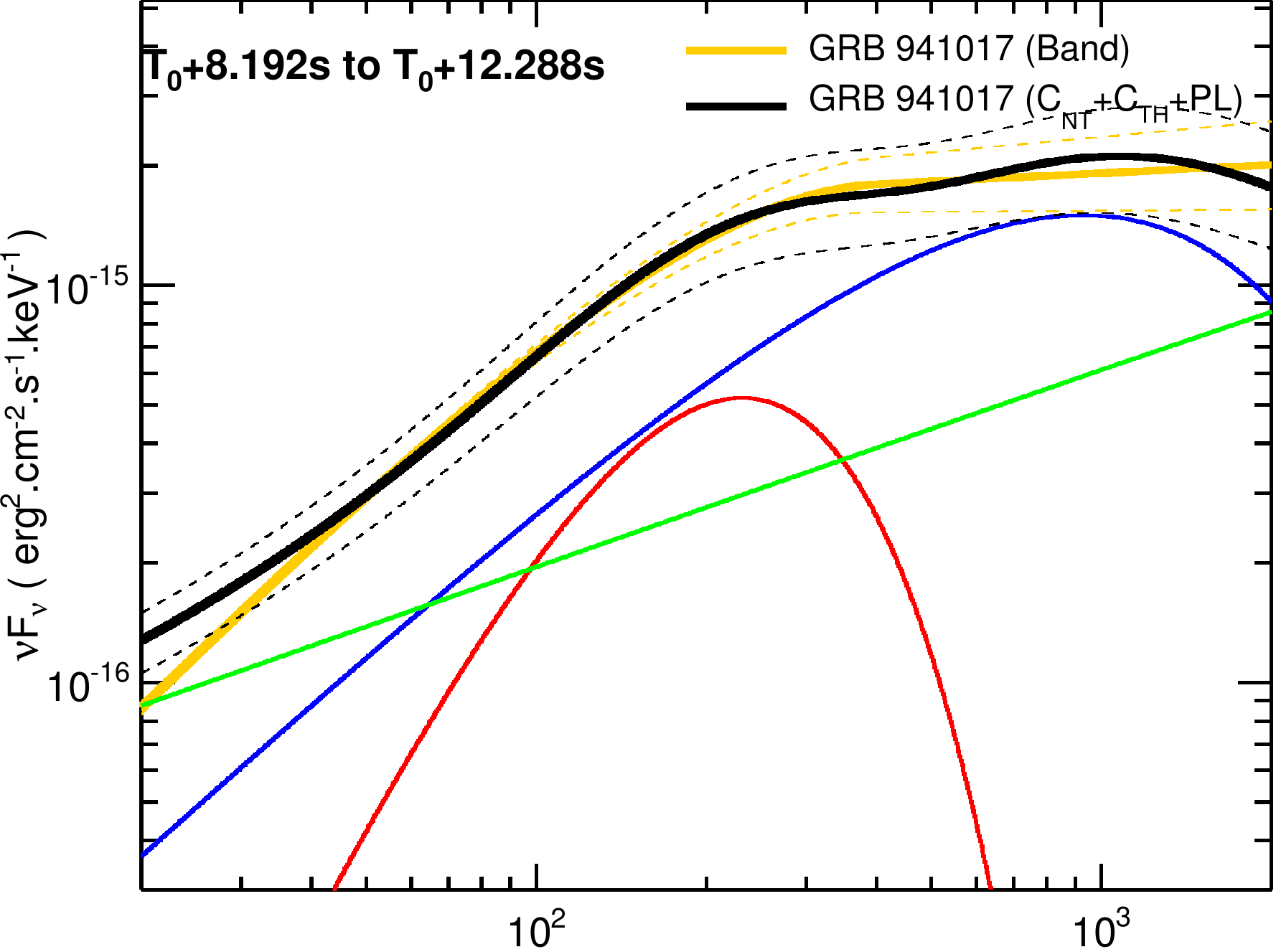}
\includegraphics[totalheight=0.185\textheight, clip]{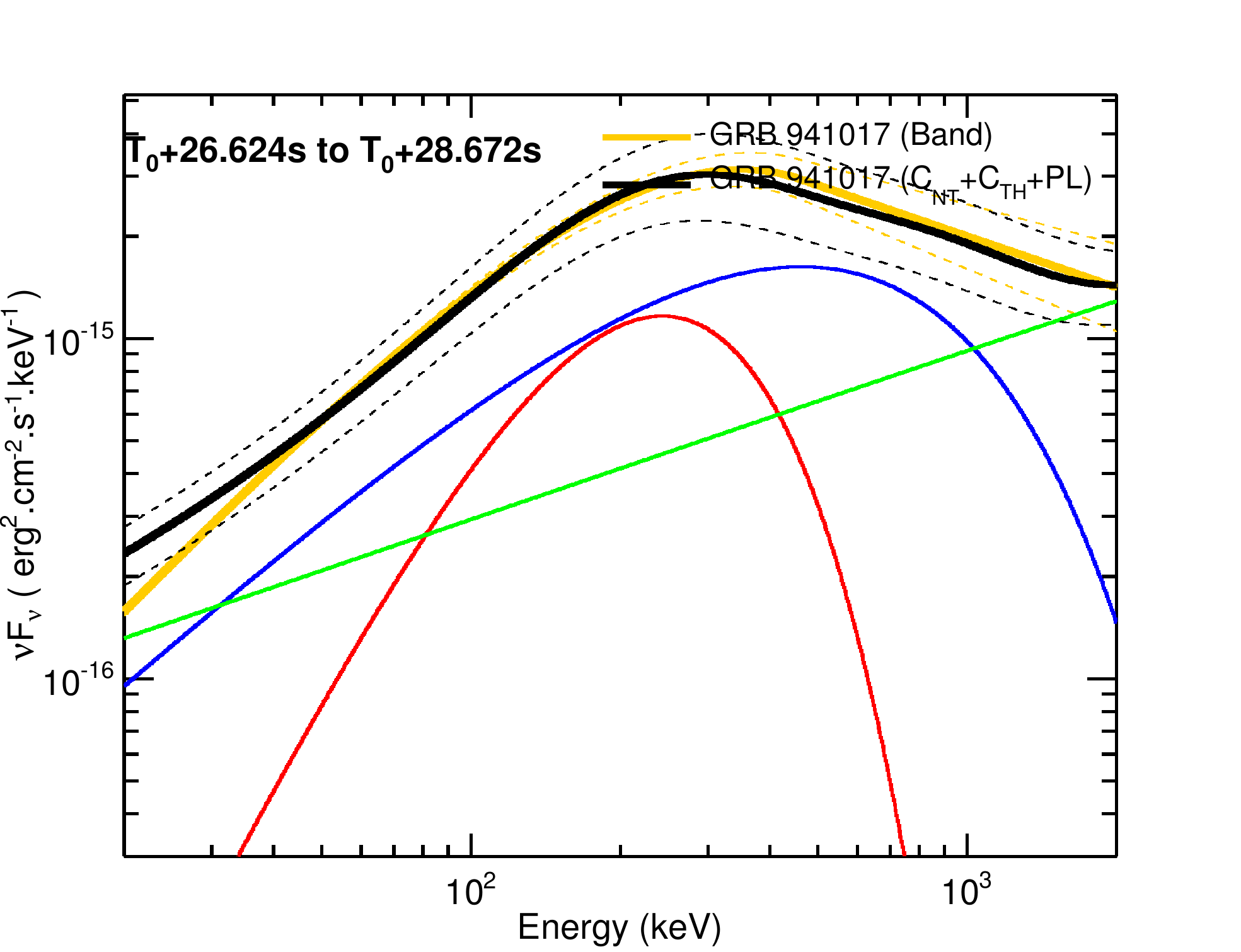}
\includegraphics[totalheight=0.185\textheight, clip]{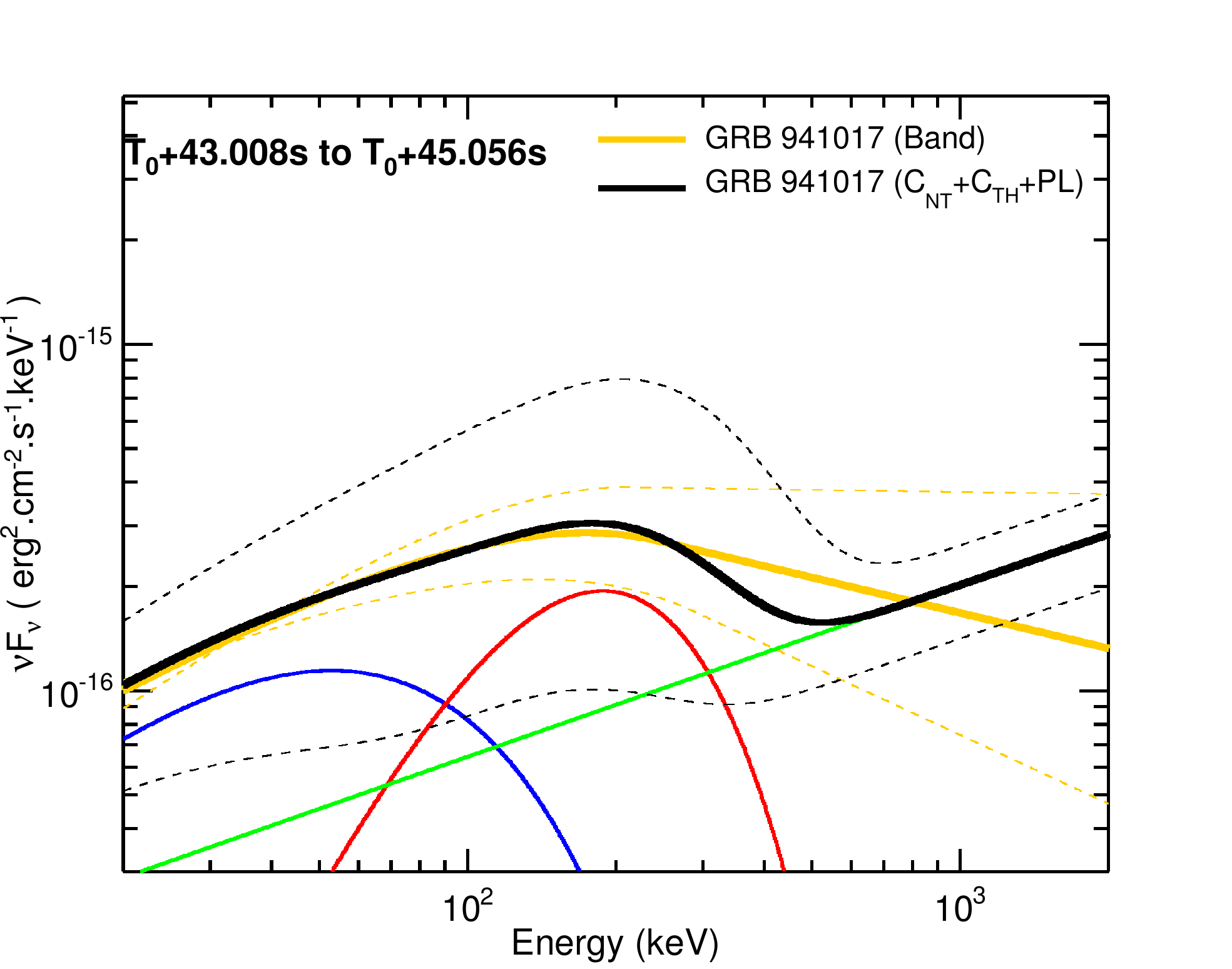}

\includegraphics[totalheight=0.185\textheight, clip]{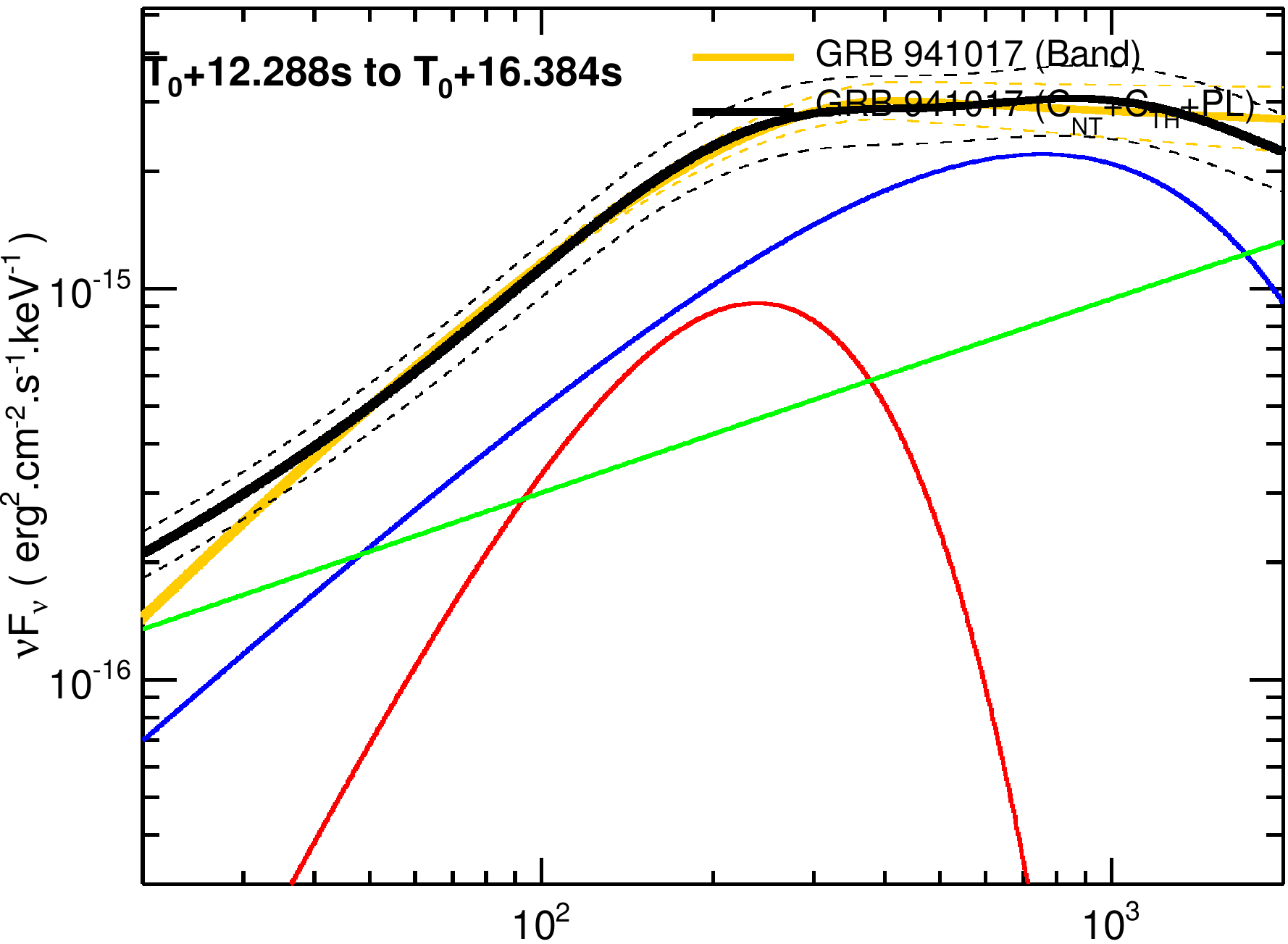}
\includegraphics[totalheight=0.185\textheight, clip]{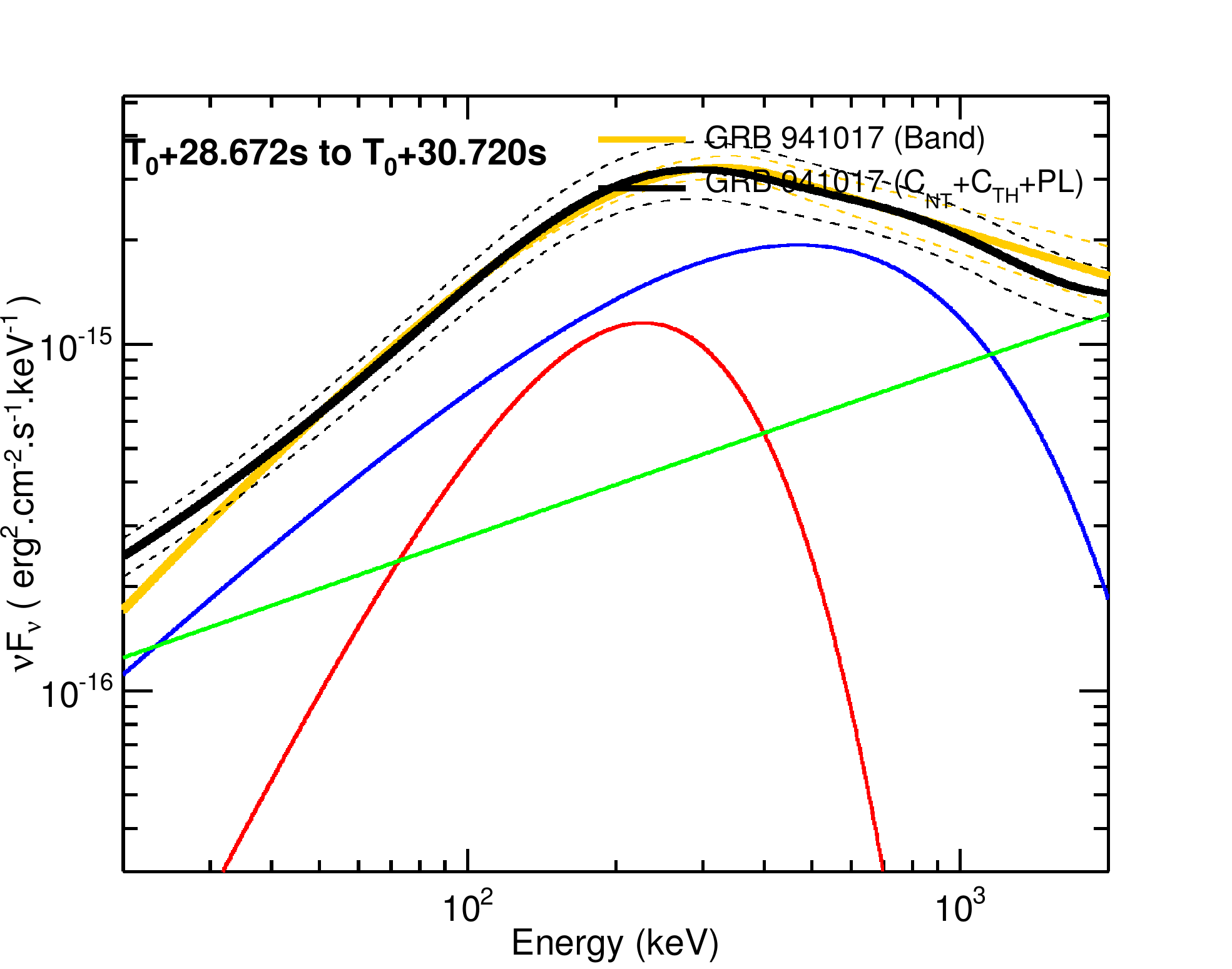}
\includegraphics[totalheight=0.185\textheight, clip]{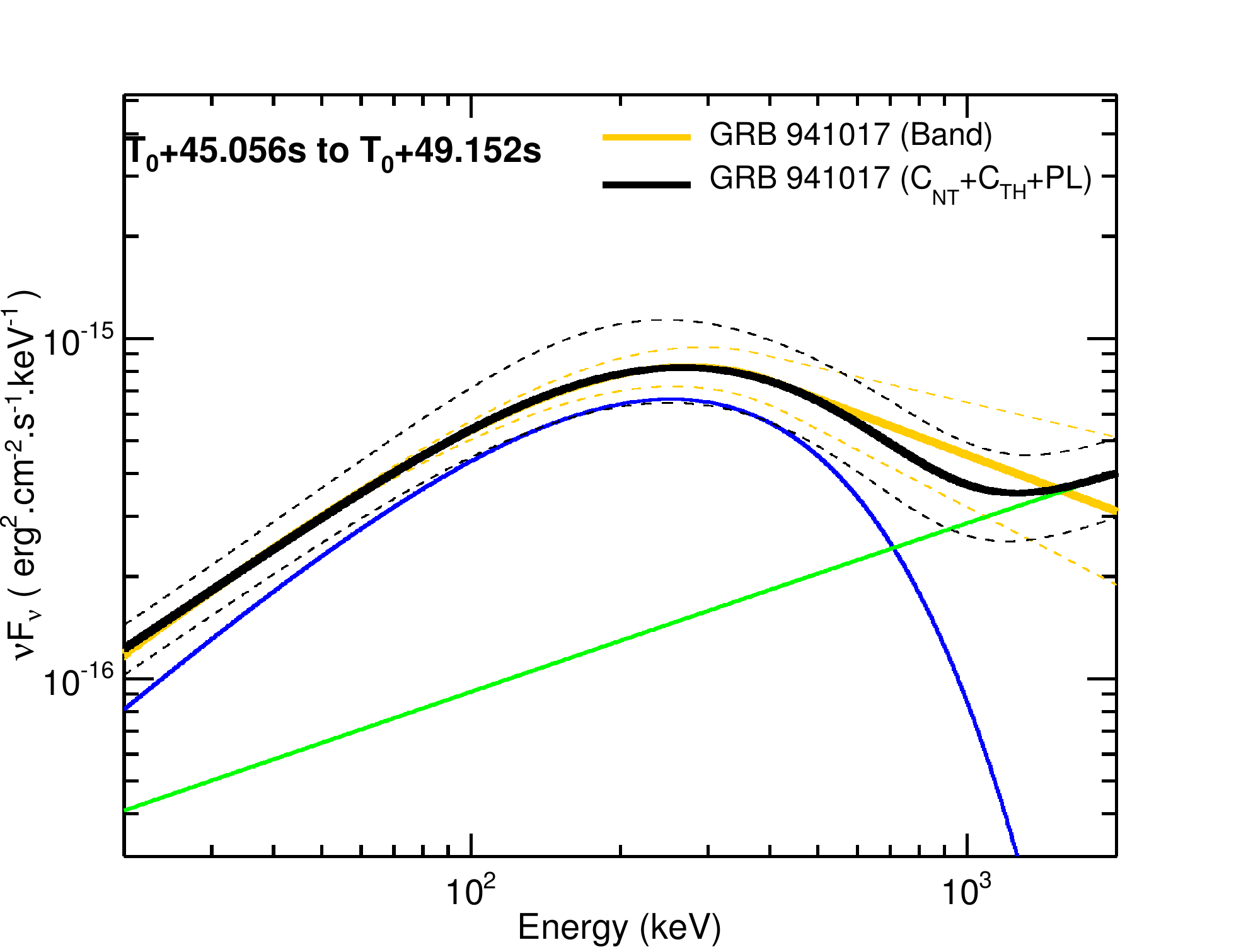}

\includegraphics[totalheight=0.185\textheight, clip]{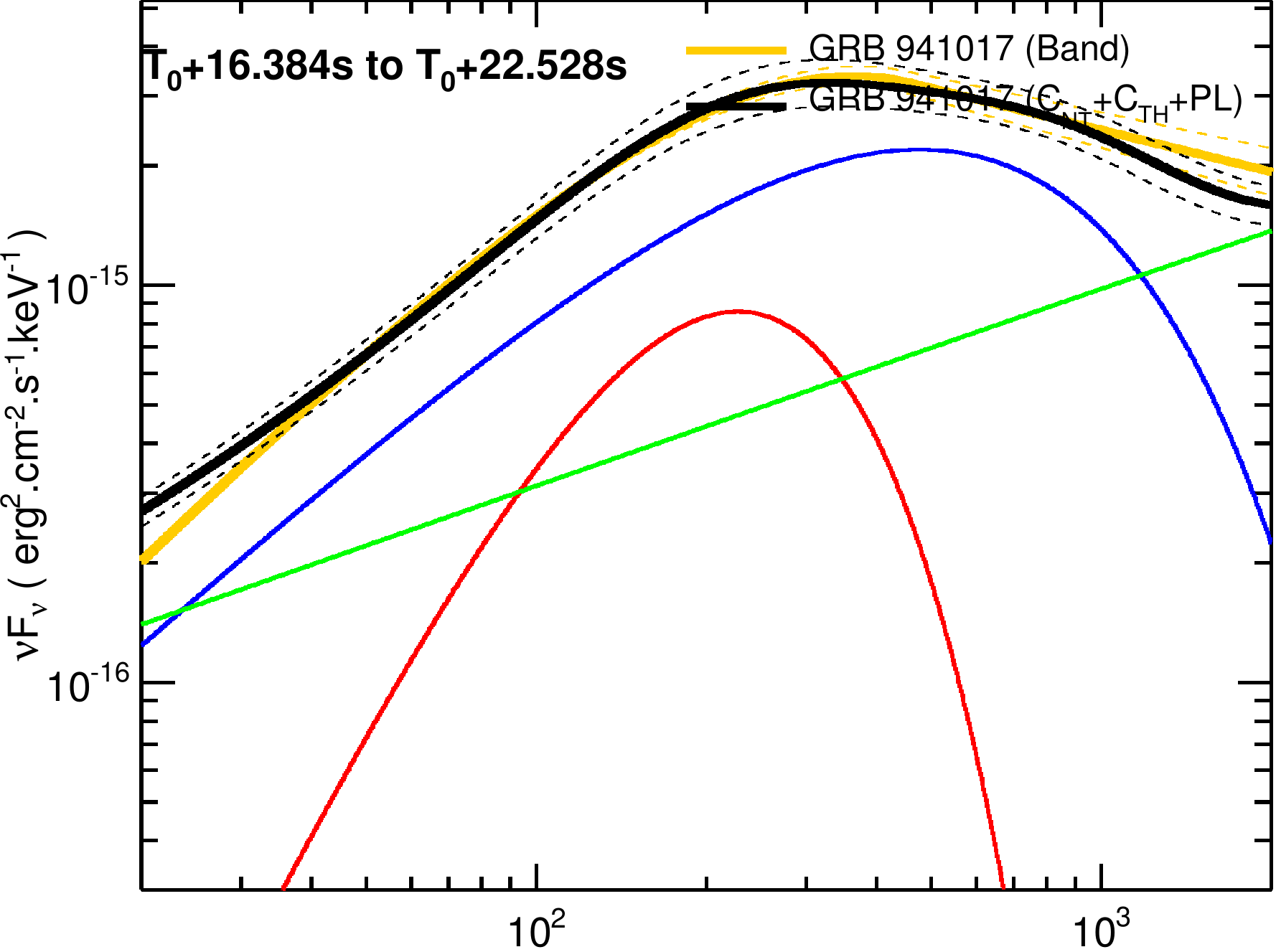}
\includegraphics[totalheight=0.185\textheight, clip]{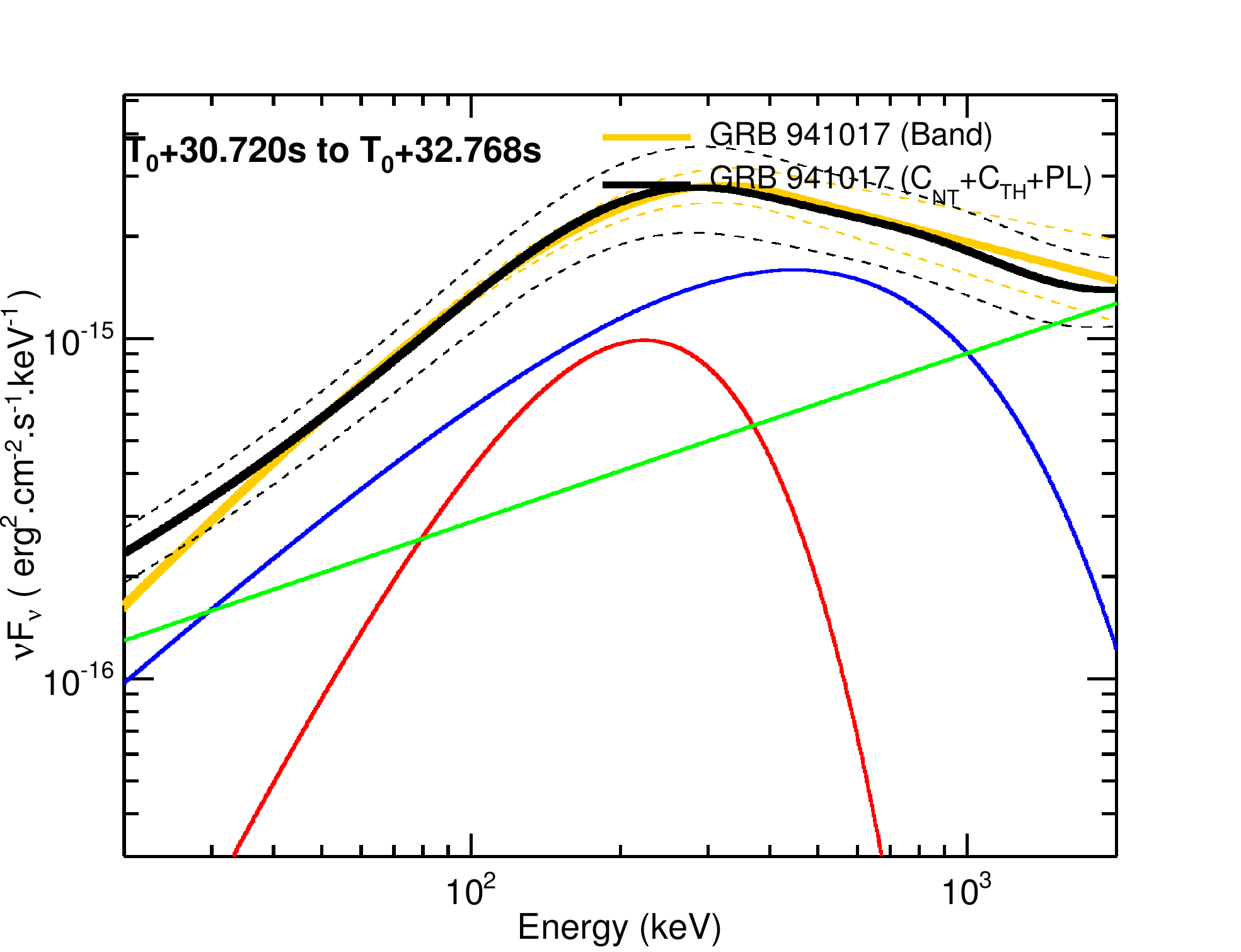}
\includegraphics[totalheight=0.185\textheight, clip]{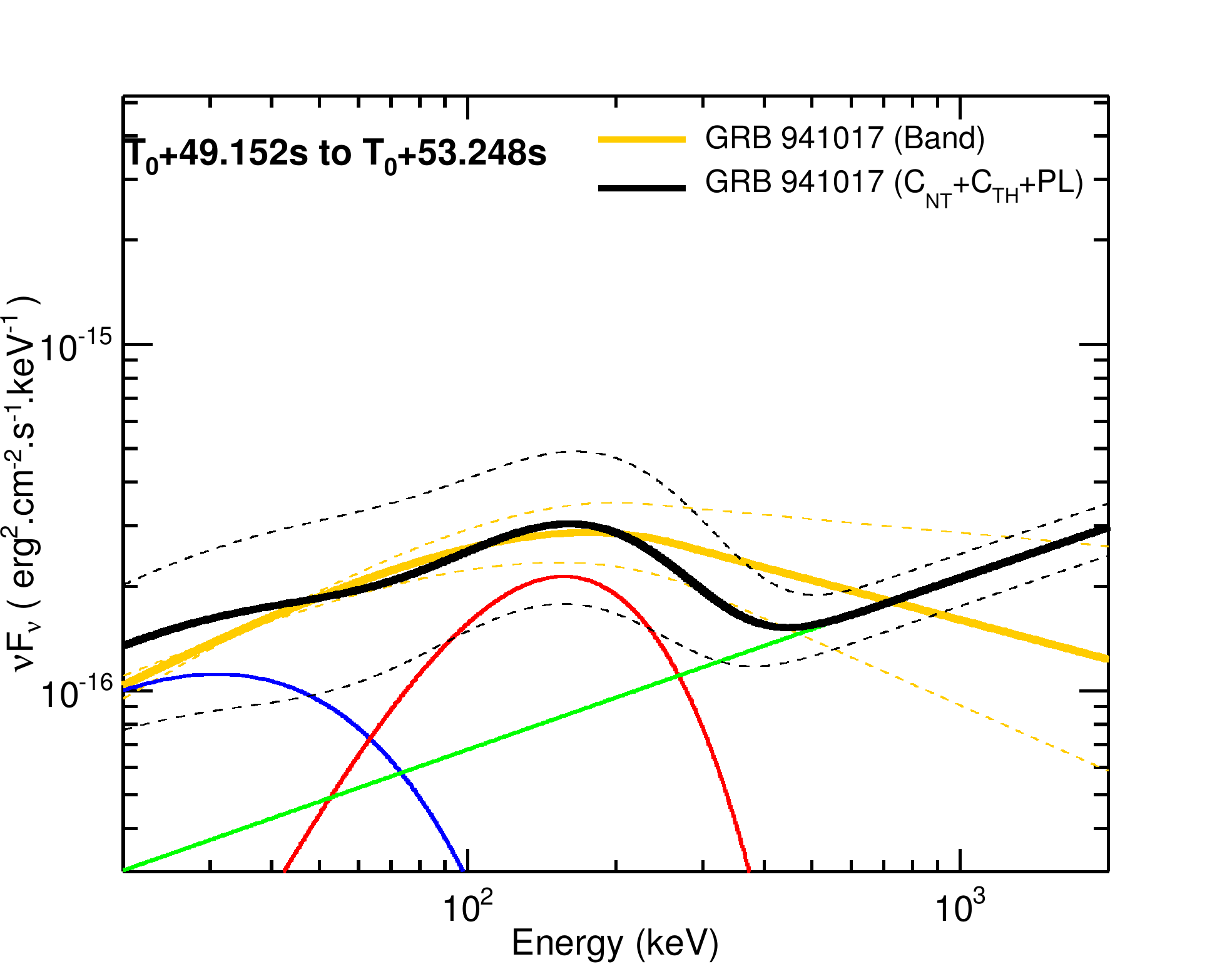}

\includegraphics[totalheight=0.195\textheight, clip]{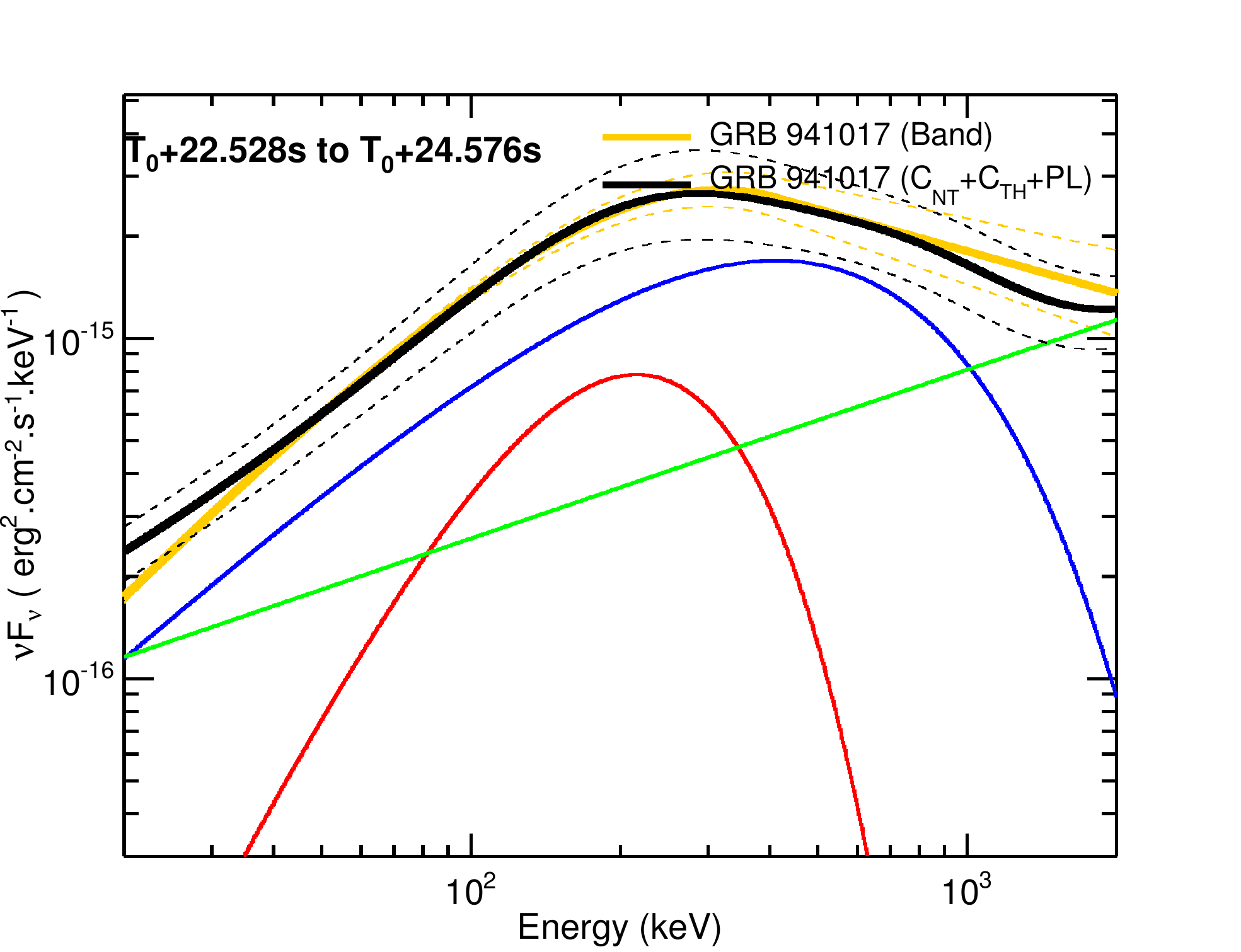}
\includegraphics[totalheight=0.195\textheight, clip]{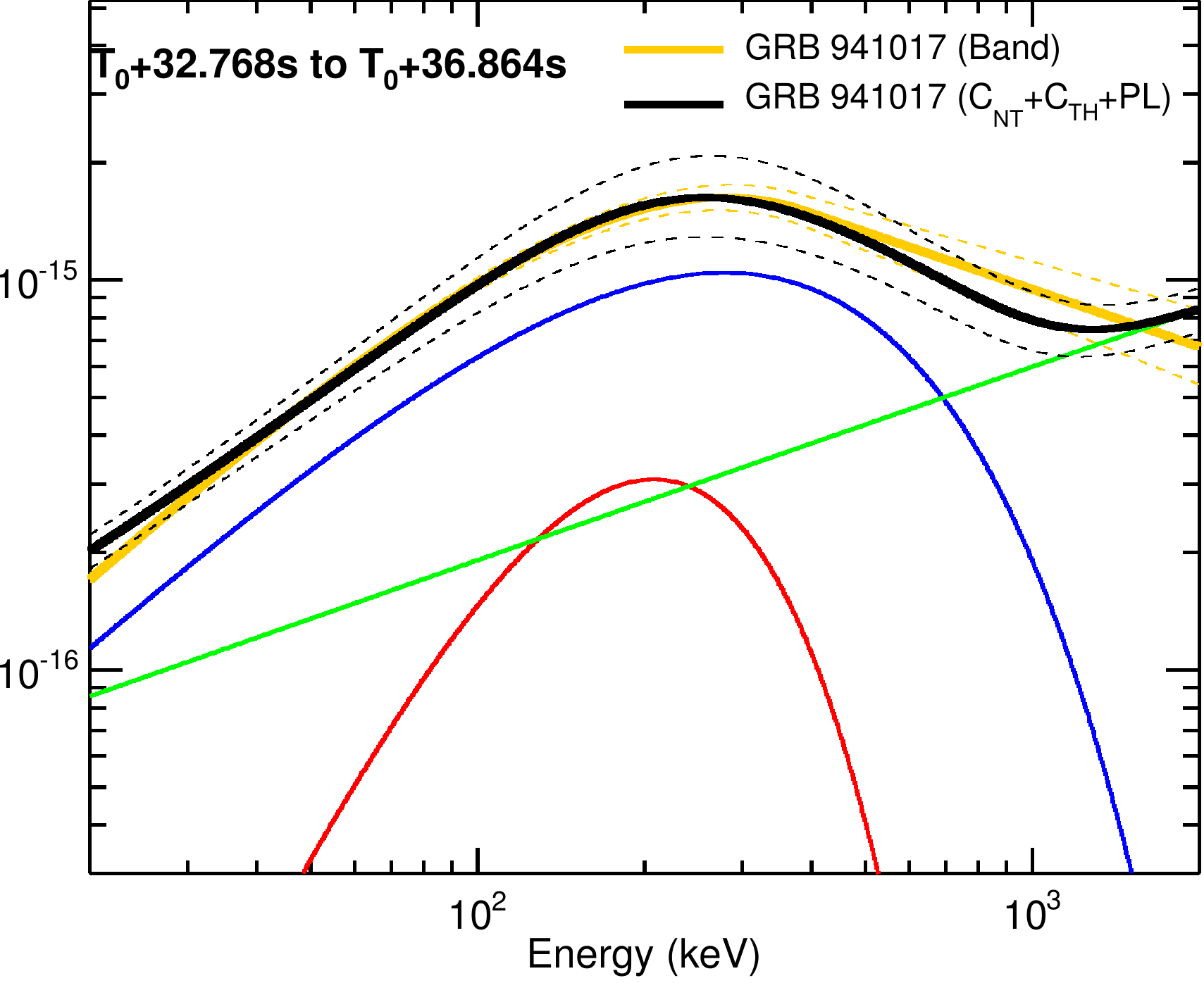}
\includegraphics[totalheight=0.195\textheight, clip]{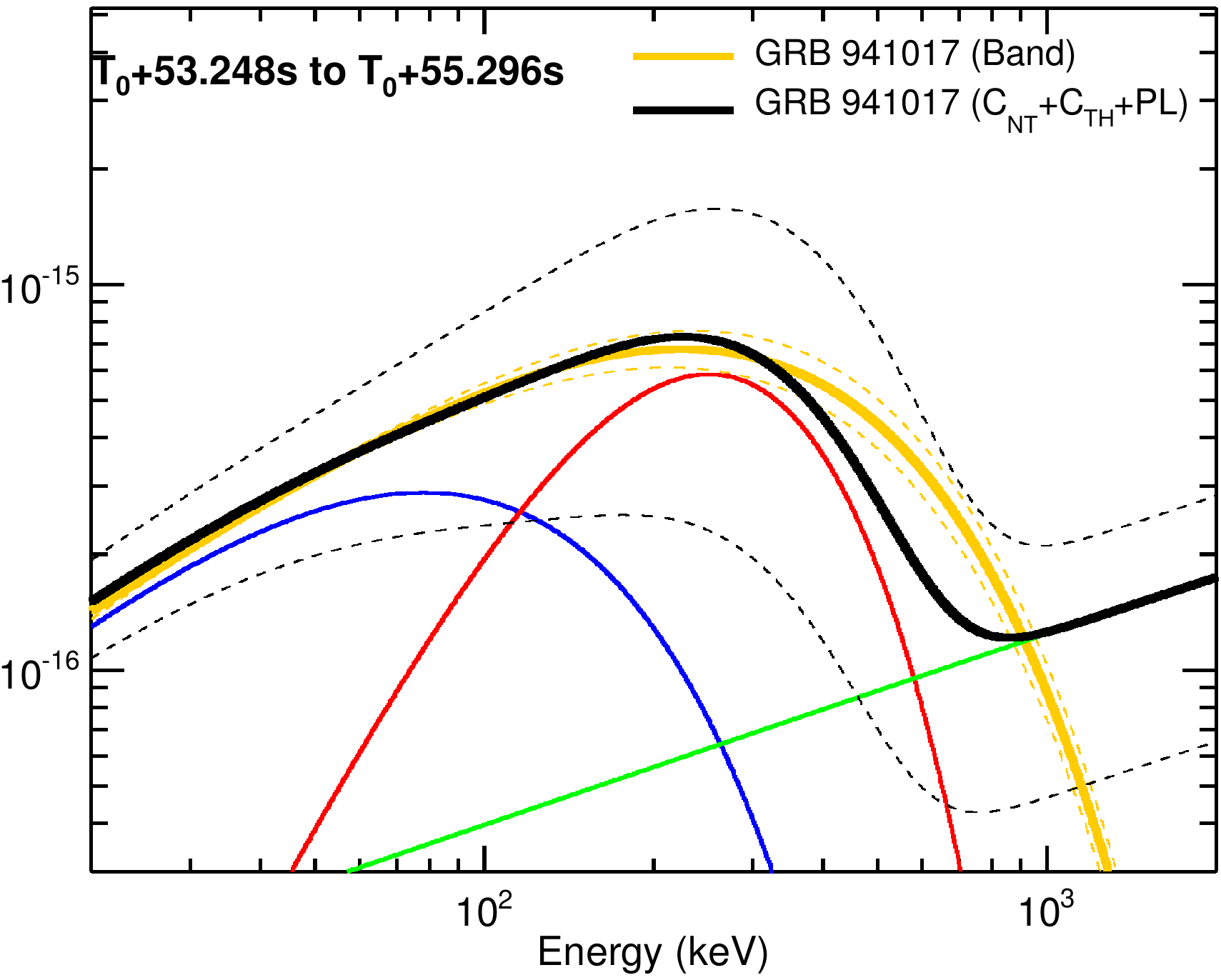}

\end{center}
\end{figure*}

\newpage

\begin{figure*}
\begin{center}
\includegraphics[totalheight=0.185\textheight, clip]{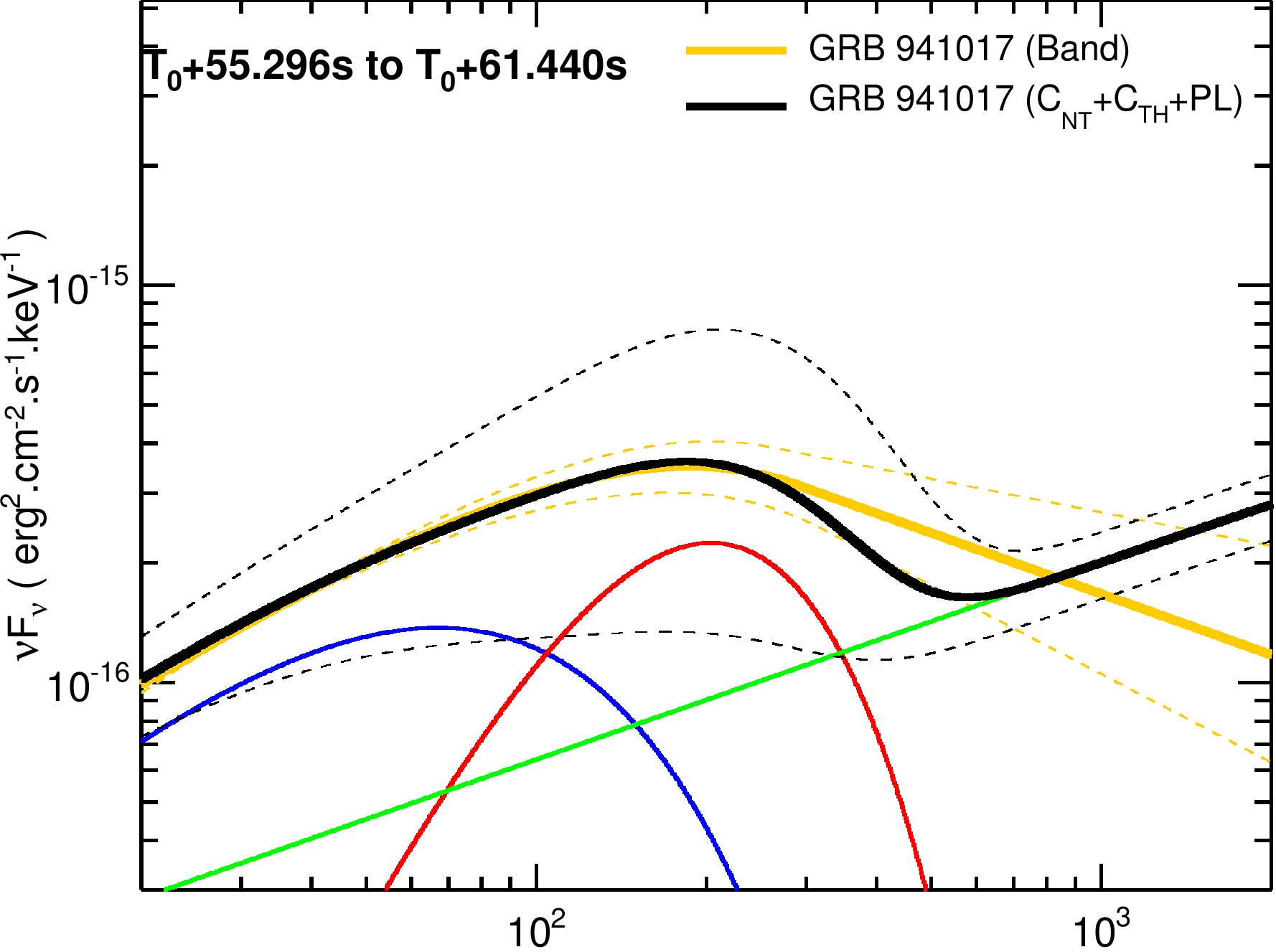}
\includegraphics[totalheight=0.185\textheight, clip]{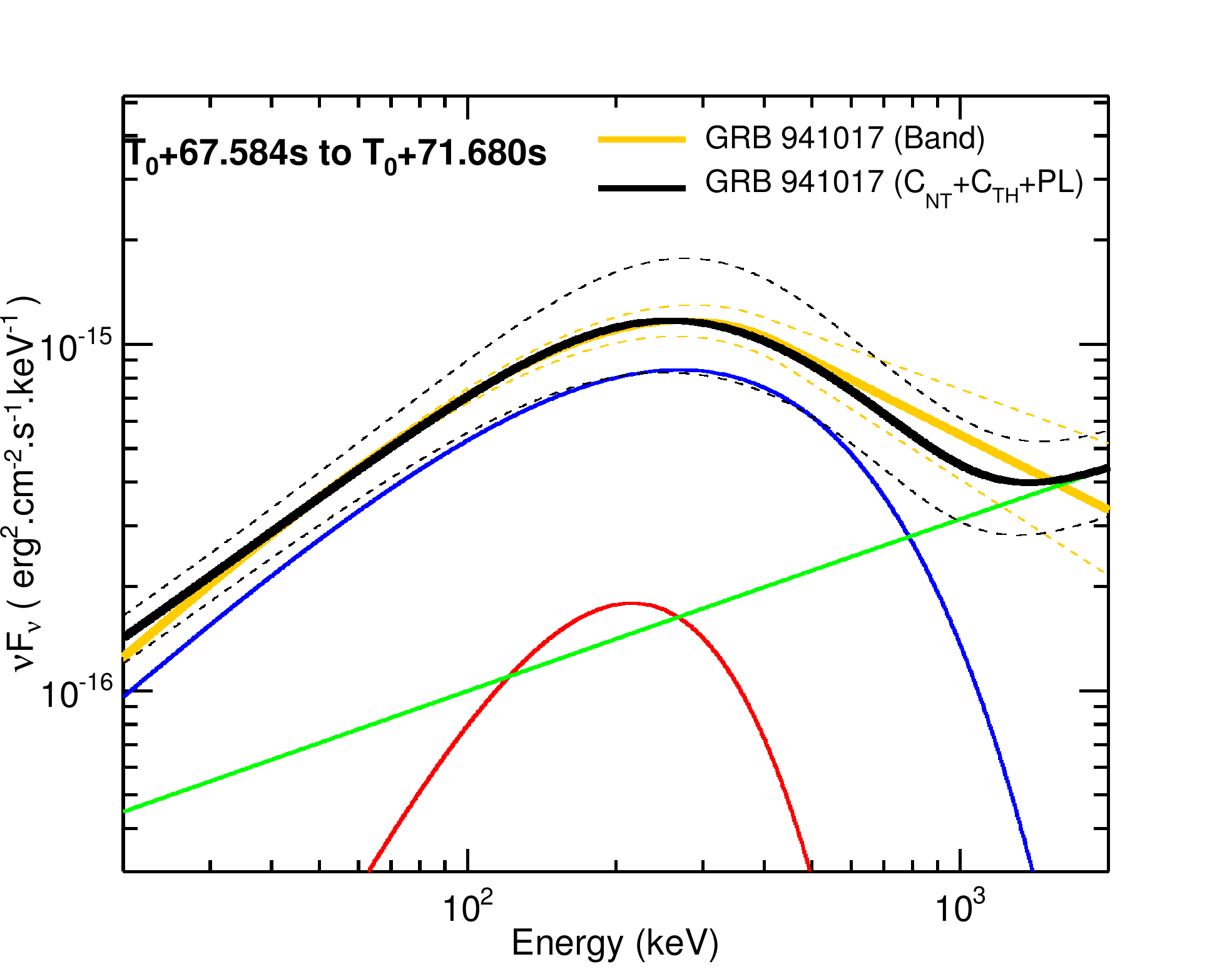}
\includegraphics[totalheight=0.185\textheight, clip]{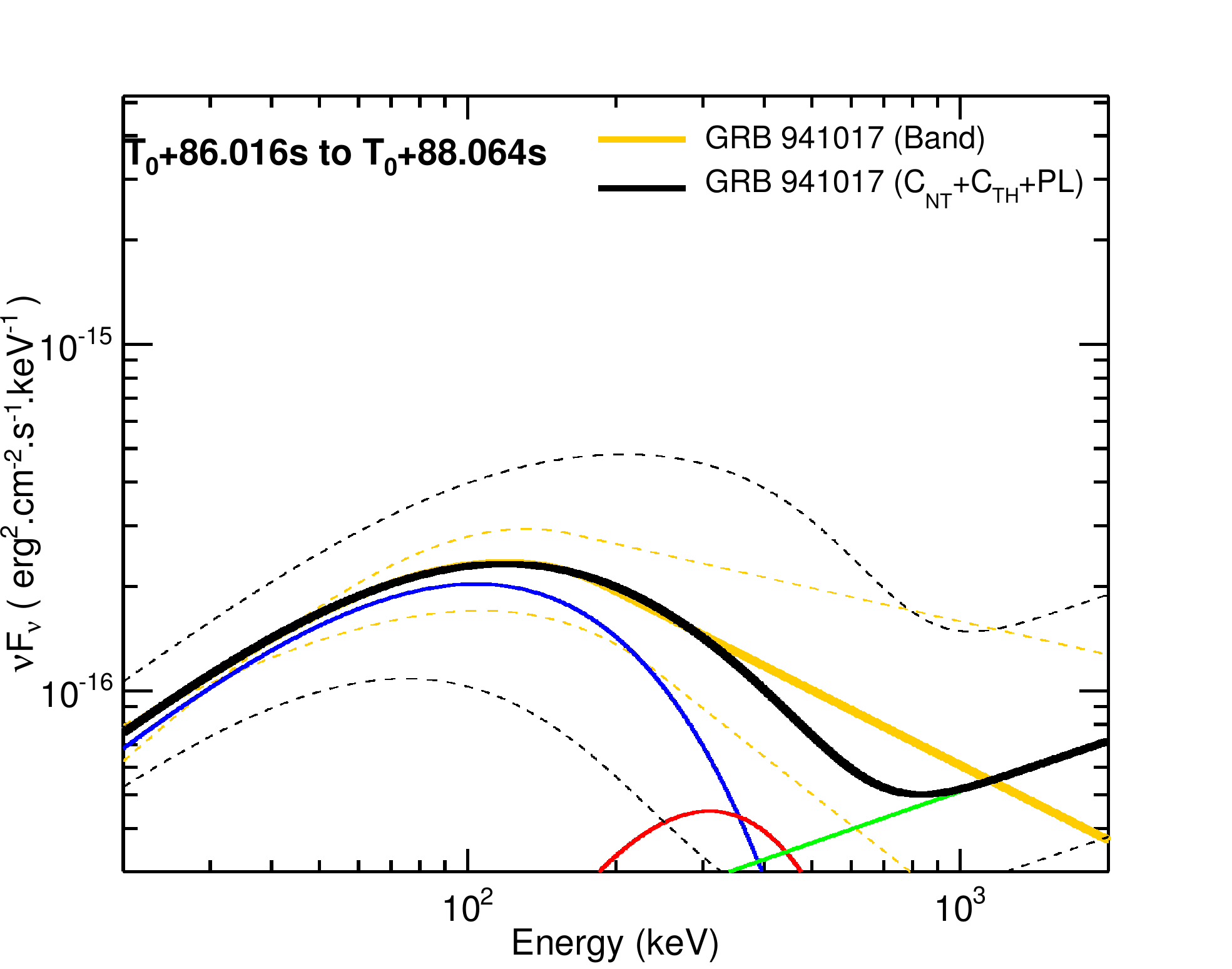}

\includegraphics[totalheight=0.185\textheight, clip]{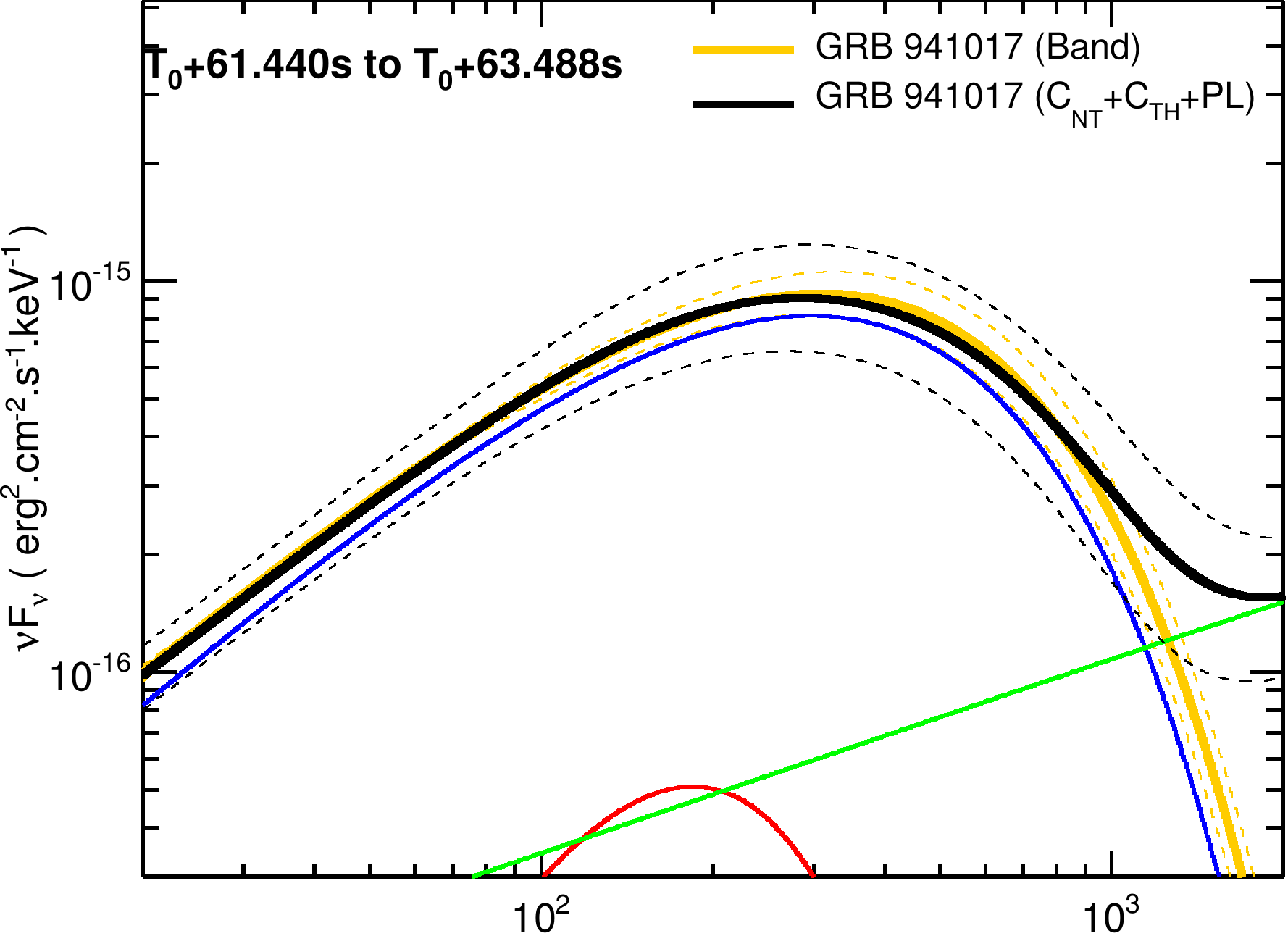}
\includegraphics[totalheight=0.185\textheight, clip]{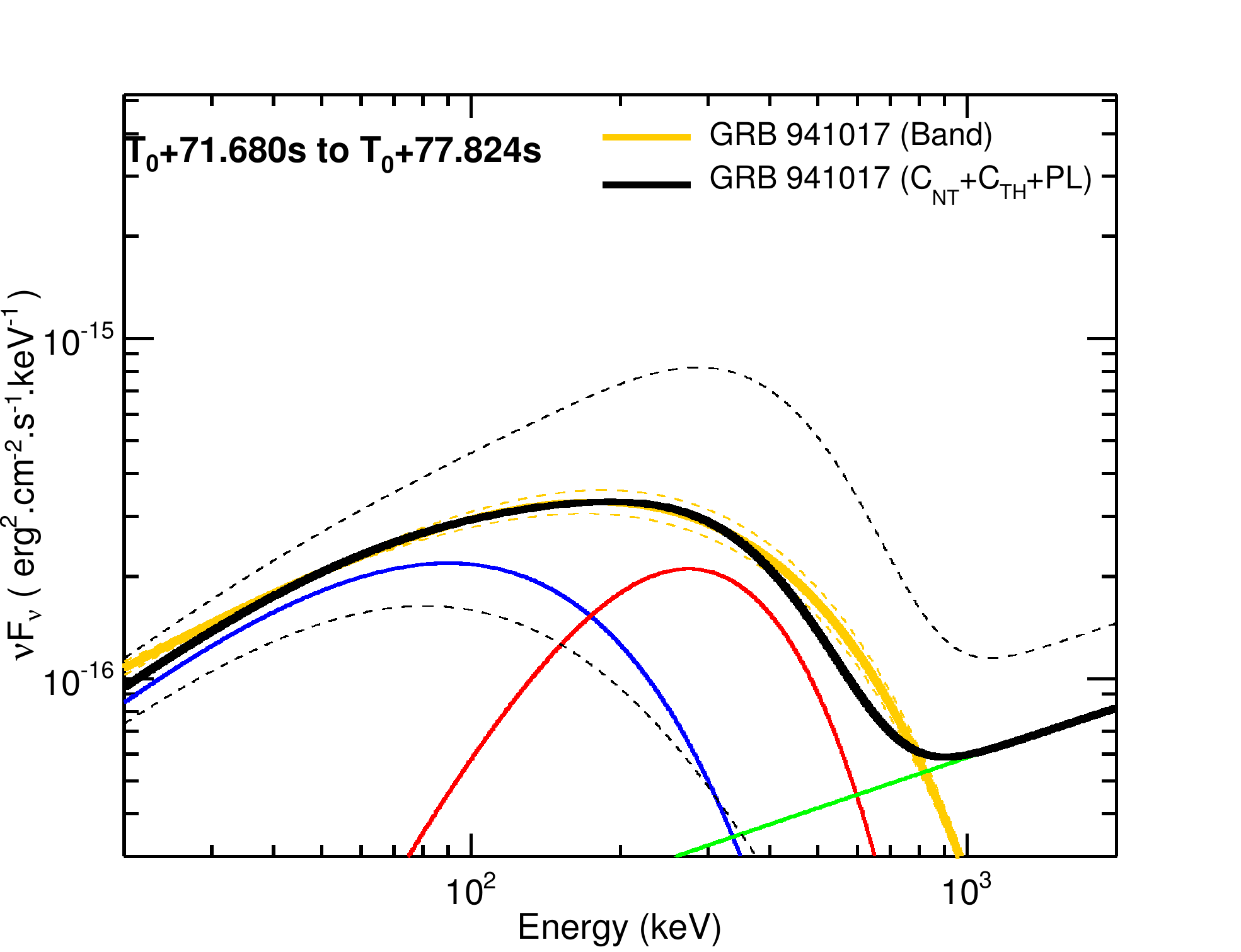}
\includegraphics[totalheight=0.185\textheight, clip]{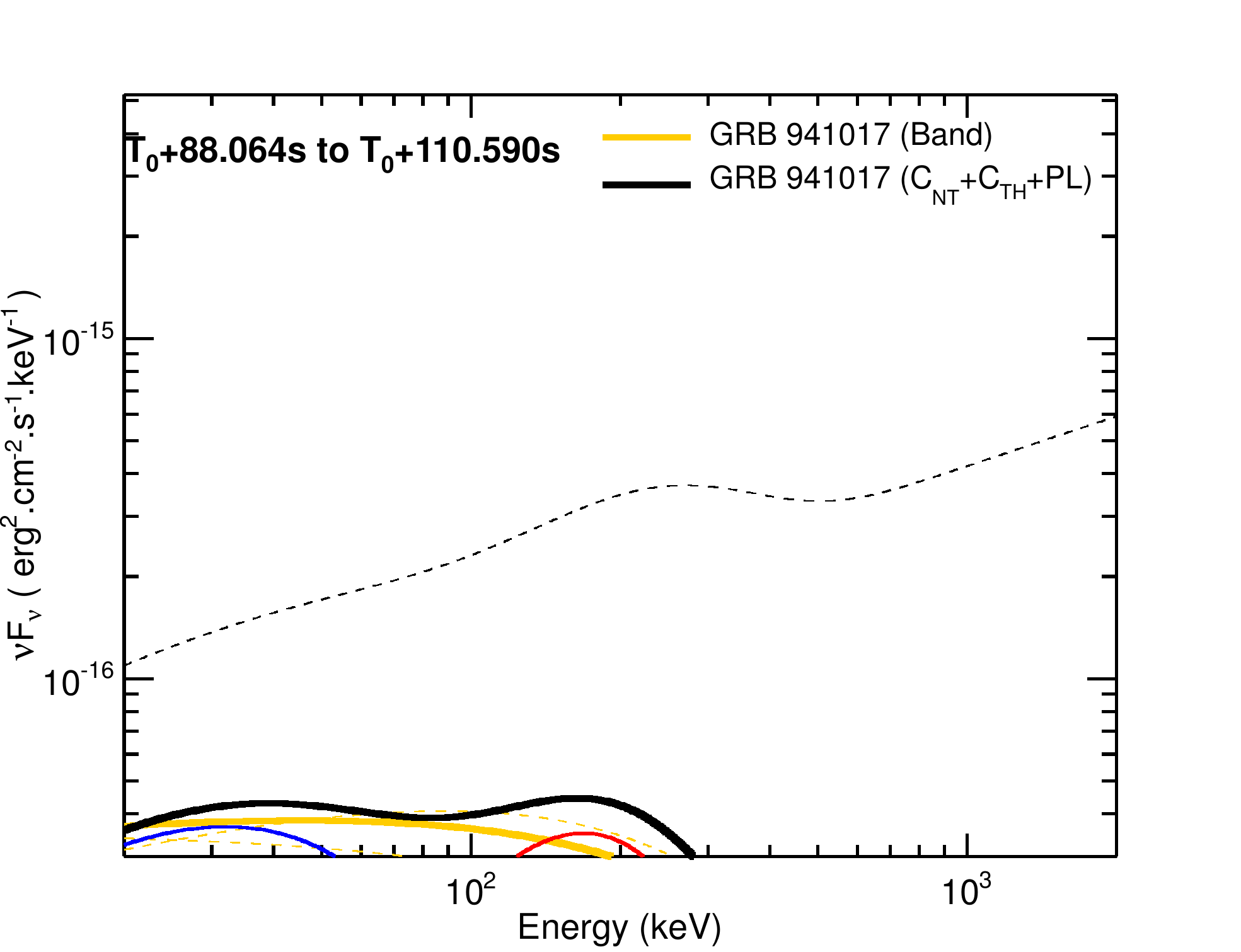}

\includegraphics[totalheight=0.185\textheight, clip]{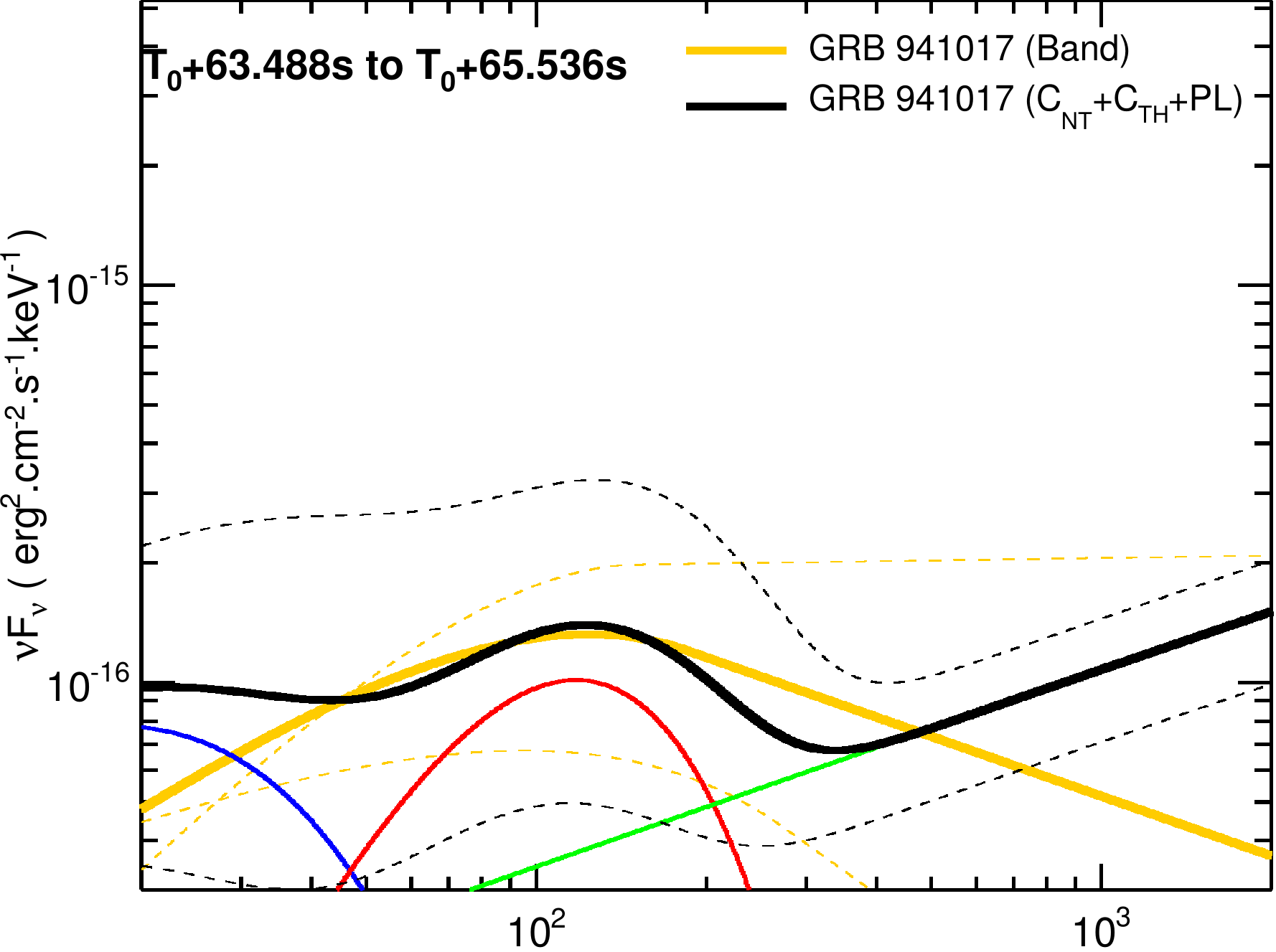}
\includegraphics[totalheight=0.185\textheight, clip]{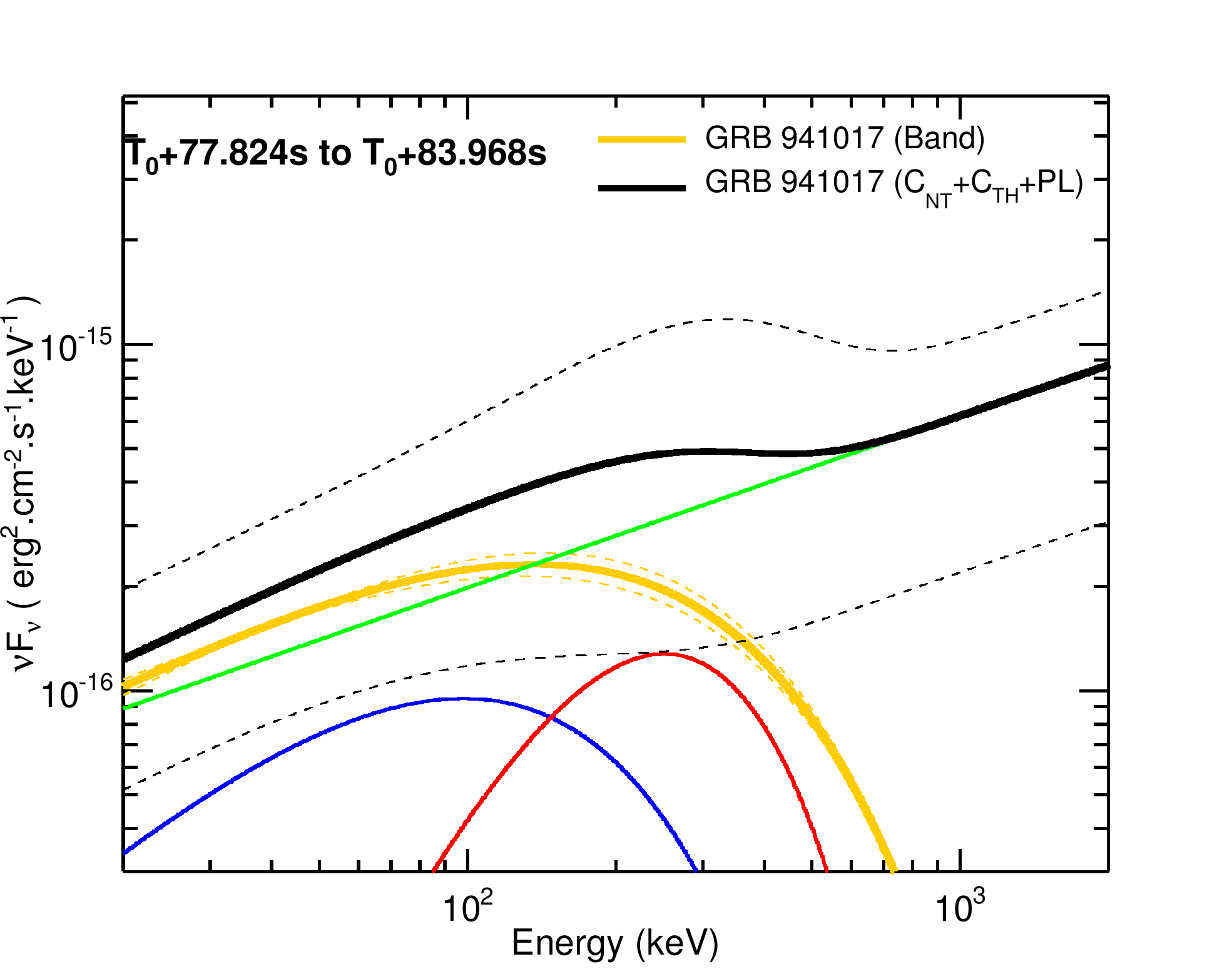}
\includegraphics[totalheight=0.185\textheight, clip]{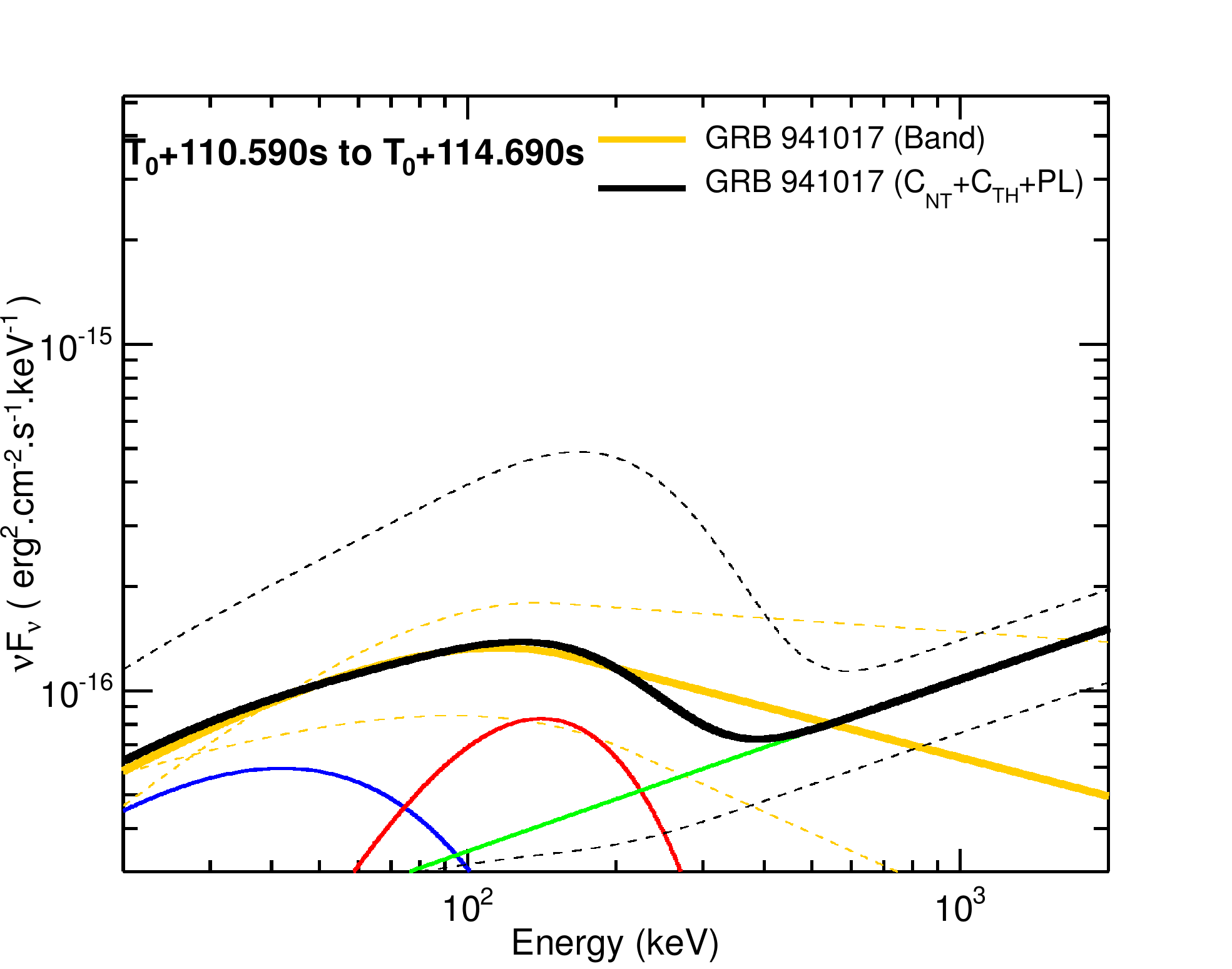}

\includegraphics[totalheight=0.195\textheight, clip]{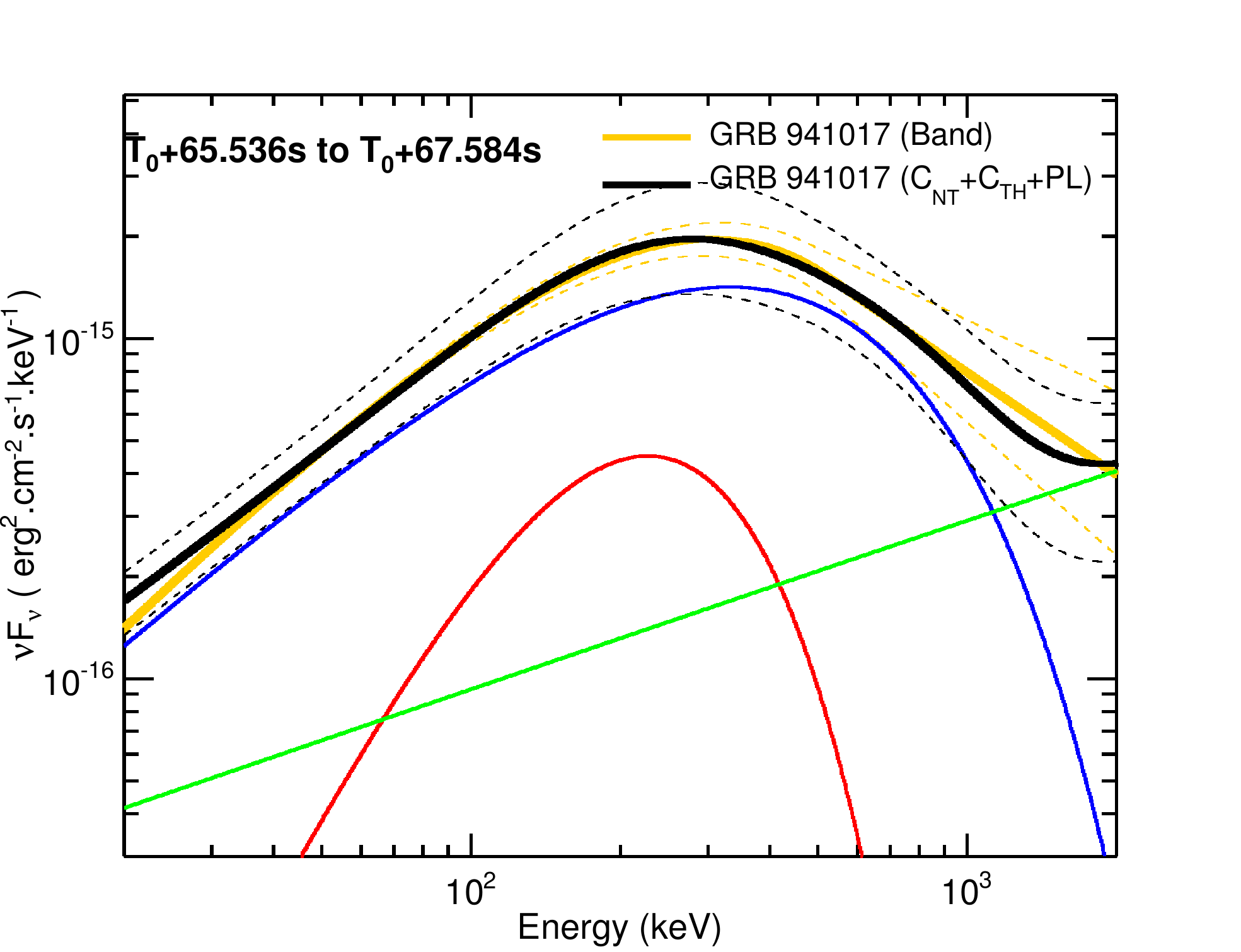}
\includegraphics[totalheight=0.195\textheight, clip]{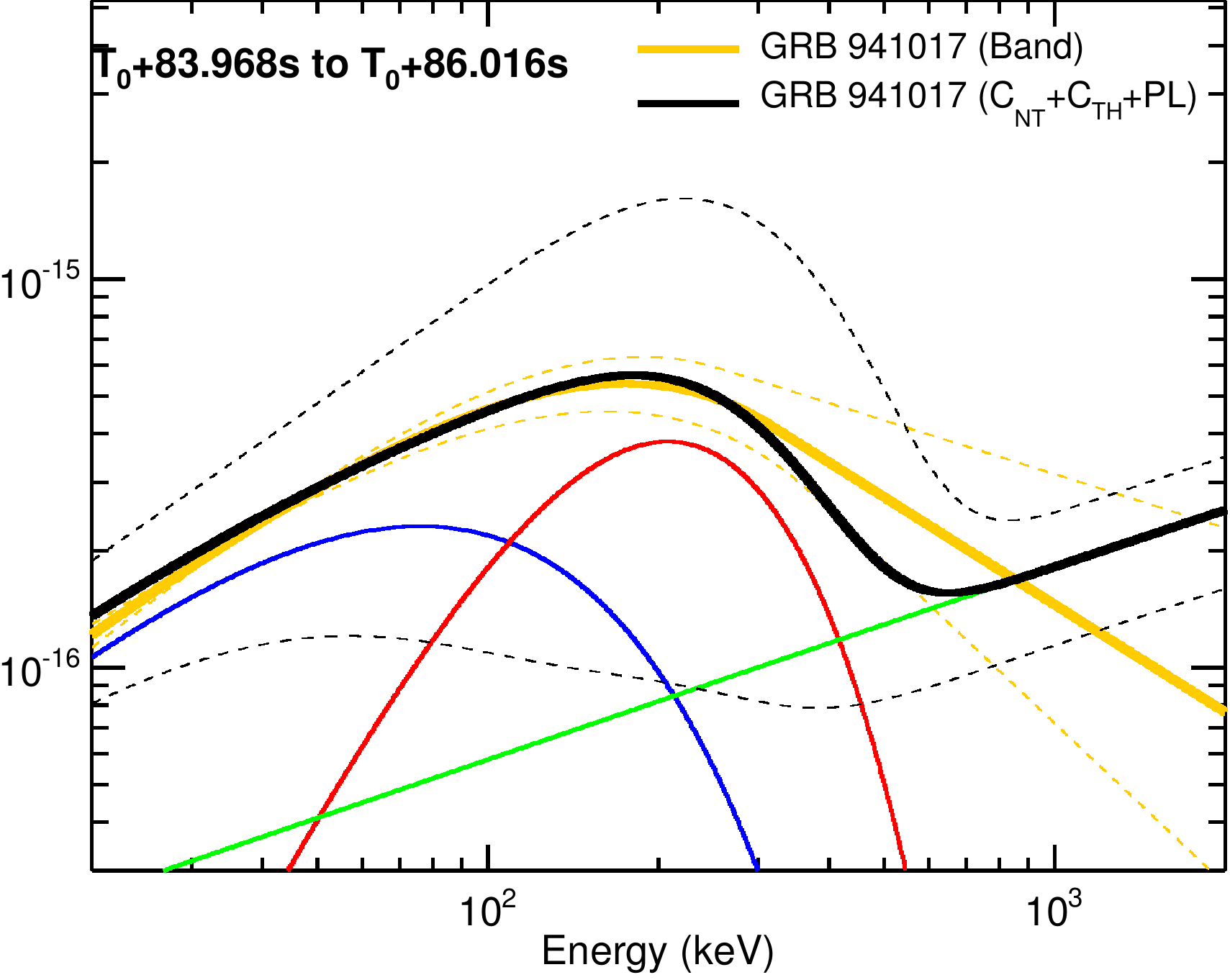}
\includegraphics[totalheight=0.195\textheight, clip]{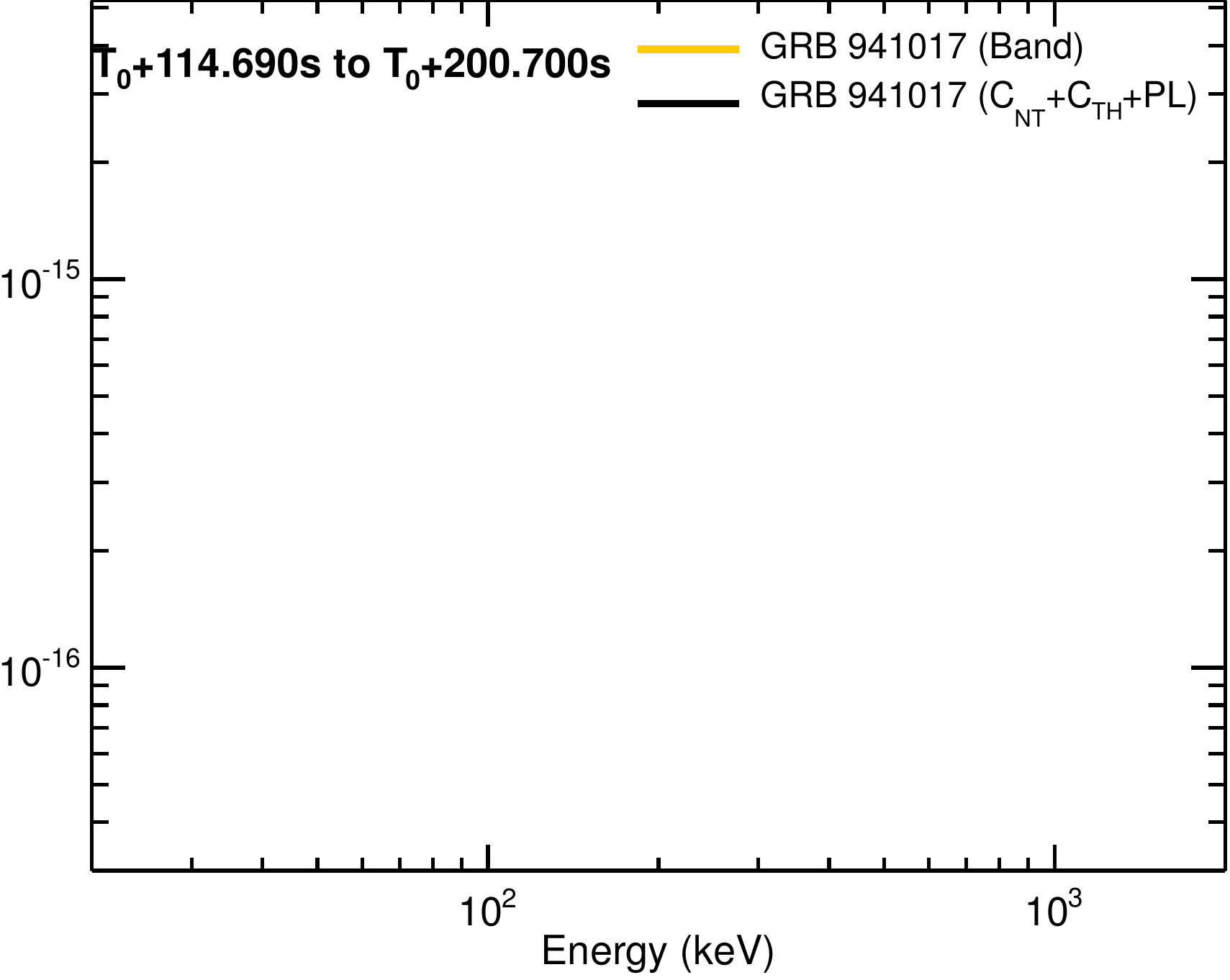}

\caption{\label{fig18}GRB~941017 : $\nu$F$_\nu$ spectra resulting from the fine-time analysis presented in Section~\ref{sec:trsa}. The solid yellow and black lines correspond to the best Band-only and C$_\mathrm{nTh}$+C$_\mathrm{Th}$+PL fits, respectively. The dashed yellow and black lines correspond to the 1--$\sigma$ confidence regions of the Band-only and C$_\mathrm{nTh}$+C$_\mathrm{Th}$+PL fits, respectively. The solid blue, red and green lines correspond to C$_\mathrm{nTh}$, C$_\mathrm{Th}$ and the additional PL resulting from the best C$_\mathrm{nTh}$+C$_\mathrm{Th}$+PL fits (i.e., solid black line) to the data, respectively.}
\end{center}
\end{figure*}

\newpage

\begin{figure*}
\begin{center}
\includegraphics[totalheight=0.185\textheight, clip]{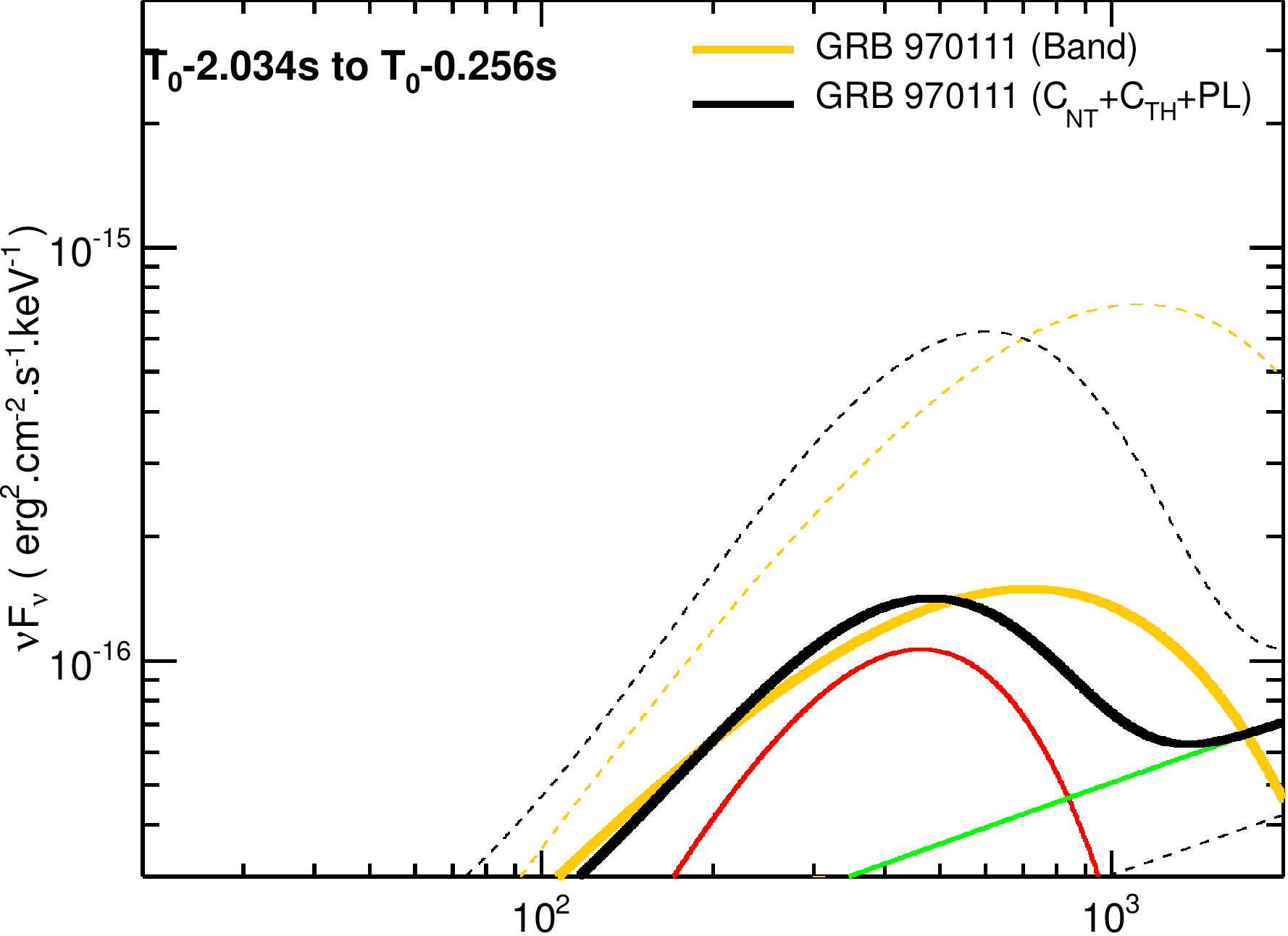}
\includegraphics[totalheight=0.185\textheight, clip]{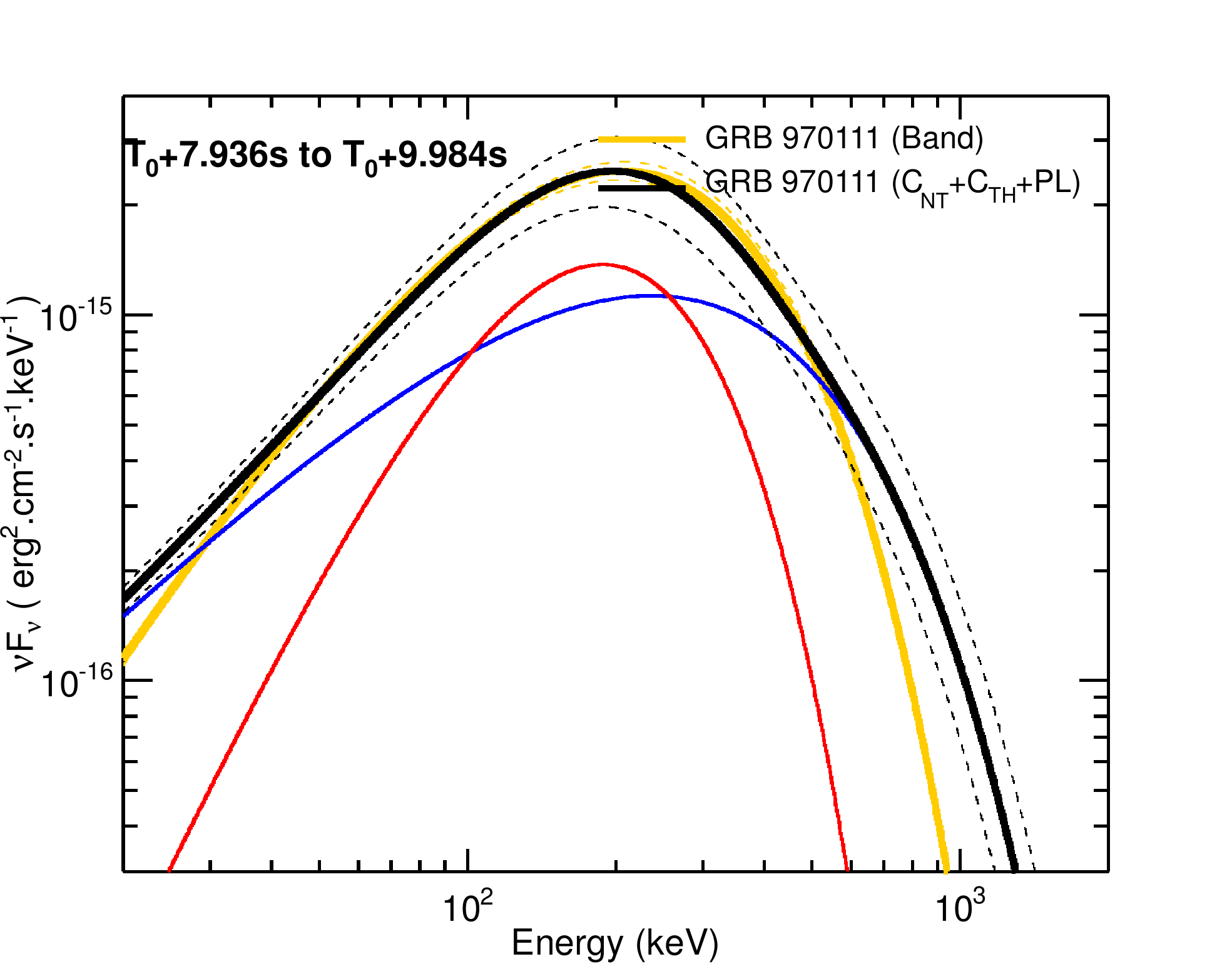}
\includegraphics[totalheight=0.185\textheight, clip]{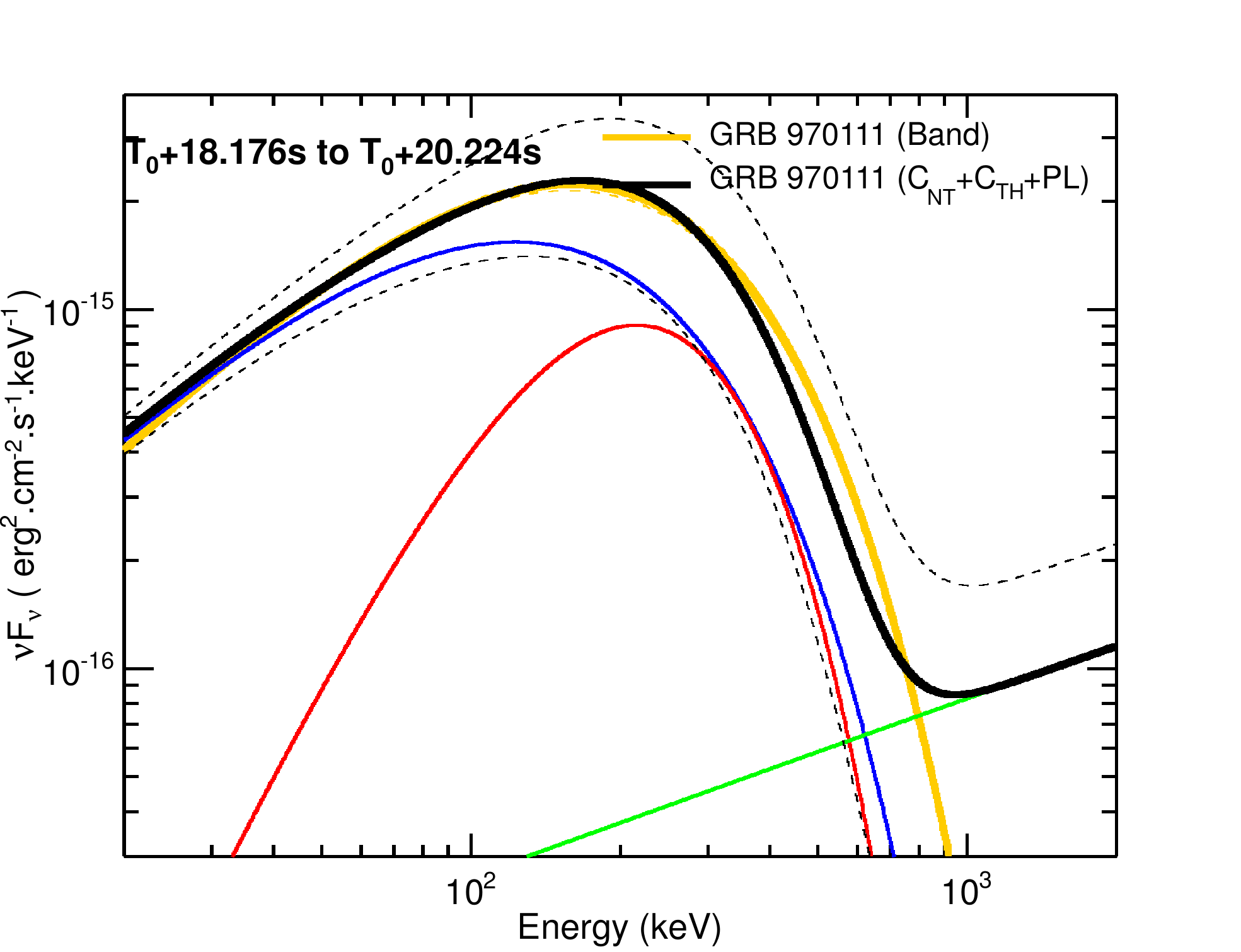}

\includegraphics[totalheight=0.185\textheight, clip]{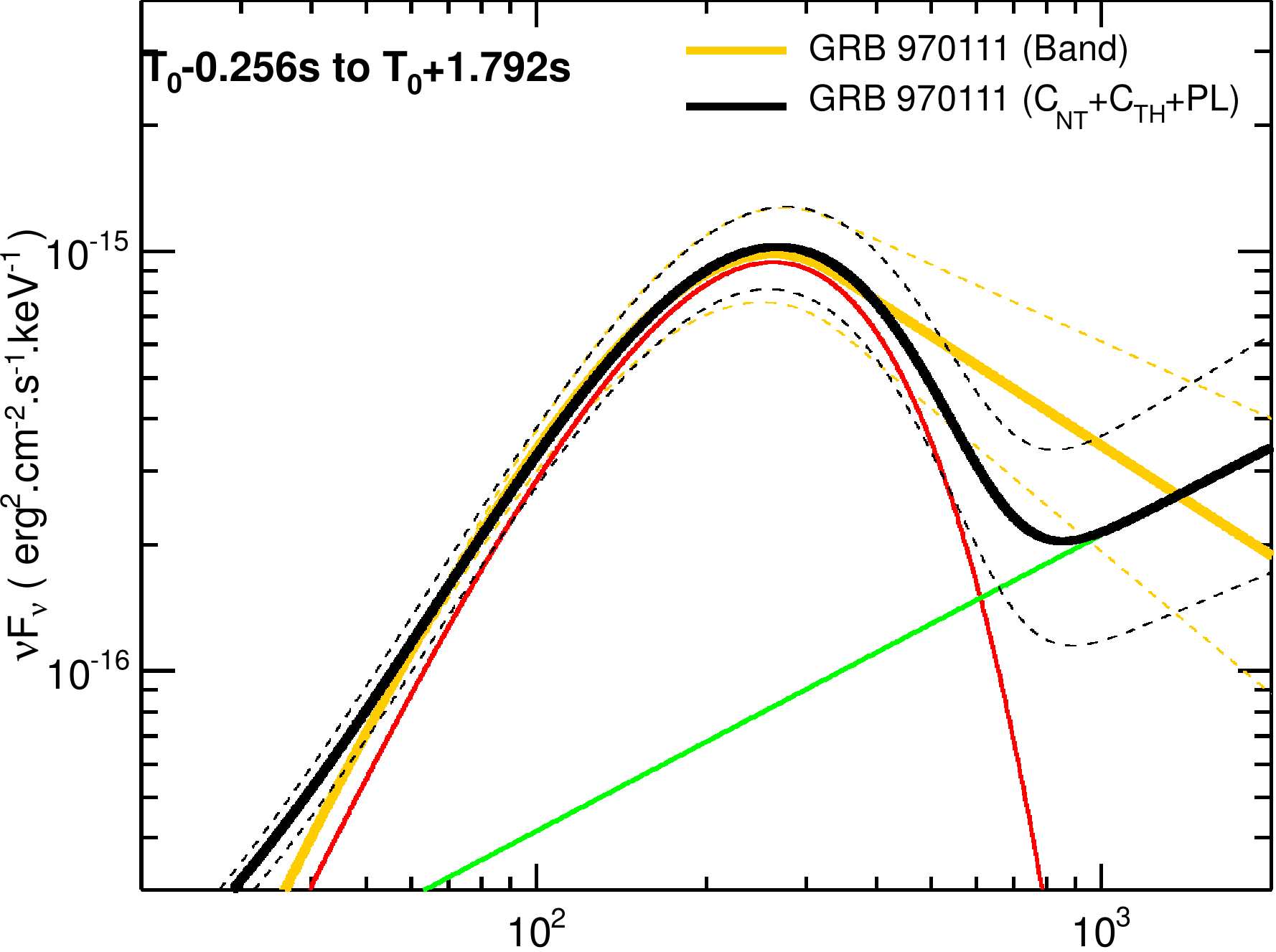}
\includegraphics[totalheight=0.185\textheight, clip]{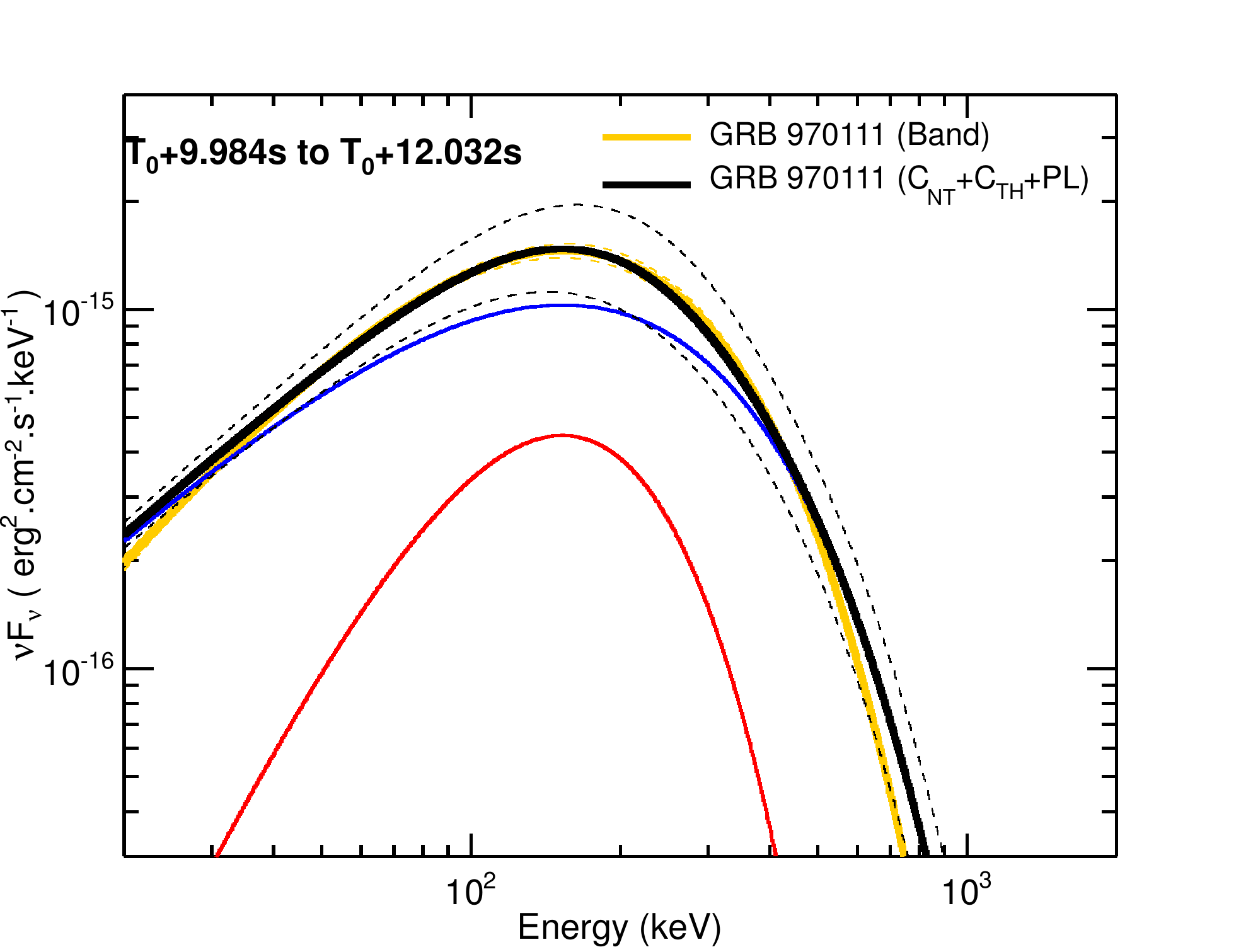}
\includegraphics[totalheight=0.185\textheight, clip]{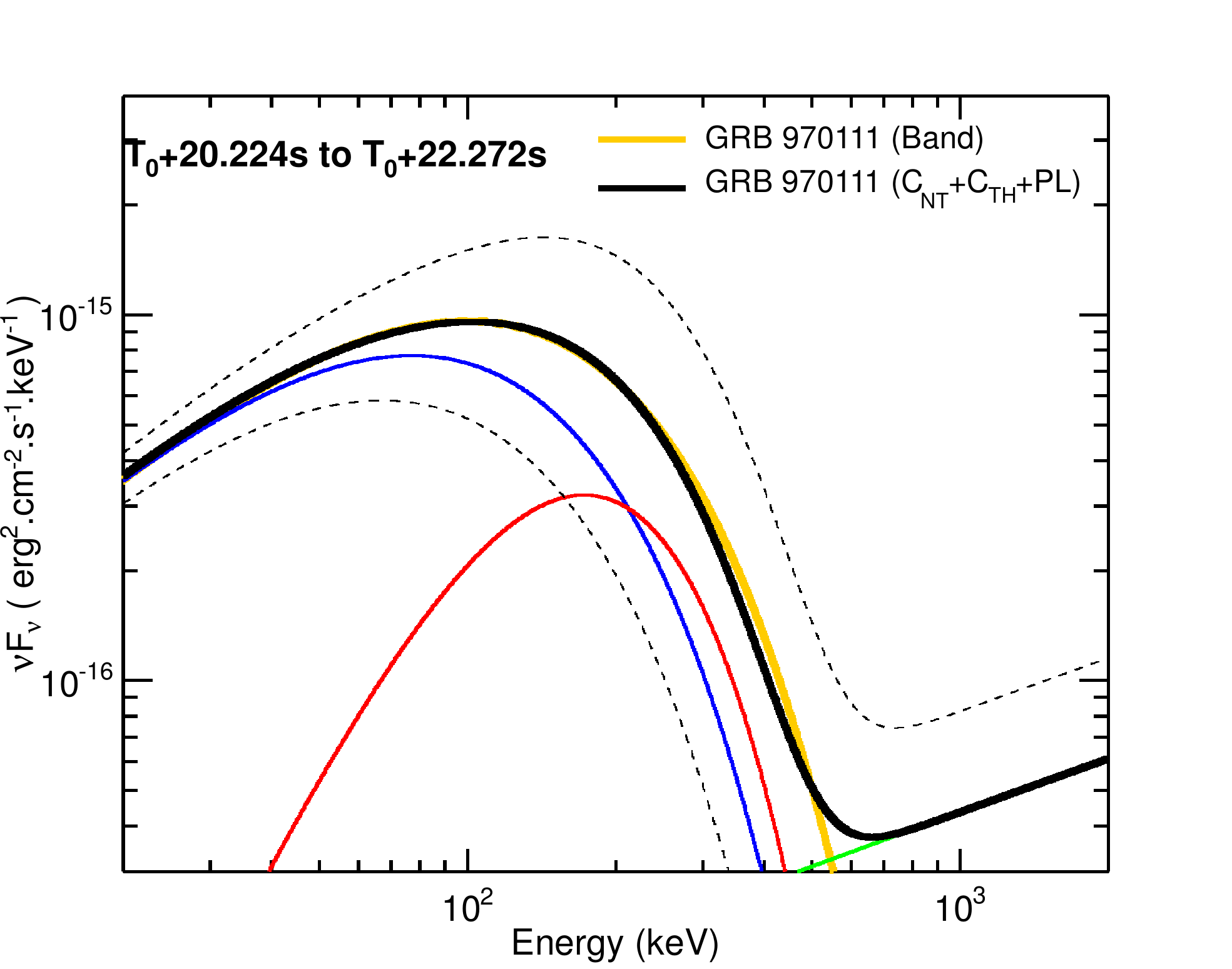}

\includegraphics[totalheight=0.185\textheight, clip]{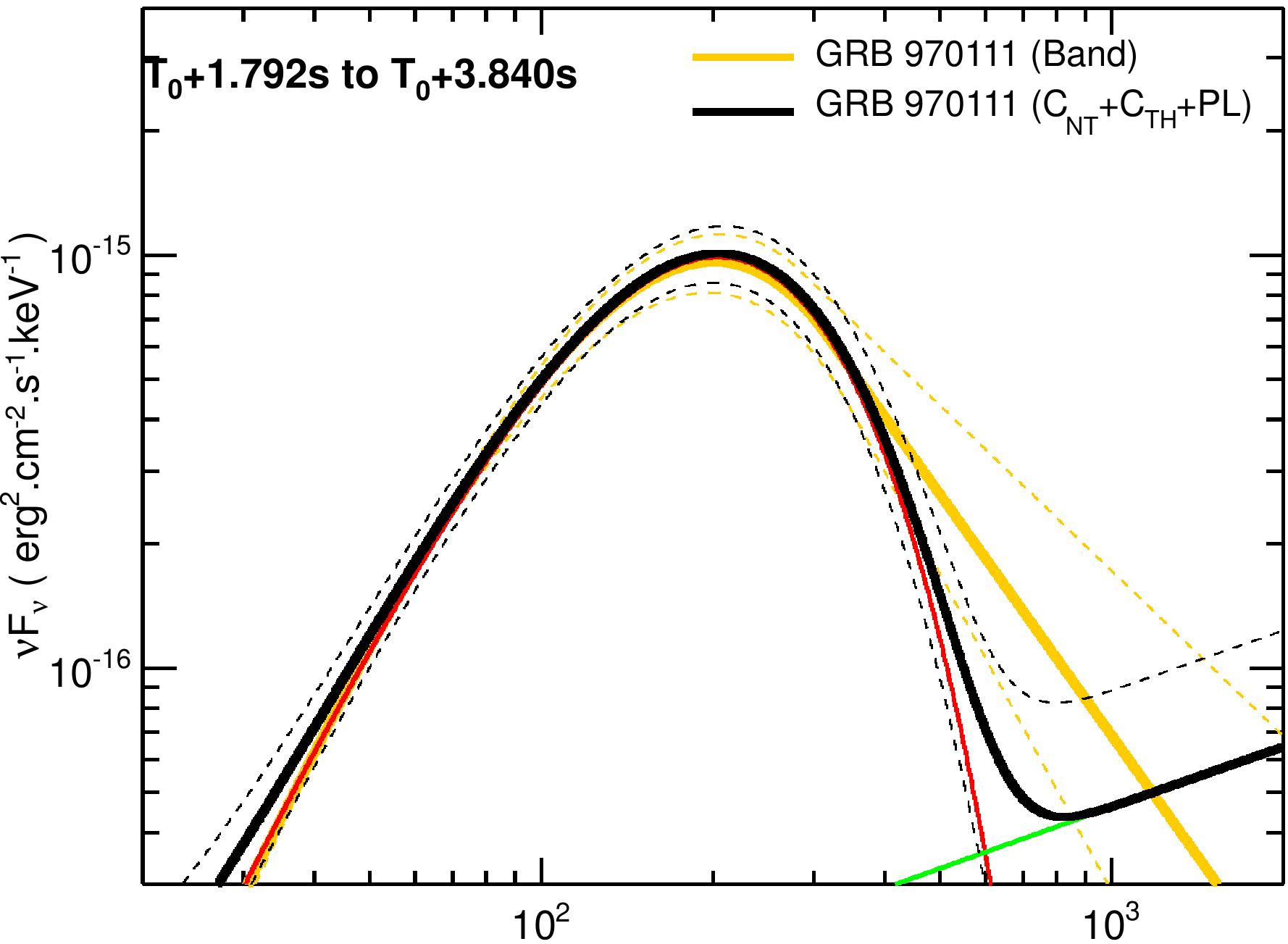}
\includegraphics[totalheight=0.185\textheight, clip]{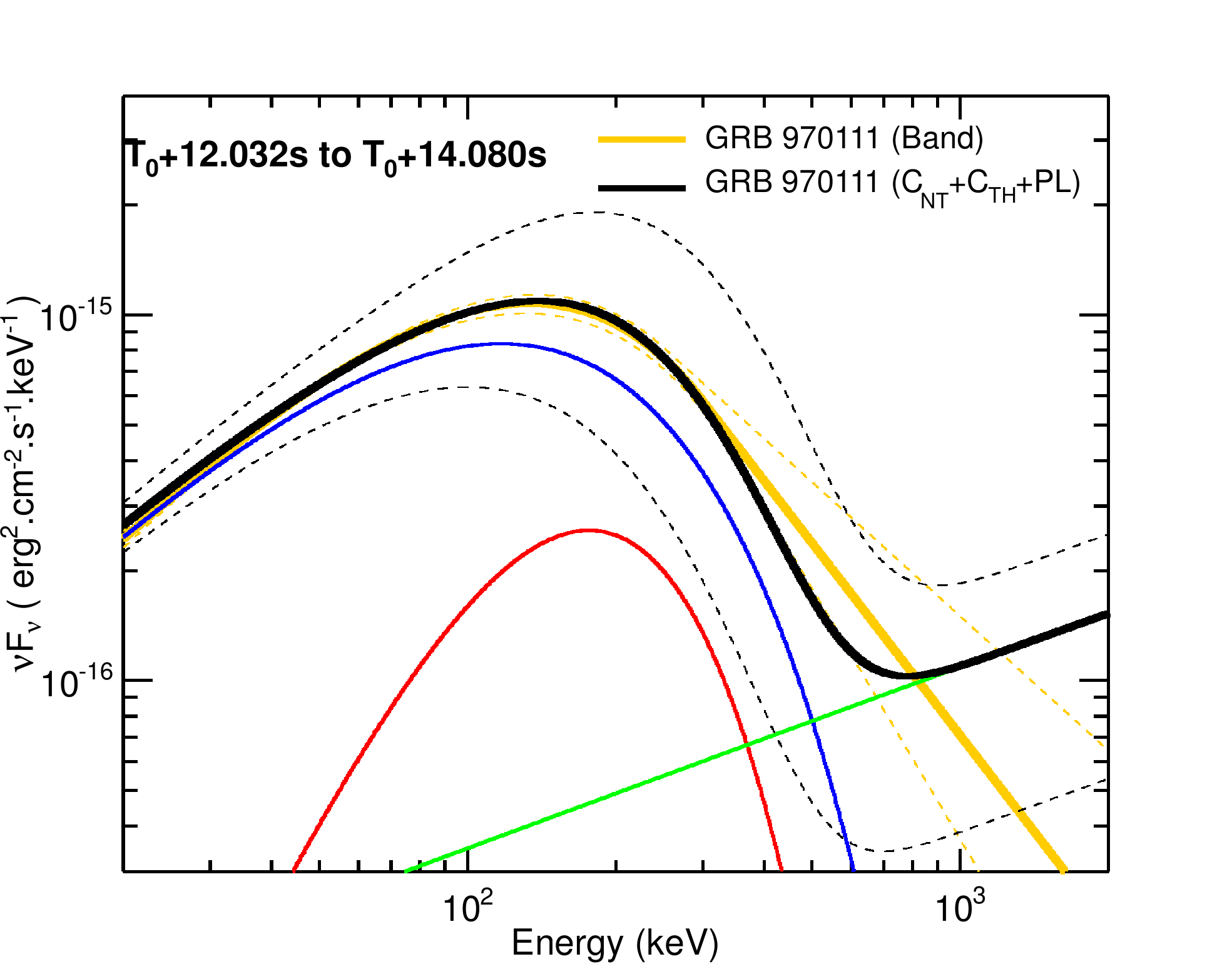}
\includegraphics[totalheight=0.185\textheight, clip]{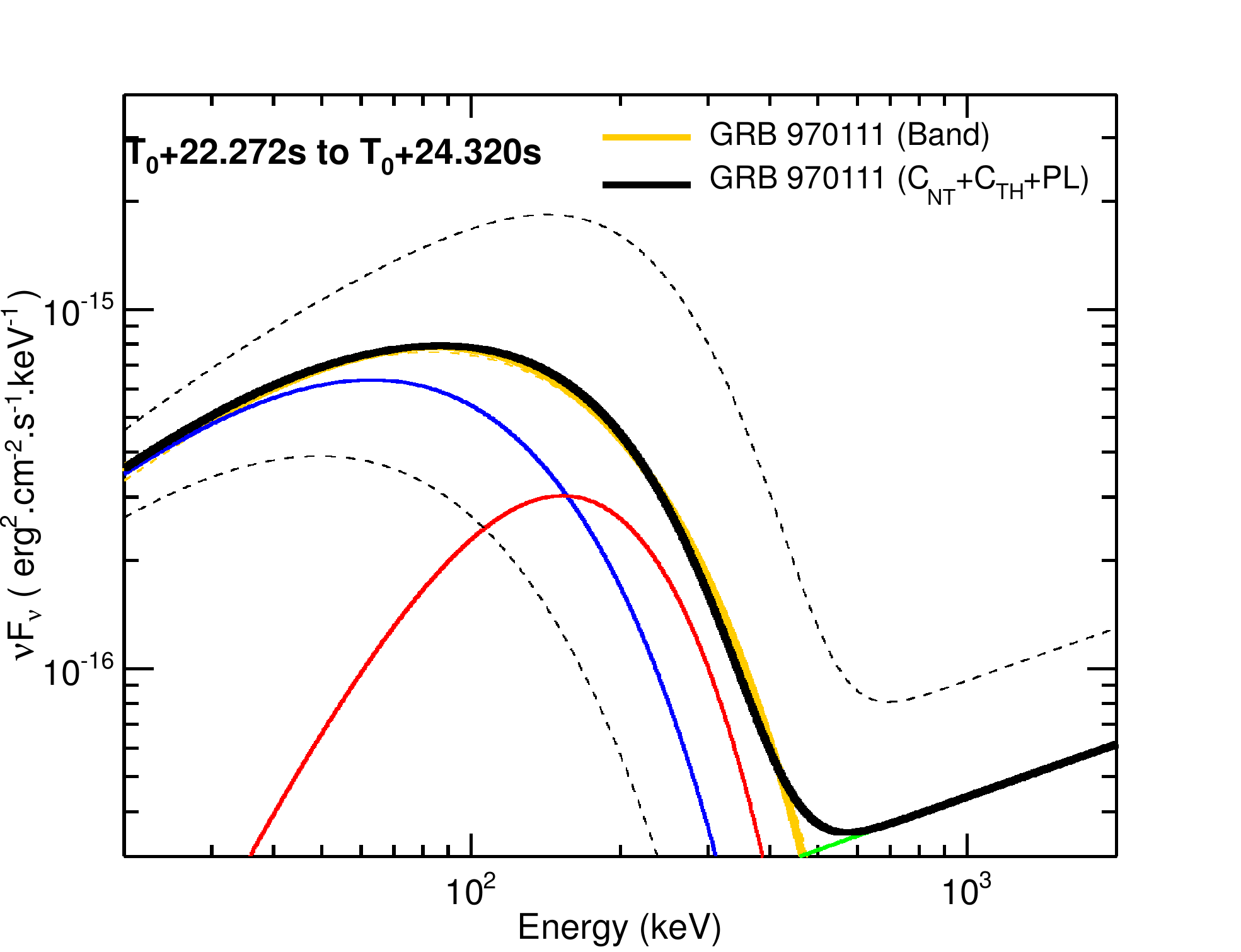}

\includegraphics[totalheight=0.185\textheight, clip]{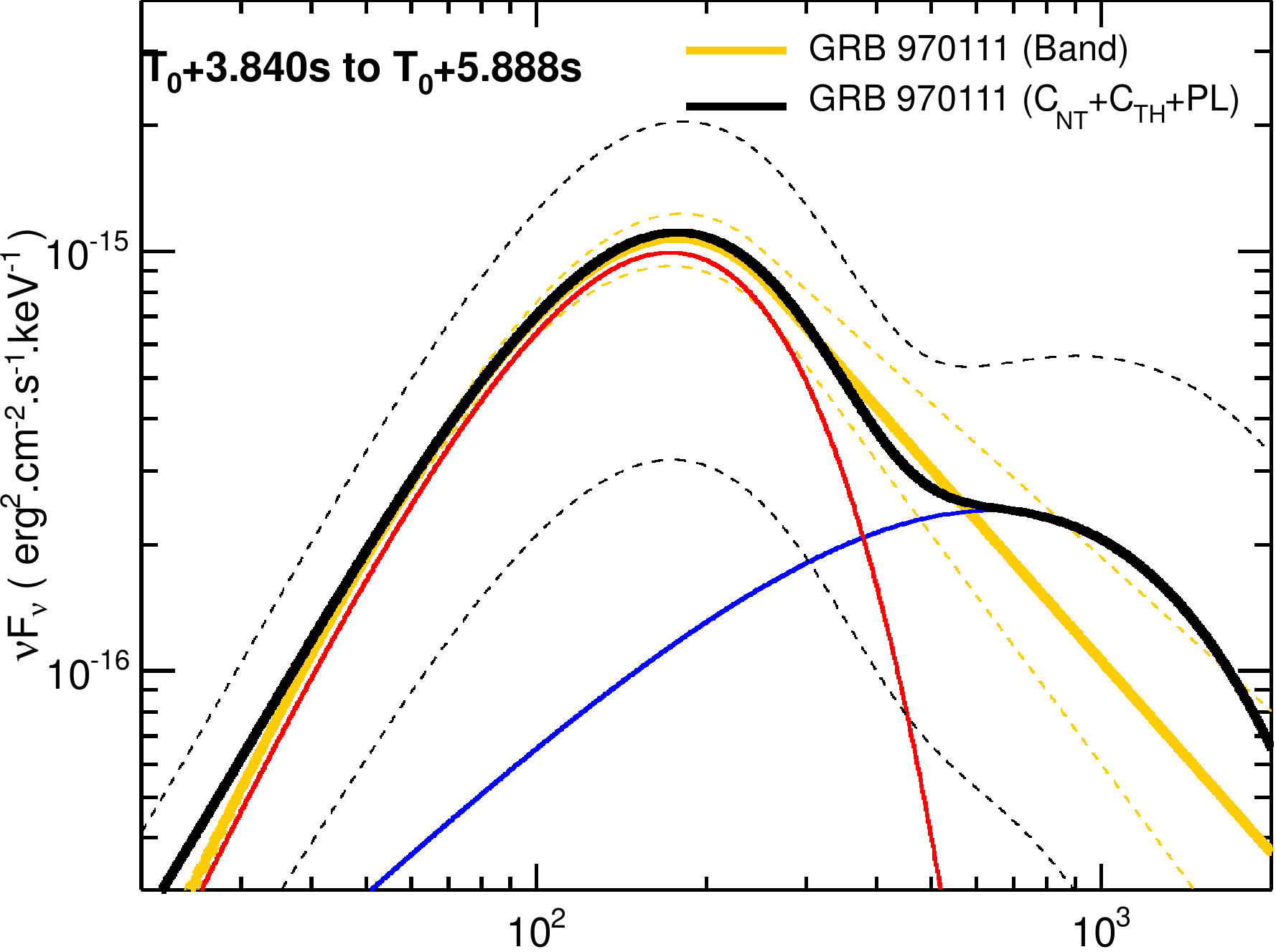}
\includegraphics[totalheight=0.185\textheight, clip]{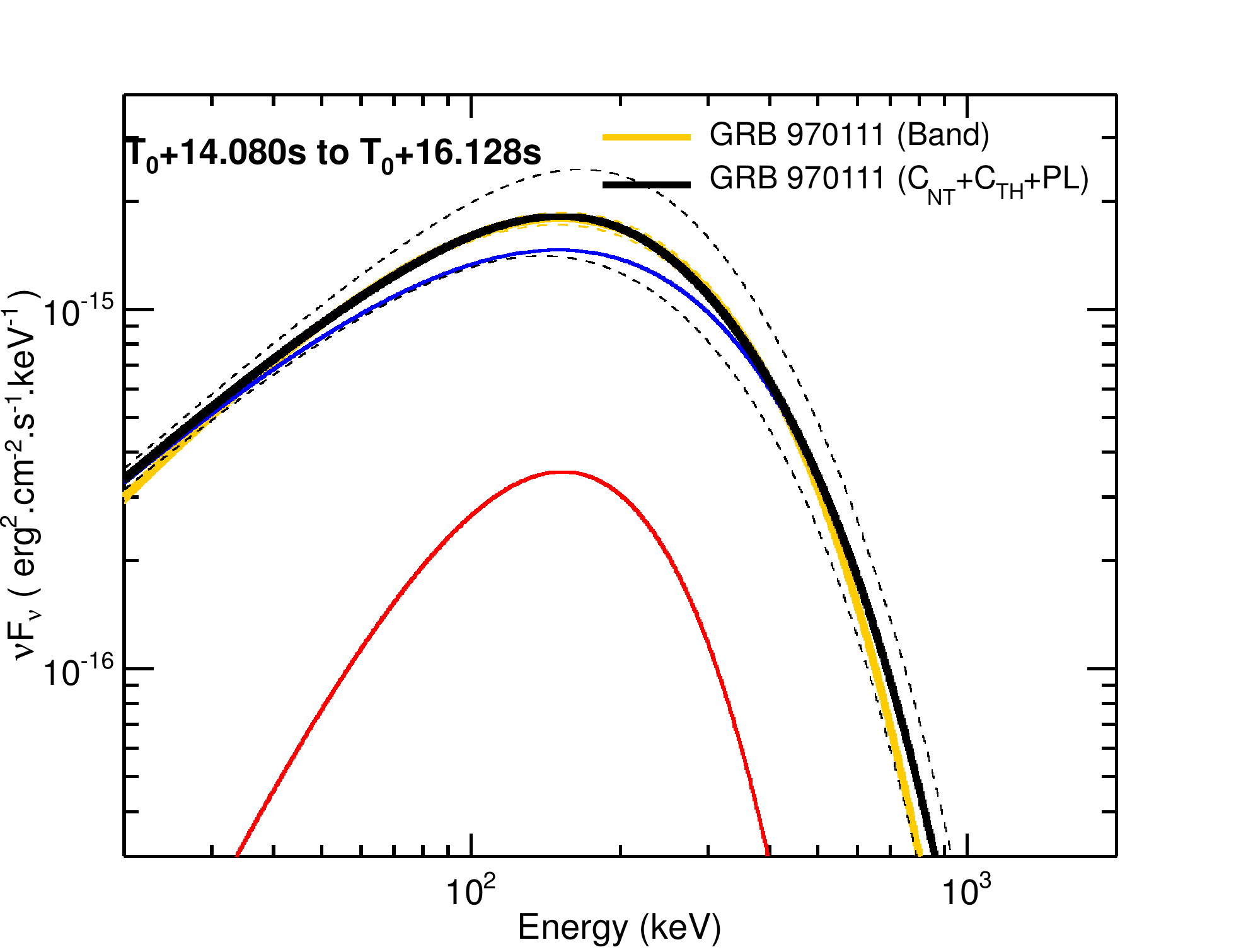}
\includegraphics[totalheight=0.185\textheight, clip]{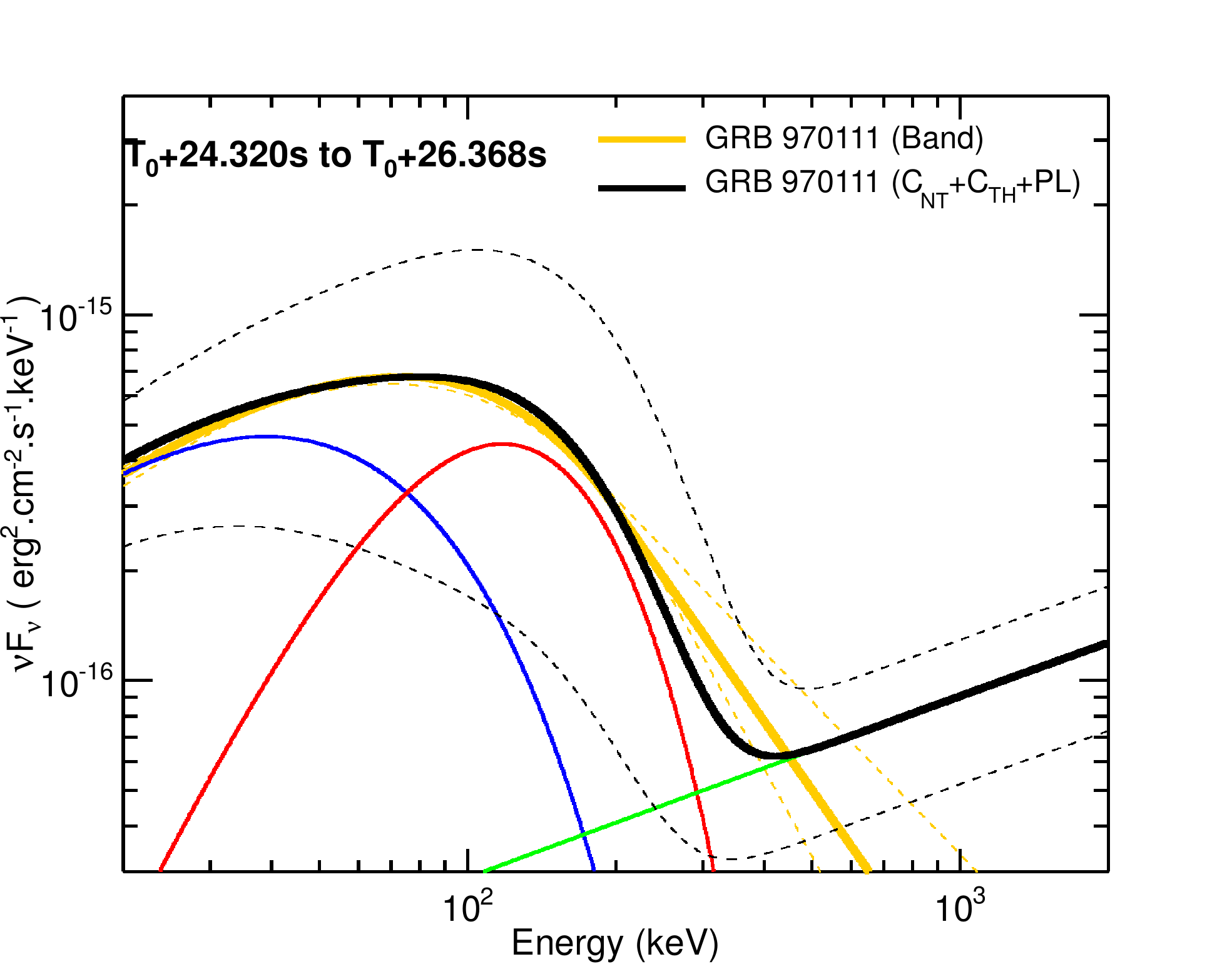}

\includegraphics[totalheight=0.195\textheight, clip]{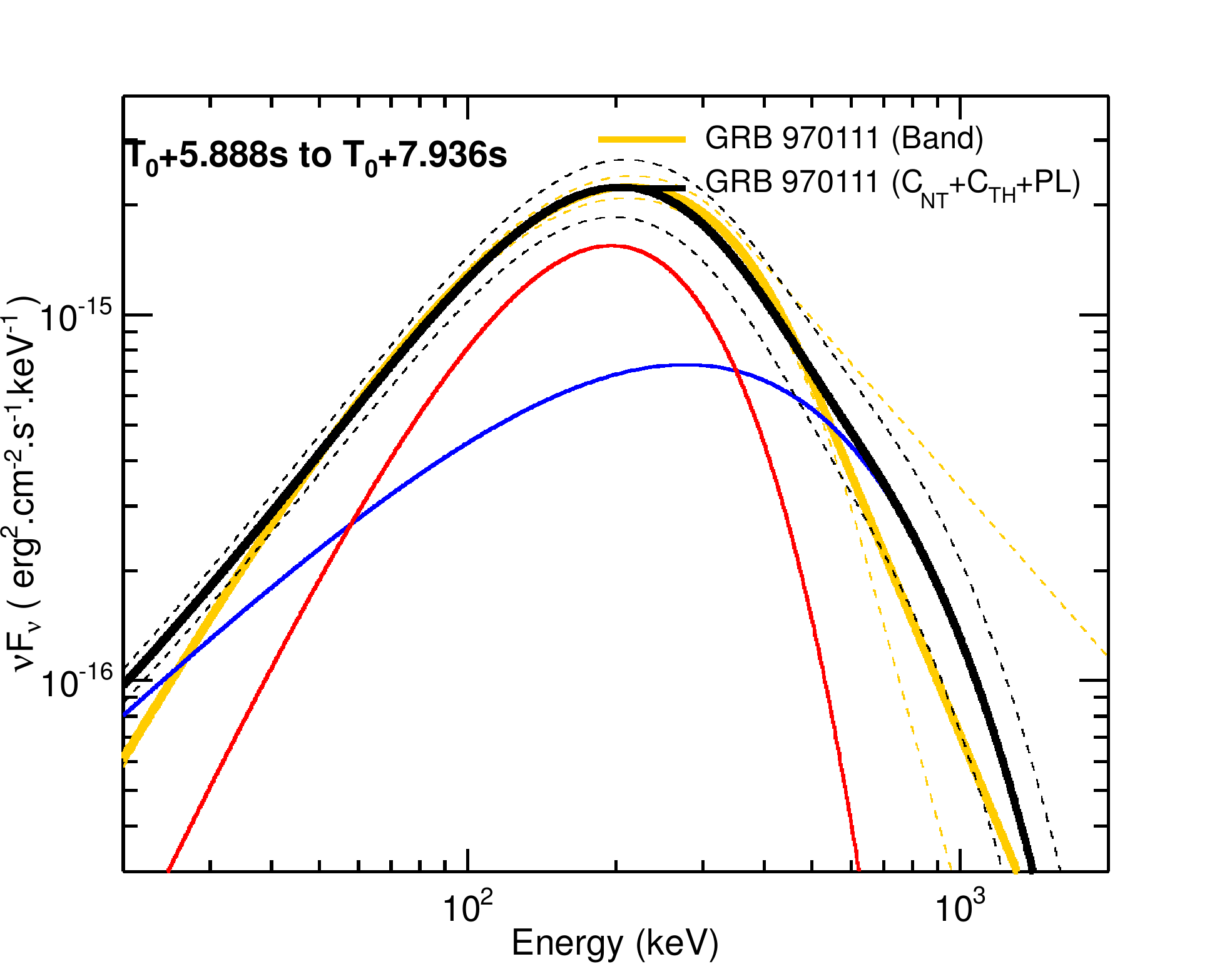}
\includegraphics[totalheight=0.195\textheight, clip]{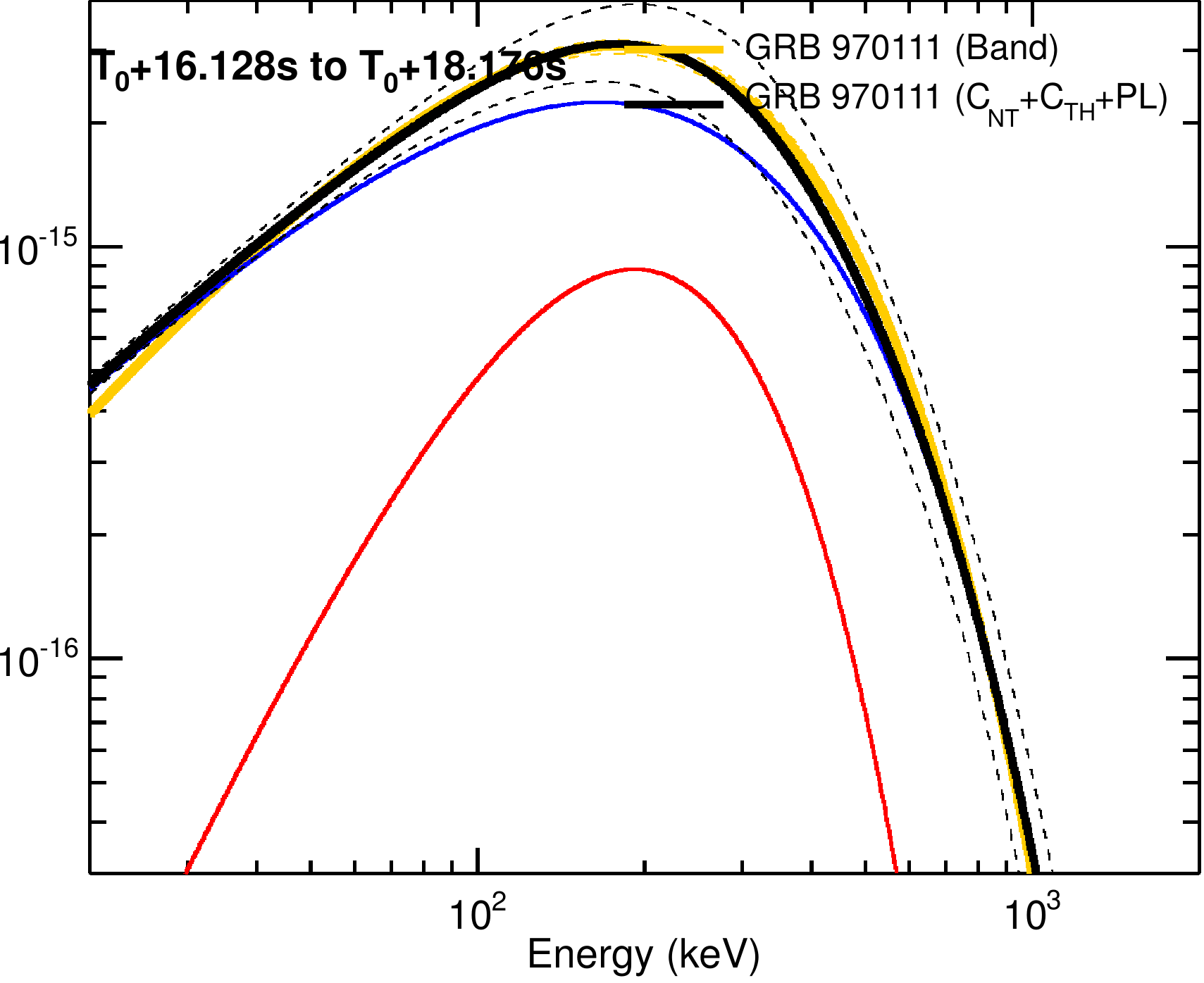}
\includegraphics[totalheight=0.195\textheight, clip]{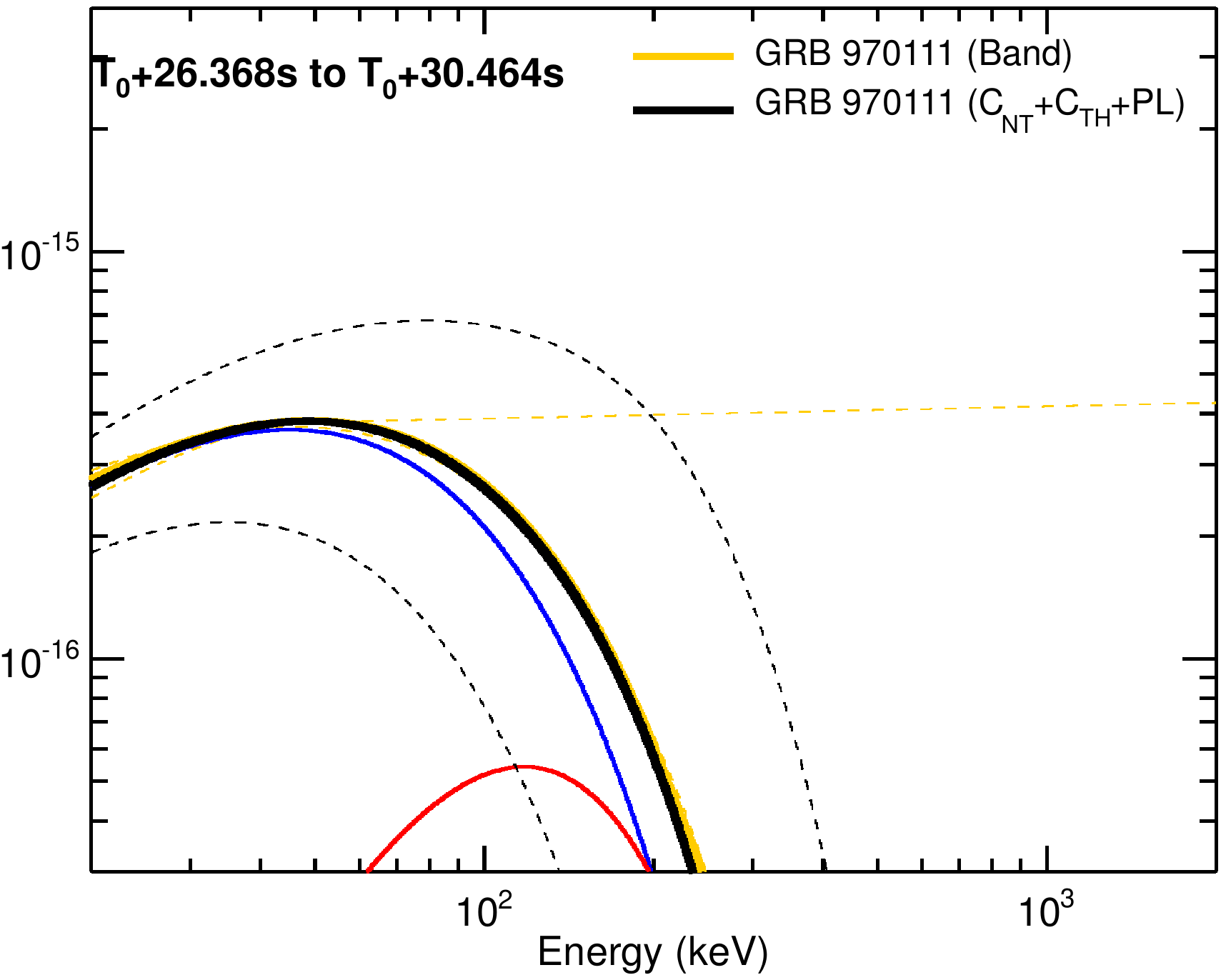}

\end{center}
\end{figure*}

\newpage

\begin{figure*}
\begin{center}
\includegraphics[totalheight=0.185\textheight, clip]{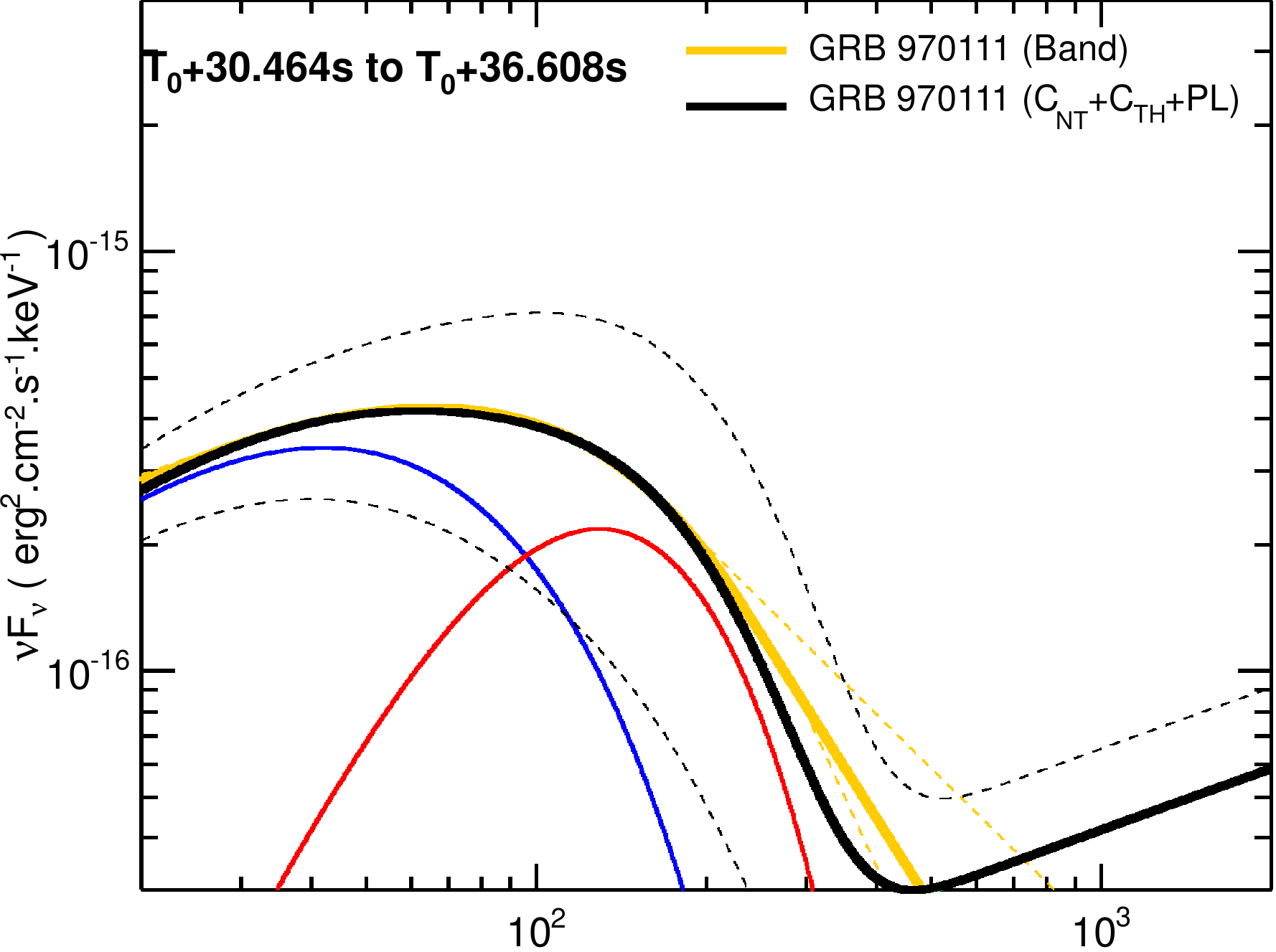}
\includegraphics[totalheight=0.185\textheight, clip]{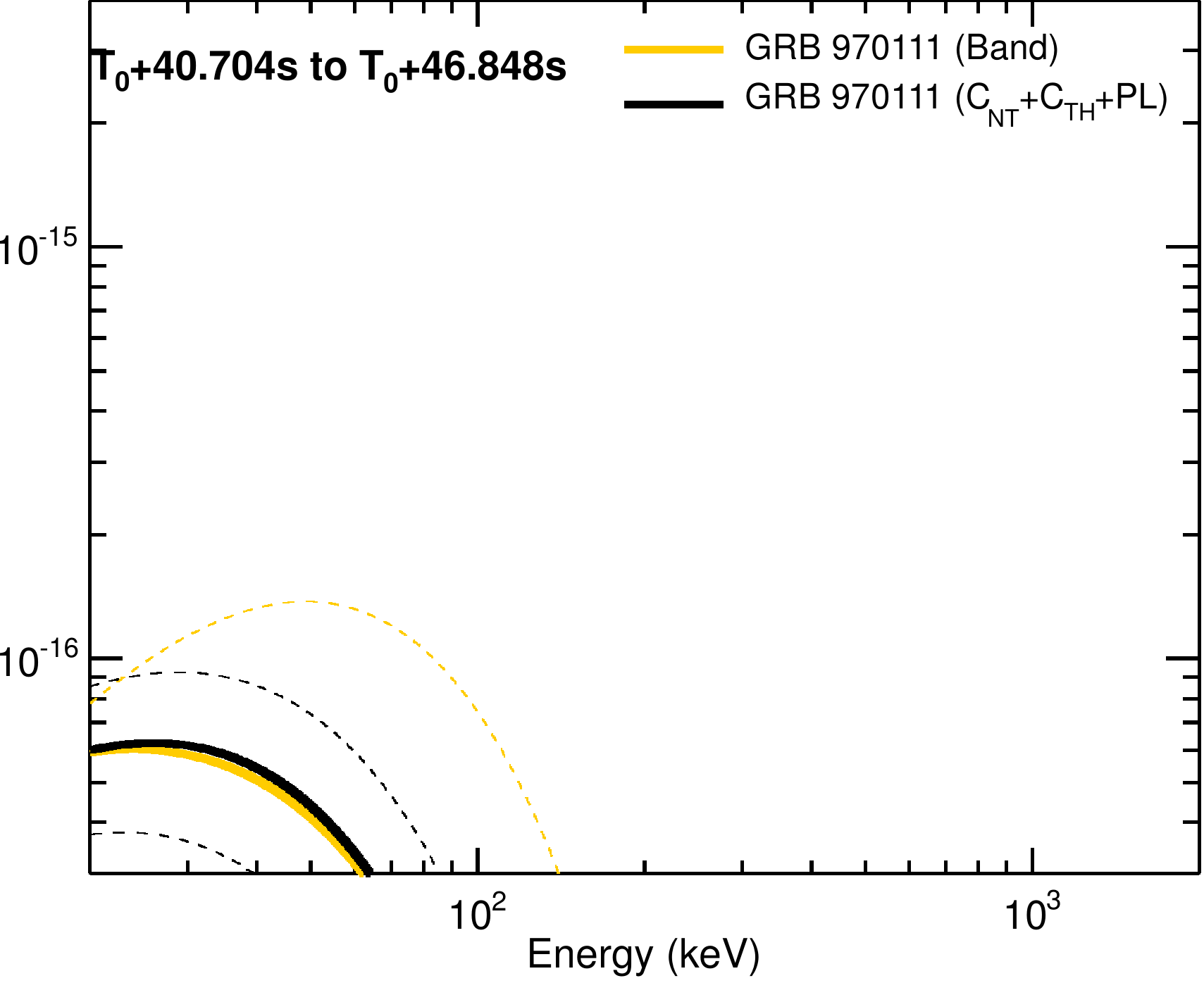}

\hspace{-6.cm}
\includegraphics[totalheight=0.195\textheight, clip]{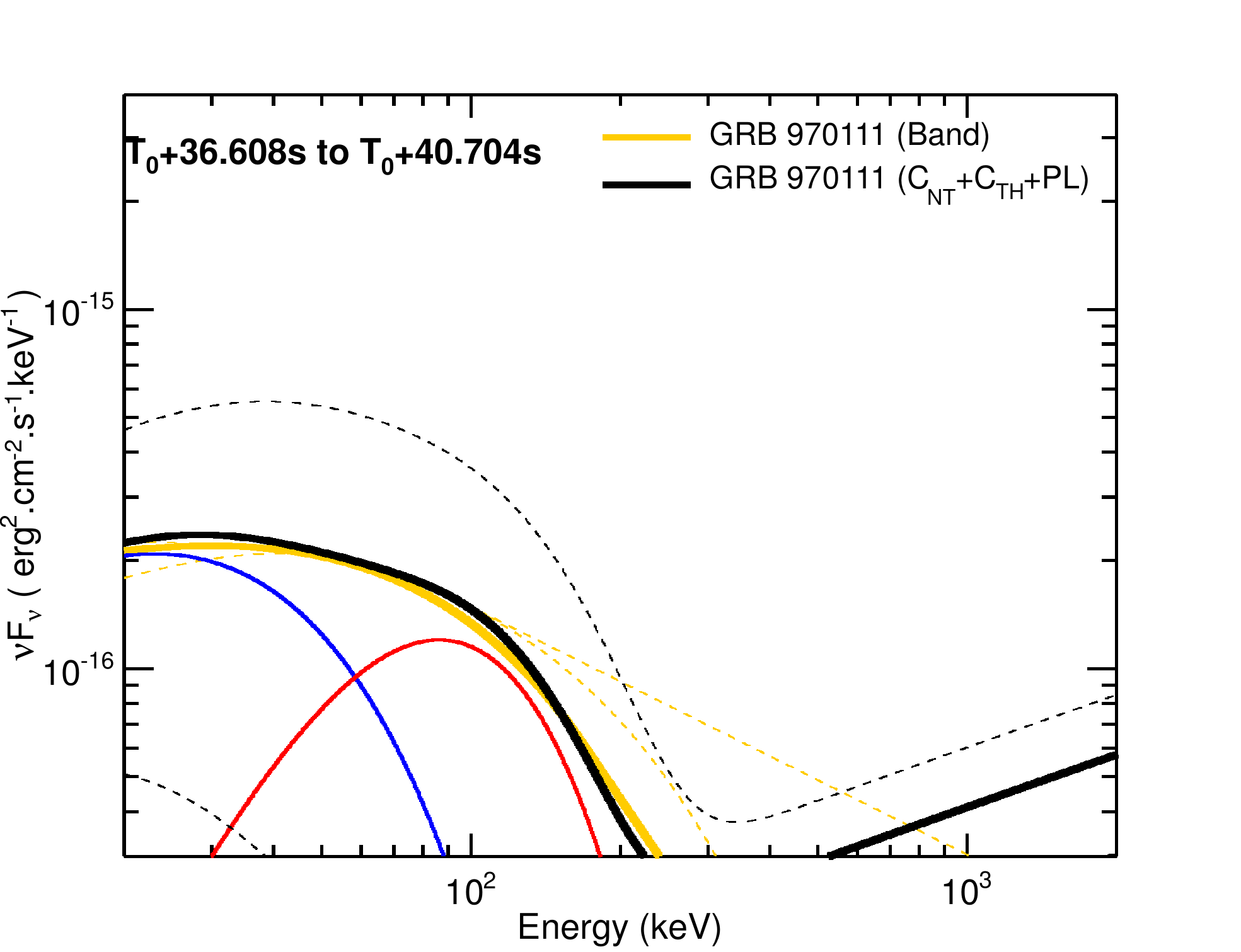}

\caption{\label{fig19}GRB~970111 : $\nu$F$_\nu$ spectra resulting from the fine-time analysis presented in Section~\ref{sec:trsa}. The solid yellow and black lines correspond to the best Band-only and C$_\mathrm{nTh}$+C$_\mathrm{Th}$+PL fits, respectively. The dashed yellow and black lines correspond to the 1--$\sigma$ confidence regions of the Band-only and C$_\mathrm{nTh}$+C$_\mathrm{Th}$+PL fits, respectively. The solid blue, red and green lines correspond to C$_\mathrm{nTh}$, C$_\mathrm{Th}$ and the additional PL resulting from the best C$_\mathrm{nTh}$+C$_\mathrm{Th}$+PL fits (i.e., solid black line) to the data, respectively.}
\end{center}
\end{figure*}

\newpage

\begin{figure*}
\begin{center}
\includegraphics[totalheight=0.185\textheight, clip]{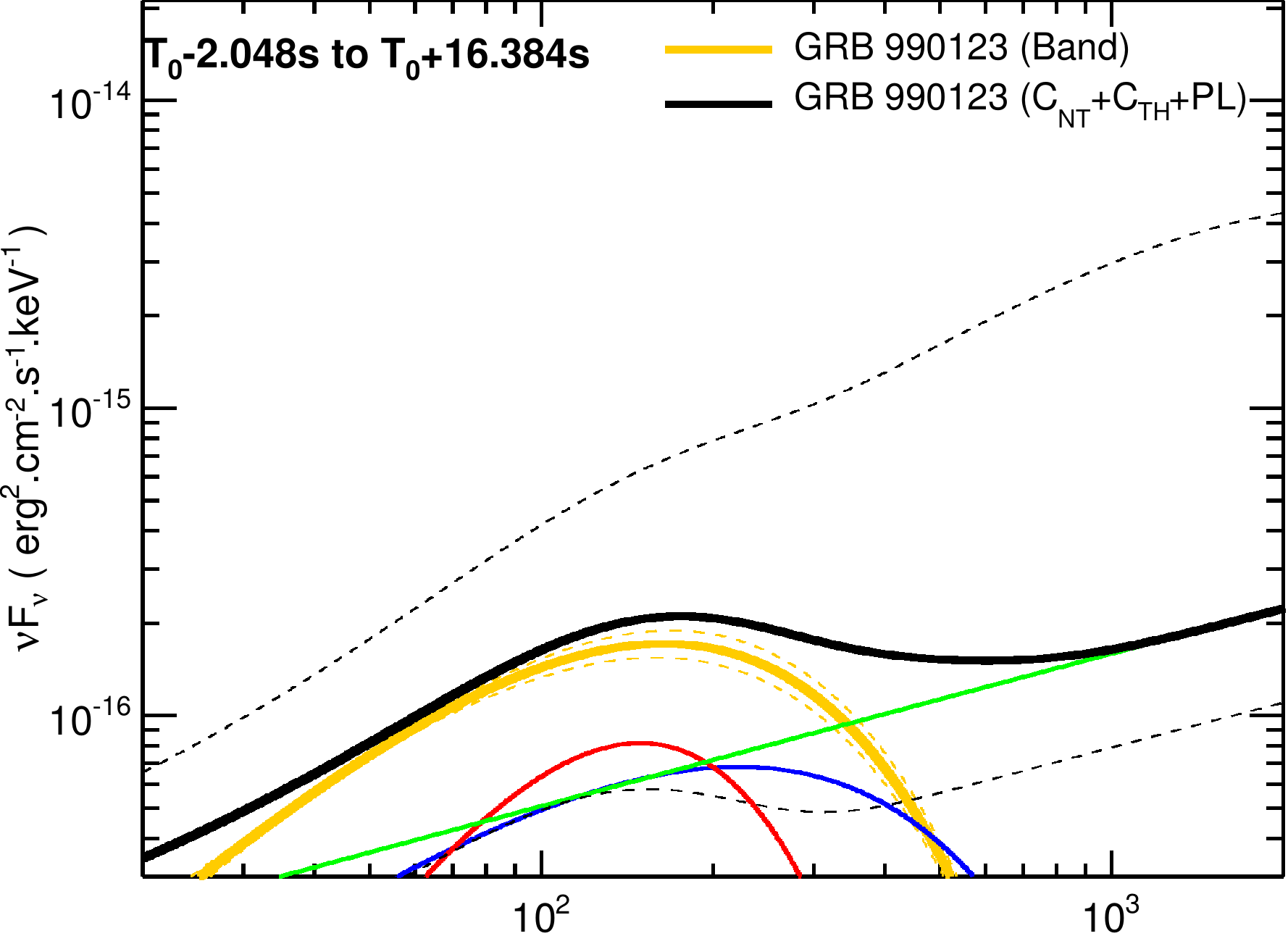}
\includegraphics[totalheight=0.185\textheight, clip]{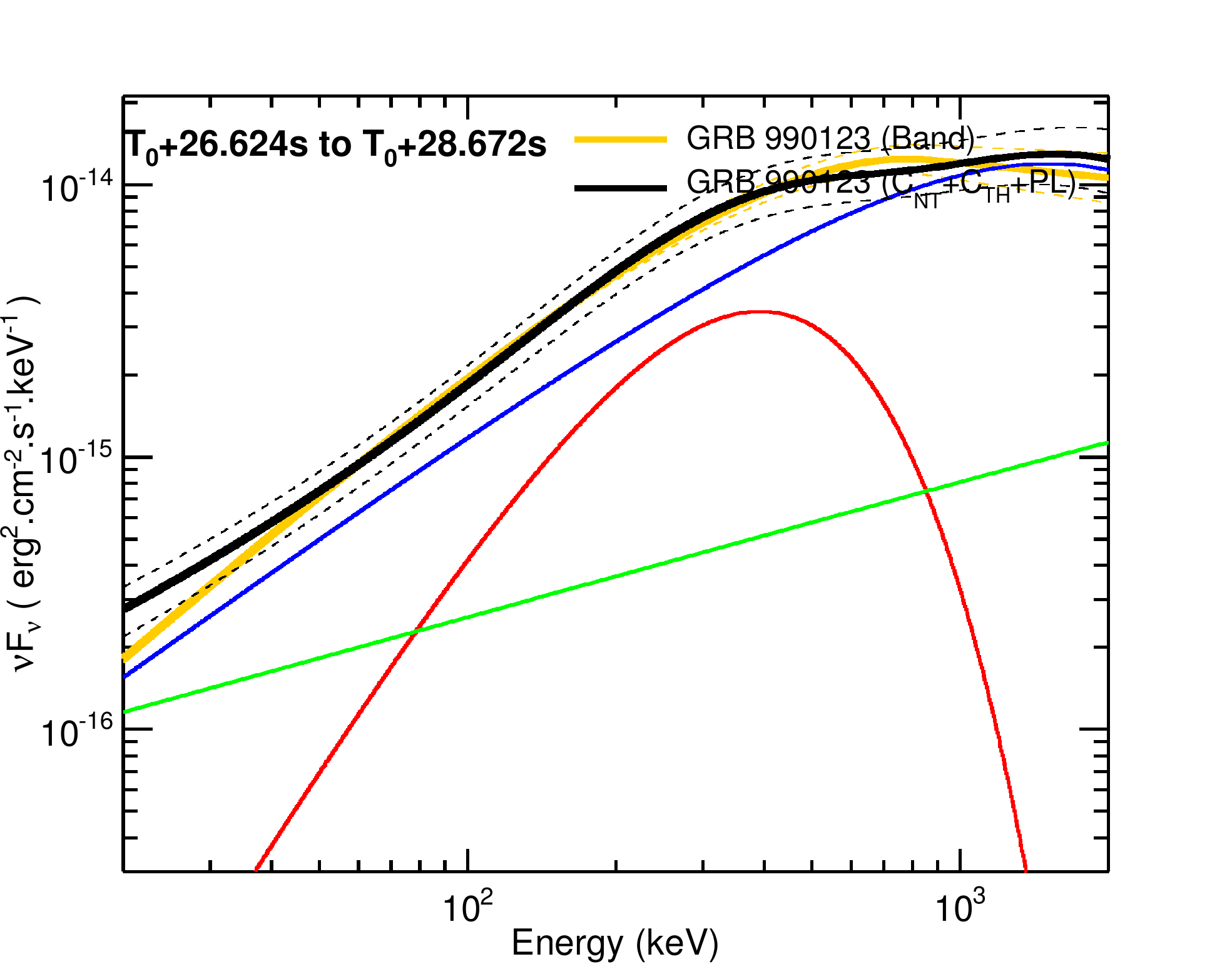}
\includegraphics[totalheight=0.185\textheight, clip]{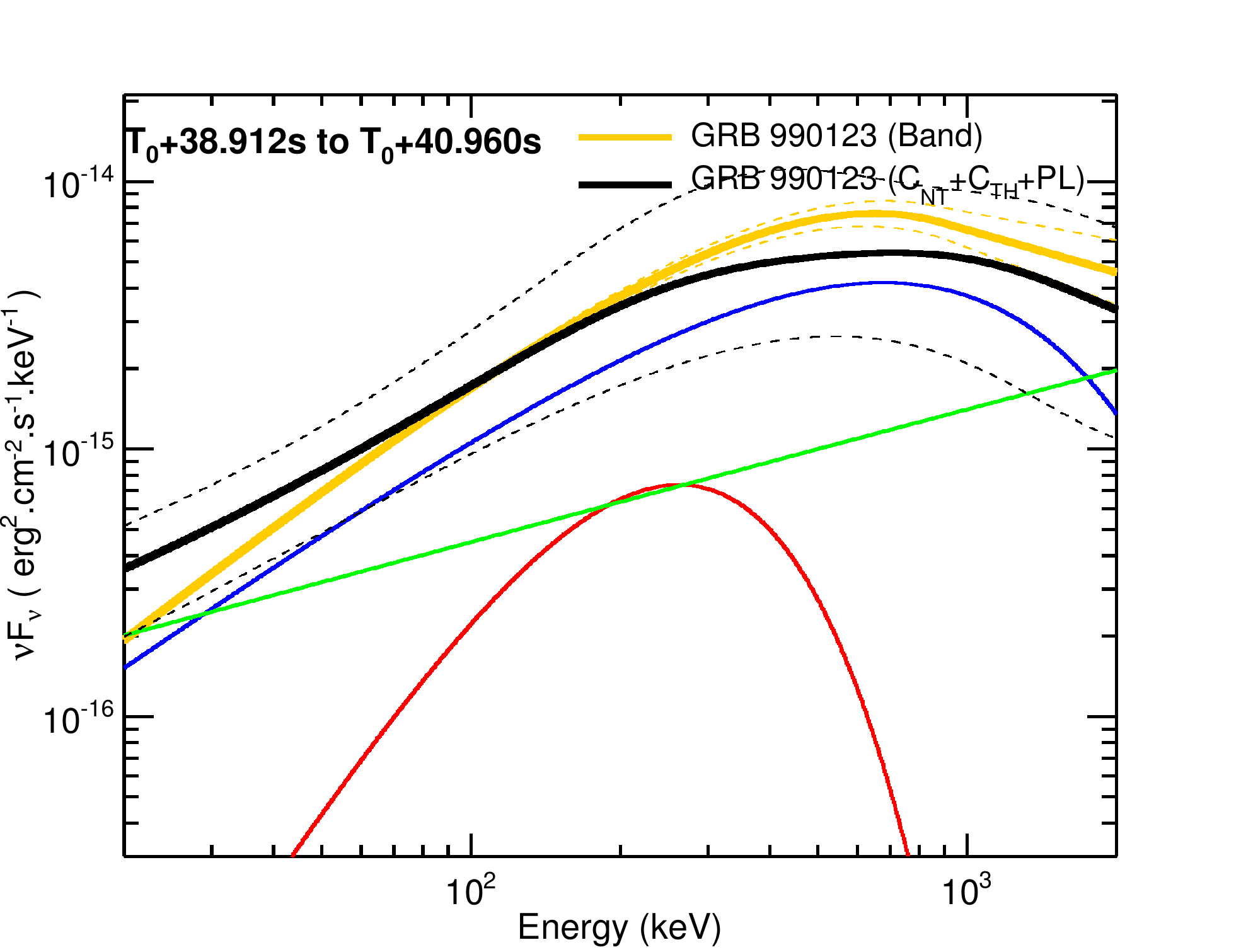}

\includegraphics[totalheight=0.185\textheight, clip]{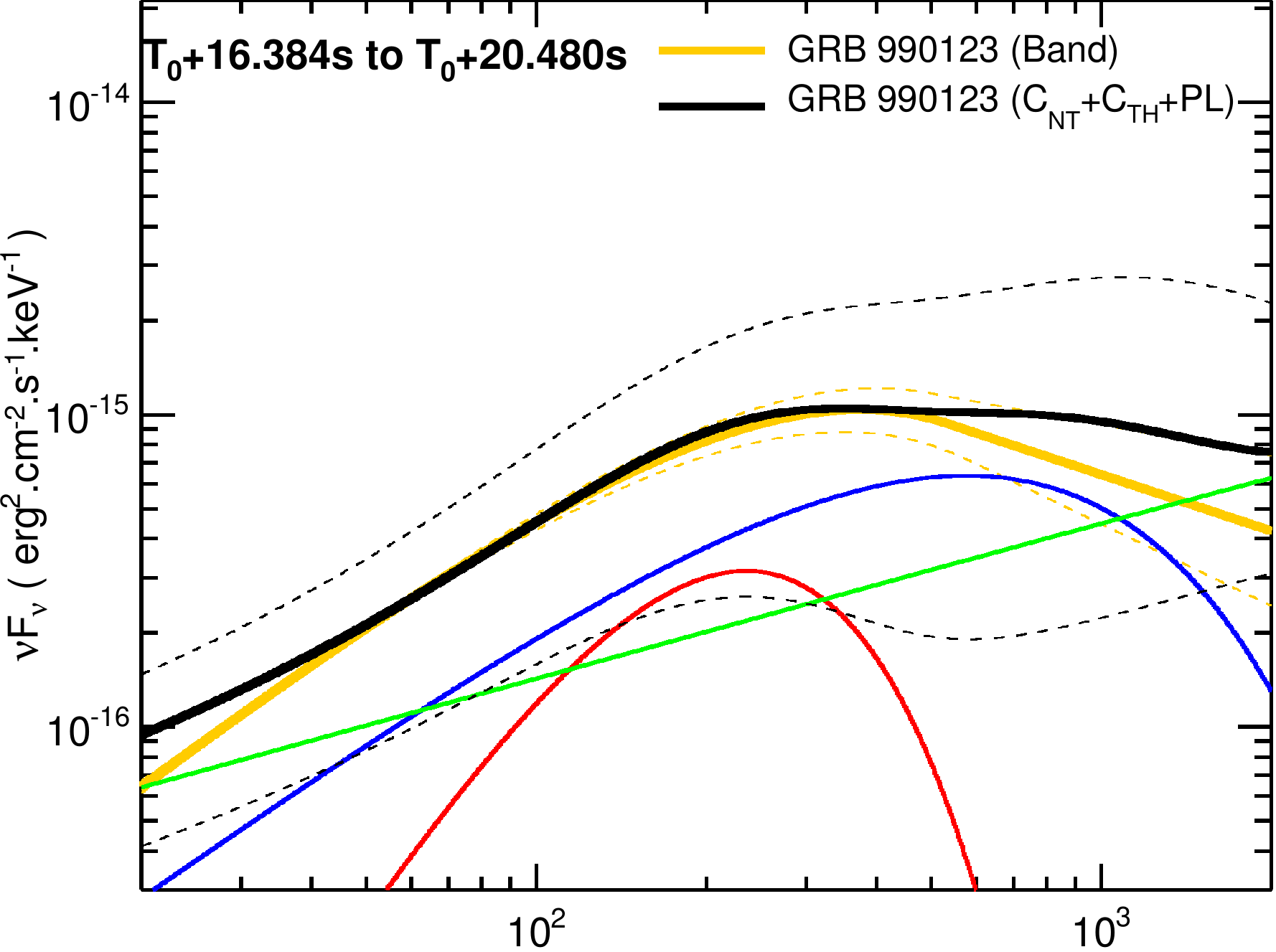}
\includegraphics[totalheight=0.185\textheight, clip]{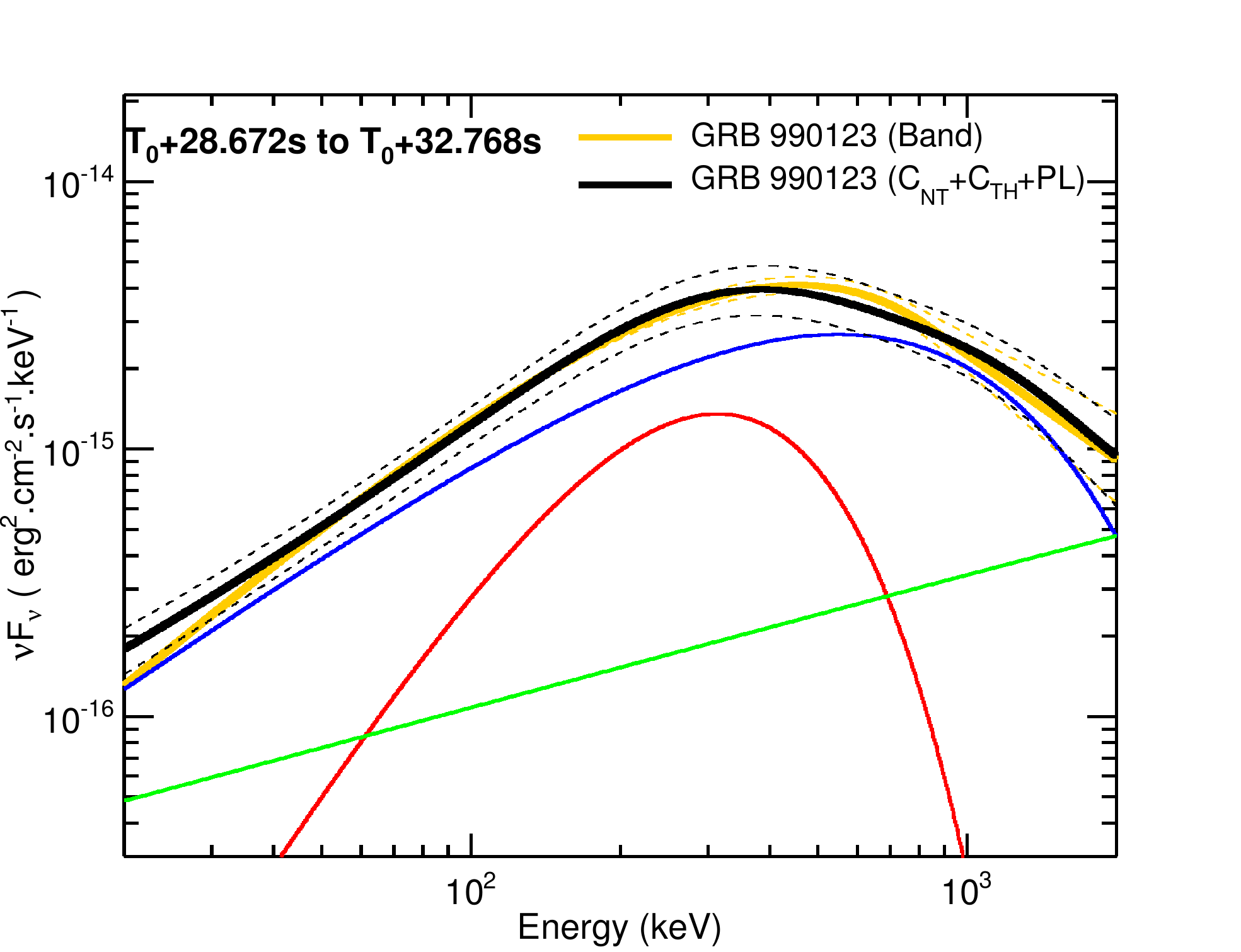}
\includegraphics[totalheight=0.185\textheight, clip]{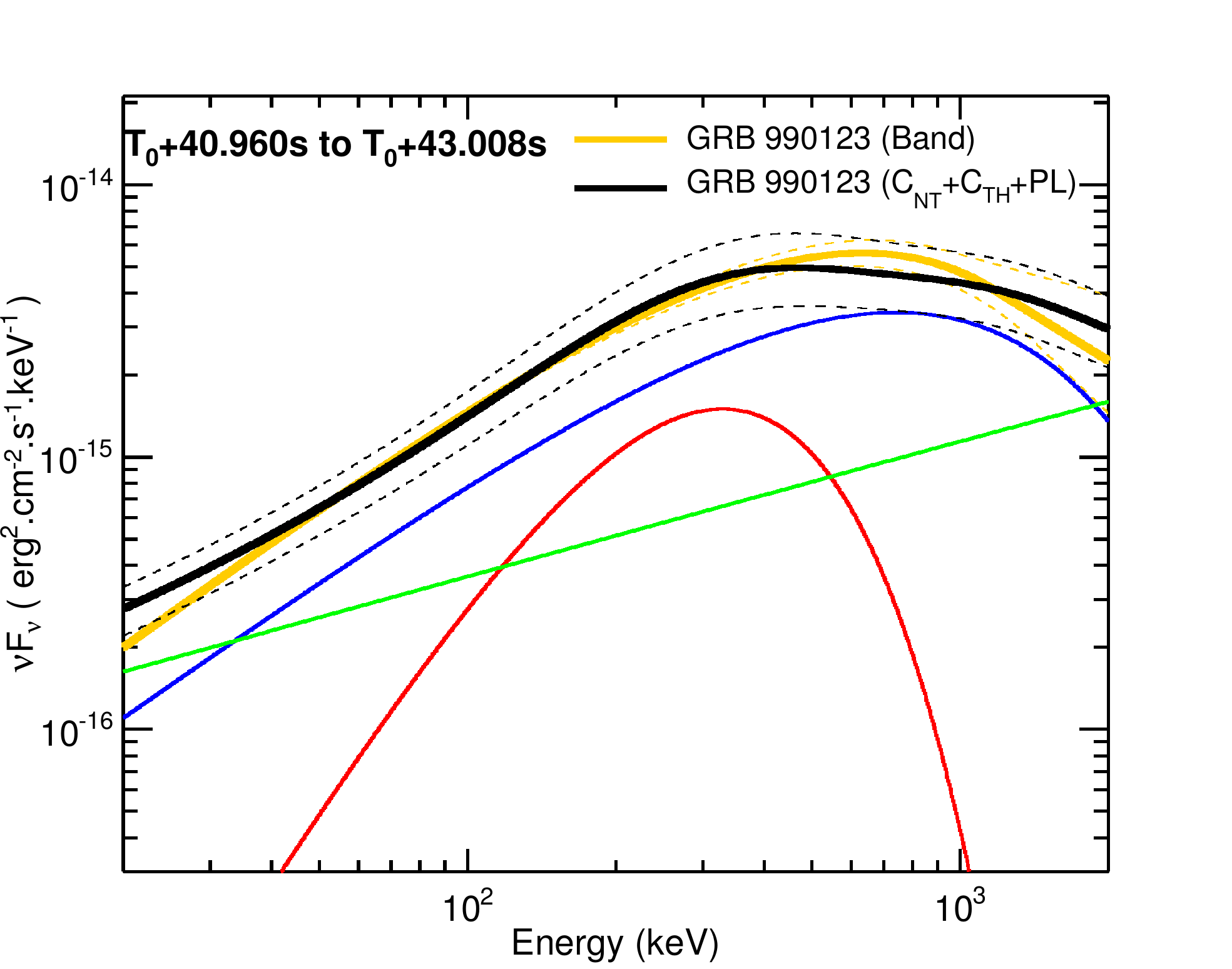}

\includegraphics[totalheight=0.185\textheight, clip]{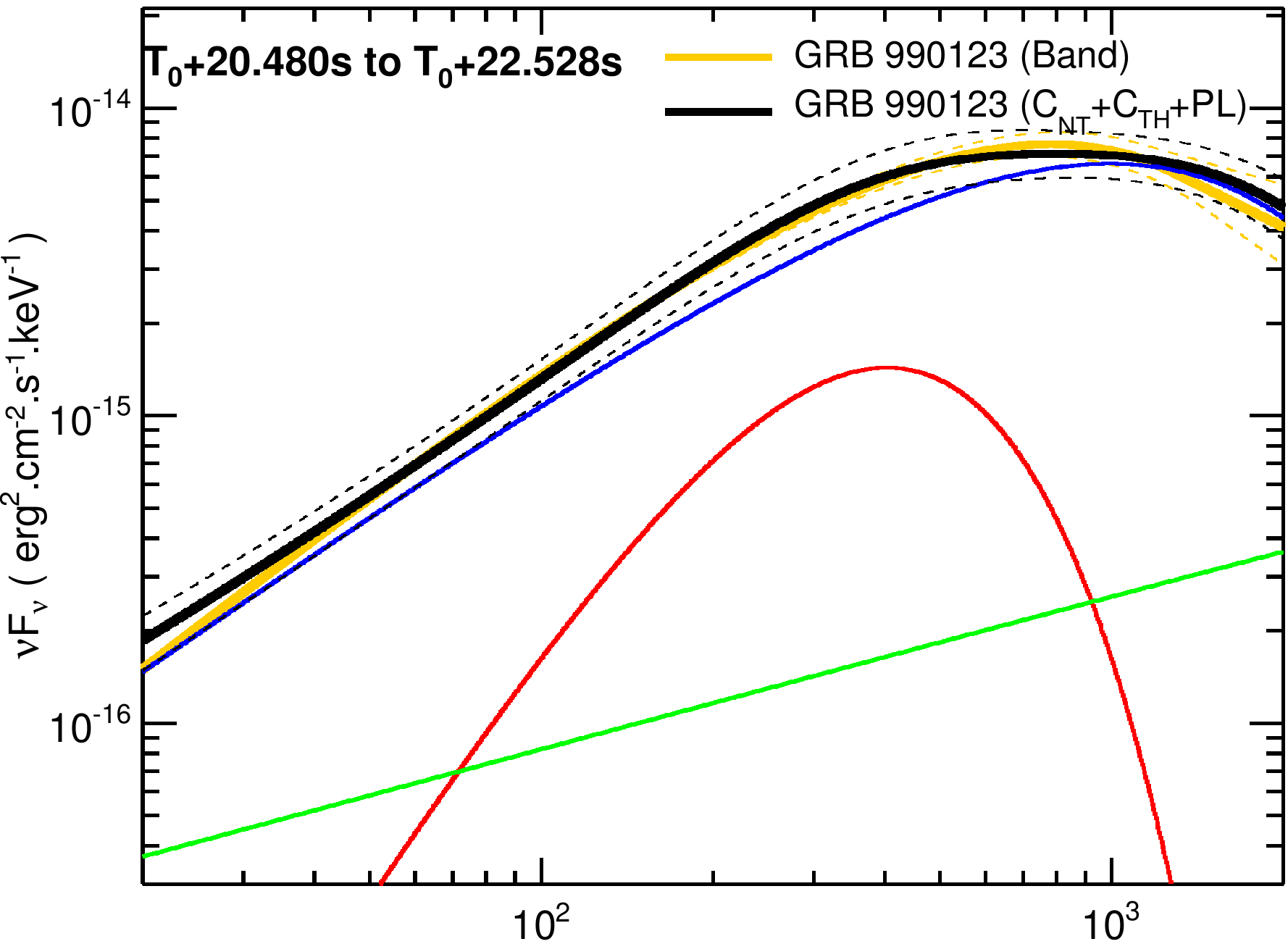}
\includegraphics[totalheight=0.185\textheight, clip]{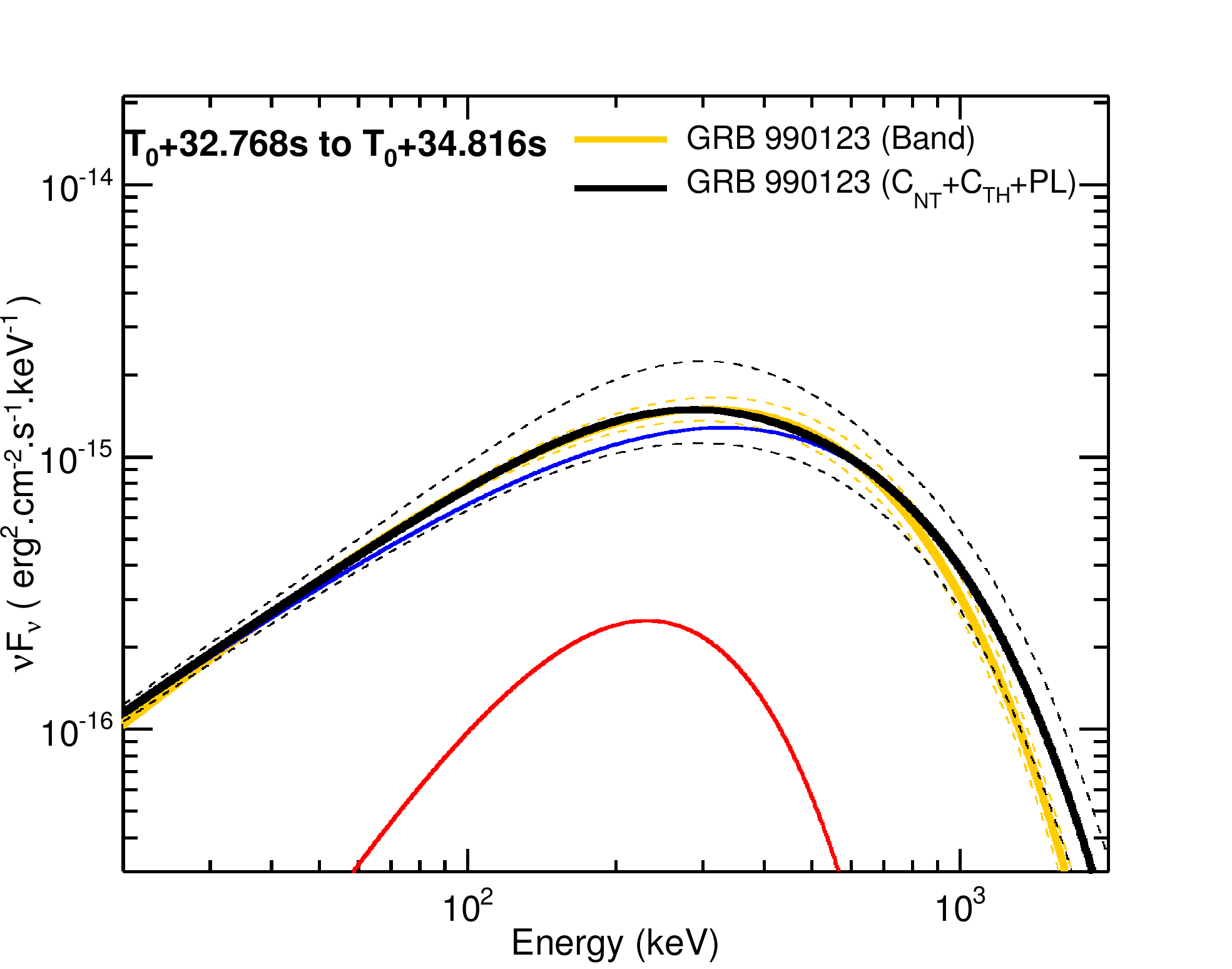}
\includegraphics[totalheight=0.185\textheight, clip]{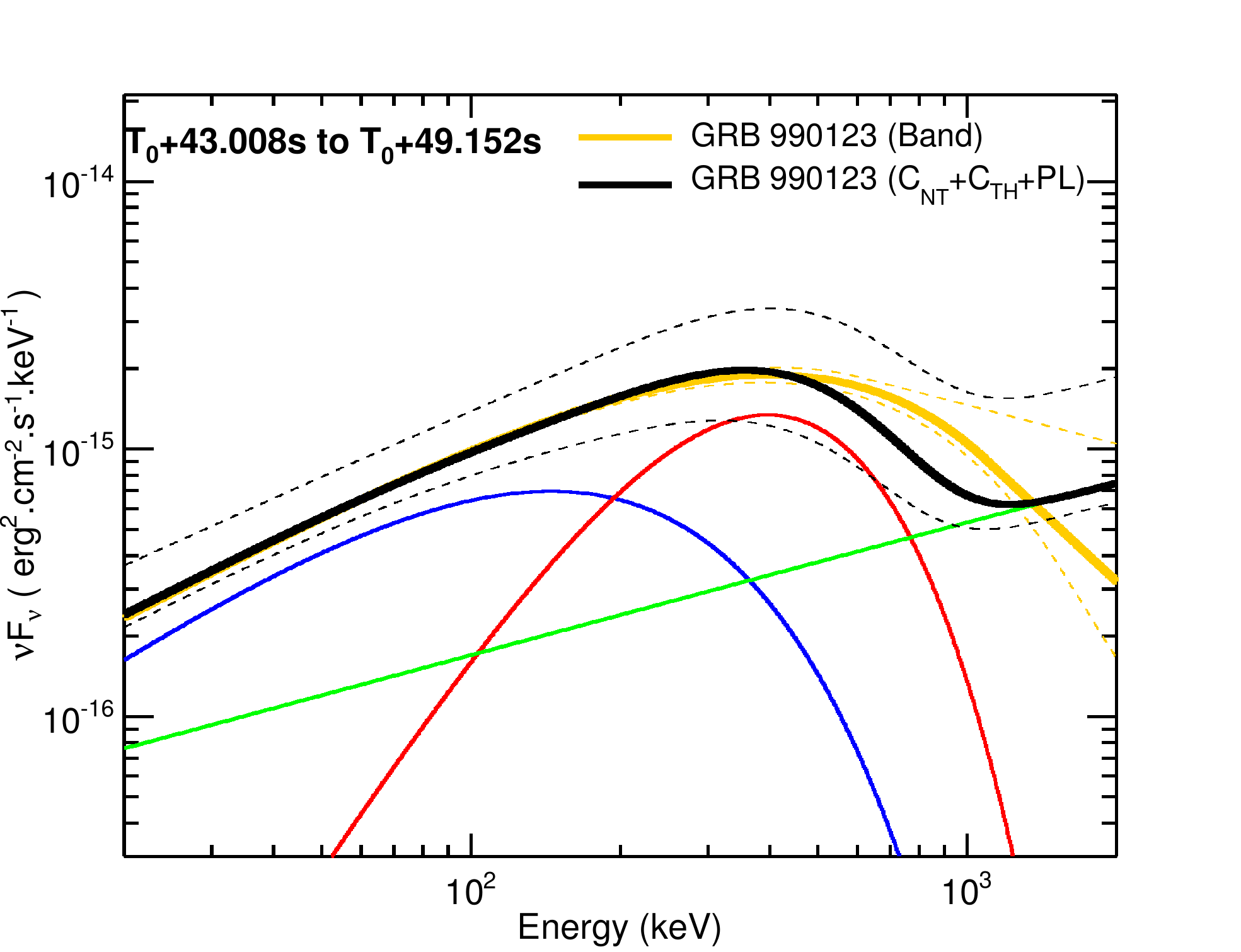}

\includegraphics[totalheight=0.185\textheight, clip]{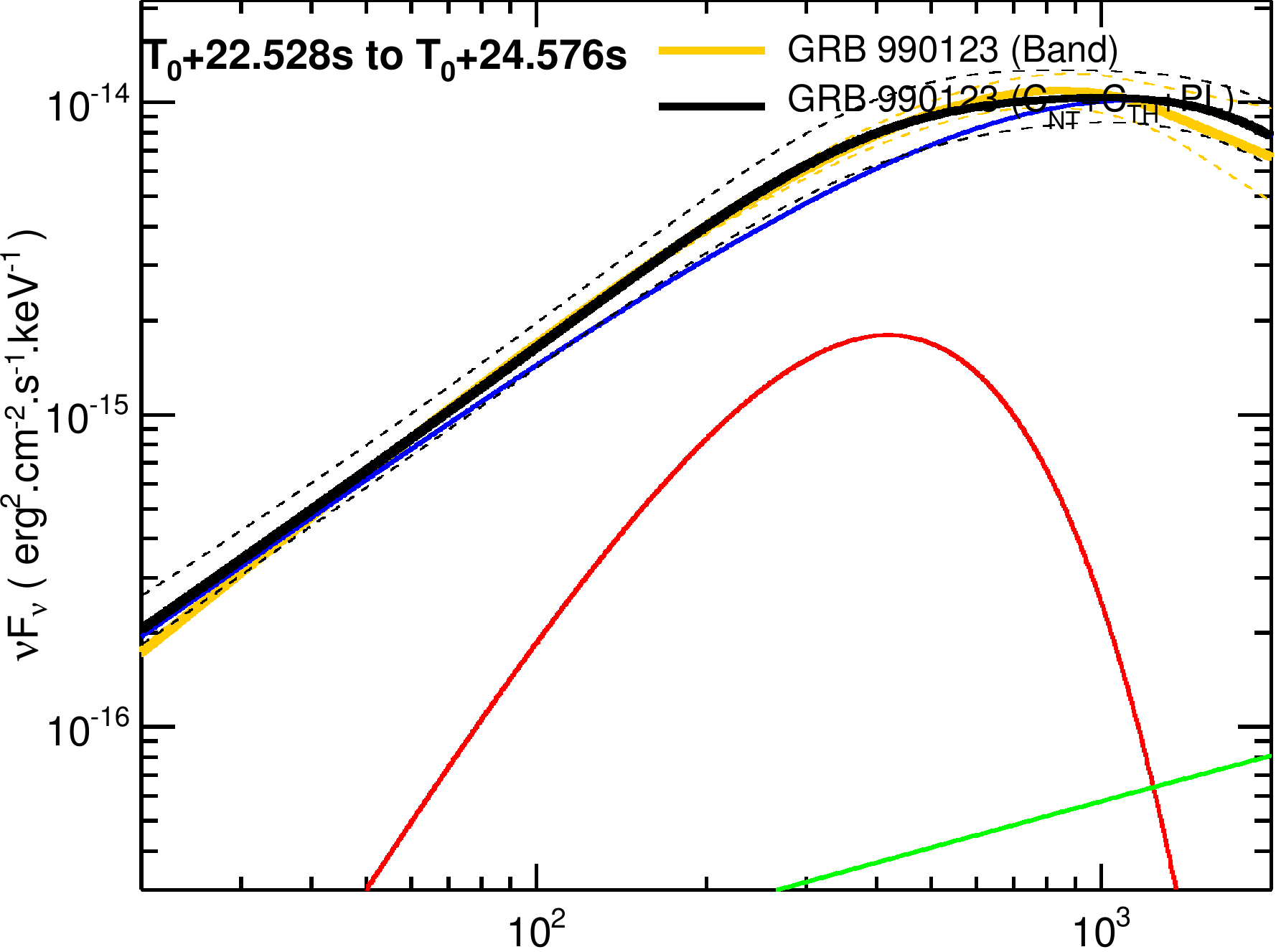}
\includegraphics[totalheight=0.185\textheight, clip]{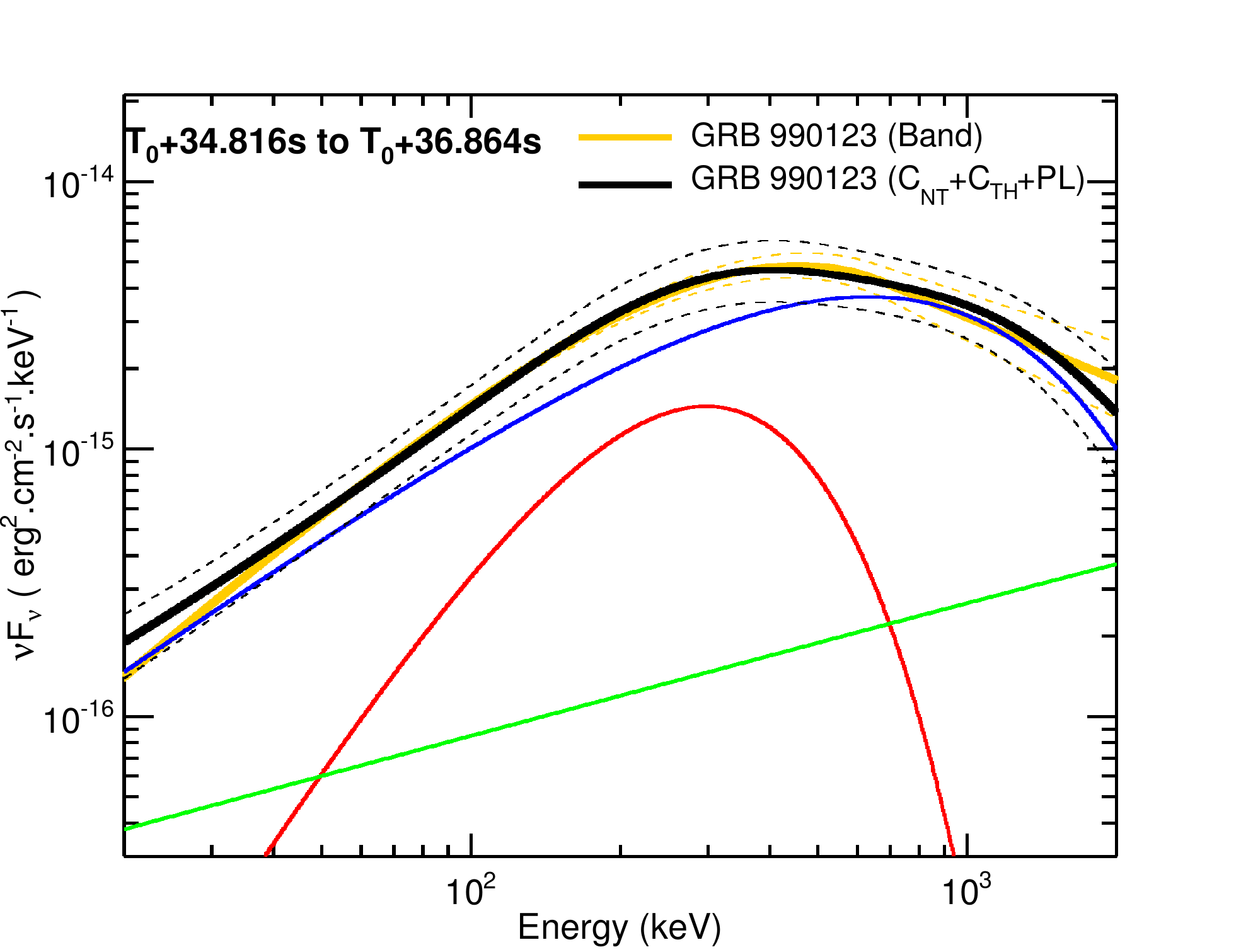}
\includegraphics[totalheight=0.185\textheight, clip]{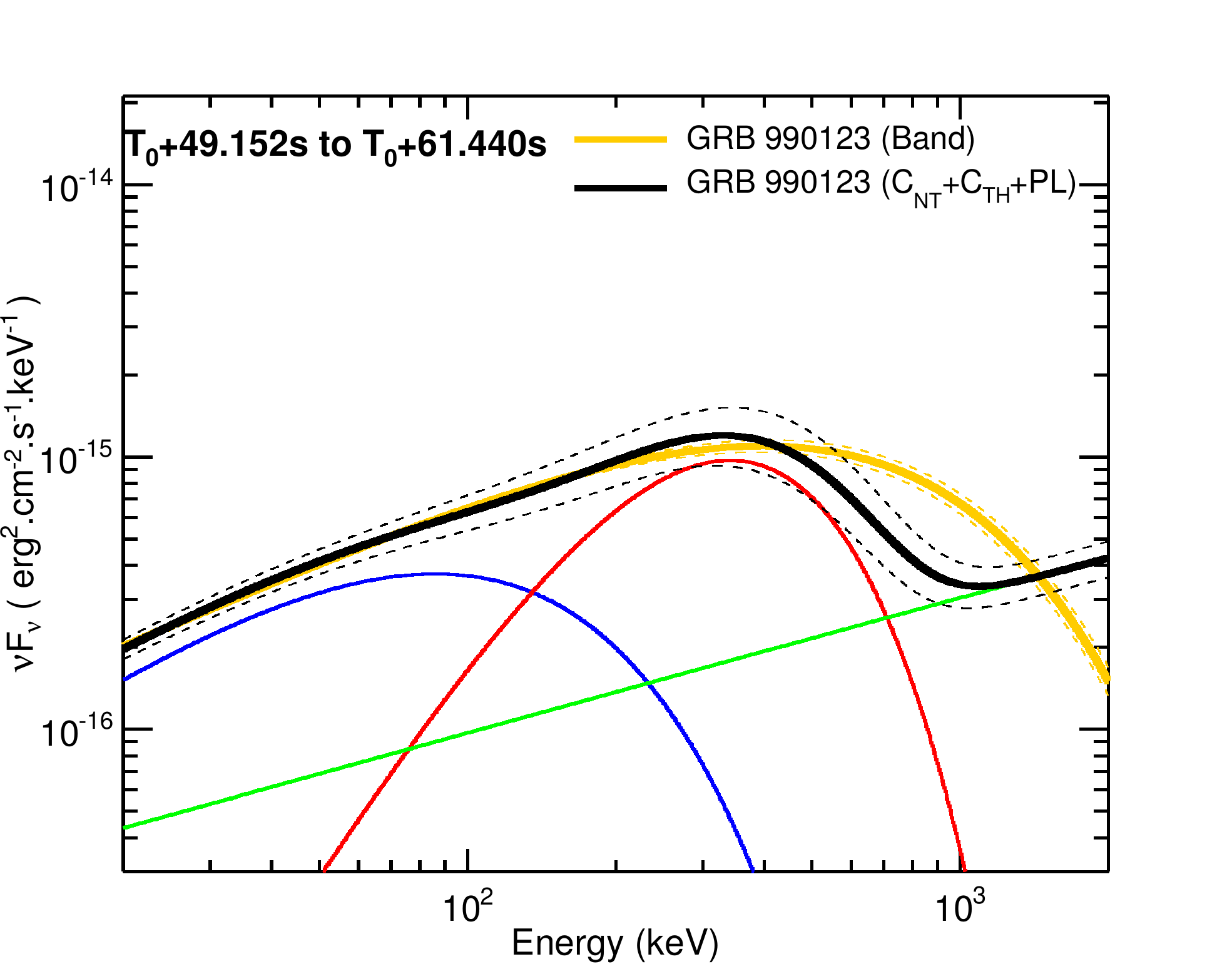}

\includegraphics[totalheight=0.195\textheight, clip]{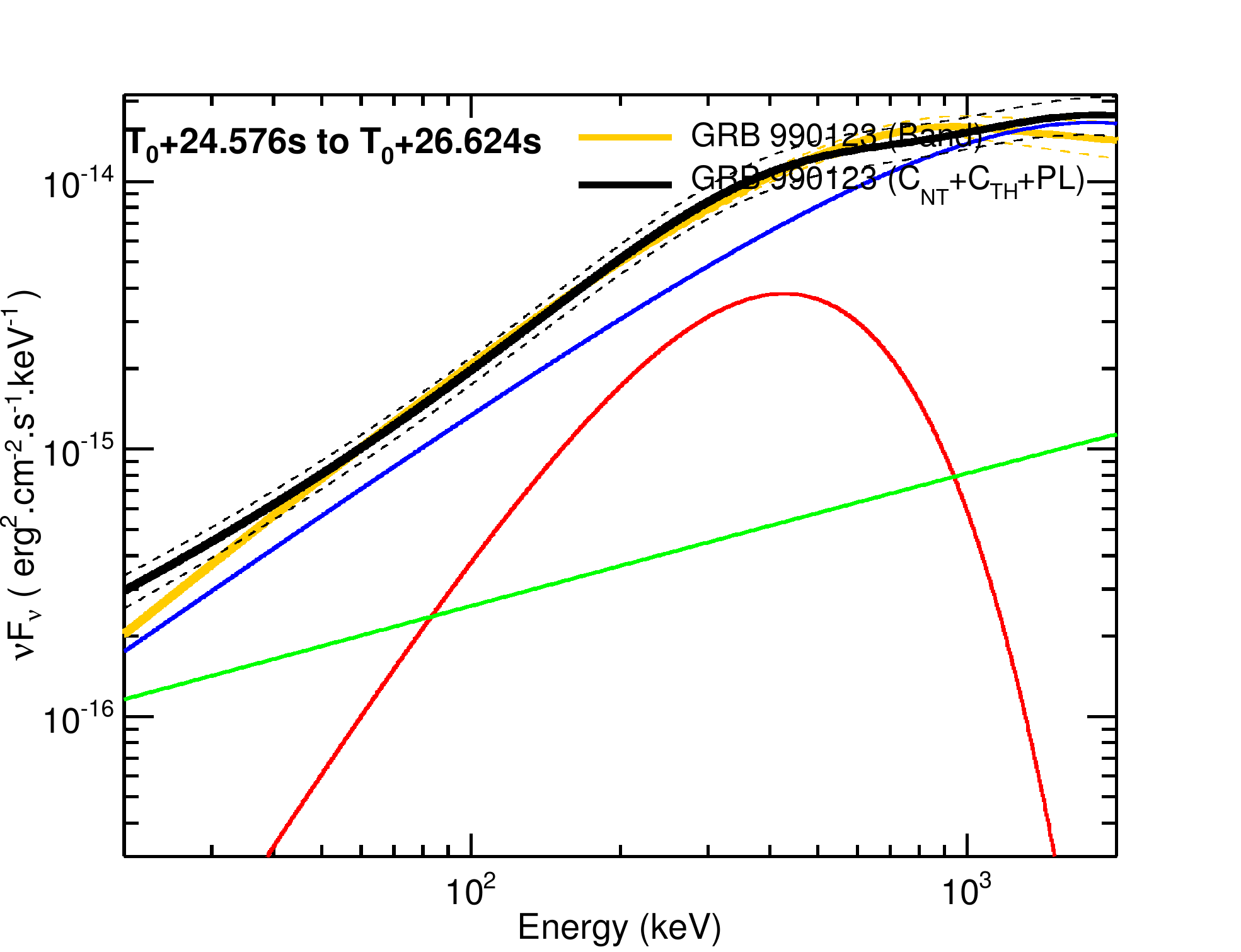}
\includegraphics[totalheight=0.195\textheight, clip]{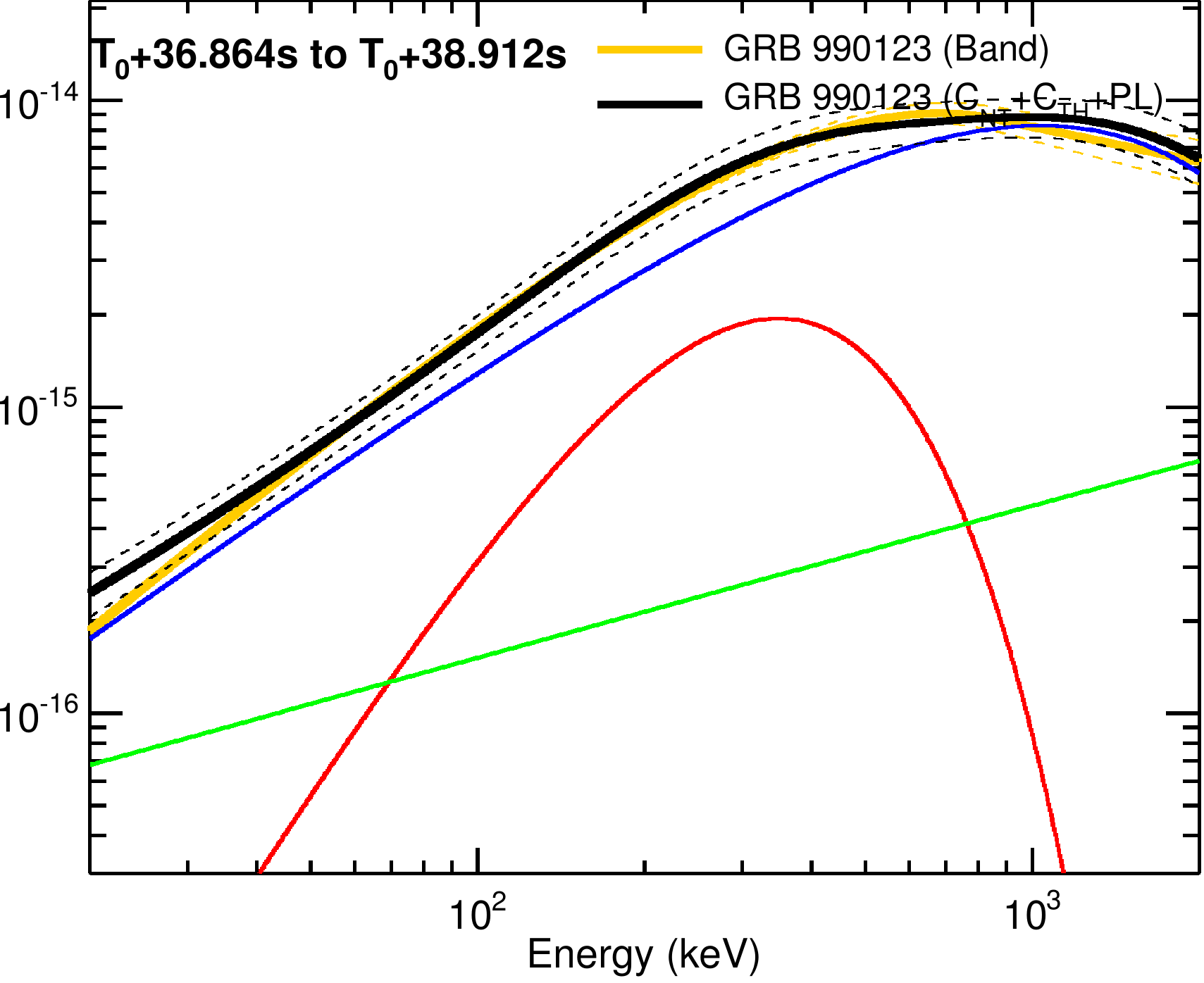}
\includegraphics[totalheight=0.195\textheight, clip]{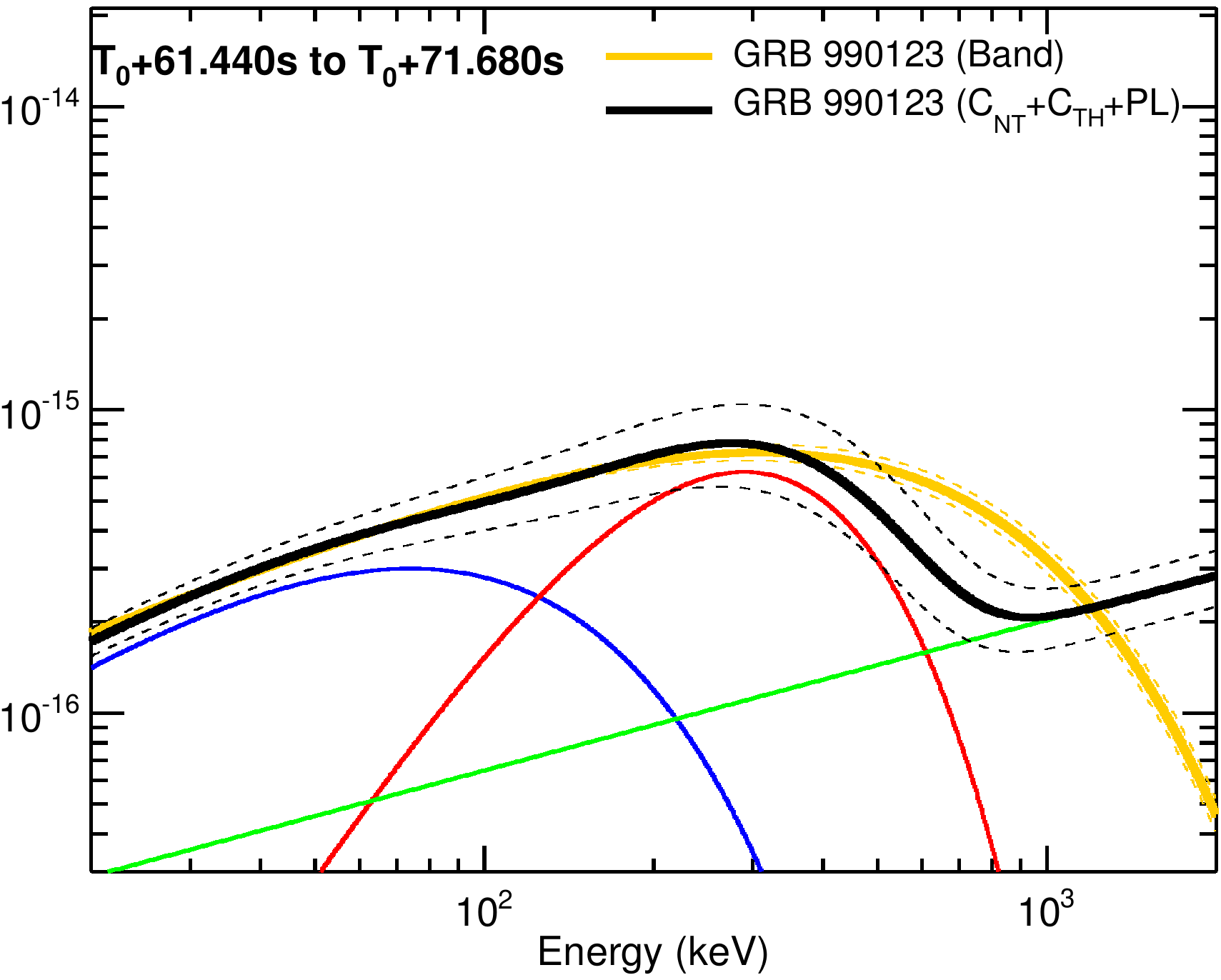}

\end{center}
\end{figure*}

\newpage

\begin{figure*}
\begin{center}
\includegraphics[totalheight=0.185\textheight, clip]{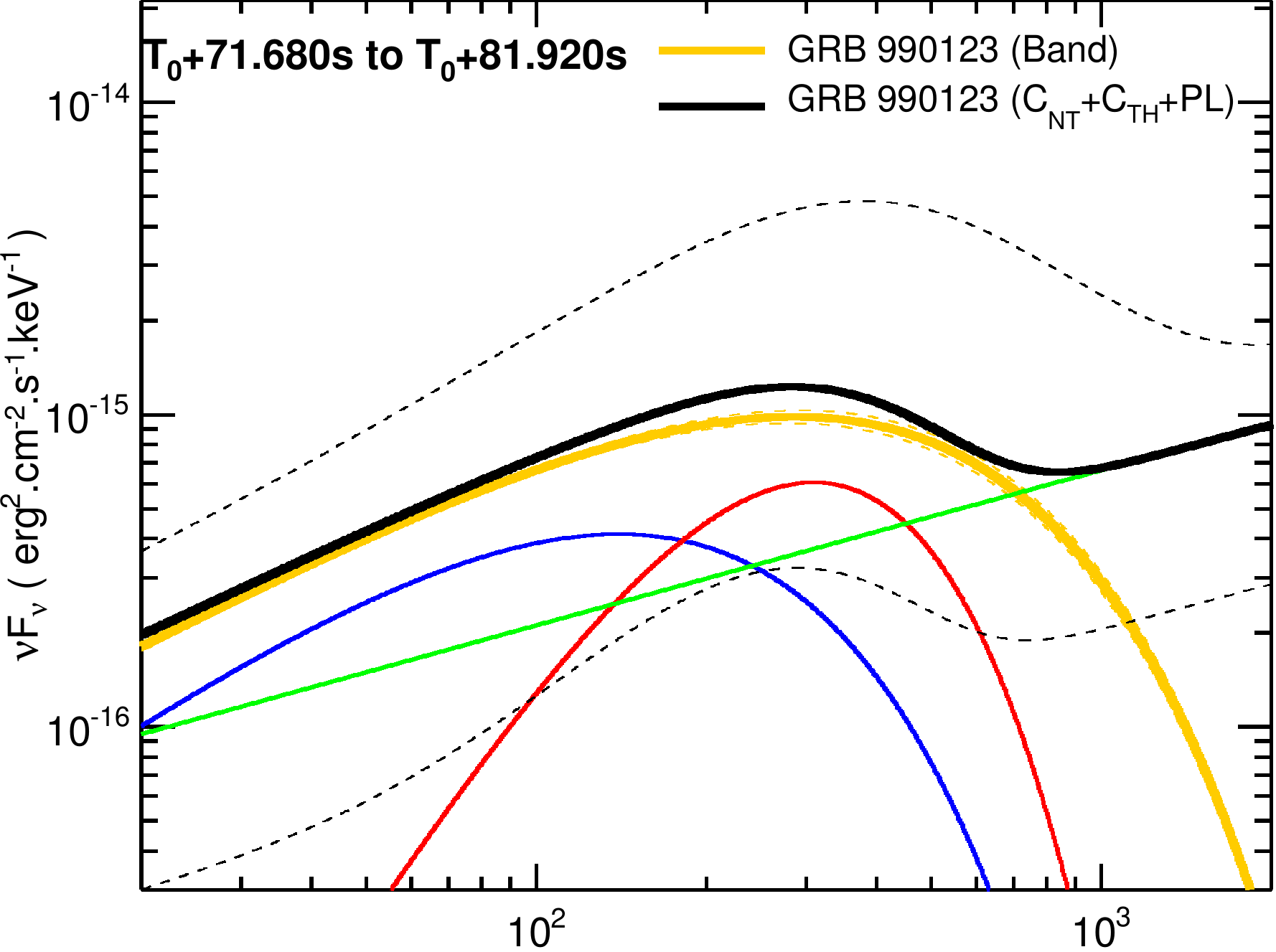}
\includegraphics[totalheight=0.185\textheight, clip]{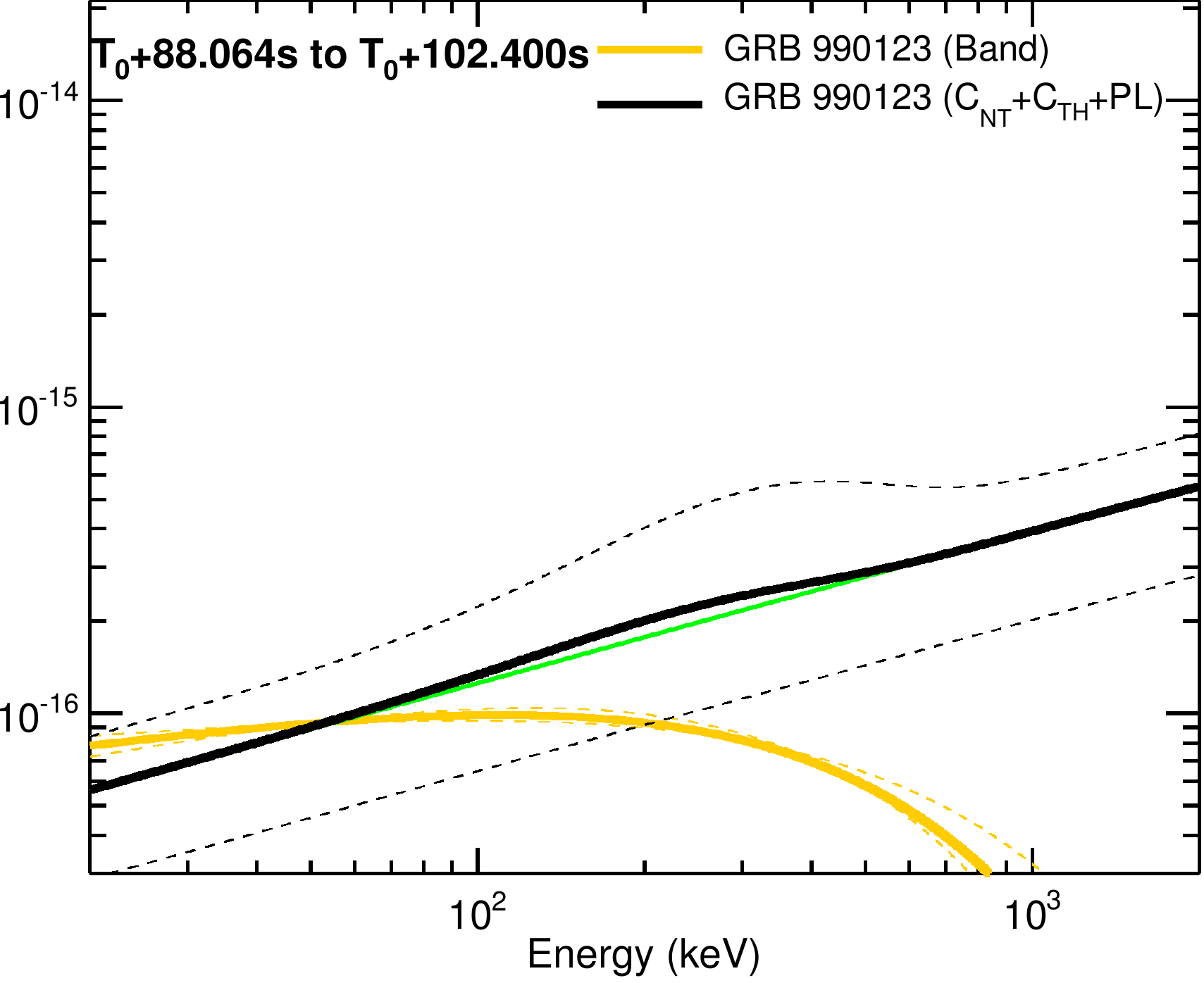}

\hspace{-6.cm}
\includegraphics[totalheight=0.195\textheight, clip]{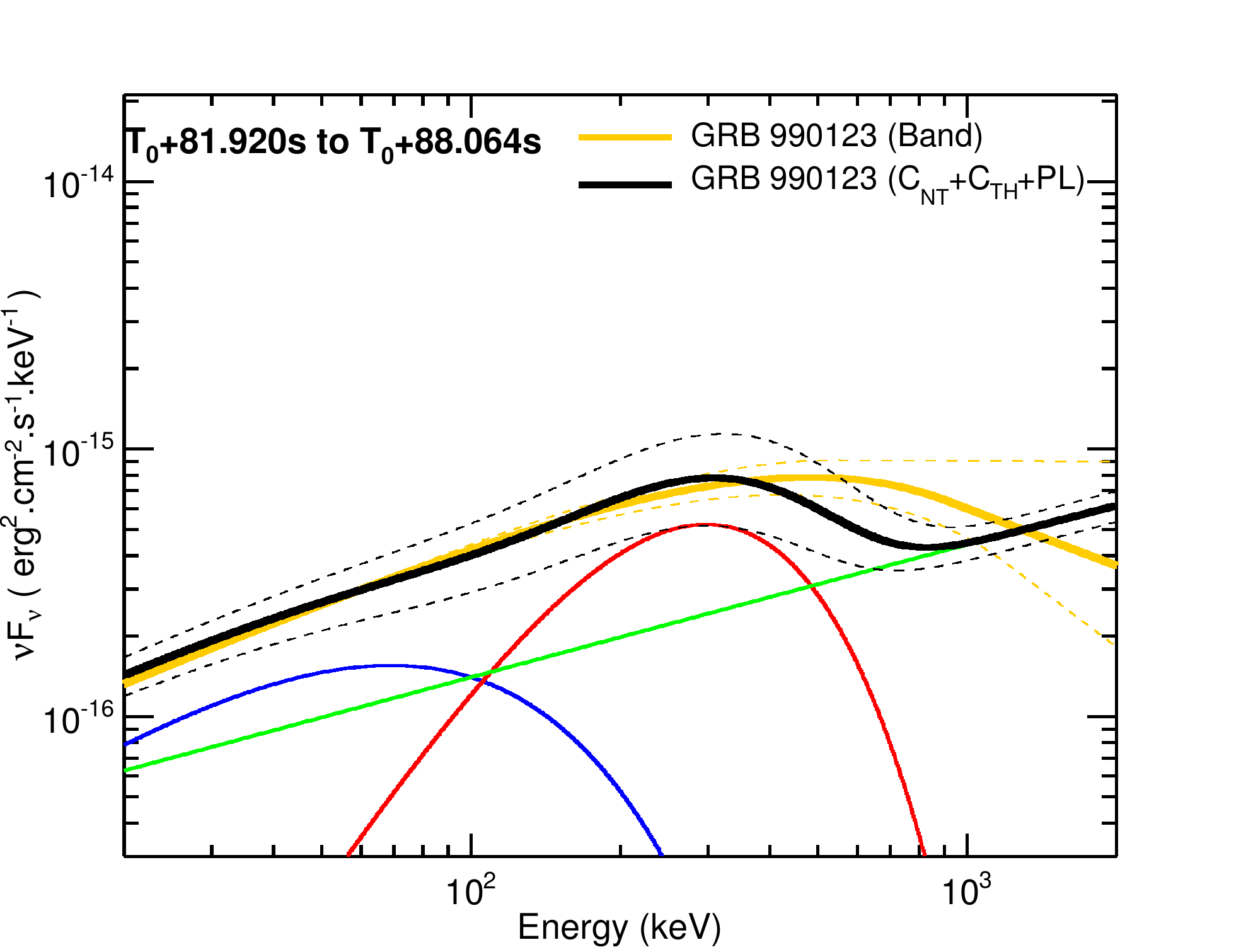}

\caption{\label{fig20}GRB~990123 : $\nu$F$_\nu$ spectra resulting from the fine-time analysis presented in Section~\ref{sec:trsa}. The solid yellow and black lines correspond to the best Band-only and C$_\mathrm{nTh}$+C$_\mathrm{Th}$+PL fits, respectively. The dashed yellow and black lines correspond to the 1--$\sigma$ confidence regions of the Band-only and C$_\mathrm{nTh}$+C$_\mathrm{Th}$+PL fits, respectively. The solid blue, red and green lines correspond to C$_\mathrm{nTh}$, C$_\mathrm{Th}$ and the additional PL resulting from the best C$_\mathrm{nTh}$+C$_\mathrm{Th}$+PL fits (i.e., solid black line) to the data, respectively.}
\end{center}
\end{figure*}

\clearpage

\setcounter{table}{0}
\renewcommand{\thetable}{A\arabic{table}}

\begin{table*}
\caption{\label{tab04}Model parameter values resulting from the fine-time spectral analysis of GRB~941017 with their 1-$\sigma$ uncertainties as presented in Section~\ref{sec:trsa}.}
\begin{center}
{\tiny
\begin{tabular}{|c|c|c|c|c|c|c|c|c|c|c|c|c|c|c|c|}
\hline
\multicolumn{2}{|c|}{Time from T$_\mathrm{0}$} & \multicolumn{1}{|c|}{Models} &\multicolumn{3}{|c|}{Base Component} & \multicolumn{2}{|c|}{Additional Components} & \multicolumn{1}{|c|}{Cstat/dof} \\
\hline
 \multicolumn{2}{|c|}{} &    &\multicolumn{3}{|c|}{CPL or Band} & \multicolumn{1}{|c|}{BB} & \multicolumn{1}{|c|}{ PL} & \multicolumn{1}{|c|}{}\\
\hline
\multicolumn{1}{|c|}{T$_\mathrm{start}$} & \multicolumn{1}{|c|}{T$_\mathrm{stop}$} & \multicolumn{1}{|c|}{Parameters} & \multicolumn{1}{|c|}{E$_{\rm peak}$} & \multicolumn{1}{|c|}{$\alpha$} & \multicolumn{1}{|c|}{$\beta$} & \multicolumn{1}{|c|}{kT} & \multicolumn{1}{|c|}{$\alpha$} & \\
\multicolumn{1}{|c|}{(s)} & \multicolumn{1}{|c|}{(s)} & \multicolumn{1}{|c|}{} & \multicolumn{1}{|c|}{(keV)} & \multicolumn{1}{|c|}{} & \multicolumn{1}{|c|}{} & \multicolumn{1}{|c|}{(keV)} & \multicolumn{1}{|c|}{} & \\
\hline
\multicolumn{9}{|c|}{ }   \\ 
-4.096 & +8.192 & Band & 456$\pm$62 & -0.76$\pm$0.09 & -2.84$\pm$1.04 & -- & -- & 20.7/9 \\
& & C$_\mathrm{nTh}$+C$_\mathrm{Th}$ & 669$\pm$224 & -1.02$\pm$0.21 & -- & 58$\pm$13 & -- & 18.7/8 \\
& & C$_\mathrm{nTh}$+C$_\mathrm{Th}$+PL& 488$\pm$69 & -0.7 (fix) & -- & 52$\pm$10 & -1.5 (fix) & 19.0/8 \\
\multicolumn{9}{|c|}{ }   \\ 
+8.192 & +12.288 & Band & 422$\pm$29 & -0.55$\pm$0.05 & -1.93$\pm$0.06 & -- & -- & 16.0/9 \\
& & C$_\mathrm{nTh}$+C$_\mathrm{Th}$ & 1391$\pm$272 & -1.08$\pm$0.06 & -- & 62$\pm$4 & -- & 5.5/8 \\
& & C$_\mathrm{nTh}$+C$_\mathrm{Th}$+PL& 929$\pm$102 & -0.7 (fix) & -- & 59$\pm$3 & -1.5 (fix) & 5.7/8 \\
\multicolumn{9}{|c|}{ }   \\ 
+12.288 & +16.384 & Band & 411$\pm$18 & -0.53$\pm$0.04 & -2.07$\pm$0.05 & -- & -- & 15.6/9 \\
& & C$_\mathrm{nTh}$+C$_\mathrm{Th}$ & 1212$\pm$161 & -1.09$\pm$0.05 & -- & 64$\pm$2 & -- & 14.6/8 \\
& & C$_\mathrm{nTh}$+C$_\mathrm{Th}$+PL& 755$\pm$48 & -0.7 (fix) & -- & 61$\pm$2 & -1.5 (fix) & 13.8/8 \\
\multicolumn{9}{|c|}{ }   \\ 
+16.384 & +22.528 & Band & 364$\pm$8 & -0.56$\pm$0.02 & -2.35$\pm$0.05 & -- & -- & 50.6/9 \\
& & C$_\mathrm{nTh}$+C$_\mathrm{Th}$ & 783$\pm$52 & -1.10$\pm$0.04 & -- & 60$\pm$1 & -- & 17.6/8 \\
& & C$_\mathrm{nTh}$+C$_\mathrm{Th}$+PL& 477$\pm$13 & -0.7 (fix) & -- & 58$\pm$2 & -1.5 (fix) & 14.9/8 \\
\multicolumn{9}{|c|}{ }   \\ 
+22.528 & +24.576 & Band & 318$\pm$14 & -0.49$\pm$0.05 & -2.41$\pm$0.10 & -- & -- & 26.5/9 \\
& & C$_\mathrm{nTh}$+C$_\mathrm{Th}$ & 722$\pm$113 & -1.16$\pm$0.08 & -- & 57$\pm$2 & -- & 10.9/8 \\
& & C$_\mathrm{nTh}$+C$_\mathrm{Th}$+PL& 412$\pm$26 & -0.7 (fix) & -- & 55$\pm$4 & -1.5 (fix) & 15.5/8 \\
\multicolumn{9}{|c|}{ }   \\ 
+24.576 & +26.624 & Band & 412$\pm$18 & -0.50$\pm$0.04 & -2.43$\pm$0.10 & -- & -- & 21.4/9 \\
& & C$_\mathrm{nTh}$+C$_\mathrm{Th}$ & 930$\pm$145 & -1.12$\pm$0.07 & -- & 71$\pm$2 & -- & 6.4/8 \\
& & C$_\mathrm{nTh}$+C$_\mathrm{Th}$+PL& 547$\pm$34 & -0.7 (fix) & -- & 70$\pm$3 & -1.5 (fix) & 5.8/8 \\
\multicolumn{9}{|c|}{ }   \\ 
+26.624 & +28.672 & Band & 350$\pm$15 & -0.44$\pm$0.04 & -2.50$\pm$0.11 & -- & -- & 39.1/9 \\
& & C$_\mathrm{nTh}$+C$_\mathrm{Th}$ & 859$\pm$161 & -1.18$\pm$0.08 & -- & 64$\pm$2 & -- & 12.1/8 \\
& & C$_\mathrm{nTh}$+C$_\mathrm{Th}$+PL& 463$\pm$34 & -0.7 (fix) & -- & 62$\pm$3 & -1.5 (fix) & 15.6/8 \\
\multicolumn{9}{|c|}{ }   \\ 
+28.672 & +30.720 & Band & 330$\pm$9 & -0.42$\pm$0.03 & -2.43$\pm$0.07 & -- & -- & 40.6/9 \\
& & C$_\mathrm{nTh}$+C$_\mathrm{Th}$ & 772$\pm$78 & -1.11$\pm$0.05 & -- & 59$\pm$2 & -- & 17.7/8 \\
& & C$_\mathrm{nTh}$+C$_\mathrm{Th}$+PL& 469$\pm$21 & -0.7 (fix) & -- & 58$\pm$2 & -1.5 (fix) & 13.6/8 \\
\multicolumn{9}{|c|}{ }   \\ 
+30.720 & +32.768 & Band & 325$\pm$14 & -0.46$\pm$0.05 & -2.38$\pm$0.09 & -- & -- & 21.4/9 \\
& & C$_\mathrm{nTh}$+C$_\mathrm{Th}$ & 832$\pm$142 & -1.17$\pm$0.08 & -- & 58$\pm$2 & -- & 8.7/8 \\
& & C$_\mathrm{nTh}$+C$_\mathrm{Th}$+PL& 447$\pm$32 & -0.7 (fix) & -- &57$\pm$3 & -1.5 (fix) & 6.7/8 \\
\multicolumn{9}{|c|}{ }   \\ 
+32.768 & +36.864 & Band & 280$\pm$9 & -0.68$\pm$0.03 & -2.49$\pm$0.08 & -- & -- & 26.0/9 \\
& & C$_\mathrm{nTh}$+C$_\mathrm{Th}$ & 614$\pm$76 & -1.31$\pm$0.06 & -- & 51$\pm$2 & -- & 11.8/8 \\
& & C$_\mathrm{nTh}$+C$_\mathrm{Th}$+PL& 279$\pm$15 & -0.7 (fix) & -- & 53$\pm$5 & -1.5 (fix) & 11.7/8 \\
\multicolumn{9}{|c|}{ }   \\ 
+36.864 & +43.008 & Band & 193$\pm$9 & -0.96$\pm$0.06 & -2.48$\pm$0.11 & -- & -- & 19.9/9 \\
& & C$_\mathrm{nTh}$+C$_\mathrm{Th}$ & 537$\pm$177 & -1.69$\pm$0.08 & -- & 41$\pm$2 & -- & 11.3/8 \\
& & C$_\mathrm{nTh}$+C$_\mathrm{Th}$+PL& 99$\pm$20 & -0.7 (fix) & -- & 61$\pm$6 & -1.5 (fix) & 7.7/8 \\
\multicolumn{9}{|c|}{ }   \\ 
+43.008 & +45.056 & Band & 174$\pm$35 & -1.18$\pm$0.20 & -2.34$\pm$0.32 & -- & -- & 8.4/9 \\
& & C$_\mathrm{nTh}$+C$_\mathrm{Th}$ & 5124$\pm$3000 & -1.88$\pm$0.17 & -- & 39$\pm$5 & -- & 8.2/8 \\
& & C$_\mathrm{nTh}$+C$_\mathrm{Th}$+PL& 53$\pm$23 & -0.7 (fix) & -- & 48$\pm$9 & -1.5 (fix) & 5.2/8 \\
\multicolumn{9}{|c|}{ }   \\ 
+45.056 & +49.152 & Band & 274$\pm$18 & -0.85$\pm$0.06 & -2.55$\pm$0.21 & -- & -- & 16.3/9 \\
& & C$_\mathrm{nTh}$+C$_\mathrm{Th}$ & 485$\pm$101 & -1.33$\pm$0.12 & -- & 48$\pm$3 & -- & 8.7/8 \\
& & C$_\mathrm{nTh}$+C$_\mathrm{Th}$+PL& 253$\pm$18 & -0.7 (fix) & -- & 45$\pm$4 & -1.5 (fix) & 7.8/8 \\
\multicolumn{9}{|c|}{ }   \\ 
+49.152 & +53.248 & Band & 176$\pm$24 & -1.21$\pm$0.14 & -2.38$\pm$0.25 & -- & -- & 25.0/9 \\
& & C$_\mathrm{nTh}$+C$_\mathrm{Th}$ & 7679$\pm$5000 & -1.95$\pm$0.14 & -- & 42$\pm$3 & -- & 14.5/8 \\
& & C$_\mathrm{nTh}$+C$_\mathrm{Th}$+PL& 31$\pm$7 & -0.7 (fix) & -- & 40$\pm$3 & -1.5 (fix) & 11.6/8 \\
\multicolumn{9}{|c|}{ }   \\ 
+53.248 & +55.296 & Band & 224$\pm$12 & -0.96$\pm$0.07 & $<$-5 & -- & -- & 13.5/9 \\
& & C$_\mathrm{nTh}$+C$_\mathrm{Th}$ & 215$\pm$49 & -1.48$\pm$0.26 & -- & 53$\pm$6 & -- & 8.0/8 \\
& & C$_\mathrm{nTh}$+C$_\mathrm{Th}$+PL& 78$\pm$18 & -0.7 (fix) & -- & 64$\pm$7 & -1.5 (fix) & 7.8/8 \\
\multicolumn{9}{|c|}{ }   \\ 
+55.296 & +61.440 & Band & 187$\pm$16 & -1.06$\pm$0.09 & -2.51$\pm$0.23 & -- & -- & 17.6/9 \\
& & C$_\mathrm{nTh}$+C$_\mathrm{Th}$ & 818$\pm$500 & -1.81$\pm$0.12 & -- & 40$\pm$3 & -- & 12.7/8 \\
& & C$_\mathrm{nTh}$+C$_\mathrm{Th}$+PL& 67$\pm$19 & -0.7 (fix) & -- & 52$\pm$6 & -1.5 (fix) & 12.3/8 \\
\multicolumn{9}{|c|}{ }   \\ 
+61.440 & +63.488 & Band & 310$\pm$17 & -0.76$\pm$0.06 & $<$-5 & -- & -- & 12.7/9 \\
& & C$_\mathrm{nTh}$+C$_\mathrm{Th}$ & 410$\pm$95 & -1.14$\pm$0.21 & -- & 53$\pm$6 & -- & 8.2/8 \\
& & C$_\mathrm{nTh}$+C$_\mathrm{Th}$+PL& 297$\pm$26 & -0.7 (fix) & -- & 47$\pm$10 & -1.5 (fix) & 6.9/8 \\
\multicolumn{9}{|c|}{ }   \\ 
+63.488 & +65.536 & Band & 124$\pm$29 & -0.97$\pm$0.49 & -2.50$\pm$0.52 & -- & -- & 11.2/9 \\
& & C$_\mathrm{nTh}$+C$_\mathrm{Th}$ & 717$\pm$600 & -2.26$\pm$0.30 & -- & 35$\pm$5 & -- & 3.7/8 \\
& & C$_\mathrm{nTh}$+C$_\mathrm{Th}$+PL& 18$^\mathrm{+10}_\mathrm{-5}$ & -0.7 (fix) & -- & 30$\pm$4 & -1.5 (fix) & 5.0/8 \\
\multicolumn{9}{|c|}{ }   \\ 
+65.536 & +67.584 & Band & 306$\pm$13 & -0.53$\pm$0.05 & -3.00$\pm$0.30 & -- & -- & 20.4/9 \\
& & C$_\mathrm{nTh}$+C$_\mathrm{Th}$ & 451$\pm$63 & -1.06$\pm$0.13 & -- & 57$\pm$3 & -- & 13.2/8 \\
& & C$_\mathrm{nTh}$+C$_\mathrm{Th}$+PL& 332$\pm$24 & -0.7 (fix) & -- & 58$\pm$7 & -1.5 (fix) & 13.0/8 \\
\multicolumn{9}{|c|}{ }   \\ 
\end{tabular}
}
\end{center}
\end{table*}

\newpage

\begin{table*}
\begin{center}
{\tiny
\begin{tabular}{|c|c|c|c|c|c|c|c|c|c|c|c|c|c|c|c|}
\multicolumn{9}{|c|}{ }   \\ 
+67.584 & +71.680 & Band & 276$\pm$12 & -0.68$\pm$0.05 & -2.72$\pm$0.20 & -- & -- & 15.2/9 \\
& & C$_\mathrm{nTh}$+C$_\mathrm{Th}$ & 461$\pm$73 & -1.25$\pm$0.11 & -- & 52$\pm$3 & -- & 6.5/8 \\
& & C$_\mathrm{nTh}$+C$_\mathrm{Th}$+PL& 269$\pm$22 & -0.7 (fix) & -- & 55$\pm$10 & -1.5 (fix) & 6.4/8 \\
\multicolumn{9}{|c|}{ }   \\ 
+71.680 & +77.824 & Band & 177$\pm$8 & -1.13$\pm$0.07 & $<$-5 & -- & -- & 7.3/9 \\
& & C$_\mathrm{nTh}$+C$_\mathrm{Th}$ & 177$\pm$8 & -1.13$\pm$0.07 & -- & 10$\pm$2 & -- & 6.0/8 \\
& & C$_\mathrm{nTh}$+C$_\mathrm{Th}$+PL& 90$\pm$15 & -0.7 (fix) & -- & 70$\pm$11 & -1.5 (fix) & 7.3/8 \\
\multicolumn{9}{|c|}{ }   \\ 
+77.824 & +83.968 & Band & 137$\pm$7 & -1.24$\pm$0.09 & $<$-5 & -- & -- & 7.3/9 \\
& & C$_\mathrm{nTh}$+C$_\mathrm{Th}$ & 137$\pm$7 & -1.24$\pm$0.09 & -- & 10$\pm$2 & -- & 6.1/8 \\
& & C$_\mathrm{nTh}$+C$_\mathrm{Th}$+PL& 98$\pm$20 & -0.7 (fix) & -- & 64$\pm$15 & -1.5 (fix) & 5.4/8 \\
\multicolumn{9}{|c|}{ }   \\ 
+83.968 & +86.016 & Band & 175$\pm$13 & -0.84$\pm$0.12 & -2.92$\pm$0.46 & -- & -- & 6.2/9 \\
& & C$_\mathrm{nTh}$+C$_\mathrm{Th}$ & 233$\pm$65 & -1.54$\pm$0.29 & -- & 40$\pm$4 & -- & 5.2/8 \\
& & C$_\mathrm{nTh}$+C$_\mathrm{Th}$+PL& 76$\pm$31 & -0.7 (fix) & -- & 53$\pm$8 & -1.5 (fix) & 4.1/8 \\
\multicolumn{9}{|c|}{ }   \\ 
+86.016 & +88.064 & Band & 119$\pm$13 & -0.81$\pm$0.30 & -2.72$<$0.40 & -- & -- & 5.0/9 \\
& & C$_\mathrm{nTh}$+C$_\mathrm{Th}$ & 130$\pm$9 & -1.01$\pm$0.18 & -- & 10$\pm$2 & -- & 4.5/8 \\
& & C$_\mathrm{nTh}$+C$_\mathrm{Th}$+PL& 104$\pm$33 & -0.7 (fix) & -- & 79$\pm$15 & -1.5 (fix) & 5.0/8 \\
\multicolumn{9}{|c|}{ }   \\ 
+88.064 & +110.590 & Band & 46$\pm$42 & -1.86$^\mathrm{+0.23}_\mathrm{-0.13}$ & $<$-5 & -- & -- & 16.5/9 \\
& & C$_\mathrm{nTh}$+C$_\mathrm{Th}$ & 16$\pm$5 & -1.83$\pm$0.06 & -- & 51$\pm$17 & -- & 13.0/8 \\
& & C$_\mathrm{nTh}$+C$_\mathrm{Th}$+PL& 32$\pm$10 & -0.7 (fix) & -- & 43$\pm$15 & -1.5 (fix) & 11.9/8 \\
\multicolumn{9}{|c|}{ }   \\ 
+110.590 & +114.690 & Band & 116$\pm$23 & -1.12$\pm$0.37 & -2.37$\pm$0.27 & -- & -- & 13.7/9 \\
& & C$_\mathrm{nTh}$+C$_\mathrm{Th}$ & 137$\pm$15 & -1.36$\pm$0.17 & -- & 10$\pm$2 & -- & 12.0/8 \\
& & C$_\mathrm{nTh}$+C$_\mathrm{Th}$+PL& 42$\pm$15 & -0.7 (fix) & -- & 36$\pm$10 & -1.5 (fix) & 11.2/8 \\
\multicolumn{9}{|c|}{ }   \\ 
+114.690 & +200.700 & Band & 91$\pm$22 & -0.80$\pm$0.89 & -2.36$\pm$0.27 & -- & -- & 7.2/9 \\
& & C$_\mathrm{nTh}$+C$_\mathrm{Th}$ & 109$\pm$13 & -1.27$\pm$0.30 & -- & 10$\pm$2 & -- & 6.5/8 \\
& & C$_\mathrm{nTh}$+C$_\mathrm{Th}$+PL& 109$\pm$13 & -1.27$\pm$0.30 & -- & 10$\pm$2 & -- & 6.5/8 \\
\multicolumn{9}{|c|}{ }   \\ 
\hline
\end{tabular}
}
\end{center}
\end{table*}

\newpage

\begin{table*}
\caption{\label{tab05}Model parameter values resulting from the fine-time spectral analysis of GRB~970111 with their 1-$\sigma$ uncertainties as presented in Section~\ref{sec:trsa}.}
\begin{center}
{\tiny
\begin{tabular}{|c|c|c|c|c|c|c|c|c|c|c|c|c|c|c|c|}
\hline
\multicolumn{2}{|c|}{Time from T$_\mathrm{0}$} & \multicolumn{1}{|c|}{Models} &\multicolumn{3}{|c|}{Base Component} & \multicolumn{2}{|c|}{Additional Components} & \multicolumn{1}{|c|}{Cstat/dof} \\
\hline
 \multicolumn{2}{|c|}{} &    &\multicolumn{3}{|c|}{CPL or Band} & \multicolumn{1}{|c|}{BB} & \multicolumn{1}{|c|}{ PL} & \multicolumn{1}{|c|}{}\\
\hline
\multicolumn{1}{|c|}{T$_\mathrm{start}$} & \multicolumn{1}{|c|}{T$_\mathrm{stop}$} & \multicolumn{1}{|c|}{Parameters} & \multicolumn{1}{|c|}{E$_{\rm peak}$} & \multicolumn{1}{|c|}{$\alpha$} & \multicolumn{1}{|c|}{$\beta$} & \multicolumn{1}{|c|}{kT} & \multicolumn{1}{|c|}{$\alpha$} & \\
\multicolumn{1}{|c|}{(s)} & \multicolumn{1}{|c|}{(s)} & \multicolumn{1}{|c|}{} & \multicolumn{1}{|c|}{(keV)} & \multicolumn{1}{|c|}{} & \multicolumn{1}{|c|}{} & \multicolumn{1}{|c|}{(keV)} & \multicolumn{1}{|c|}{} & \\
\hline
\multicolumn{9}{|c|}{ }   \\ 
-2.034 & -0.256 & Band & 715$\pm$403 & -0.47$\pm$0.49 & $<$-5 & -- & -- & 13.0/9 \\
& & C$_\mathrm{nTh}$+C$_\mathrm{Th}$ & 903$\pm$369 & -0.7 (fix) & -- & 20$\pm$2 & -- & 12.0/9 \\
& & C$_\mathrm{nTh}$+C$_\mathrm{Th}$+PL& -- & -- & -- & 118$\pm$34 & -1.5 (fix) & 12.6/10 \\
\multicolumn{9}{|c|}{ }   \\ 
-0.256 & +1.792 & Band & 263$\pm$12 & +1.09$\pm$0.17 & -2.87$\pm$0.26 & -- & -- & 7.6/9 \\
& & C$_\mathrm{nTh}$+C$_\mathrm{Th}$ & 1019$\pm$391 & -0.7 (fix) & -- & 65$\pm$2 & -- & 7.7/9 \\
& & C$_\mathrm{nTh}$+C$_\mathrm{Th}$+PL& -- & -- & -- & 67$\pm$2 & -1.5 (fix) & 12.6/10 \\
\multicolumn{9}{|c|}{ }   \\ 
+1.792 & +3.840 & Band & 201$\pm$5 & +1.36$\pm$0.16 & -3.94$\pm$0.61 & -- & -- & 11.3/9 \\
& & C$_\mathrm{nTh}$+C$_\mathrm{Th}$ & 2565$\pm$400 & -0.7 (fix) & -- & 51$\pm$1 & -- & 9.8/9 \\
& & C$_\mathrm{nTh}$+C$_\mathrm{Th}$+PL& -- & -- & -- & 51$\pm$1 & -1.5 (fix) & 12.7/10 \\
\multicolumn{9}{|c|}{ }   \\ 
+3.840 & +5.888 & Band & 177$\pm$4 & +1.20$\pm$0.14 & -3.52$\pm$0.30 & -- & -- & 11.4/9 \\
& & C$_\mathrm{nTh}$+C$_\mathrm{Th}$ & 540$\pm$237 & -0.19$\pm$0.92 & -- & 43$\pm$2 & -- & 10.4/8 \\
& & C$_\mathrm{nTh}$+C$_\mathrm{Th}$+PL& 635$\pm$269 & -0.7 (fix) & -- & 44$\pm$1 & -- & 11.1/9 \\
\multicolumn{9}{|c|}{ }   \\ 
+5.888 & +7.936 & Band & 213$\pm$3 & +0.47$\pm$0.06 & -5.22$\pm$1.68 & -- & -- & 2.5/9 \\
& & C$_\mathrm{nTh}$+C$_\mathrm{Th}$ & 228$\pm$23 & +0.08$\pm$0.43 & -- & 49$\pm$4 & -- & 2.5/8 \\
& & C$_\mathrm{nTh}$+C$_\mathrm{Th}$+PL& 277$\pm$26 & -0.7 (fix) & -- & 50$\pm$1 & -- & 6.3/9 \\
\multicolumn{9}{|c|}{ }   \\ 
+7.936 & +9.984 & Band & 206$\pm$3 & +0.15$\pm$0.05 & $<$-5 & -- & -- & 12.7/9 \\
& & C$_\mathrm{nTh}$+C$_\mathrm{Th}$ & 215$\pm$15 & -0.22$\pm$0.30 & -- & 48$\pm$4 & -- & 10.9/8 \\
& & C$_\mathrm{nTh}$+C$_\mathrm{Th}$+PL& 236$\pm$16 & -0.7 (fix) & -- & 48$\pm$2 & -- & 14.0/9 \\
\multicolumn{9}{|c|}{ }   \\ 
+9.984 & +12.032 & Band & 154$\pm$2 & -0.29$\pm$0.06 & $<$-5 & -- & -- & 14.5/9 \\
& & C$_\mathrm{nTh}$+C$_\mathrm{Th}$ & 153$\pm$9 & -0.64$\pm$0.32 & -- & 39$\pm$4 & -- & 13.2/8 \\
& & C$_\mathrm{nTh}$+C$_\mathrm{Th}$+PL& 153$\pm$9 & -0.7 (fix) & -- & 39$\pm$3 & -- & 13.2/9 \\
\multicolumn{9}{|c|}{ }   \\ 
+12.032 & +14.080 & Band & 134$\pm$3 & -0.59$\pm$0.08 & -3.75$\pm$0.53 & -- & -- & 12.5/9 \\
& & C$_\mathrm{nTh}$+C$_\mathrm{Th}$ & 131$\pm$11 & -1.18$\pm$0.31 & -- & 36$\pm$3 & -- & 10.8/8 \\
& & C$_\mathrm{nTh}$+C$_\mathrm{Th}$+PL& 117$\pm$23 & -0.7 (fix) & -- & 45$\pm$8 & -1.5 (fix) & 10.8/8 \\
\multicolumn{9}{|c|}{ }   \\ 
+14.080 & +16.128 & Band & 151$\pm$2 & -0.45$\pm$0.05 & $<$-5 & -- & -- & 14.3/9 \\
& & C$_\mathrm{nTh}$+C$_\mathrm{Th}$ & 150$\pm$8 & -0.95$\pm$0.24 & -- & 38$\pm$2 & -- & 9.2/8 \\
& & C$_\mathrm{nTh}$+C$_\mathrm{Th}$+PL& 150$\pm$9 & -0.7 (fix) & -- & 39$\pm$5 & -1.5 (fix) & 10.6/8 \\
\multicolumn{9}{|c|}{ }   \\ 
+16.128 & +18.176 & Band & 177$\pm$2 & -0.41$\pm$0.03 & $<$-5 & -- & -- & 13.2/9 \\
& & C$_\mathrm{nTh}$+C$_\mathrm{Th}$ & 166$\pm$9 & -0.68$\pm$0.16 & -- & 50$\pm$6 & -- & 4.9/8 \\
& & C$_\mathrm{nTh}$+C$_\mathrm{Th}$+PL& 166$\pm$9 & -0.7 (fix) & -- & 49$\pm$4 & -- & 4.9/8 \\
\multicolumn{9}{|c|}{ }   \\ 
+18.176 & +20.224 & Band & 159$\pm$2 & -0.58$\pm$0.04 & $<$-5 & -- & -- & 15.8/9 \\
& & C$_\mathrm{nTh}$+C$_\mathrm{Th}$ & 144$\pm$9 & -0.95$\pm$0.18 & -- & 46$\pm$4 & -- & 5.1/8 \\
& & C$_\mathrm{nTh}$+C$_\mathrm{Th}$+PL& 123$\pm$21 & -0.7 (fix) & -- & 55$\pm$2 & -1.5 (fix) & 6.4/8 \\
\multicolumn{9}{|c|}{ }   \\ 
+20.224 & +22.272 & Band & 100$\pm$1 & -0.77$\pm$0.05 & $<$-5 & -- & -- & 10.3/9 \\
& & C$_\mathrm{nTh}$+C$_\mathrm{Th}$ & 100$\pm$17 & -0.77$\pm$0.20 & -- & 45$\pm$10 & -- & 10.4/8 \\
& & C$_\mathrm{nTh}$+C$_\mathrm{Th}$+PL& 77$\pm$12 & -0.7 (fix) & -- & 44$\pm$4 & -1.5 (fix) & 7.6/8 \\
\multicolumn{9}{|c|}{ }   \\ 
+22.272 & +24.320 & Band & 85$\pm$2 & -0.83$\pm$0.09 & $<$-5 & -- & -- & 13.8/9 \\
& & C$_\mathrm{nTh}$+C$_\mathrm{Th}$ & 67$\pm$45 & -0.75$\pm$0.83 & -- & 41$\pm$6 & -- & 12.7/8 \\
& & C$_\mathrm{nTh}$+C$_\mathrm{Th}$+PL& 63$\pm$15 & -0.7 (fix) & -- & 39$\pm$6 & -1.5 (fix) & 11.8/8 \\
\multicolumn{9}{|c|}{ }   \\ 
+24.320 & +26.368 & Band & 72$\pm$2 & -0.90$\pm$0.13 & -3.93$\pm$0.54 & -- & -- & 16.8/9 \\
& & C$_\mathrm{nTh}$+C$_\mathrm{Th}$ & 67$\pm$14 & -1.06$\pm$0.54 & -- & 28$\pm$15 & -- & 19.1/8 \\
& & C$_\mathrm{nTh}$+C$_\mathrm{Th}$+PL& 39$\pm$10 & -0.7 (fix) & -- & 30$\pm$3 & -1.5 (fix) & 15.8/8 \\
\multicolumn{9}{|c|}{ }   \\ 
+26.368 & +30.464 & Band & 49$\pm$3 & -0.90$\pm$0.17 & $<$-5 & -- & -- & 9.1/9 \\
& & C$_\mathrm{nTh}$+C$_\mathrm{Th}$ & 46$\pm$30 & -0.79$\pm$1.00 & -- & 31$\pm$18 & -- & 9.0/8 \\
& & C$_\mathrm{nTh}$+C$_\mathrm{Th}$+PL& 45$\pm$10 & -0.7 (fix) & -- & 30$\pm$10 & -- & 9.0/9 \\
\multicolumn{9}{|c|}{ }   \\ 
+30.464 & +36.608 & Band & 65$\pm$2 & -1.16$\pm$0.09 & -4.14$\pm$0.81 & -- & -- & 6.8/9 \\
& & C$_\mathrm{nTh}$+C$_\mathrm{Th}$ & 62$\pm$10 & -1.21$\pm$0.32 & -- & 32$\pm$12 & -- & 7.0/8 \\
& & C$_\mathrm{nTh}$+C$_\mathrm{Th}$+PL& 42$\pm$6 & -0.7 (fix) & -- & 33$\pm$3 & -1.5 (fix) & 6.2/8 \\
\multicolumn{9}{|c|}{ }   \\ 
+36.608 & +40.704 & Band & 30$\pm$12 & -1.56$\pm$0.29 & -4.04$\pm$1.35 & -- & -- & 12.2/9 \\
& & C$_\mathrm{nTh}$+C$_\mathrm{Th}$ & 28$\pm$17 & -0.73$\pm$0.80 & -- & 26$\pm$8 & -- & 12.4/8 \\
& & C$_\mathrm{nTh}$+C$_\mathrm{Th}$+PL& 23$\pm$8 & -0.7 (fix) & -- & 22$\pm$3 & -1.5 (fix) & 7.0/8 \\
\multicolumn{9}{|c|}{ }   \\ 
+40.704 & +46.848 & Band & 24$\pm$23 & -0.92$\pm$0.80 & $<$-5 & -- & -- & 23.1/9 \\
& & C$_\mathrm{nTh}$+C$_\mathrm{Th}$ & 24$\pm$20 & -0.92$\pm$0.80 & -- & 26$\pm$8 & -- & 20.0/8 \\
& & C$_\mathrm{nTh}$+C$_\mathrm{Th}$+PL& 26$\pm$3 & -0.7 (fix) & -- & -- & -- & 21.8/8 \\
\multicolumn{9}{|c|}{ }   \\ 
\hline
\end{tabular}
}
\end{center}
\end{table*}

\newpage

\begin{table*}
\caption{\label{tab06}Model parameter values resulting from the fine-time spectral analysis of GRB~990123 with their 1-$\sigma$ uncertainties as presented in Section~\ref{sec:trsa}.}
\begin{center}
{\tiny
\begin{tabular}{|c|c|c|c|c|c|c|c|c|c|c|c|c|c|c|c|}
\hline
\multicolumn{2}{|c|}{Time from T$_\mathrm{0}$} & \multicolumn{1}{|c|}{Models} &\multicolumn{3}{|c|}{Base Component} & \multicolumn{2}{|c|}{Additional Components} & \multicolumn{1}{|c|}{Cstat/dof} \\
\hline
 \multicolumn{2}{|c|}{} &    &\multicolumn{3}{|c|}{CPL or Band} & \multicolumn{1}{|c|}{BB} & \multicolumn{1}{|c|}{ PL} & \multicolumn{1}{|c|}{}\\
\hline
\multicolumn{1}{|c|}{T$_\mathrm{start}$} & \multicolumn{1}{|c|}{T$_\mathrm{stop}$} & \multicolumn{1}{|c|}{Parameters} & \multicolumn{1}{|c|}{E$_{\rm peak}$} & \multicolumn{1}{|c|}{$\alpha$} & \multicolumn{1}{|c|}{$\beta$} & \multicolumn{1}{|c|}{kT} & \multicolumn{1}{|c|}{$\alpha$} & \\
\multicolumn{1}{|c|}{(s)} & \multicolumn{1}{|c|}{(s)} & \multicolumn{1}{|c|}{} & \multicolumn{1}{|c|}{(keV)} & \multicolumn{1}{|c|}{} & \multicolumn{1}{|c|}{} & \multicolumn{1}{|c|}{(keV)} & \multicolumn{1}{|c|}{} & \\
\hline
\multicolumn{9}{|c|}{ }   \\ 
-2.048 & +16.384 & Band & 164$\pm$5 & -0.30$\pm$0.11 & $<$-5 & -- & -- & 3.7/9 \\
& & C$_\mathrm{nTh}$+C$_\mathrm{Th}$ & 177$\pm$35 & -0.98$\pm$0.60 & -- & 39$\pm4$ & -- & 2.2/8 \\
& & C$_\mathrm{nTh}$+C$_\mathrm{Th}$+PL & 632$^\mathrm{+238}_\mathrm{-155}$ & -0.7 (fix) & -- & 44.8$^\mathrm{+20.5}_\mathrm{-11.4}$ & -1.5 (fix) & 2.2/8 \\
\multicolumn{9}{|c|}{ }   \\ 
+16.384 & +20.480 & Band & 375$\pm$25 & -0.60$\pm$0.06 & -2.60$\pm$0.27 & -- & -- & 18.0/9 \\
& & C$_\mathrm{nTh}$+C$_\mathrm{Th}$ & 645$\pm$142 & -1.09$\pm$0.14 & -- & 63$\pm$5 & -- & 11.3/8 \\
& & C$_\mathrm{nTh}$+C$_\mathrm{Th}$+PL& 577$\pm$392 & -0.7 (fix) & -- & 60$\pm$5 & -1.5 (fix) &  12.9/8\\
\multicolumn{9}{|c|}{ }   \\ 
+20.480 & +22.528 & Band & 783$\pm$30 & -0.54$\pm$0.02 & -2.92$\pm$0.39 & -- & -- & 67.6/9 \\
& & C$_\mathrm{nTh}$+C$_\mathrm{Th}$ & 1059$\pm$85 & -0.78$\pm$0.05 & -- & 106$\pm5$ & -- & 49.2/8 \\
& & C$_\mathrm{nTh}$+C$_\mathrm{Th}$+PL& 994$\pm$38 & -0.7 (fix) & -- & 103$\pm$5 & -1.5 (fix) &  49.0/8\\
\multicolumn{9}{|c|}{ }   \\ 
+22.528 & +24.576 & Band & 841$\pm$42 & -0.50$\pm$0.03 & -2.76$\pm$0.40 & -- & -- & 38.8/9 \\
& & C$_\mathrm{nTh}$+C$_\mathrm{Th}$ & 1101$\pm$104 & -0.70$\pm$0.07 & -- & 108$\pm$8 & -- & 31.6/8 \\
& & C$_\mathrm{nTh}$+C$_\mathrm{Th}$+PL& 1115$\pm$55 & -0.7 (fix) & -- & 107$\pm$8 & -1.5 (fix) &  31.6/8\\
\multicolumn{9}{|c|}{ }   \\ 
+24.576 & +26.624 & Band & 944$\pm$35 & -0.48$\pm$0.02 & -2.17$\pm$0.09 & -- & -- & 91.6/9 \\
& & C$_\mathrm{nTh}$+C$_\mathrm{Th}$ & 2053$\pm$231 & -0.84$\pm$0.04 & -- & 115$\pm$3 & -- & 31.1/8 \\
& & C$_\mathrm{nTh}$+C$_\mathrm{Th}$+PL& 1789$\pm$106 & -0.7 (fix) & -- & 109$\pm$3 & -1.5 (fix) &  25.2/8\\
\multicolumn{9}{|c|}{ }   \\ 
+26.624 & +28.672 & Band & 783$\pm$36 & -0.43$\pm$0.03 & -2.18$\pm$0.10 & -- & -- & 54.5/9 \\
& & C$_\mathrm{nTh}$+C$_\mathrm{Th}$ & 1756$\pm$267 & -0.86$\pm$0.06 & -- & 105$\pm$4 & -- & 24.6/8 \\
& & C$_\mathrm{nTh}$+C$_\mathrm{Th}$+PL& 1516$\pm$123 & -0.7 (fix) & -- & 100$\pm$3 & -1.5 (fix) & 22.3/8 \\
\multicolumn{9}{|c|}{ }   \\ 
+28.672 & +32.768 & Band & 456$\pm$12 & -0.41$\pm$0.03 & -3.31$\pm$0.32 & -- & -- & 46.8/9 \\
& & C$_\mathrm{nTh}$+C$_\mathrm{Th}$ & 649$\pm$54 & -0.89$\pm$0.07 & -- & 80$\pm$2 & -- & 15.4/8 \\
& & C$_\mathrm{nTh}$+C$_\mathrm{Th}$+PL& 550$\pm$21 & -0.7 (fix) & -- & 80$\pm$3 & -1.5 (fix) & 16.4/8 \\
\multicolumn{9}{|c|}{ }   \\ 
+32.768 & +34.816 & Band & 308$\pm$11 & -0.52$\pm$0.05 & $<$-5 & -- & -- & 6.3/9 \\
& & C$_\mathrm{nTh}$+C$_\mathrm{Th}$ & 342$\pm$44 & -0.78$\pm$0.21 & -- & 59$\pm$8 & -- & 4.7/8 \\
& & C$_\mathrm{nTh}$+C$_\mathrm{Th}$+PL& 331$\pm$23 & -0.7 (fix) & -- & 59$\pm$11 & -- & 4.9/8 \\
\multicolumn{9}{|c|}{ }   \\ 
+34.816 & +36.864 & Band & 457$\pm$17 & -0.37$\pm$0.04 & -2.78$\pm$0.18 & -- & -- & 19.8/9 \\
& & C$_\mathrm{nTh}$+C$_\mathrm{Th}$ & 687$\pm$67 & -0.80$\pm$0.09 & -- & 76$\pm$4 & -- & 14.1/8 \\
& & C$_\mathrm{nTh}$+C$_\mathrm{Th}$+PL& 632$\pm$27 & -0.7 (fix) & -- & 76$\pm$4 & -1.5 (fix) & 13.8/8 \\
\multicolumn{9}{|c|}{ }   \\ 
+36.864 & +38.912 & Band & 682$\pm$21 & -0.48$\pm$0.02 & -2.38$\pm$0.09 & -- & -- & 54.8/9 \\
& & C$_\mathrm{nTh}$+C$_\mathrm{Th}$ & 1119$\pm$81 & -0.80$\pm$0.04 & -- & 92$\pm$4 & -- & 33.8/8 \\
& & C$_\mathrm{nTh}$+C$_\mathrm{Th}$+PL& 1034$\pm$39 & -0.7 (fix) & -- & 89$\pm$3 & -1.5 (fix) & 32.3/8 \\
\multicolumn{9}{|c|}{ }   \\ 
+38.912 & +40.960 & Band & 652$\pm$29 & -0.54$\pm$0.03 & -2.53$\pm$0.18 & -- & -- & 27.7/9 \\
& & C$_\mathrm{nTh}$+C$_\mathrm{Th}$ & 941$\pm$86 & -0.79$\pm$0.06 & -- & 86$\pm$6 & -- & 21.1/8 \\
& & C$_\mathrm{nTh}$+C$_\mathrm{Th}$+PL& 677$\pm$171 & -0.7 (fix) & -- & 67$\pm$23 & -1.5 (fix) & 20.8/8 \\
\multicolumn{9}{|c|}{ }   \\ 
+40.960 & +43.008 & Band & 635$\pm$31 & -0.66$\pm$0.03 & -3.12$\pm$0.62 & -- & -- & 59.9/9 \\
& & C$_\mathrm{nTh}$+C$_\mathrm{Th}$ & 810$\pm$83 & -0.89$\pm$0.08 & -- & 88$\pm$7 & -- & 41.6/8 \\
& & C$_\mathrm{nTh}$+C$_\mathrm{Th}$+PL& 740$\pm$47 & -0.7 (fix) & -- &84$\pm$5 & -1.5 (fix) & 30.9/8 \\
\multicolumn{9}{|c|}{ }   \\ 
+43.008 & +49.152 & Band & 402$\pm$15 & -0.98$\pm$0.02 & -3.73$\pm$1.25 & -- & -- & 32.7/9 \\
& & C$_\mathrm{nTh}$+C$_\mathrm{Th}$ & 468$\pm$39 & -1.25$\pm$0.06 & -- & 71$\pm$4 & -- & 13.3/8 \\
& & C$_\mathrm{nTh}$+C$_\mathrm{Th}$+PL& 145$\pm$15 & -0.7 (fix) & -- & 101$\pm$6 & -1.5 (fix) & 30.6/8 \\
\multicolumn{9}{|c|}{ }   \\ 
+49.152 & +61.440 & Band & 399$\pm$14 & -1.17$\pm$0.02 & $<$-5 & -- & -- & 57.6/9 \\
& & C$_\mathrm{nTh}$+C$_\mathrm{Th}$ & 373$\pm$36 & -1.43$\pm$0.05 & -- & 81$\pm$5 & -- & 11.3/8 \\
& & C$_\mathrm{nTh}$+C$_\mathrm{Th}$+PL& 86$\pm$5 & -0.7 (fix) & -- & 87$\pm$3 & -1.5 (fix) & 42.2/8 \\
\multicolumn{9}{|c|}{ }   \\ 
+61.440 & +71.680 & Band & 308$\pm$13 & -1.24$\pm$0.03 & $<$-5 & -- & -- & 24.0/9 \\
& & C$_\mathrm{nTh}$+C$_\mathrm{Th}$ & 295$\pm$36 & -1.50$\pm$0.08 & -- & 66$\pm$5 & -- & 5.5/8 \\
& & C$_\mathrm{nTh}$+C$_\mathrm{Th}$+PL& 74$\pm$5 & -0.7 (fix) & -- & 74$\pm$3 & -1.5 (fix) & 28.6/8 \\
\multicolumn{9}{|c|}{ }   \\ 
+71.680 & +81.920 & Band & 286$\pm$7 & -1.02$\pm$0.03 & $<$-5 & -- & -- & 36.9/9 \\
& & C$_\mathrm{nTh}$+C$_\mathrm{Th}$ & 299$\pm$26 & -1.38$\pm$0.08 & -- & 60$\pm$3 & -- & 10.0/8 \\
& & C$_\mathrm{nTh}$+C$_\mathrm{Th}$+PL& 140$^\mathrm{+249}_\mathrm{-139}$ & -0.7 (fix) & -- & 79$\pm$10 & -1.5 (fix) & 10.2/8 \\
\multicolumn{9}{|c|}{ }   \\ 
+81.920 & +88.064 & Band & 482$\pm$55 & -1.20$\pm$0.04 & -2.71$\pm$0.70 & -- & -- & 22.8/9 \\
& & C$_\mathrm{nTh}$+C$_\mathrm{Th}$ & 997$\pm$500 & -1.59$\pm$0.09 & -- & 71$\pm$5 & -- & 6.4/8 \\
& & C$_\mathrm{nTh}$+C$_\mathrm{Th}$+PL& 69$\pm$10 & -0.7 (fix) & -- & 76$\pm$5 & -1.5 (fix) & 12.0/8 \\
\multicolumn{9}{|c|}{ }   \\ 
+88.064 & +102.400 & Band & 110$\pm$15 & -1.74$\pm$0.11 & $<$-5 & -- & -- & 15.1/9 \\
& & C$_\mathrm{nTh}$+C$_\mathrm{Th}$ & 110$\pm$15 & -1.74$\pm$0.11 & -- & 50$\pm$1 & -- & 13.5/8 \\
& & C$_\mathrm{nTh}$+C$_\mathrm{Th}$+PL& 10$\pm$1 & -0.7 (fix) & -- & 66$\pm$20 & -1.5 (fix) & 12.0/8 \\
\multicolumn{9}{|c|}{ }   \\ 
\hline
\end{tabular}
}
\end{center}
\end{table*}

\end{document}